\documentclass[12pt]{article}
\pdfoutput=1
\usepackage[]{hyperref}
\usepackage{xcolor}
\usepackage[]{putex}
\usepackage[utf8]{inputenc}
\usepackage{subcaption}
\usepackage{mathtools}
\usepackage{tikz}
\usetikzlibrary{decorations.markings,arrows}
\usepackage{amsmath,amsthm}

\DeclareMathOperator*{\res}{Res}

\theoremstyle{plain}
\newtheorem{theorem}{Theorem}

\newtheorem{ex}{Definition}

\theoremstyle{plain}
\newtheorem{prescription}[theorem]{Prescription}
\theoremstyle{definition}
\newtheorem{example}[ex]{Example}

\makeatother
\hypersetup{
  colorlinks=true,
  linkcolor=blue,
  citecolor=green!60!black,
  urlcolor=cyan!70!black,
  linktoc=all
  }

\numberwithin{equation}{section}

\input{glyphtounicode}
\definecolor{light-gray}{gray}{0.85}

\tikzset{->-/.style={decoration={
  markings,
  mark=at position .6 with {\arrow[>=stealth]{>}}},postaction={decorate}}}

\begin{document}

\preprint{PUPT-2577}

\title{Recursion Relations in $p$-adic Mellin Space}
\authors{Christian Baadsgaard Jepsen$^1$\footnote{\tt cjepsen@princeton.edu} \& Sarthak Parikh$^2$\footnote{\tt sparikh@caltech.edu}}
\institution{PU}{$^1$Joseph Henry Laboratories, Princeton University, Princeton, NJ 08544, USA}
\institution{Caltech}{$^2$Division of Physics, Mathematics and Astronomy,\cr\hskip0.06in California Institute of Technology, Pasadena, CA 91125, USA}

\abstract{
In this work, we formulate a set of rules for writing down $p$-adic Mellin amplitudes at tree-level. The rules lead to closed-form expressions for Mellin amplitudes for arbitrary scalar bulk diagrams. The prescription is recursive in nature, with two different physical interpretations: one as a recursion on the number of internal lines in the diagram, and the other as reminiscent of on-shell BCFW recursion for flat-space amplitudes, especially when viewed in auxiliary momentum space.  The prescriptions are proven in full generality, and their close connection with Feynman rules for real Mellin amplitudes is explained. We also show that the integrands in the Mellin-Barnes representation of both real and $p$-adic Mellin amplitudes, the so-called pre-amplitudes, can be constructed according to virtually identical rules, and that these pre-amplitudes themselves may be re-expressed as products of particular Mellin amplitudes with complexified conformal dimensions.
}

\date{\today}

\maketitle

{\hypersetup{linkcolor=black}
\tableofcontents
}

\section{Introduction and Summary}
\label{INTRODUCTION}

Mellin amplitudes share many properties with flat space scattering amplitudes and have been especially useful in studying holographic CFTs~\cite{Penedones:2010ue}. Until very recently, this story has remained restricted to the study of amplitudes in field theories defined over real-valued spacetimes. In this paper, we continue the holographic investigation of Mellin amplitudes in the context of $p$-adic AdS/CFT~\cite{Gubser:2016guj,Heydeman:2016ldy} initiated in Ref.~\cite{Jepsen:2018dqp}. 
These so-called $p$-adic Mellin amplitudes were shown to have analytic properties similar to those of traditional real Mellin amplitudes for several classes of bulk diagrams~\cite{Jepsen:2018dqp}. 
At the same time, $p$-adic Mellin amplitudes were found to be significantly simpler than their real counterparts owing to the absence of descendants~\cite{Melzer:1988he} in the corresponding $p$-adic CFTs, thus providing a new technically simpler arena for studying generic features of Mellin amplitudes. 
In this paper, we  establish systematically the analytic similarities between the real and $p$-adic amplitudes for \emph{all} tree-level bulk diagrams involving scalars.
The main questions we answer are: What is the $p$-adic Mellin amplitude of an arbitrary tree-level diagram, and what recursion relations do the amplitudes obey? We establish three prescriptions for writing down arbitrary Mellin amplitudes, two of which are recursive, and proceed to establish the precise connections with analogous prescriptions satisfied by corresponding real Mellin amplitudes.

For real Mellin amplitudes, the final expressions, obtained via the Feynman rules for Mellin amplitudes, are usually written in terms of unevaluated infinite sums over terms corresponding to the exchange of descendant fields in the intermediate channels. For scalar operators, the amplitudes for tree-level bulk diagrams take the following schematic form, which we refer to as the series representation of the Mellin amplitudes~\cite{Fitzpatrick:2011ia,Paulos:2011ie,Nandan:2011wc},
\eqn{seriesForm}
{
\mathcal{M} \sim \sum_{m_{i_1}, \ldots, m_{i_K} =0}^\infty \left( \prod_{i \in I} {1 \over s_{i}-\Delta_{i}-2m_{i}} \right) \left( \prod_{\text{vertices}}V(\{m_I\})\right) \qquad I=\{i_1,\ldots,i_K\}\,,
}
where the set $I$ labels the internal lines of the diagram, each admitting the exchange  of a single trace operator of conformal dimension $\Delta_i$ for $i \in I$ (with non-zero $m_i$ corresponding to the exchange of descendants from its conformal family), and each contact vertex in the diagram has an associated factor $V$ which depends  on the conformal dimensions of the external operator insertions incident at that vertex, as well as the dimensions $\Delta_i$ and the associated integers $m_i$ of operators running along internal lines incident on the same vertex. For each internal line $i \in I$ there is a ``propagator factor'', which depends on a particular Mandelstam-like variable $s_i$, which itself depends on the Mellin variables. To obtain the full Mellin amplitude, one must sum over all integers $m_i$ for $i\in I$; typically these sums are hard to evaluate analytically.  This series representation makes the pole structure of the Mellin amplitude manifest --- the Mandelstam variables pick up poles in the infinite sum \eno{seriesForm} when fields propagating along the internal lines go on-shell, signaling the exchange of single-trace operators and their descendants in the intermediate channels, and the residue at each pole factorizes into a product of left and right sub-amplitudes \cite{Fitzpatrick:2011ia,Paulos:2011ie,Goncalves:2014rfa}.

Alternatively, real (and $p$-adic) Mellin amplitudes also admit a contour representation, which we will refer to as the Mellin-Barnes integral representation. Schematically, it takes the form 
\eqn{contourForm}
{
\mathcal{M} \sim  \left( \prod_{i \in I}\,\int_{-i\infty}^{i\infty} {dc_i \over 2\pi i}\,f_{\Delta_i}(c_i)\right) \widetilde{\mathcal{M}}(\{c_I\},\{\gamma_{ij}\}) \qquad I=\{i_1,\ldots,i_K\}\,.
}
where the weight $f_{\Delta_i}(c_i)$ depends on the dimension $\Delta_i$ of the internal operator exchanged along the internal line $i \in I$ as well as the spacetime dimension, and takes as argument a complex parameter $c_i$, while $\widetilde{\mathcal{M}}$, which following Refs.~\cite{Yuan:2017vgp,Yuan:2018qva} we refer to as the ``pre-amplitude'', is a function that is significantly simpler in form than the full, integrated amplitude $\mathcal{M}$ and depends on the Mellin variables $\gamma_{ij}$, the complex parameters $c_I$, and the dimensions solely of external operators.

While in principle the series and Mellin-Barnes representations solve the problem of obtaining real tree-level scalar correlators, they do not yield a closed-form expression for the amplitudes (except in a few simple cases such as the three- and four-point functions). 
And while Mellin amplitudes for special kinds of one-loop diagrams were already available in the early days of the Mellin program~\cite{Penedones:2010ue}, and recent progress has been made in studying bulk-diagrams at one-loop, both in position space~\cite{Cardona:2017tsw,Bertan:2018khc} and Mellin space \cite{Yuan:2017vgp,Yuan:2018qva}, the current technology at loop-level leaves more to be desired when compared with the state-of-the-art for flat-space amplitudes. The flat-space amplitudes program has benefited from many extremely powerful tools and techniques which go beyond summing Feynman diagrams, such as BCFW recursion~\cite{Britto:2005fq}, but these have so far remained elusive in Mellin space and serve as one of the motivations for this work.\footnote{\label{fn:lit}It should be mentioned, though, that methods based on general consistency and symmetry have been successful in a variety of settings, e.g.\ conformal bootstrap for higher dimensional CFTs in position space~\cite{Rattazzi:2008pe,Poland:2011ey,Kos:2013tga,Kos:2014bka,El-Showk:2014dwa,Paulos:2014vya,Simmons-Duffin:2015qma} and Mellin space~\cite{Gopakumar:2016cpb}, use of crossing symmetry to constrain 1-loop Mellin amplitudes~\cite{Aharony:2016dwx,Alday:2017gde,Alday:2017xua}, as well as techniques utilizing superconformal Ward identities to obtain total on-shell amplitudes without resorting to summing individual bulk diagrams in e.g.\ type IIB supergravity in both position and Mellin space~\cite{Rastelli:2016nze,Rastelli:2017udc}, to name a few. See also Refs.~\cite{Raju:2010by,Raju:2011mp} for work on BCFW recursion relations  for bulk diagrams in {\it momentum space}.
}

Turning to $p$-adic Mellin amplitudes (see Ref.~\cite{Jepsen:2018dqp} for a detailed introduction and section~\ref{setup} for a quick overview, definitions and conventions), explicit computations of the amplitudes for tree-diagrams with up to three internal lines~\cite{Jepsen:2018dqp} hint at the existence of recursion relations obeyed by the $p$-adic Mellin amplitudes of arbitrary tree-level bulk diagrams. In section \ref{recursionSection}, we present such a relation which is recursive in the number of internal lines in the diagram, with a proof in appendix~\ref{AmpProof}.
The prescription consists of assigning factors to each vertex and internal line of a given bulk diagram and then writing down the Mellin amplitude by taking a product over all these factors, but also subtracting off all possible diagrams obtained by collapsing in the original diagram every possible subset of internal lines. The expressions for the diagrams to be subtracted off are obtained by a recursive application of this procedure. In the end the entire Mellin amplitude is expressible in closed-form in terms of the factors associated with contact interaction vertices and internal lines. Explicit examples are also provided, and in section~\ref{recursionSectionReal} we present the Feynman rules for real Mellin amplitudes for comparison.

 As pointed out above, real and $p$-adic Mellin amplitudes admit a Mellin-Barnes representation in terms of multi-contour integrals. 
 The reason such a representation exists is that by applying the split representation to all the bulk-to-bulk propagators of the diagram, a Mellin amplitude can be written as a multi-dimensional contour integral over what is referred to as the \emph{pre-amplitude}.
  In section \ref{preAmpSec} we formulate a set of  rules for constructing any tree-level pre-amplitude. 
   Curiously, this prescription  applies universally to \emph{both} $p$-adic and real Mellin amplitudes. 
   
    In the same section we also describe an interesting connection  between pre-amplitudes and Mellin amplitudes, which holds true for both $p$-adic and real Mellin amplitudes: 
  We show that pre-amplitudes may be obtained by taking products of diagrams which appear in the recursive prescription for Mellin amplitudes,\footnote{In section~\ref{recursionSection} we refer to these diagrams as ``undressed diagrams'', and they are essentially the same as Mellin amplitudes up to simple overall factors associated with each vertex. These diagrams can be obtained via the prescriptions of sections~\ref{recursionSection} and~\ref{BCFW} over $p$-adics and the Feynman rules~\cite{Fitzpatrick:2011ia,Paulos:2011ie,Nandan:2011wc} over reals.} except with the dimensions in the diagrams assigned specific complex values that depend on the complex variables of the contour integrals. This prescription is proven in section \ref{PreAmpProof} via an inductive argument, and explicit examples are provided in section~\ref{PreampExamples}.
  
 We emphasize that the claim of the previous paragraph essentially is that the {\it integrand} in the Mellin-Barnes integral representation of the Mellin amplitude secretly takes the same functional form as the {\it integral}, i.e.\  the full Mellin amplitude. 
 Thus the pre-amplitude is given simply by a product of a particular set of full Mellin amplitudes with the dimensions of certain operators set to special values. 
  Although the series and Mellin-Barnes integral representations of the real Mellin amplitudes obfuscate this property, it holds true even over the reals as discussed in section~\ref{PreampRecursion}.

 In section \ref{BCFW} we move on to present a different type of recursion relation obeyed by $p$-adic Mellin amplitudes, which makes the factorization property of $p$-adic Mellin amplitudes manifest.  This prescription allows for an arbitrary tree-level bulk diagram to be expressed in terms of lower-point bulk diagrams with shifted Mandelstam-like invariants, constructed out of the same interaction vertices as the original diagram.
We prove in section \ref{BCFW} that this  prescription follows from the recursive prescription from section \ref{recursionSection}.
We also argue that an adaptation of this prescription applies to real Mellin amplitudes and in fact gives back the Feynman rules prescription. 
In appendix \ref{BCFWMOM} we illustrate with the help of explicit examples, how this prescription originates from an application of Cauchy's residue theorem to complex deformed Mellin amplitudes obtained via complex shifting (auxiliary) momenta, {\it \`a la} on-shell BCFW recursion relations in flat space~\cite{Britto:2005fq}. 
 Thus this prescription is closer in form to  on-shell recursive relations in flat space  since  it provides a decomposition of Mellin amplitudes into products of lower-point sub-amplitudes with all external legs ``on-shell'', joined together by ``propagator'' factors, reminiscent of BCFW recursion. 
We stress, however, that the recursive relation  applies at the level of {\it individual} diagrams, and not at the level of full amplitudes.
We close with final comments and future directions in section \ref{DISCUSSION}.

\section{Mellin Amplitudes and Pre-Amplitudes}
\label{setup}

In this section we recall some basic facts and results for $p$-adic holography and Mellin amplitudes and provide relevant definitions and normalizations which will be used throughout.\footnote{For recent developments in $p$-adic holography, see Refs.~\cite{Gubser:2016guj,Heydeman:2016ldy,Jepsen:2018dqp,Gubser:2017tsi,Gubser:2016htz,Gubser:2017vgc,Gubser:2017qed,Bhattacharyya:2017aly,Marcolli:2018ohd,Dutta:2018qrv,Bhowmick:2018bmn,Qu:2018ned,Gubser:2018cha,Stoica:2018zmi,Heydeman:2018qty,Hung:2018mcn}.}  More details can be found in Ref.~\cite{Jepsen:2018dqp} (see also Refs.~\cite{Gubser:2016guj,Gubser:2017tsi}).
Many results in $p$-adic AdS/CFT closely resemble their real counterparts, not just in physical interpretation but also in their mathematical formulation.
To facilitate comparison and emphasize the striking similarities, we will often present a $p$-adic result or definition along with its real counterpart from conventional continuum AdS/CFT.
In such cases, we will label the equations with a ``$\mathbb{Q}_{p^n}$'' or an ``$\mathbb{R}^n$''  depending on whether they hold in $p$-adic AdS/CFT or real Euclidean AdS${}_{n+1}$/CFT${}_n$.

In the $p$-adic setup, the bulk geometry is given by the Bruhat--Tits tree ${\cal T}_{p^n}$, which is an infinite $(p^n+1)$-regular graph without cycles, and the bulk scalar degrees of freedom live on the vertices of the tree. The boundary of the tree, $\partial {\cal T}_{p^n}$, is the projective line over the degree $n$ unramified extension of the $p$-adic numbers, denoted $\mathbb{P}^1(\mathbb{Q}_{p^n})$. $\mathbb{Q}_{p^n}$ may also be thought of as an $n$-dimensional vector-space over the base field $\mathbb{Q}_p$, thus we will refer to $n$ as the dimension of the boundary field theory. 
The vertices of the tree can be labelled by bulk coordinates $(z_0,z)$ where $z_0$, the bulk depth coordinate, is an integral power of $p$ and $z\in \mathbb{Q}_{p^n}$ represents the boundary direction. 
The vertices are not labelled uniquely by this coordinate representation: two coordinate pairs $(z_0,z)$ and $(z'_0,z')$ label the same vertex iff $z_0=z'_0$ and $z'-z \in z_0 \mathbb{Z}_{p^n}$, where $ \mathbb{Z}_{p^n}=\{x\in \mathbb{Q}_{p^n} \,\big|\, |x|_p \leq 1\}$. 
$|\cdot|_p$ denotes the ultrametric $p$-adic norm, $|\cdot|_p : \mathbb{Q}_{p^n} \to \mathbb{R}^{\geq 0}$. On the real side, the bulk geometry is $(n+1)$-dimensional Euclidean anti-de Sitter space. In the two formalisms, the free bulk actions are given by
\eqn{}{
&\mathbb{Q}_{p^n}\big)
\hspace{10mm}
S_{\text{kin}}=\sum_{\left<(z_0,z)(w_0,w)\right>}{1 \over 2}(\phi_{i(z_0,z)}-\phi_{i(w_0,w)})^2+\sum_{(z_0,z) \in \mathcal{T}_{p^n}}
\frac{1}{2} m^2_{\Delta_i}\phi_{i(z_0,z)}^2\,,
\cr
\cr
&\mathbb{R}^n\big)
\hspace{12mm}
S_{\text{kin}}=\int_{\text{AdS}_{n+1}}dZ 
\left[
\frac{1}{2}\big(\nabla\phi_i(Z)\big)^2+\frac{1}{2}m_{\Delta_i}^2\phi_i^2(Z)
\right],
}
where we have expressed the real action in embedding space coordinates such that the bulk coordinate $Z$ lives in an $(n+2)$-dimensional Minkowski space satisfying $Z^2 \equiv Z\cdot Z =-1$.
The bulk scalar fields $\phi_i$ have masses $m_{\Delta_i}^2$ and scaling dimensions $\Delta_i$, related to each other via
\eqn{mDef}
{
\mathbb{Q}_{p^n}\big)
\hspace{10mm}
m_\Delta^2 = \frac{-1}{\zeta_p(-\Delta)\zeta_p(\Delta-n)}\,,
\qquad \qquad 
\mathbb{R}^n\big)
\hspace{12mm}
m_\Delta^2 = \Delta(\Delta-n)\,,
}
where $\zeta_v: \mathbb{C}\rightarrow \mathbb{C}$, which will be called the local zeta function, is a meromorphic function defined for $v=p$ and $v=\infty$ by
\eqn{zetaDef}
{
\zeta_p(z) = \frac{1}{1-p^{-z}}\,, \qquad \qquad  \qquad \zeta_\infty(z)=\pi^{-\frac{z}{2}}\Gamma\left(\frac{z}{2}\right).
}
The bulk-to-boundary propagators for a bulk scalar of dimension $\Delta$ are given by
\eqn{KDef}
{
&
\mathbb{Q}_{p^n}\big)
\hspace{10mm}
K_\Delta(z_0,z;x)=\zeta_p(2\Delta)\frac{|z_0|_p^{\Delta}}{|z_0,z-x|_s^{2\Delta}}
\,,
\cr \cr
&
\mathbb{R}^n\big)
\hspace{12mm}
K_\Delta(Z,P)
=
\frac{\zeta_\infty(2\Delta)}{(-2P\cdot Z)^\Delta}
\,,
}
where $|\cdot,\cdot|_s$ denotes the supremum norm, 
\eqn{}{
|x,y|_s \equiv \sup\{|x|_p,|y|_p\}\,,
}
and the real bulk-to-boundary propagator is expressed in the embedding space formalism, with the boundary coordinate satisfying $P^2=0$.

The $p$-adic and real bulk-to-bulk propagators,  each of which admits  a \emph{split representation} in terms of integrals over bulk-to-boundary propagators, are given by
\eqn{Gsplit}
{
&
\mathbb{Q}_{p^n}\big)
\hspace{10mm}
G_\Delta(z_0,z;w_0,w)=\zeta_p(2\Delta)p^{-\Delta d[z_0,z;w_0,w]} 
\cr
& \hspace{46mm}
=
 \int_{-\frac{i\pi}{\log p}}^{\frac{i\pi}{\log p}}\, {dc \over 2\pi i}\, f_\Delta(c) \int_{\partial {\cal T}_{p^n}}dx\,K_{h-c}(z_0,z;x)K_{h+c}(w_0,w;x)\,,
\cr
\cr
&
\mathbb{R}^n\big)
\hspace{12mm}
G_\Delta(Z,W)=\frac{\zeta_\infty(2\Delta)}{(Z-W)^{2\Delta}} \: {}_2F_1\left(\Delta,\Delta-h+\frac{1}{2};2\Delta-2h+1;-\frac{4}{(Z-W)^2}\right)
\cr
&
\hspace{38mm}
=
{1\over 2}
 \int_{-i\infty}^{i\infty}\, {dc \over 2\pi i} f_{\Delta}(c) \int_{\partial {\rm AdS}}dP\,K_{h-c}(Z,P)K_{h+c}(W,P)
\,,}
where we have defined 
\eqn{hDef}{ 
h \equiv \frac{n}{2}\,,
}
with
\eqn{fDef}{
&
\mathbb{Q}_{p^n}\big)\hspace{10mm}
f_\Delta(c) \equiv  {\nu_\Delta \over m^2_{\Delta}-m^2_{h-c}} \frac{ \zeta_p(2\Delta-2h)}{\zeta_p(2c)\zeta_p(-2c)} \log p \,, \cr
\cr
&
\mathbb{R}^n\big)\hspace{12mm}
f_\Delta(c) \equiv  {\nu_\Delta \over m^2_{\Delta}-m^2_{h-c}}  \frac{\zeta_\infty(2\Delta-2h)}{\zeta_\infty(2c)\zeta_\infty(-2c)}\,,
}
and
\eqn{nuDef}{
\mathbb{Q}_{p^n}\big) \hspace{10mm} 2\nu_\Delta \equiv p^{\Delta_+}-p^{\Delta_-} = {p^\Delta \over \zeta_p(2\Delta-n)}\,, \qquad \quad \mathbb{R}^n\big) \hspace{12mm}  2\nu_\Delta \equiv \Delta_+-\Delta_- = 2\Delta-n\,,
}
and $d[z_0,z;w_0,w]$ denotes the graph distance between two nodes $(z_0,z)$ and $(w_0,w)$ on the Bruhat--Tits tree, while $(Z-W)^2$ is related to the chordal distance in real continuum AdS.

We note that consistent with Ref.~\cite{Jepsen:2018dqp}, we are using non-standard normalizations for the bulk-to-boundary and bulk-to-bulk propagators, so that the $p$-adic and real bulk-to-bulk propagators satisfy the equations\footnote{The relative sign mismatch in \eno{GEOM} is merely cosmetic, and an artifact of the sign convention used in defining the Laplacian operator $\square$ on the Bruhat--Tits tree~\cite{Gubser:2016guj}.}
\eqn{GEOM}{
&\mathbb{Q}_{p^n}\big) \hspace{10mm}
\left( \square_{z_0,z} + m^2_\Delta \right) G_{\Delta}(z_0,z;w_0,w) = 2\nu_\Delta\: \zeta_p(2\Delta-2h) \delta(z_0,z;w_0,w)\,, \cr \cr
&\mathbb{R}^n\big) \hspace{20mm} 
\left(-\nabla_Z^2 +m^2_\Delta \right) G_{\Delta}(Z,W) = 2\nu_\Delta\: \zeta_\infty(2\Delta-2h) \delta(Z-W)\,.
}
We will consider bulk contact-interactions of the type 
\eqn{}{
\mathbb{Q}_{p^n}\big)
\hspace{10mm}
\sum_{(z_0,z)\in\mathcal{T}_{p^n}}\prod_{i=1}^\mathcal{N}\phi_{i(z_0,z)}\,,
\hspace{30mm}
\mathbb{R}^n\big)
\hspace{12mm}
\int_{\text{AdS}_{n+1}}dX\prod_{i=1}^\mathcal{N}\phi_i(X)\,,
}
and study boundary correlators $\langle {\cal O}_{\Delta_1}(x_1) \ldots {\cal O}_{\Delta_{\cal N}}(x_{\cal N}) \rangle$, which are functions of boundary insertion points $x_i$, and are built holographically from  ${\cal N}$-point position space amplitudes  ${\cal A}(\{x_i\})$. These amplitudes are given by products of bulk-to-boundary and bulk-to-bulk propagators with bulk vertices integrated over the entire bulk, and they depend on the boundary insertion points only via the ${\cal N}({\cal N}-3)/2$ independent conformally invariant cross-ratios constructed out of the boundary points.\footnote{To obtain  this counting, we assume $n+1 \geq {\cal N}$.}

\emph{Mellin amplitudes} are defined via multi-dimensional inverse-Mellin transforms of the position space amplitudes,
\eqn{pMel}
{
&
\mathbb{Q}_{p^n}\big)
\hspace{10mm}
 \mathcal{A}(\{x_i\})=\int[d\gamma]\,\mathcal{M}(\{\gamma_{ij}\})\prod_{1\leq i < j \leq \mathcal{N}}\frac{\zeta_p(2\gamma_{ij})}{|x_i-x_j|_p^{2\gamma_{ij}}}\,,
\cr
 \cr
&
\mathbb{R}^n\big)
\hspace{12mm}
\mathcal{A}(\{x_i\})=\int[d\gamma]\,\mathcal{M}(\{\gamma_{ij}\})\prod_{1\leq i < j \leq \mathcal{N}}\frac{\zeta_\infty(2\gamma_{ij})}{|x_i-x_j|^{2\gamma_{ij}}}\,.
}
The definition of Mellin amplitudes over the reals differs from the ones usually found in the literature only in the choice of  normalization of ${\cal M}$. As will become clear shortly, this choice makes the parallels between real and $p$-adic results the crispest.
The  complex $\gamma_{ij}$ variables above are the Mellin variables, which satisfy
\eqn{MellinVarConstraints}{
\gamma_{ij}=\gamma_{ji}\,, \qquad   \sum_{j=1}^\mathcal{N}\gamma_{ij}=0 \quad {\rm and }\quad  \gamma_{ii}=-\Delta_i \quad {\rm (no\ sum\ over\ } i) \qquad i=1,\ldots, \mathcal{N}\,,
}
leading also to ${\cal N}({\cal N}-3)/2$ independent components. The integration measure in \eno{pMel} is over the independent Mellin variables,
\eqn{}{
&
\mathbb{Q}_{p^n}\big)
\hspace{10mm}
[d\gamma]\equiv\left(\frac{2\log p}{2\pi i}\right)^{\frac{\mathcal{N}(\mathcal{N}-3)}{2}}\left[\prod_{1\leq i < j \leq \mathcal{N}}d\gamma_{ij}\right]\left[\prod_{i=1}^\mathcal{N}\delta\big(\sum_{j=1}^\mathcal{N}\gamma_{ij}\big)\right],
\cr
\cr
&
\mathbb{R}^n\big)
\hspace{12mm}
[d\gamma]\equiv\left(\frac{1}{2\pi i}\right)^{\frac{\mathcal{N}(\mathcal{N}-3)}{2}}\left[\prod_{1\leq i < j \leq \mathcal{N}}d\gamma_{ij}\right]\left[\prod_{i=1}^\mathcal{N}\delta\big(\sum_{j=1}^\mathcal{N}\gamma_{ij}\big)\right],
}
and the precise contour prescription is similar in both cases, with the main distinction being that the Mellin variables in $p$-adic Mellin space live on a ``complex cylinder'', $\mathbb{R} \times S^1$ where the imaginary direction is periodically identified: $\gamma_{ij} \sim  \gamma_{ij}+\frac{i\pi}{\log p}$~\cite{Jepsen:2018dqp}. This distinction arises for the following reason. Mellin variables may be identified as the conjugate of the dilatation parameter. However the dilatation group for $p$-adic coordinates is discrete, owing to the discreteness of the $p$-adic norm, and the Pontryagin dual of a discrete group is compact.\footnote{We thank Brian Trundy for this observation.}

The constraints in \eno{MellinVarConstraints} can be solved in auxiliary momentum space, provided the fictitious momenta $k_i$ satisfy the following constraints,
 \eqn{momConstraints}{
 k_i\cdot k_j = \gamma_{ij}\,, \qquad \sum_{i=1}^\mathcal{N} k_i = 0\,, \qquad i,j \in \{1,\ldots,\mathcal{N}\}\,,
 }
 and the on-shell condition,
  \eqn{OnshellMom}{
k_i^2 = k_i \cdot k_i = -\Delta_i\,, \qquad i\in \{1,\ldots,\mathcal{N}\}\,.
 }
Mandelstam-like variables can be defined via
 \eqn{MandelstamDef}
{
s_{i_1 \ldots i_K} \equiv -\left(\sum_{i\in S} k_i\right)^2 = \sum_{i\in S}\Delta_i - 2\sum_{\substack{i,j \in S,\\ i<j}}\gamma_{ij}\,,
}
where $S = \{i_1,\ldots,i_K\}  \subset \{1,\ldots, {\cal N}\}$ is a subset of the set of all external legs. For large $N$ CFTs, Mellin amplitudes are meromorphic functions of such Mandelstam-like variables, built out of the local zeta function $\zeta_p$ or $\zeta_\infty$, with the poles corresponding to the exchange of single-trace operators (and their descendants in $\mathbb{R}^n$). 

\begin{example}
The simplest Mellin amplitudes are the contact amplitudes, represented diagrammatically thus:
 \eqn{}
{
\mathcal{M}^{\text{contact}} =  \begin{matrix}
\text{
 \begin{tikzpicture}
\draw[thick] (0,0) ellipse (1.3cm and 1.3cm);
\draw[thick,fill=black] (0,0) ellipse (0.05cm and 0.05cm);
\draw[very thick] (0,0) -- (0,1.3);
\node at (0,1.6) {2}; 
\draw[very thick] (0,0) -- (1.23637,0.401722);
\node at (1.5,0.5) {3}; 
\draw[very thick] (0,0) -- (-1.23637,0.401722);
\node at (-1.5,0.5) {1}; 
\draw[very thick] (0,0) -- (0.764121,-1.05172);
\node at (0.95,-1.25) {4}; 
\draw[very thick] (0,0) -- (-0.764121,-1.05172);
\node at (-1.05,-1.35) {$\mathcal{N}$}; 
\node at (0,-0.8) {...}; 
\end{tikzpicture}
}
  \end{matrix}\,.
}
In our normalization, the Mellin amplitudes for the contact diagram are given by
\eqn{}{
&
\mathbb{Q}_{p^n}\big)
\hspace{10mm}
\mathcal{M}^{\text{contact}}=
\zeta_p\big(\sum_{i=1}^\mathcal{N}\Delta_i-n\big)\,,
\cr
&
\mathbb{R}^n\big)
\hspace{12mm}
\mathcal{M}^{\text{contact}}=\frac{1}{2}
\zeta_\infty\big(\sum_{i=1}^\mathcal{N}\Delta_i-n\big)\,.
}
\end{example}

For diagrams with internal lines corresponding to exchange of single-trace operators, it is convenient to express Mellin amplitudes as contour integrals over the so-called \emph{pre-amplitudes} (denoted with a tilde):
\eqn{preampDef}{
&
\mathbb{Q}_{p^n}\big)
\hspace{10mm}
\mathcal{M}=
\left[\prod_{I}
\int_{-\frac{i\pi}{\log p}}^{\frac{i\pi}{\log p}}\, \frac{dc_I}{ 2\pi i}\, f_{\Delta_I}(c_I) \right]
\widetilde{\mathcal{M}}\,,
\cr
\cr
&
\mathbb{R}^n\big)
\hspace{12mm}
\mathcal{M}= 
\left[\prod_{I}
\int_{-i\infty}^{i\infty} \frac{dc_I}{2\pi i}\, f_{\Delta_I}(c_I)
\right]
\widetilde{\mathcal{M}}\,,
}
 where $f_\Delta(c)$ is defined in \eno{fDef}. The index $I$ runs over all internal lines in the diagram.
Like Mellin amplitudes, pre-amplitudes depend on Mellin variables solely via Mandelstam-like variables, but unlike Mellin amplitudes, pre-amplitudes depend only on external scaling dimensions, not internal ones.

\begin{example}
The amplitude with exactly one internal line, that is the exchange amplitude, is depicted thus:
 \eqn{}
{
{\mathcal{M}}^{\text{exchange}} = \begin{matrix}
\text{
\begin{tikzpicture}
\draw[thick] (0,0) ellipse (2.cm and 1.3cm);
\draw[thick,fill=black] (-0.8,0) ellipse (0.05cm and 0.05cm);
\draw[thick,fill=black] (0.8,0) ellipse (0.05cm and 0.05cm);
\draw[very thick] (0.8,0)--(-0.8,0);
\draw[very thick] (0.8,0)--(1.73,0.63);
\draw[very thick] (0.8,0)--(1.73,-0.63);
\draw[very thick] (-0.8,0)--(-1.73,0.63);
\draw[very thick] (-0.8,0)--(-1.73,-0.63);
\node at (0,0.3) {$s$}; 
\node at (0,-0.4) {$\Delta$}; 
\node at (1.6,0.1) {$\vdots$}; 
\node at (-1.6,0.1) {$\vdots$}; 
\node at (2.4,0) {$i_R$}; 
\node at (-2.35,0) {$i_L$}; 
\end{tikzpicture}
}
  \end{matrix}\,.
}
The associated Mandelstam invariant $s$ is given by
\eqn{sExchange}{
 s=  \sum\Delta_{i_L} - 2\sum \gamma_{i_L j_L} = \sum\Delta_{i_R}  -2\sum \gamma_{i_R j_R}\,,
 }
 where  the limits of the various sums have been left implicit, but it should be clear from the context what indices are being summed over. For example, $\sum \Delta_{i_L} = \sum_{i_L} \Delta_{i_L}$ is a sum over all possible values that $i_L$ takes, and $\sum \gamma_{i_L j_L} = \sum_{i_L < j_L} \gamma_{i_L j_L}$ where $i_L, j_L$ are summed over all possible indices that label the external legs to the left  of the internal line.
The $p$-adic and real pre-amplitudes for the exchange diagram are then given by
\eqn{ExchPreamp}{
\mathbb{Q}_{p^n}\big)
\hspace{10mm}
\widetilde{\mathcal{M}}^{\text{exchange}} =&\,
\zeta_p(\sum\Delta_{i_L}+c-h)
\zeta_p(\sum\Delta_{i_R}-c-h)
\cr
&
\times
\beta_p\!\left(h+c-s,\sum\Delta_{i_L}-h-c\right) \beta_p\!\left(h-c-s,\sum\Delta_{i_R}-h+c\right),
\cr
\cr
\mathbb{R}^n\big)
\hspace{12mm}
\widetilde{\mathcal{M}}^{\text{exchange}} =&\,
\frac{1}{4}
\zeta_\infty(\sum\Delta_{i_L}+c-h)
\zeta_\infty(\sum\Delta_{i_R}-c-h)
\cr
&
\times
\beta_\infty\!\left(h+c-s,\sum\Delta_{i_L}-h-c\right) \beta_\infty\!\left(h-c-s,\sum\Delta_{i_R}-h+c\right),
}
where we have defined
\eqn{BetaPDef}{
\mathbb{Q}_{p^n}\big) \hspace{10mm} \beta_p(s,t) \equiv {\zeta_p(s) \zeta_p(t) \over \zeta_p(s+t)} = \zeta_p(s) - \zeta_p(-t) \,, \qquad \mathbb{R}^n\big) \hspace{12mm} \beta_\infty(s,t) \equiv {\zeta_\infty(s) \zeta_\infty(t) \over \zeta_\infty(s+t)}\,.
}
\end{example}
The functional similarities between the real and $p$-adic results reviewed in this section hint at a framework in which perhaps both real and $p$-adic computations can be performed in a unified manner. Indeed as stated in the introduction, we will present in this paper prescriptions for constructing arbitrary-point tree-level Mellin amplitudes for scalars which apply to both the real and $p$-adic AdS/CFT frameworks simultaneously, demonstrating the close ties between the formalisms and providing further evidence for a larger universal framework in which the real and $p$-adic formulations are treated on the same footing.

\section{Construction of Mellin Amplitudes}
\label{recursionSection}

\subsection{Recursive prescription for $p$-adic Mellin amplitudes}
\label{recursionSectionPadic}

In Ref.~\cite{Jepsen:2018dqp}, we computed the $p$-adic Mellin amplitudes for a general scalar contact diagram, as well as for arbitrary-point tree-level bulk diagrams containing one, two or three internal lines. 
 A careful comparison of these amplitudes hints at a recursive structure. In this section, we provide a precise formulation of this structure and a prescription for recovering all previously computed Mellin amplitudes from Ref.~\cite{Jepsen:2018dqp}. 
In appendices \ref{PreAmpProof} and \ref{AmpProof}, we prove that this recursive prescription extends to arbitrary tree-level bulk diagrams, and thus can be used to directly write down closed-form expressions for Mellin amplitudes corresponding to tree-level bulk diagrams with an arbitrary number of internal lines admitting scalar exchanges,  helping bypass tedious bulk computations.

\vspace{1em}
It is well known that in position space,  any tree-level bulk diagram, also referred to as a Witten diagram, is built out of three ingredients: external legs corresponding to bulk-to-boundary propagators, internal lines corresponding to bulk-to-bulk propagators, and interaction vertices corresponding to coupling constants. 
The same continues to hold true in $p$-adic AdS/CFT~\cite{Gubser:2016guj}.
The full ($p$-adic) position space amplitude is obtained by performing bulk integration over all bulk vertices of an integrand given by the product over all factors associated with the individual ingredients going into the diagram.

In contrast, in Mellin space the closed-form expressions for Mellin amplitudes from Ref.~\cite{Jepsen:2018dqp} suggest that $p$-adic Mellin amplitudes can be written in terms of two basic building blocks, which which we represent graphically as:
\begin{itemize}
\item Contact vertex: 
\eqn{vertexFactorDef}{
\begin{matrix}
\includegraphics[height=14ex]{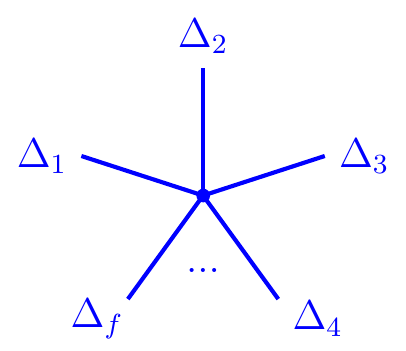}
 \end{matrix}
 \quad  
 \equiv
 \quad  
  \zeta_p\!\left(\sum_{i=1}^f\Delta_i-n\right).
}
\item Internal line:  
\eqn{internalPropagator}
{\begin{matrix}
\includegraphics[height=6ex]{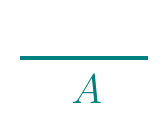}
 \end{matrix}
 \quad   
 \equiv 
 \quad  
 \zeta_p(s_A-\Delta_A)\,.
}
\end{itemize}

We present two examples.
\begin{example}
\label{ex:exch}
Using the graphical representations above, we can rewrite the closed-form $p$-adic Mellin amplitude for the arbitrary-point exchange diagram as 
\eqn{singleAmp}{
\begin{matrix}
\includegraphics[height=12ex]{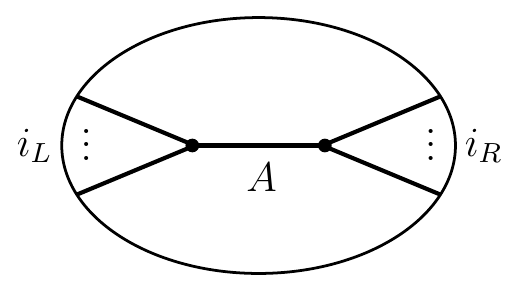}
 \end{matrix}
 =
 \left(
\begin{matrix}
\includegraphics[height=5ex]{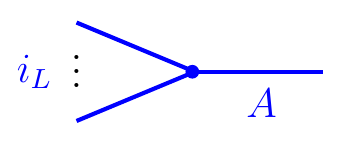}
 \end{matrix}
 \right)
\left(
\begin{matrix}
\includegraphics[height=5ex]{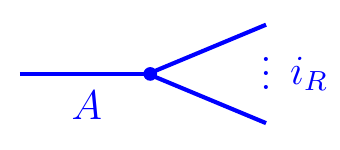}
 \end{matrix}
 \right)
 \left(
\begin{matrix}
\includegraphics[height=5ex]{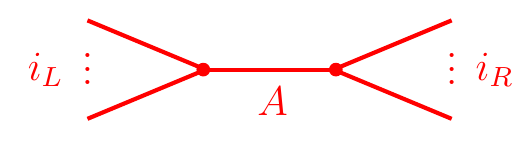}
 \end{matrix}
 \right),
}
where we have defined
\eqn{singleAmpUndressed}{
\begin{matrix}
\includegraphics[height=5ex]{figures/singleRed.pdf}
 \end{matrix}
 \equiv
 (-1)^1
 \left\{
 \left(
 \begin{matrix}
\includegraphics[height=5ex]{figures/propA.pdf}
 \end{matrix}
 \right)
-
   \left(
 \begin{matrix}
\includegraphics[height=5ex]{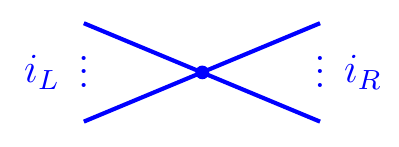}
 \end{matrix}
 \right)
 \right\}
  .
}
 It is straightforward to verify that this is simply a diagrammatic rewriting of equation (4.42) of Ref.~\cite{Jepsen:2018dqp} provided we identify the Mandelstam invariant $s_A$ with the expression in \eno{sExchange}.
\end{example}

\begin{example}
\label{ex:double}
Similar to the previous example, the $p$-adic Mellin amplitude for a bulk diagram with two internal lines (but arbitrary external insertions) is given by 
\eqn{doubleAmp}{
\begin{matrix}
\includegraphics[height=14ex]{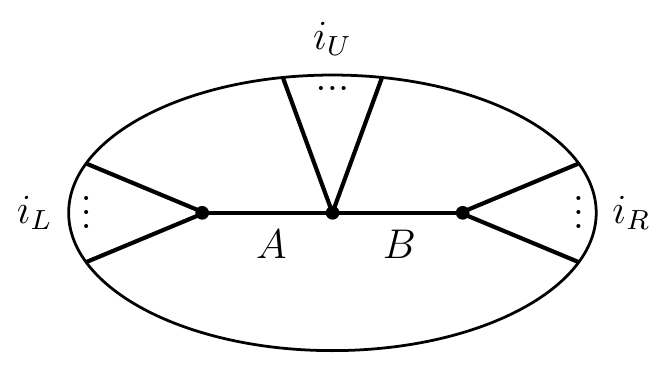}
 \end{matrix}
 =
 \cr
 \left(
\begin{matrix}
\includegraphics[height=5ex]{figures/leftBlue.pdf}
 \end{matrix}
 \right)
 &
 \left(
 \begin{matrix}
\includegraphics[height=10ex]{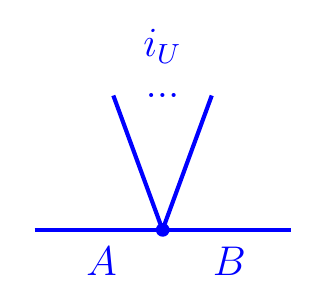}
 \end{matrix}
 \right)
\left(
\begin{matrix}
\includegraphics[height=5ex]{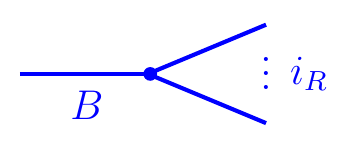}
 \end{matrix}
 \right)
 \left(
\begin{matrix}
\includegraphics[height=11ex]{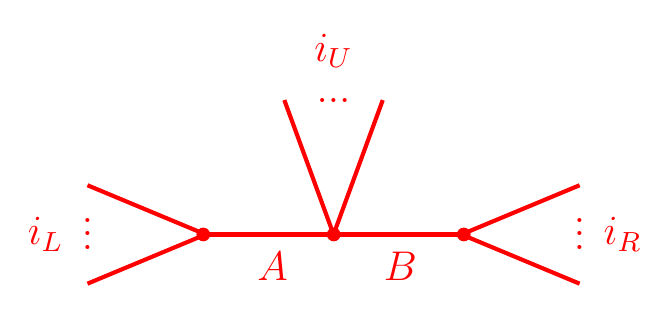}
 \end{matrix}
 \right),
}
where we have defined 
\eqn{doubleAmpUndressed}{
& \begin{matrix}
\includegraphics[height=11ex]{figures/doubleRed.pdf}
 \end{matrix} \equiv \cr 
 &
  (-1)^2 \left\{
  \left(
 \begin{matrix}
\includegraphics[height=5ex]{figures/propA.pdf}
 \end{matrix}
 \right)
  \left(
 \begin{matrix}
\includegraphics[height=5ex]{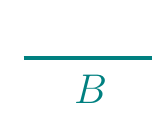}
 \end{matrix}
 \right)
 -
 \left(
\begin{matrix}
\includegraphics[height=10ex]{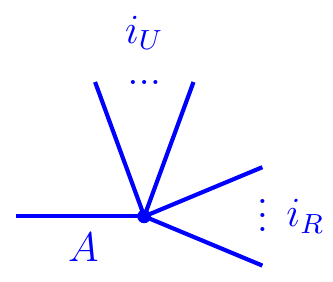}
 \end{matrix}
 \right)
 (-1)^1
 \left(
\begin{matrix}
\includegraphics[height=10ex]{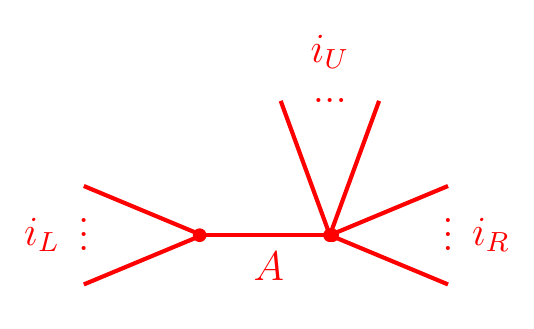}
 \end{matrix}
 \right)
 \right.
 \cr
 & \left. 
 -\left(
\begin{matrix}
\includegraphics[height=10ex]{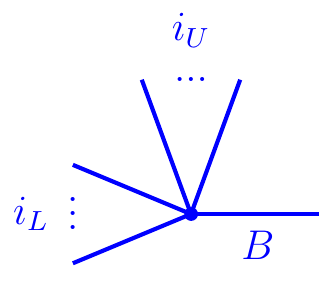}
 \end{matrix}
 \right)
 (-1)^1
 \left(
\begin{matrix}
\includegraphics[height=10ex]{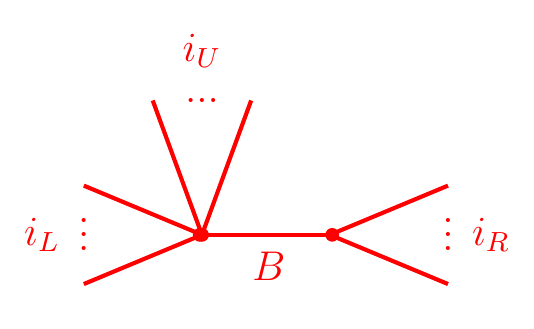}
 \end{matrix}
 \right)
- \left(
 \begin{matrix}
\includegraphics[height=10ex]{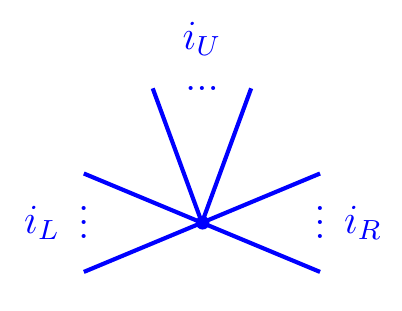}
 \end{matrix}
 \right)
 \right\}.
}
This reproduces the amplitude given in equation (4.56) of Ref.~\cite{Jepsen:2018dqp} if we set the Mandelstam-like variables $s_A, s_B$ to
\eqn{}{
s_A = \sum \Delta_{i_L} - 2\sum\gamma_{i_Lj_L}\,, \qquad 
 s_B = \sum \Delta_{i_R} - 2\sum\gamma_{i_Rj_R}\,,
}
and use \eno{singleAmpUndressed} to further reduce \eno{doubleAmpUndressed} to an expression given entirely in terms of vertex and internal line  factors \eno{vertexFactorDef}-\eno{internalPropagator}.
\end{example}

The two examples above are suggestive of the recursive procedure for reconstructing $p$-adic Mellin amplitudes which we will now describe.
In this prescription, no integration is necessary and the basic building blocks are the  bulk contact-interaction vertices and internal lines, \eno{vertexFactorDef} and \eno{internalPropagator} respectively. 
This recursive procedure will be proven in the appendices. 

\begin{prescription}[$p$-adic Mellin amplitudes]
\label{pres:padicMellin}
The Mellin amplitude for a particular bulk diagram is given by the product over a vertex factor \eno{vertexFactorDef} for each contact vertex in the diagram, times what we will refer to as an \emph{``undressed Mellin amplitude''} (which we have depicted in red in \eno{singleAmp} and \eno{doubleAmp}).
The undressed amplitudes are constructed \emph{recursively} as follows:
\begin{itemize}
\item
The undressed amplitude for a bulk diagram with \emph{no} internal lines is equal to one.
\item
The undressed amplitude for a bulk diagram with one or more internal lines  is given by 
 an overall factor of minus one raised to the number of internal lines times the following combination:
the product over all internal line factors \eno{internalPropagator} associated with the bulk diagram, minus all possible undressed diagrams associated with bulk diagrams obtained by collapsing all possible subsets of internal lines in the original bulk diagram,
with each such subtracted term weighted by minus one raised to the number of \emph{remaining} internal lines and also weighted by the vertex factor(s) \eno{vertexFactorDef} associated with any \emph{new} contact vertex (or vertices) generated upon affecting such a collapse.
\end{itemize}
\end{prescription}

Schematically, this prescription takes the form:
\eqn{decomp1}
{
\text{(undressed diagram)} &= (-1)^{\#\text{ internal lines}} \Bigg\{
\prod\big(\text{internal line factors}\big) \cr 
& \qquad  -\sum \Big[ (-1)^{\#\text{remaining internal lines}}\big(\text{new vertex factor(s)} \big) \ \cr  
& \quad \qquad \qquad   \times
\big(\text{reduced undressed diagram}\big) \Big] \Bigg\}.
}

\clearpage

We make two remarks about the prescription described above:
\begin{itemize}
\item[$\star$] Each vertex factor is given by the $p$-adic local zeta function \eno{vertexFactorDef} whose argument is given by the sum over the scaling dimensions of all (internal and external) operators incident on the vertex, minus the dimension of the boundary field theory $n$.

\item[$\star$] In each factor \eno{internalPropagator} associated with an internal line, the Mandelstam variable carried by the line is given by a combination of Mellin variables $\gamma_{ij}$ and external scaling dimensions $\Delta_i$ as defined in \eno{MandelstamDef}, with the subset $S$ in \eno{MandelstamDef} taken to be all the external indices on one side of the internal line in question.
\end{itemize}
 As is clear from the recursive prescription for the undressed amplitudes, the recursion is on the number of internal lines (exchanges), with each undressed amplitude expressible in terms of undressed amplitudes containing fewer internal lines (exchange channels). 

\vspace{0.5em}
It will be often convenient to represent this recursive procedure diagrammatically. To distinguish between the full Mellin amplitudes and undressed Mellin amplitudes, we will use the following convention:
\begin{itemize}
    \item[$\star$] Mellin amplitudes will be represented like standard Witten diagrams, drawn in black with the Poincar\'{e} disk shown explicitly.
    \item[$\star$]  Undressed amplitudes will be represented in red with no Poincar\'{e} disk shown, and we will refer to these as ``undressed diagrams''. 
\end{itemize}
 
For illustrative purposes,  we apply this procedure to two non-trivial diagrams, namely the two topologically distinct types of diagrams involving three internal lines.

\clearpage

\begin{example} 
\label{ex:sixptseries}
For the diagram with three internal lines arranged in a series,
 the recursive procedure leads to the following Mellin amplitude. 
\eqn{tripleLineAmp}{
&
\begin{matrix}
\includegraphics[height=16ex]{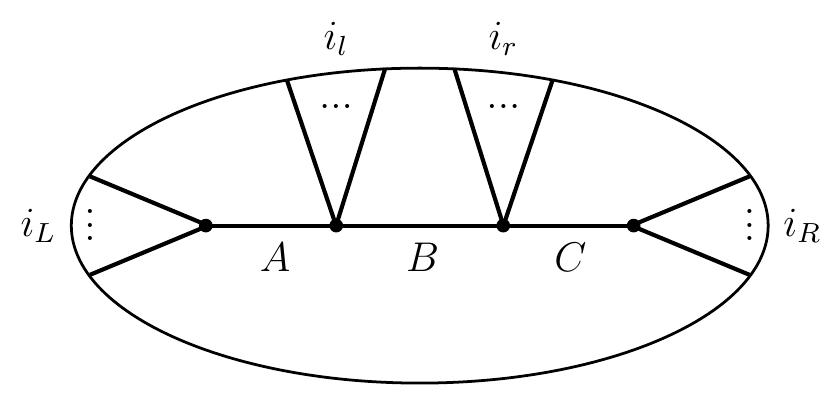}
 \end{matrix}
 =
 \cr
 \left(
 \begin{matrix}
\includegraphics[height=5ex]{figures/leftBlue.pdf}
 \end{matrix}
 \right)
 &
 \left(
 \begin{matrix}
\includegraphics[height=9ex]{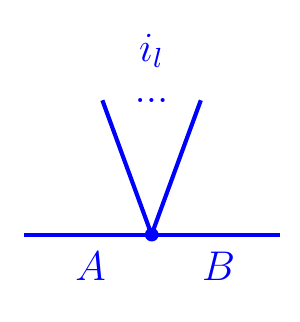}
 \end{matrix}
 \right)
 \left(
 \begin{matrix}
\includegraphics[height=9ex]{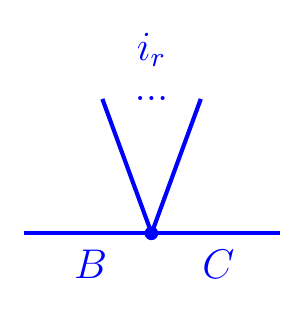}
 \end{matrix}
 \right)
 \left(
 \begin{matrix}
\includegraphics[height=4ex]{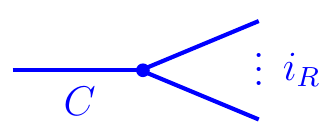}
 \end{matrix}
 \right)
 \left(
 \begin{matrix}
\includegraphics[height=11ex]{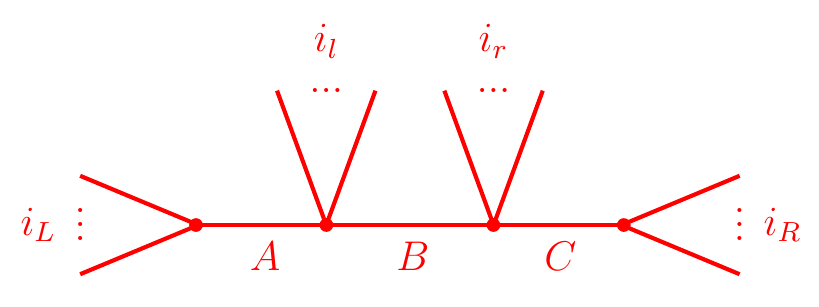}
 \end{matrix}
 \right),
}
where  the undressed diagram is given by
\eqn{seriesUndressed}{
&
\begin{matrix}
\includegraphics[height=11ex]{figures/tripleLineRed.pdf}
 \end{matrix}
 =
 \cr
 &
 (-1)^3 \left\{
   \left(
 \begin{matrix}
\includegraphics[height=4ex]{figures/propA.pdf}
 \end{matrix}
 \right)
  \left(
 \begin{matrix}
\includegraphics[height=4ex]{figures/propB.pdf}
 \end{matrix}
 \right)
  \left(
 \begin{matrix}
\includegraphics[height=4ex]{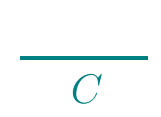}
 \end{matrix}
 \right)
 -
  \left(
 \begin{matrix}
\includegraphics[height=8ex]{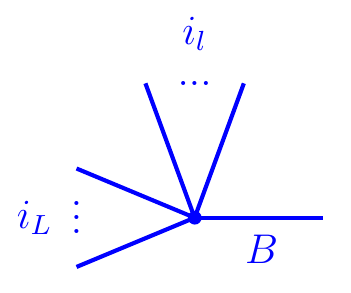}
 \end{matrix}
 \right) 
 (-1)^2
   \left(
 \begin{matrix}
\includegraphics[height=11ex]{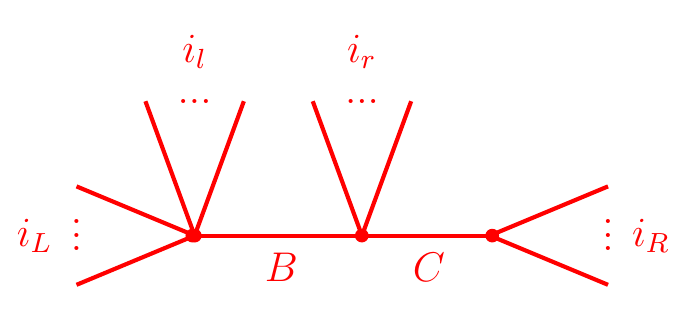}
 \end{matrix}
 \right) \right.
 \cr
  &-
  \left(
 \begin{matrix}
\includegraphics[height=13ex]{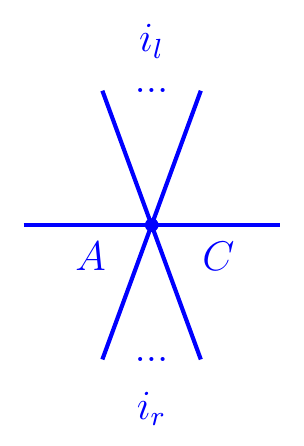}
 \end{matrix}
 \right) 
 (-1)^2
   \left(
 \begin{matrix}
\includegraphics[height=14ex]{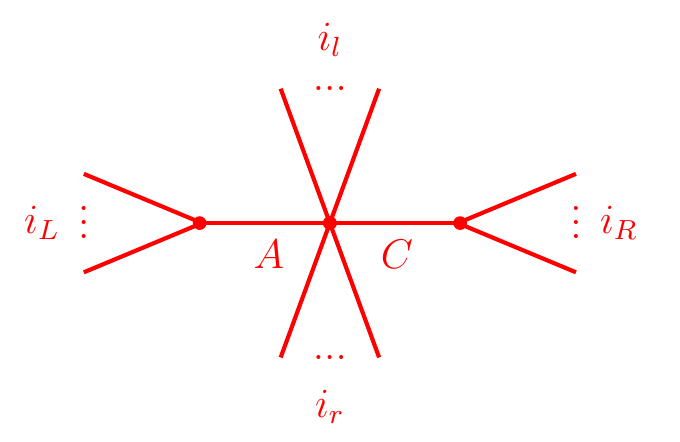}
 \end{matrix}
 \right) 
 -
  \left(
 \begin{matrix}
\includegraphics[height=8ex]{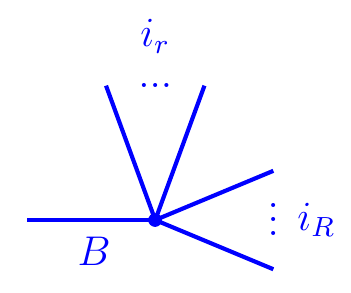}
 \end{matrix}
 \right) 
 (-1)^2
   \left(
 \begin{matrix}
\includegraphics[height=11ex]{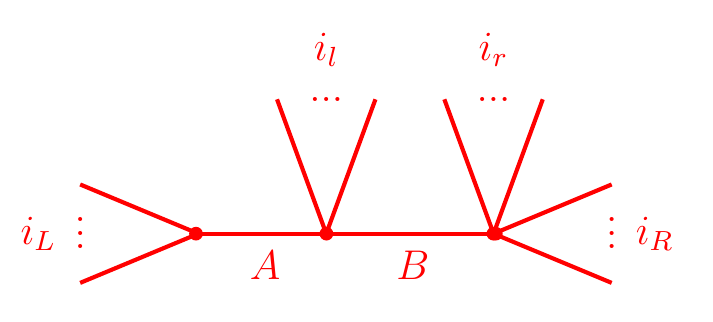}
 \end{matrix}
 \right)
 \cr
  &-
  \left(
 \begin{matrix}
\includegraphics[height=13ex]{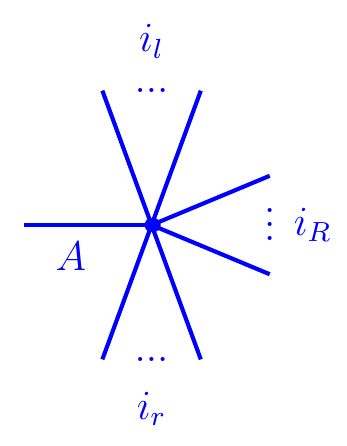}
 \end{matrix}
 \right) 
 (-1)^1
   \left(
 \begin{matrix}
\includegraphics[height=14ex]{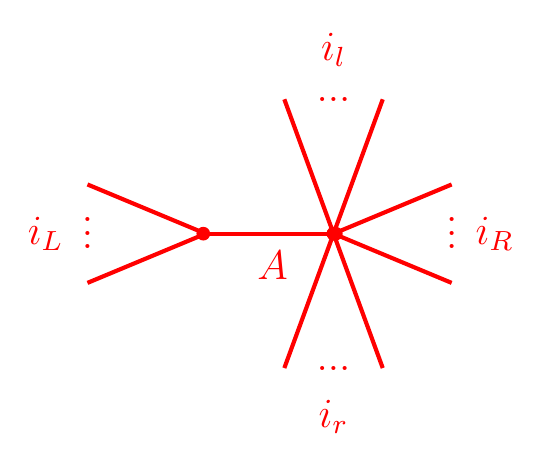}
 \end{matrix}
 \right) 
 -
  \left(
 \begin{matrix}
\includegraphics[height=13ex]{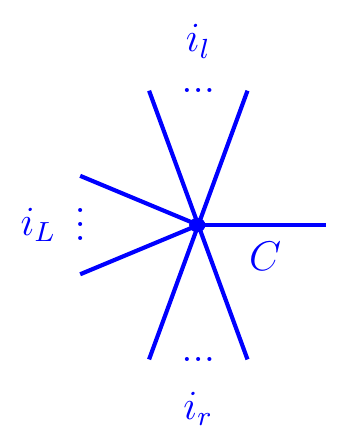}
 \end{matrix}
 \right) 
 (-1)^1
   \left(
 \begin{matrix}
\includegraphics[height=14ex]{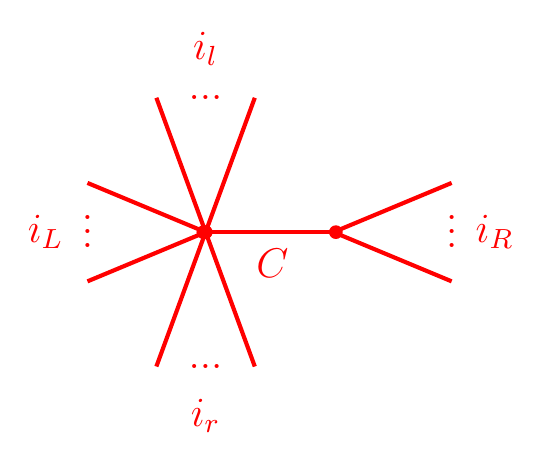}
 \end{matrix}
 \right)
 \cr
  & \left. 
  -
  \left(
 \begin{matrix}
\includegraphics[height=8ex]{figures/tripleLineV5.pdf}
 \end{matrix}
 \right) 
  \left(
 \begin{matrix}
\includegraphics[height=8ex]{figures/tripleLineV4.pdf}
 \end{matrix}
 \right) 
  (-1)^1
   \left(
 \begin{matrix}
\includegraphics[height=11ex]{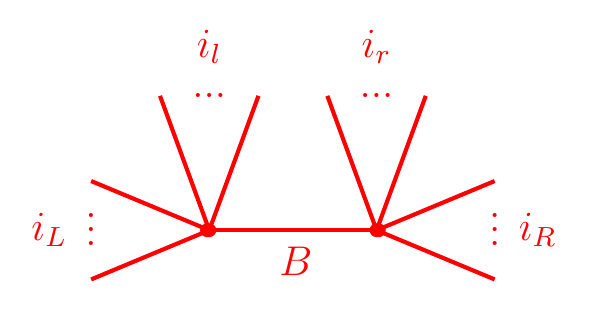}
 \end{matrix}
 \right) 
 -
  \left(
 \begin{matrix}
\includegraphics[height=13ex]{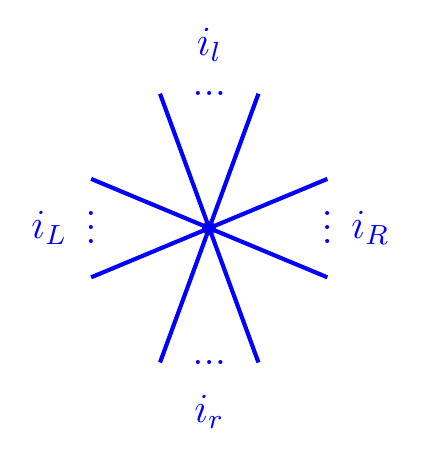}
 \end{matrix}
 \right) \right\}.
}
It should be clear in the diagrammatic representation above which internal lines were collapsed in each of the terms subtracted off from the first term, the product over all internal line factors.
Further, according to the prescription, the Mandelstam-like variables which feature in this example are given by
\eqn{}{ 
s_A &= \sum \Delta_{i_L} -2\sum \gamma_{i_L j_L}\,, \cr
s_B &= \sum \Delta_{i_L} + \sum \Delta_{i_l} - 2\sum \gamma_{i_L j_L} -2\sum \gamma_{i_l j_l} - 2\sum \gamma_{i_L j_l}\,, \cr 
s_C &= \sum \Delta_{i_R} -2\sum \gamma_{i_R j_R}\,.
}
Note that there is no restriction on the indices in the sum, $\sum \gamma_{i_L j_l} = \sum_{i_L, j_l} \gamma_{i_L j_l}$, while in other sums there is one, e.g.\ $\sum \gamma_{i_L j_L} = \sum_{i_L<j_L} \gamma_{i_L j_L}$. The undressed diagrams in \eno{seriesUndressed} can be reduced further using previous results \eno{singleAmpUndressed} and \eno{doubleAmpUndressed}, at which point the final closed-form Mellin amplitude obtained via the recursive procedure of this section directly matches the  result of the first principles computation given in equation (4.66) of Ref.~\cite{Jepsen:2018dqp}.
\end{example}

As a second example, we now consider the other type of bulk configuration with three internal lines.
\begin{example} 
\label{ex:sixptstar}
 According to the recursive prescription, the Mellin amplitude for the star diagram, which is a bulk diagram where three internal lines meet at a central bulk vertex, takes the form
\eqn{starAmp}{
&
\begin{matrix}
\includegraphics[height=20ex]{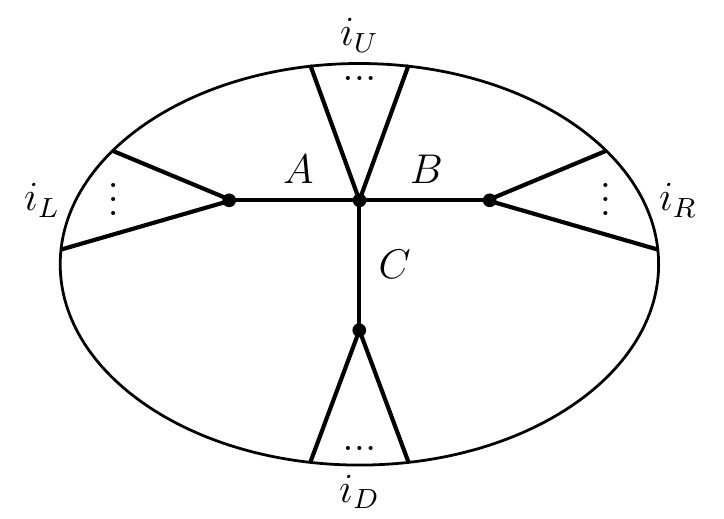} 
 \end{matrix} =
 \cr
 \left(
 \begin{matrix}
\includegraphics[height=5ex]{figures/leftBlue.pdf}
 \end{matrix}
 \right)
 &
 \begin{pmatrix}
\includegraphics[height=13ex]{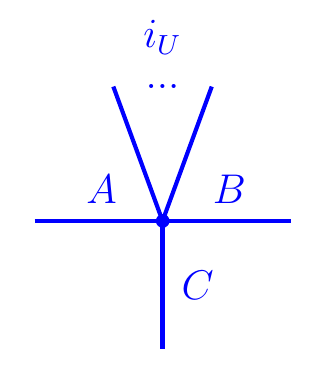}
 \end{pmatrix}
 \left(
 \begin{matrix}
\includegraphics[height=15ex]{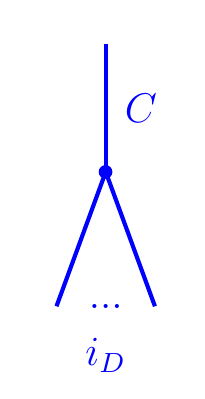}
 \end{matrix}
 \right)
 \left(
 \begin{matrix}
\includegraphics[height=5ex]{figures/RightBlue2.pdf}
 \end{matrix}
 \right)
 \left(
 \begin{matrix}
\includegraphics[height=20ex]{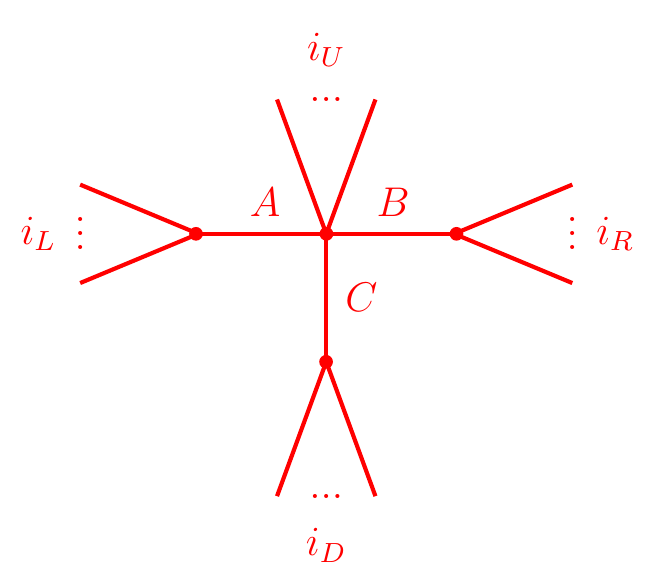}
 \end{matrix}
 \right),
}
where the undressed diagram is given by
\eqn{starAmpB}{
&
 \begin{matrix}
\includegraphics[height=20ex]{figures/triangleRed.pdf}
 \end{matrix}
 =
 \cr
 &
 (-1)^3\left\{
  \left(
 \begin{matrix}
\includegraphics[height=5ex]{figures/propA.pdf}
 \end{matrix}
 \right)
  \left(
 \begin{matrix}
\includegraphics[height=5ex]{figures/propB.pdf}
 \end{matrix}
 \right)
  \left(
 \begin{matrix}
\includegraphics[height=7ex]{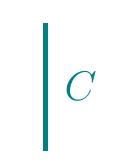}
 \end{matrix}
 \right) 
 -
   \left(
 \begin{matrix}
\includegraphics[height=12ex]{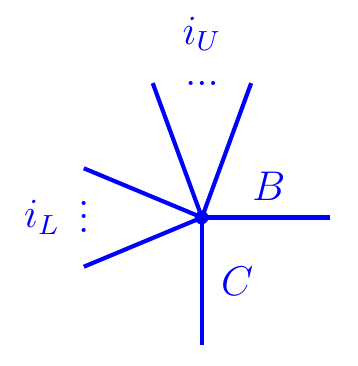}
 \end{matrix}
 \right)
  \right. 
 (-1)^2
    \left(
 \begin{matrix}
\includegraphics[height=18ex]{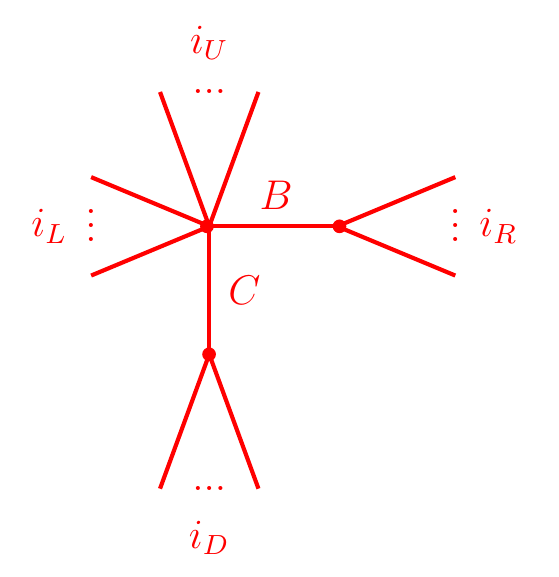}
 \end{matrix}
 \right)
 \cr
 &
 -
  \left(
 \begin{matrix}
\includegraphics[height=11ex]{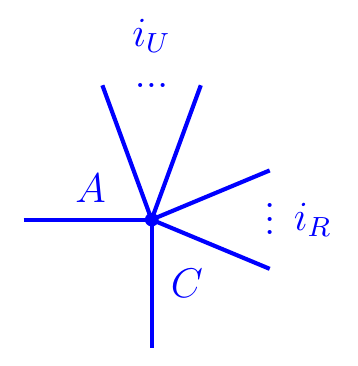}
 \end{matrix}
 \right)
 (-1)^2
     \left(
 \begin{matrix}
\includegraphics[height=18ex]{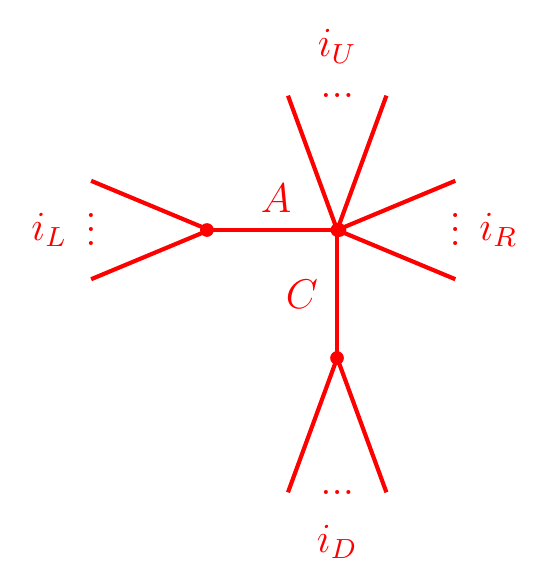}
 \end{matrix}
 \right)
-
  \left(
 \begin{matrix}
\includegraphics[height=12ex]{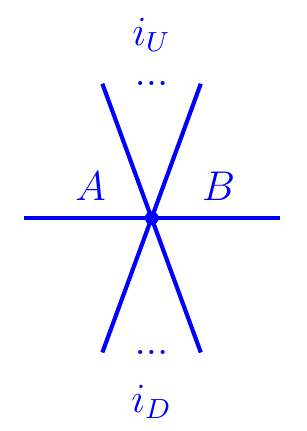}
 \end{matrix}
 \right)
 (-1)^2
     \left(
 \begin{matrix}
\includegraphics[height=14ex]{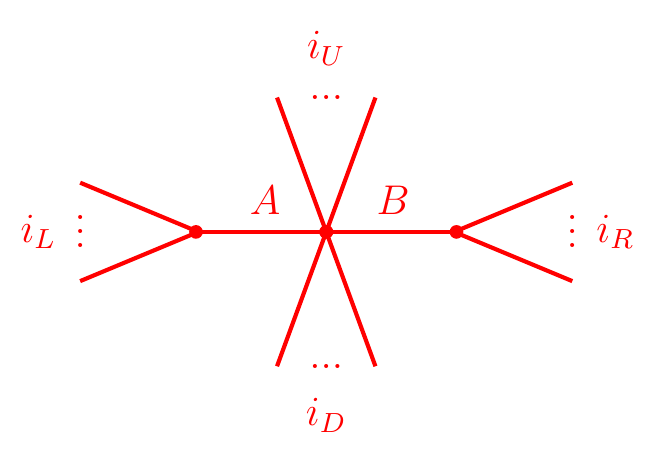}
 \end{matrix}
 \right)
  \cr
 &-
  \left(
 \begin{matrix}
\includegraphics[height=12ex]{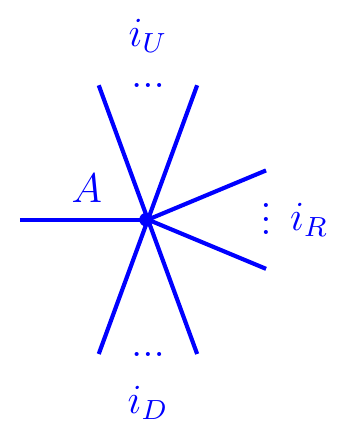}
 \end{matrix}
 \right)
 (-1)^1
     \left(
 \begin{matrix}
\includegraphics[height=14ex]{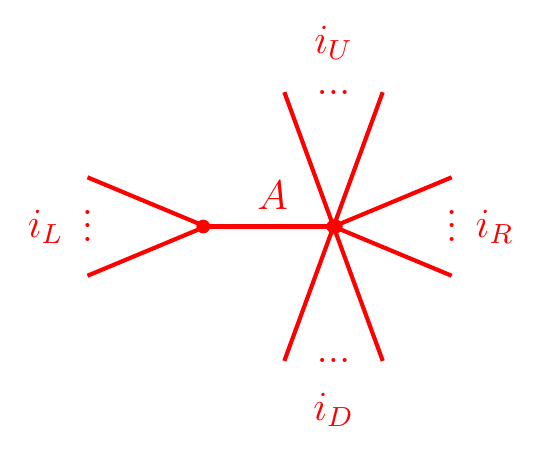}
 \end{matrix}
 \right)
-
  \left(
 \begin{matrix}
\includegraphics[height=13ex]{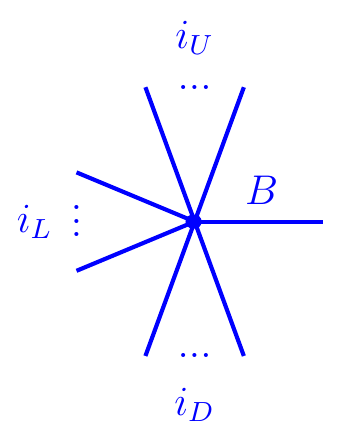}
 \end{matrix}
 \right)
 (-1)^1
     \left(
 \begin{matrix}
\includegraphics[height=14ex]{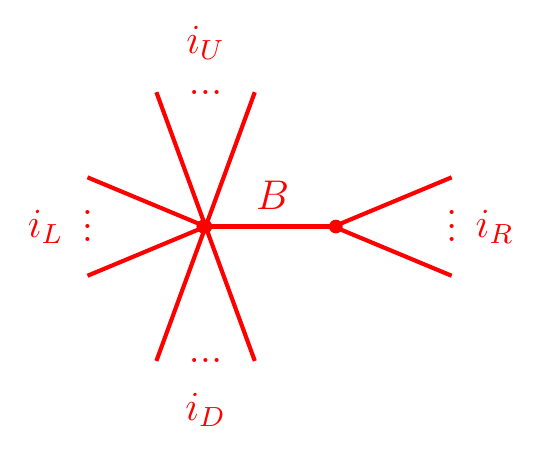}
 \end{matrix}
 \right)
  \cr
 &
 -
  \left(
 \begin{matrix}
\includegraphics[height=11ex]{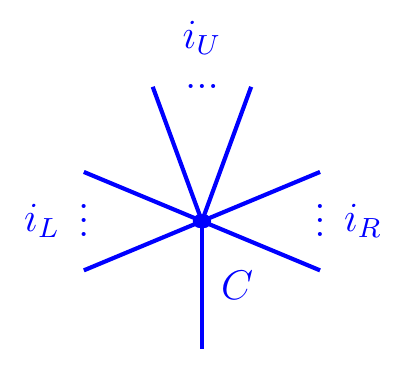}
 \end{matrix}
 \right)
 (-1)^1
     \left(
 \begin{matrix}
\includegraphics[height=18ex]{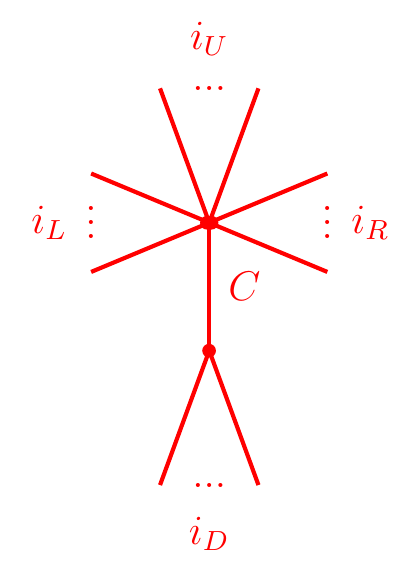}
 \end{matrix}
 \right)
-
   \left.
  \left(
 \begin{matrix}
\includegraphics[height=13ex]{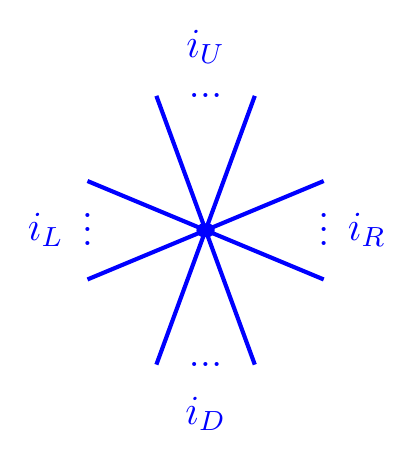}
 \end{matrix}
 \right)\right\},
}
with the Mandelstam-like variables given by
\eqn{}{ 
s_A &= \sum \Delta_{i_L} - 2\sum \gamma_{i_Lj_L}\,, \cr
s_B &= \sum \Delta_{i_R} - 2\sum \gamma_{i_Rj_R}\,, \cr
s_C &= \sum \Delta_{i_D} - 2\sum \gamma_{i_Dj_D}\,.
}
Just like in the previous example, the undressed diagram in  \eno{starAmpB} may recursively be reduced further using expressions \eno{singleAmpUndressed} and \eno{doubleAmpUndressed}, to give an amplitude expressed solely in terms of the internal line and contact vertex factors. It is again easily verified that this result matches the previously computed amplitude given in equation (4.78) of Ref.~\cite{Jepsen:2018dqp}.
\end{example}

\subsection{Feynman rules for  real Mellin amplitudes}
\label{recursionSectionReal}

It is interesting to compare the recursive procedure describe above with the so-called Feynman rules for real Mellin amplitudes. 

\begin{prescription}[Feynman rules for real Mellin amplitudes~\cite{Fitzpatrick:2011ia,Paulos:2011ie,Nandan:2011wc}]
\label{pres:Feyn}
Given a tree-level bulk diagram, (adapted to the conventions and normalization described in section \ref{setup}) the Feynman rules are:

\begin{itemize}

\item Label the internal lines with an index $j$ running from 1 to $v-1$ and associate to each internal line an integer $n_j$ and a factor of
\eqn{lineFactorFeyn}{
L_j(s_j,\Delta_j,n_j) = \frac{1}{n_j!}\frac{(1+\Delta_j-h)_{n_j}}{(\Delta_j-s_j)/2+n_j}\,,
}
where $s_j$ is the Mandelstam invariant associated to internal leg $j$, and $(a)_{n_j}\equiv \frac{\Gamma(a+n_j)}{\Gamma(a)}$ is the Pochhammer symbol.

\item Label the vertices with an index $j$ running from one to $v$. For a vertex $j$ connected to $L$ legs  with scaling dimensions $\Delta_1$ to $\Delta_L$, label the internal legs by indices 1 to $l$ and the external legs by indices $l+1$ to $L$, let $n_1$ to $n_l$ be the integers associated with the internal legs, and associate to the vertex $j$ a factor of
\eqn{vertexFactorFeyn}{
V_j(\{\Delta_i\},&\{n_i\})=\,
\frac{1}{2}\zeta_\infty\big(\sum_{i=1}^L\Delta_i-n\big)
\cr
&
\times
F_A^{(l)}\left(\frac{1}{2}\sum_{i=1}^L\Delta_i-h;\{-n_1,\ldots,-n_l\};\{1+\Delta_1-h,\ldots,1+\Delta_l-h\};1,\ldots,1\right),
}
where $F_A^{(l)}$ is the Lauricella hypergeometric series~\cite{Lauricella1893,Srivastava1985book,AartsMathWorld,Paulos:2011ie}, defined as
\eqn{LauricellaDef}
{
F_A^{(l)}\left(g;\{a_1,\ldots,a_l\};\{b_1,\ldots,b_l\};x_1,\ldots,x_l\right)
=\left[\prod_{i=1}^l\sum_{n_i=0}^\infty \right](g)_{\sum_{i=1}^l n_i}\prod_{i=1}^l \frac{(a_i)_{n_i}}{(b_i)_{n_i}}\frac{x_i^{n_i}}{n_i!}\,.
}

\item The Mellin amplitude is given by taking the product of all line and vertex factors and then summing over the integers associated with the internal legs:
\eqn{}{
\mathcal{M}(\{\gamma_{ij}\},\{\Delta_i\})= 
\left[\prod_{i=1}^{v-1} \sum_{n_i=0}^\infty \right]
\left[\prod_{j=1}^{v-1} L_j(s_j,\Delta_j,n_j) \right]
\left[\prod_{j=1}^{v} V_j(\{\Delta_i\},\{n_i\}) \right].
}

\end{itemize}
\end{prescription}
These rules are arguably more compact and straightforward than the prescription for obtaining the $p$-adic Mellin amplitudes, but this apparent simplicity is deceptive and merely due to the fact that the real Mellin amplitude is left in the form of  unevaluated infinite sums, while the prescription in the $p$-adic case fully reduces the Mellin amplitudes to finite sums of elementary functions. For this reason, prescriptions~\ref{pres:padicMellin} and~\ref{pres:Feyn} do not highlight the similarity in the structure of real and $p$-adic Mellin amplitudes. For a unified perspective on real and $p$-adic amplitudes, it is useful to turn instead to pre-amplitudes which will be the subject of the next section. However, we do note  at this point that as in the $p$-adic case, we can decompose both real and $p$-adic Mellin amplitudes into a product of contact vertices and an ``undressed diagram'':
\eqn{MContUndress}
{
\mathcal{M}(\{\gamma_{ij}\},\{\Delta_i\})= (\text{all contact vertices }) \times (\text{undressed diagram})\,,
}
where in the real case the vertex factors are defined as
\begin{itemize}
\item Contact vertex: 
\eqn{vertexFactorDefreal}{
\begin{matrix}
\includegraphics[height=14ex]{figures/GenericVertex.pdf}
 \end{matrix}
 \quad  
 \equiv
 \quad  
  \frac{1}{2}\zeta_\infty\left(\sum_{i=1}^f\Delta_i-n\right).
}
\end{itemize}
With this definition in place, equations \eno{singleAmp}, \eno{doubleAmp}, \eno{tripleLineAmp}, and \eno{starAmp} continue to apply for real Mellin amplitudes, with the explicit expressions for the undressed diagrams over the $p$-adics (given by prescription~\ref{pres:padicMellin}) getting replaced by infinite sums over products of Lauricella functions over the reals (given by prescription~\ref{pres:Feyn}). 
Equation \eno{MContUndress} serves as the definition of undressed diagrams over reals.
We will show in the next section that the undressed diagrams over the reals and the $p$-adics are in fact closely related to each other via a concrete relation.

\section{Construction of Pre-Amplitudes}
\label{preAmpSec}

Mellin amplitudes $\mathcal{M}$, real and $p$-adic, admit  a representation  called the Mellin-Barnes representation  in terms of a contour integral over a pre-amplitude $\widetilde{\mathcal{M}}$, defined in~\eno{preampDef}. Compared to Mellin amplitudes, pre-amplitudes are markedly simpler: they do not depend on internal scaling dimensions, and for any given bulk diagram the corresponding pre-amplitude is simply a product over factors associated with each vertex of the bulk diagram. Furthermore, pre-amplitudes of real and $p$-adic AdS/CFT take identical forms, and for this reason we can simultaneously present the prescription for either type of pre-amplitude. In the real case, the prescription essentially follows from Ref.~\cite{Nandan:2011wc} after adjusting overall coefficients to account for our normalization conventions described in section~\ref{setup}. For the $p$-adic case, we present an inductive proof of the prescription in appendix~\ref{PreAmpProof}.

\begin{prescription}[Pre-amplitudes]
\label{pres:Preamp}
For a diagram with $v$ vertices labelled by indices $j
\in \{1,\ldots,v\}$, the \emph{pre-amplitude} is constructed as follows:
\begin{itemize}
\item Associate to each internal leg an orientation and a complex variable $c_i$.
\item 
For a vertex $j$ connected to $L$ legs with scaling dimensions $\Delta_1$ to $\Delta_L$, let indices 1 to $l$ label internal legs and let indices $l+1$ to $L$ label external leg. Futhermore, define scaling dimensions $\widetilde{\Delta}_i$ as follows: if $i\in\{1,\ldots,l\}$ so that $i$ labels an internal leg, define a complex scaling dimension $\widetilde{\Delta}_i=h \pm c_i$, where the plus sign is chosen if the orientation of the internal leg is incoming and the minus sign is chosen if the internal leg is outgoing with respect to the vertex; if $i\in\{l+1,\ldots,L\}$ so that $i$ labels an external leg, define  $\widetilde{\Delta}_i=\Delta_i$. Also associate to each vertex a factor of:
\eqn{intForms}
{
\mathbb{Q}_{p^n}\big)
\hspace{10mm}
&\widetilde{V}_j(\{s_i\},\{\widetilde{\Delta}_i\})
=
\cr
\zeta_p\big(\sum_{i=1}^L & \widetilde{\Delta}_i - n\big)
 \bigg[\prod_{i=1}^l {2\,\zeta_p(1) \over |2|_p}\int_{\mathbb{Q}_p^2}\frac{dx_i}{|x_i|_p}\,|x_i|_p^{\widetilde{\Delta}_i-s_i \over 2} |1,x_i|^{\widetilde{\Delta}_i+s_i-n \over 2}_s\bigg] \left|1,x_1,\ldots,x_l\right|_s^{n-\sum_{i=1}^L\widetilde{\Delta}_i \over 2},
\cr
\cr
\mathbb{R}^n\big)
\hspace{12mm}
&\widetilde{V}_j(\{s_i\},\{\widetilde{\Delta}_i\})
=
\cr
\frac{1}{2}
\zeta_\infty\big(\sum_{i=1}^L & \widetilde{\Delta}_i - n\big)
\left[\prod_{i=1}^l\int_0^\infty \frac{dx_i}{|x_i|}|x_i|^{\frac{\widetilde{\Delta}_i-s_i}{2}}|1+x_i|^{\frac{\widetilde{\Delta}_i+s_i-n}{2}}\right]\big|1+\sum_{i=1}^l x_i\big|^{\frac{n-\sum_{i=1}^L\widetilde{\Delta}_i}{2}}\,.
}
Here $\{s_i\}$ is the set of Mandelstam-like variables associated with the internal lines and $\mathbb{Q}_p^2$ denotes the set of $p$-adic squares, see \eno{square}.

\item
The \emph{Mellin pre-amplitude} is given by the product of these vertex factors:
\eqn{preampVertex}{
\widetilde{\mathcal{M}}(\{\gamma_{ij}\},\{\Delta_{\text{\emph{ext,}}i}\},\{c_i\})=\prod_{j=1}^v
\widetilde{V}_j(\{s_{i}\},\{\widetilde{\Delta}_i\})\,.
}

\end{itemize}
\end{prescription}

We point out that the integral representation of the pre-amplitude vertex factor in~\eno{intForms} over the $p$-adics and the reals is related to  undressed diagrams as defined in \eno{MContUndress}.
     Thus these integrals admit a recursive construction via prescription~\ref{pres:padicMellin} for the $p$-adics and prescription~\ref{pres:Feyn} for the reals. 
This point is explained in more detail next.

\subsection{Obtaining pre-amplitudes from Mellin amplitudes}
\label{PreampRecursion}

The reader might have noticed that the contact vertices defined in \eno{vertexFactorDef} and \eno{vertexFactorDefreal} appear in the prescription above for pre-amplitude vertex factors, and as we commented above the remaining factors in \eno{intForms} can be identified with undressed diagrams. Essentially the claim is that pre-amplitudes decompose into products of Mellin amplitudes with a special assignment of internal dimensions, so that the Mellin-Barnes integral representation expresses any Mellin amplitude as a contour integral over simpler Mellin amplitudes. 

More precisely, the pre-amplitude vertex factor $\widetilde{V}_j(\{s_i\},\{\widetilde{\Delta}_i\})$ from \eno{intForms} associated with a given vertex in a bulk diagram, with  internal lines assigned orientations according to prescription \ref{pres:Preamp}  as displayed in black in \eno{PreampRuleDiag}, is expressible as a product of a contact vertex and an undressed diagram associated to a full Mellin amplitude, as shown:
\eqn{PreampRuleDiag}{
&\hspace{45mm}
\left(
\begin{matrix}
\text{
\begin{tikzpicture}
\draw[thick,fill=black] (0,0) ellipse (0.05cm and 0.05cm);
\draw[thick,fill=black] (1.2,0.5) ellipse (0.05cm and 0.05cm);
\draw[thick,fill=black] (1.2,-0.5) ellipse (0.05cm and 0.05cm);
\draw[very thick,->-] (1.2,0.5) to (0,0);
\draw[very thick,->-] (1.2,-0.5) to (0,0);
\draw[very thick] (-1.1*1.2,-1.1*0.5) to (0,0);
\draw[very thick] (-1.1*1.2,1.1*0.5) to (0,0);
\node at (-1.5*1.2,-1.4*0.5) {$\Delta_{l+1}$};
\node at (-1.5*1.2,1.4*0.5) {$\Delta_{L}$};
\node at (1,0.1) {$\vdots$}; 
\node at (-1.1,0.1) {$\vdots$}; 
\node at (0.5,0.7) {$s_1, \Delta_1$}; 
\node at (0.5,-0.7) {$s_l, \Delta_l$}; 
\end{tikzpicture}
}
 \end{matrix}
 \right):
\cr
&
\widetilde{V}_j(\{s_i\},\{\widetilde{\Delta}_i\})=\left(
 \begin{matrix}
\text{
\begin{tikzpicture}
\draw[blue,thick,fill=blue] (0,0) ellipse (0.05cm and 0.05cm);
\draw[blue,very thick] (1.2,0.5) to (0,0);
\draw[blue,very thick] (1.2,-0.5) to (0,0);
\draw[blue,very thick] (-1.1*1.2,-1.1*0.5) to (0,0);
\draw[blue,very thick] (-1.1*1.2,1.1*0.5) to (0,0);
\node at (-1.5*1.2,-1.4*0.5) {\textcolor{blue}{$\Delta_{l+1}$}};
\node at (-1.5*1.2,1.4*0.5) {\textcolor{blue}{$\Delta_{L}$}};
\node at (1,0.1) {\textcolor{blue}{$\vdots$}}; 
\node at (-1.1,0.1) {\textcolor{blue}{$\vdots$}}; 
\node at (1.1,0.8) {\textcolor{blue}{$h+c_1$}}; 
\node at (1.1,-0.8) {\textcolor{blue}{$h+c_l$}}; 
\end{tikzpicture}
}
 \end{matrix}
 \right)
  \left(
 \begin{matrix}
 \text{
 \begin{tikzpicture}
\draw[red,thick,fill=red] (0,0) ellipse (0.05cm and 0.05cm);
\draw[red,thick,fill=red] (1.2,0.5) ellipse (0.05cm and 0.05cm);
\draw[red,thick,fill=red] (1.2,-0.5) ellipse (0.05cm and 0.05cm);
\draw[red,very thick] (2*1.2,2*0.5) to (0,0);
\draw[red,very thick] (2*1.2,-2*0.5) to (0,0);
\draw[red,very thick] (-1.1*1.2,-1.1*0.5) to (0,0);
\draw[red,very thick] (-1.1*1.2,1.1*0.5) to (0,0);
\node at (-1.5*1.2,-1.4*0.5) {\textcolor{red}{$\Delta_{l+1}$}};
\node at (-1.5*1.2,1.4*0.5) {\textcolor{red}{$\Delta_{L}$}};
\node at (1,0.1) {\textcolor{red}{$\vdots$}}; 
\node at (-1.1,0.1) {\textcolor{red}{$\vdots$}}; 
\node at (0.5,1.2) {\textcolor{red}{$s_1,$}}; 
\node at (0.5,0.75) {\textcolor{red}{$h+c_1$}}; 
\node at (0.5,-0.7) {\textcolor{red}{$h+c_l,$}}; 
\node at (0.5,-1.25) {\textcolor{red}{$s_l$}}; 
\node at (1.8,1.27) {\textcolor{red}{$h-c_1$}}; 
\node at (1.8,-1.23) {\textcolor{red}{$h-c_l$}}; 
\end{tikzpicture}
 }
 \end{matrix}
 \right),
 }
where the contact vertex factor (shown in blue in \eno{PreampRuleDiag}) is given by \eno{vertexFactorDef} over $p$-adics and \eno{vertexFactorDefreal} over reals, and the undressed diagram  (shown in red in \eno{PreampRuleDiag}) can be explicitly constructed over the $p$-adics using the recursive prescription \ref{pres:padicMellin}, while over the reals it is constructed using the Feynman rule prescription \ref{pres:Feyn}. 

We point out that the undressed diagram in \eno{PreampRuleDiag} has $l$ internal lines, with Mandelstam invariants $s_i$ and propagating complex dimensions $h+c_i$.
Furthermore, the $l$ vertices at which operators of dimensions $h-c_i$ are incident should formally be thought of as contact interaction vertices, each of which has precisely one leg incident on another bulk vertex, while all other legs connect to external operator insertions.

We now prove this non-trivial claim, beginning with the real case.
From the Feynman rules and definitions in section \ref{recursionSectionReal}, it follows that 
\eqn{}{
&
\mathbb{R}^n\big)
\hspace{40mm}
 \begin{matrix}
 \text{
 \begin{tikzpicture}
\draw[red,thick,fill=red] (0,0) ellipse (0.05cm and 0.05cm);
\draw[red,thick,fill=red] (1.2,0.5) ellipse (0.05cm and 0.05cm);
\draw[red,thick,fill=red] (1.2,-0.5) ellipse (0.05cm and 0.05cm);
\draw[red,very thick] (2*1.2,2*0.5) to (0,0);
\draw[red,very thick] (2*1.2,-2*0.5) to (0,0);
\draw[red,very thick] (-1.1*1.2,-1.1*0.5) to (0,0);
\draw[red,very thick] (-1.1*1.2,1.1*0.5) to (0,0);
\node at (-1.5*1.2,-1.4*0.5) {\textcolor{red}{$\Delta_{l+1}$}};
\node at (-1.5*1.2,1.4*0.5) {\textcolor{red}{$\Delta_{L}$}};
\node at (1,0.1) {\textcolor{red}{$\vdots$}}; 
\node at (-1.1,0.1) {\textcolor{red}{$\vdots$}}; 
\node at (0.5,1.2) {\textcolor{red}{$s_1,$}}; 
\node at (0.5,0.75) {\textcolor{red}{$h+c_1$}}; 
\node at (0.5,-0.7) {\textcolor{red}{$h+c_l,$}}; 
\node at (0.5,-1.25) {\textcolor{red}{$s_l$}}; 
\node at (1.8,1.27) {\textcolor{red}{$h-c_1$}}; 
\node at (1.8,-1.23) {\textcolor{red}{$h-c_l$}}; 
\end{tikzpicture}
 }
 \end{matrix}
\cr
 &= \left[\prod_{i=1}^l\sum_{n_i=0}^\infty \right]
  \left[ \prod_{j=1}^l \frac{1}{n_j!} \frac{(1+\widetilde{\Delta}_j-h)_{n_j}}{(\widetilde{\Delta}_j-s_j)/2+n_j}
  \,
  F_A^{(1)}\left(\frac{\widetilde{\Delta}_j+h-c_j}{2}-h;\{-n_j\};\{1+c_j\};1\right)
  \right]
 \cr
 & \quad\times 
 F_A^{(l)}
 \left(\frac{1}{2}\sum_{i=1}^L\widetilde{\Delta}_i-h;\{-n_1,\ldots,-n_l\};\{1+c_1,\ldots,1+c_l\};1,\ldots,1\right).
}
Since $\widetilde{\Delta}_j=h+c_j$ for $j\in\{1,\ldots,l\}$, the Lauricella functions $F_A^{(1)}$ in the above expression are all equal to unity. Using the definition \eno{LauricellaDef} for the Lauricella function as a multiple sum, changing the order of summation between this multiple sum and the sums over $n_i$, and applying the identity 
\eqn{}
{
\sum_{n=0}^\infty \frac{(-n)_m(1+c)_n}{n!(A+n)}=\frac{\Gamma(A+m)\Gamma(-c)}{\Gamma(A-c)}=(A)_m\,\beta_\infty(2A,-2c)\,,
}
where $\beta_\infty(\cdot,\cdot)$ is related to the usual Euler Beta function and was defined in \eno{BetaPDef}, one can  carry out the sums over the integers $n_i$ to find, 
\eqn{realIntForm}{ 
&
\mathbb{R}^n\big)
\hspace{40mm}
 \begin{matrix}
 \text{
 \begin{tikzpicture}
\draw[red,thick,fill=red] (0,0) ellipse (0.05cm and 0.05cm);
\draw[red,thick,fill=red] (1.2,0.5) ellipse (0.05cm and 0.05cm);
\draw[red,thick,fill=red] (1.2,-0.5) ellipse (0.05cm and 0.05cm);
\draw[red,very thick] (2*1.2,2*0.5) to (0,0);
\draw[red,very thick] (2*1.2,-2*0.5) to (0,0);
\draw[red,very thick] (-1.1*1.2,-1.1*0.5) to (0,0);
\draw[red,very thick] (-1.1*1.2,1.1*0.5) to (0,0);
\node at (-1.5*1.2,-1.4*0.5) {\textcolor{red}{$\Delta_{l+1}$}};
\node at (-1.5*1.2,1.4*0.5) {\textcolor{red}{$\Delta_{L}$}};
\node at (1,0.1) {\textcolor{red}{$\vdots$}}; 
\node at (-1.1,0.1) {\textcolor{red}{$\vdots$}}; 
\node at (0.5,1.2) {\textcolor{red}{$s_1,$}}; 
\node at (0.5,0.75) {\textcolor{red}{$h+c_1$}}; 
\node at (0.5,-0.7) {\textcolor{red}{$h+c_l,$}}; 
\node at (0.5,-1.25) {\textcolor{red}{$s_l$}}; 
\node at (1.8,1.27) {\textcolor{red}{$h-c_1$}}; 
\node at (1.8,-1.23) {\textcolor{red}{$h-c_l$}}; 
\end{tikzpicture}
 }
 \end{matrix}
\cr
&=
F_A^{(l)}\!\left(\frac{\sum_{i=1}^L\widetilde{\Delta}_i-n}{2};\left\{\frac{\widetilde{\Delta}_1-s_1}{2},\ldots,\frac{\widetilde{\Delta}_l-s_l}{2}\right\};\{1+c_1,\ldots,1+c_l\};1,\ldots,1\right)
\prod_{i=1}^l \beta_\infty\big(\widetilde{\Delta}_i-s_i,-2c_i\big)
\cr
&=
\left[\prod_{i=1}^l\int_0^\infty \frac{dx_i}{|x_i|}|x_i|^{\frac{\widetilde{\Delta}_i-s_i}{2}}|1+x_i|^{\frac{\widetilde{\Delta}_i+s_i-n}{2}}\right]\big|1+\sum_{i=1}^l x_i\big|^{\frac{n-\sum_{i=1}^L\widetilde{\Delta}_i}{2}}\,,
}
where the last equality arises from a formal substitution justified via an appropriate analytic continuation; we refer the reader to equation (3.30) of Ref.~\cite{Nandan:2011wc}. 
This proves \eno{PreampRuleDiag} over $\mathbb{R}^n$.

\vspace{1em}
We now turn to the $p$-adic case, where we need to prove the identity
\eqn{toShowPadic}{ 
&
\mathbb{Q}_{p^n}\big)
\hspace{30mm}
 \begin{matrix}
 \text{
 \begin{tikzpicture}
\draw[red,thick,fill=red] (0,0) ellipse (0.05cm and 0.05cm);
\draw[red,thick,fill=red] (1.2,0.5) ellipse (0.05cm and 0.05cm);
\draw[red,thick,fill=red] (1.2,-0.5) ellipse (0.05cm and 0.05cm);
\draw[red,very thick] (2*1.2,2*0.5) to (0,0);
\draw[red,very thick] (2*1.2,-2*0.5) to (0,0);
\draw[red,very thick] (-1.1*1.2,-1.1*0.5) to (0,0);
\draw[red,very thick] (-1.1*1.2,1.1*0.5) to (0,0);
\node at (-1.5*1.2,-1.4*0.5) {\textcolor{red}{$\Delta_{l+1}$}};
\node at (-1.5*1.2,1.4*0.5) {\textcolor{red}{$\Delta_{L}$}};
\node at (1,0.1) {\textcolor{red}{$\vdots$}}; 
\node at (-1.1,0.1) {\textcolor{red}{$\vdots$}}; 
\node at (0.5,1.2) {\textcolor{red}{$s_1,$}}; 
\node at (0.5,0.75) {\textcolor{red}{$h+c_1$}}; 
\node at (0.5,-0.7) {\textcolor{red}{$h+c_l,$}}; 
\node at (0.5,-1.25) {\textcolor{red}{$s_l$}}; 
\node at (1.8,1.27) {\textcolor{red}{$h-c_1$}}; 
\node at (1.8,-1.23) {\textcolor{red}{$h-c_l$}}; 
\end{tikzpicture}
 }
 \end{matrix}
\cr
\stackrel{!}{=}&\,
 \bigg[\prod_{i=1}^l {2\,\zeta_p(1) \over |2|_p}\int_{\mathbb{Q}_p^2}\frac{dx_i}{|x_i|_p}\,|x_i|_p^{c_i+h-s_i \over 2} |1,x_i|^{c_i+s_i-h \over 2}_s\bigg] \left|1,x_1,\ldots,x_l\right|_s^{2h-\Delta-\sum_{i=1}^l(c_i+h) \over 2}\equiv \mathcal{D}\,,
}
where $\Delta\equiv\sum_{i=l+1}^L\Delta_i$ and the left-hand side of the equation is obtained using prescription~\ref{pres:padicMellin} in section~\ref{recursionSectionPadic} and we have introduced the symbol $\mathcal{D}$ as a convenient shorthand for the right-hand side. As a first step, we  change to variables $t_i \in \mathbb{Q}_p$ satisfying $x_i=t_i^2$. Rather than picking a specific root for each $x_i$, we can include both roots in the domain of $t_i$ at the cost of a factor of $2/|2|_p$ per variable. We get that
\eqn{recipint}
{
\mathcal{D}=\prod_{i=1}^l\bigg[\zeta_p(1)\int_{\mathbb{Q}_p}\frac{dt_i}{|t_i|_p}\,|t_i|_p^{h+c_i-s_i}|1,t_i|^{c_i+s_i-h}_s\bigg]\left|1,t_1,\ldots,t_l\right|_s^{2h-\Delta-\sum_{i=1}^l(c_i+h)}\,.
}
Starting from this expression one can verify \eno{toShowPadic} when $l=1$:
\eqn{}
{
{\cal D}\big|_{l=1} = \zeta_p(1)\int_{\mathbb{Q}_p}\frac{dt}{|t|_p}\,|t|_p^{h+c_1-s_1}|1,t|^{s_1-\Delta}_s & =\zeta_p(\Delta-c_1-h)-\zeta_p(s_1-c_1-h) \cr 
&= \begin{matrix}
\text{
\begin{tikzpicture}
\draw[red,thick,fill=red] (-0.8,0) ellipse (0.05cm and 0.05cm);
\draw[red,thick,fill=red] (0.6,0) ellipse (0.05cm and 0.05cm);
\draw[red,very thick] (2.0,0)--(-0.8,0);
\draw[red,very thick] (-0.8,0)--(-1.73,0.63);
\draw[red,very thick] (-0.8,0)--(-1.73,-0.63);
\node at (-0.1,-0.4) {\textcolor{red}{$h+c_1$}};
\node at (-0.1,0.3) {\textcolor{red}{$s_1$}}; 
\node at (1.3,-0.4) {\textcolor{red}{$h-c_1$}}; 
\node at (-1.6,0.1) {\textcolor{red}{$\vdots$}}; 
\node at (-1.5*1.4,-1.4*0.5) {\textcolor{red}{$\Delta_{2}$}};
\node at (-1.5*1.4,1.4*0.5) {\textcolor{red}{$\Delta_{L}$}};
\end{tikzpicture}
}
 \end{matrix}
\,,
}
where we refer the reader to \eno{supInt} for the second equality, and in the final equality we used the expression for the undressed diagram with a single internal leg, given in \eno{singleAmpUndressed}.

Having established the base case, we  now proceed to prove \eno{toShowPadic} inductively by showing that \eno{recipint} satisfies the same recursion relation as the undressed diagram in \eno{toShowPadic}. This can be achieved by partitioning the domain of the multi-dimensional integral in \eno{recipint} into the following sub-domains:
\begin{enumerate}
\item the part of the domain where $|t_1|_p,|t_2|_p,\ldots,|t_l|_p\leq 1$,
\item the part of the domain where $|t_1|_p=|t_2|_p=\cdots=|t_l|_p>1$, and finally
\item for each proper subset $\{u_1,u_2,\ldots,u_g\}\subset\{1,2,\ldots,l\}$, with complementary set labeled $\{u_{g+1},u_{g+1},\ldots,u_l\}$, the part of the domain where $|t_{u_1}|_p=|t_{u_2}|_p=\cdots=|t_{u_g}|_p > \{1,|t_{u_{g+1}}|_p,|t_{u_{g+2}}|_p,\ldots,|t_{u_l}|_p\}$. 
\end{enumerate}
For the first part of the domain, the integral in $\mathcal{D}$ evaluates to
\eqn{subDomain1}
{
\mathcal{D}\big|_{\text{sub-domain 1}}=&
 \prod_{i=1}^l\bigg[\zeta_p(1)\int_{\mathbb{Q}_p}\frac{dt_i}{|t_i|_p}\,|t_i|_p^{-s_i+h+c_i}\gamma_p(t_i)\bigg]
 =\prod_{i=1}^l\zeta_p\left(-s_i+h+c_i\right),
}
where $\gamma_p$ is the characteristic function of $\mathbb{Z}_{p}$, defined in \eno{characteristicfn} and we used identity \eno{Schwinger} to obtain the second equality.
The contribution to $\mathcal{D}$ coming from the second sub-domain evaluates to 
\eqn{subDomain2}
{\mathcal{D}\big|_{\text{sub-domain 2}} &=
 \zeta_p(1)\int_{\mathbb{Q}_p}\frac{dt}{|t|_p}|t|_p^{2h-\Delta-\sum_{i=1}^l(-c_i+h)}\gamma_p\left(\frac{1}{pt}\right) 
= -\zeta_p\big(2h-\Delta-\sum_{i=1}^l(-c_i+h)\big).
}
For the last type of sub-domain, the contribution  to $\mathcal{D}$ arising from the subset $\{u_1,...,u_g\}$ evaluates to 
\eqn{sD3}
{\mathcal{D}\big|_{ \substack{\{u_1,...,u_g\} \text{ part} \\ \text{of sub-domain 3}} } &=
 \prod_{i={g+1}}^l\left[\zeta_p(1)\int_{\mathbb{Q}_p}\frac{dt_{u_i}}{|t_{u_i}|_p}\,|t_{u_{i}}|_p^{-s_{u_i}+h+c_{u_i}}|1,t_{u_{i}}|^{s_{u_i}-h+c_{u_i}}_s\right]
\cr
& \quad \times \zeta_p(1)
 \int_{\mathbb{Q}_p}\frac{dt}{|t|_p}|t|_p^{2h-\Delta-\sum_{i=1}^g(h-c_{u_i})-\sum_{i=g+1}^l(h+c_{u_i})}
 \gamma_p\left(\frac{(1,t_{u_{g+1}},\ldots,t_{u_l})_s}{pt}\right).
}
By changing variable from $t$ to $u\equiv t/(1,t_{u_{g+1}},\ldots,t_{u_f})_s$, the $u$ integral factors out and can immediately be evaluated. One obtains 
\eqn{subDomain3}
{
\mathcal{D}\big|_{ \substack{\{u_1,...,u_g\} \text{ part} \\ \text{of sub-domain 3}} } =& 
-
\zeta_p\left(2h-\Delta-\sum_{i=1}^g(h-c_{u_i})-\sum_{i=g+1}^l(h+c_{u_i})\right)
\cr
&\times
\prod_{i={g+1}}^l\bigg[\zeta_p(1)\int_{\mathbb{Q}_p}\frac{dt_{u_i}}{|t_{u_i}|_p}\,|t_{u_{i}}|_p^{-s_{u_i}+h+c_{u_i}}|1,t_{u_{i}}|^{s_{u_i}-h+c_{u_i}}_s\bigg] \cr 
& \times |1,t_{u_{g+1}},\ldots,t_{u_l}|_s^{2h-\Delta-\sum_{i=1}^g(h-c_{u_i})-\sum_{i=g+1}^l(h+c_{u_i})}\,.
}
We notice that the above expression is equal to a local zeta function times an integral of the same form as the original integral \eqref{recipint} except with $l-g$ instead of $l$ internal legs and with $\Delta-\sum_{i=1}^g(h-c_{u_i})$ substituted for $\Delta$. Via an inductive reasoning, we assume that this remaining integral obeys \eno{toShowPadic}, that is it is equal to the undressed diagram obtained by collapsing legs indexed by $u_1$ to $u_g$ in the undressed diagram on the right-hand side of \eno{toShowPadic}, and now proceed to show that this implies the full diagram obeys \eno{toShowPadic} too.

In section~\ref{recursionSectionPadic} we defined undressed diagrams as satisfying  the recursion relation \eno{decomp1}. From the decomposition of $\mathcal{D}$ into sub-domains as described above, we have in fact recovered the same recursive structure except  we did not pick up explicit factors of $(-1)^{\rm \#internal\ lines}$ and the local zeta functions in 
 \eno{subDomain1}, \eno{subDomain2}, and \eno{subDomain3} appear with minus the argument with which they would have appeared  in the vertex factors and internal leg factors as prescribed by \eno{decomp1}. That is, compared to \eno{decomp1}, we find here
\eqn{decomp2}
{
\mathcal{D} &= \prod\text{(internal line factors w/ opposite sign arguments)}
\cr
&-\sum\text{(new vertex factors w/ opposite sign arguments)}\times
\text{(reduced undressed diagram)}
\,.
}
However, the two decompositions \eno{decomp1} and \eno{decomp2} are in fact equivalent, by virtue of the $p$-adic local zeta function identity
\eqn{}
{\zeta_p(x)+\zeta_p(-x)=1\,,
}
using which one can show that if one flips the sign of 1) the degree of extension $n$, 2) all the scaling dimensions $\Delta_i$, and 3) all the Mandelstam invariants $s_i$ inside the arguments of the local zeta functions, then any undressed diagram transform into itself times  a factor of $(-1)^{\text{\#internal legs}}$. The equivalence of \eno{decomp1} and \eno{decomp2} follows directly from this, completing the proof of \eno{toShowPadic}.

\subsection{Pre-amplitude examples}
\label{PreampExamples}

 In light of the identities for undressed diagrams derived in the previous subsection, the prescription for real and $p$-adic pre-amplitudes can be recast in a diagrammatic form that applies to both cases at once. 
In the contour integrals which follow, it will be understood that in the real case the contours run parallel to the imaginary axis, with limits from $-i\infty$ to $i\infty$, while in the $p$-adic case the contour wraps  once around the complex cylinder, so that the limits are from $-\frac{i\pi}{\log p}$ to $\frac{i\pi}{\log p}$.
With these conventions in place, let us consider some explicit examples for illustrative purposes.

\begin{example}
The exchange diagram Mellin amplitude and pre-amplitude take the following form:
\eqn{singlePreAmp}{
\begin{matrix}
\text{
\scalebox{0.75}{
\begin{tikzpicture}
\draw[thick] (0,0) ellipse (2.cm and 1.3cm);
\draw[thick,fill=black] (-0.8,0) ellipse (0.05cm and 0.05cm);
\draw[thick,fill=black] (0.8,0) ellipse (0.05cm and 0.05cm);
\draw[very thick] (0.8,0)--(-0.8,0);
\draw[very thick] (0.8,0)--(1.73,0.63);
\draw[very thick] (0.8,0)--(1.73,-0.63);
\draw[very thick] (-0.8,0)--(-1.73,0.63);
\draw[very thick] (-0.8,0)--(-1.73,-0.63);
\node at (0,0.3) {$s$}; 
\node at (0,-0.4) {$\Delta$}; 
\node at (1.6,0.1) {$\vdots$}; 
\node at (-1.6,0.1) {$\vdots$}; 
\node at (2.4,0) {$i_R$}; 
\node at (-2.35,0) {$i_L$}; 
\end{tikzpicture}}
}
\end{matrix}
=&
 \left(
 \begin{matrix}
\text{
\scalebox{0.75}{
\begin{tikzpicture}
\draw[blue,thick,fill=blue] (-0.8,0) ellipse (0.05cm and 0.05cm);
\draw[blue,very thick] (0.6,0)--(-0.8,0);
\draw[blue,very thick] (-0.8,0)--(-1.73,0.63);
\draw[blue,very thick] (-0.8,0)--(-1.73,-0.63);
\node at (-0.1,-0.4) {\textcolor{blue}{$\Delta$}}; 
\node at (-1.6,0.1) {\textcolor{blue}{$\vdots$}}; 
\node at (-2.1,0) {\textcolor{blue}{$i_L$}}; 
\end{tikzpicture}}
}
 \end{matrix}
 \right)
   \left(
 \begin{matrix}
\text{
\scalebox{0.75}{
\begin{tikzpicture}
\draw[blue,thick,fill=blue] (0.8,0) ellipse (0.05cm and 0.05cm);
\draw[blue,very thick] (-0.6,0)--(0.8,0);
\draw[blue,very thick] (0.8,0)--(1.73,0.63);
\draw[blue,very thick] (0.8,0)--(1.73,-0.63);
\node at (0.1,-0.4) {\textcolor{blue}{$\Delta$}}; 
\node at (1.6,0.1) {\textcolor{blue}{$\vdots$}}; 
\node at (2.1,0) {\textcolor{blue}{$i_R$}}; 
\end{tikzpicture}}
} 
\end{matrix}
 \right)
  \left(
 \begin{matrix}
\text{
\scalebox{0.75}{
\begin{tikzpicture}
\draw[red,thick,fill=red] (-0.8,0) ellipse (0.05cm and 0.05cm);
\draw[red,thick,fill=red] (0.8,0) ellipse (0.05cm and 0.05cm);
\draw[red,very thick] (-0.8,0)--(-1.73,0.63);
\draw[red,very thick] (-0.8,0)--(-1.73,-0.63);
\draw[red,very thick] (-0.8,0)--(0.8,0);
\draw[red,very thick] (0.8,0)--(1.73,0.63);
\draw[red,very thick] (0.8,0)--(1.73,-0.63);
\node at (-0.1,-0.4) {\textcolor{red}{$\Delta$}};
\node at (-0.1,0.3) {\textcolor{red}{$s$}}; 
\node at (-1.6,0.1) {\textcolor{red}{$\vdots$}}; 
\node at (-2.1,0) {\textcolor{red}{$i_L$}}; 
\node at (1.6,0.1) {\textcolor{red}{$\vdots$}}; 
\node at (2.1,0) {\textcolor{red}{$i_R$}}; 
\end{tikzpicture}}
}
 \end{matrix}
 \right)
 \cr
 =&
 \int\frac{dc}{2\pi i}\,f_\Delta(c)\,
 \left(
 \begin{matrix}
\text{
\scalebox{0.75}{
\begin{tikzpicture}
\draw[blue,thick,fill=blue] (-0.8,0) ellipse (0.05cm and 0.05cm);
\draw[blue,very thick] (0.6,0)--(-0.8,0);
\draw[blue,very thick] (-0.8,0)--(-1.73,0.63);
\draw[blue,very thick] (-0.8,0)--(-1.73,-0.63);
\node at (-0.1,-0.4) {\textcolor{blue}{$h+c$}}; 
\node at (-1.6,0.1) {\textcolor{blue}{$\vdots$}}; 
\node at (-2.1,0) {\textcolor{blue}{$i_L$}}; 
\end{tikzpicture}}
}
 \end{matrix}
 \right)
 \left(
 \begin{matrix}
\text{
\scalebox{0.75}{
\begin{tikzpicture}
\draw[red,thick,fill=red] (-0.8,0) ellipse (0.05cm and 0.05cm);
\draw[red,thick,fill=red] (0.6,0) ellipse (0.05cm and 0.05cm);
\draw[red,very thick] (2.0,0)--(-0.8,0);
\draw[red,very thick] (-0.8,0)--(-1.73,0.63);
\draw[red,very thick] (-0.8,0)--(-1.73,-0.63);
\node at (-0.1,-0.4) {\textcolor{red}{$h+c$}};
\node at (-0.1,0.3) {\textcolor{red}{$s$}}; 
\node at (1.3,-0.4) {\textcolor{red}{$h-c$}}; 
\node at (-1.6,0.1) {\textcolor{red}{$\vdots$}}; 
\node at (-2.1,0) {\textcolor{red}{$i_L$}}; 
\end{tikzpicture}}
}
 \end{matrix}
 \right)
 \cr
 &\hspace{19mm}
 \times
  \left(
 \begin{matrix}
\text{
\scalebox{0.75}{
\begin{tikzpicture}
\draw[blue,thick,fill=blue] (0.8,0) ellipse (0.05cm and 0.05cm);
\draw[blue,very thick] (-0.6,0)--(0.8,0);
\draw[blue,very thick] (0.8,0)--(1.73,0.63);
\draw[blue,very thick] (0.8,0)--(1.73,-0.63);
\node at (0.1,-0.4) {\textcolor{blue}{$h-c$}}; 
\node at (1.6,0.1) {\textcolor{blue}{$\vdots$}}; 
\node at (2.1,0) {\textcolor{blue}{$i_R$}}; 
\end{tikzpicture}}
} 
\end{matrix}
 \right)
   \left(
 \begin{matrix}
\text{
\scalebox{0.75}{
\begin{tikzpicture}
\draw[red,thick,fill=red] (0.8,0) ellipse (0.05cm and 0.05cm);
\draw[red,thick,fill=red] (-0.6,0) ellipse (0.05cm and 0.05cm);
\draw[red,very thick] (-2.0,0)--(0.8,0);
\draw[red,very thick] (0.8,0)--(1.73,0.63);
\draw[red,very thick] (0.8,0)--(1.73,-0.63);
\node at (0.1,-0.4) {\textcolor{red}{$h-c$}}; 
\node at (-1.3,-0.4) {\textcolor{red}{$h+c$}}; 
\node at (0.1,0.3) {\textcolor{red}{$s$}}; 
\node at (1.6,0.1) {\textcolor{red}{$\vdots$}}; 
\node at (2.1,0) {\textcolor{red}{$i_R$}}; 
\end{tikzpicture}}
}
 \end{matrix}
 \right),
 }
where we remind the reader that $f_\Delta(c)$ was defined in \eno{fDef}, the vertex factors are given by \eno{vertexFactorDef} over $p$-adics and \eno{vertexFactorDefreal} over reals, and the undressed diagrams are obtained from prescription \ref{pres:padicMellin} over $p$-adics and prescription \ref{pres:Feyn} over reals. It can be easily checked that this reproduces the pre-amplitudes in \eno{ExchPreamp}.
 \end{example}
 
 \clearpage
 
 \begin{example}
The Mellin amplitude and pre-amplitude for the diagram with two internal lines, \eno{doubleAmp}, are:
\eqn{doublePreAmp}{
& \begin{matrix}
\text{
\scalebox{0.75}{
\begin{tikzpicture}
\draw[thick] (0,0) ellipse (2.6cm and 1.4cm);
\draw[thick,fill=black] (0,0) ellipse (0.05cm and 0.05cm);
\draw[thick,fill=black] (-1.4,0) ellipse (0.05cm and 0.05cm);
\draw[thick,fill=black] (1.4,0) ellipse (0.05cm and 0.05cm);
\draw[very thick] (0,0)--(0.6,1.36);
\draw[very thick] (0,0)--(-0.6,1.36);
\node at (0,1.1) {...}; 
\node at (0,1.75) {$i_U$}; 
\draw[very thick] (1.4,0)--(-1.4,0);
\draw[very thick] (1.4,0)--(2.33,0.63);
\draw[very thick] (1.4,0)--(2.33,-0.63);
\draw[very thick] (-1.4,0)--(-2.33,0.63);
\draw[very thick] (-1.4,0)--(-2.33,-0.63);
\node at (-0.8,0.3) {$s_A$}; 
\node at (-0.8,-0.4) {$\Delta_A$};
\node at (0.8,0.3) {$s_B$}; 
\node at (0.8,-0.4) {$\Delta_B$}; 
\node at (2.2,0.1) {$\vdots$}; 
\node at (-2.2,0.1) {$\vdots$}; 
\node at (3.,0) {$i_R$}; 
\node at (-2.95,0) {$i_L$}; 
\end{tikzpicture}}
}
\end{matrix}
=  
  \left(
 \begin{matrix}
\text{
\scalebox{0.75}{
\begin{tikzpicture}
\draw[blue,thick,fill=blue] (-0.8,0) ellipse (0.05cm and 0.05cm);
\draw[blue,very thick] (0.6,0)--(-0.8,0);
\draw[blue,very thick] (-0.8,0)--(-1.73,0.63);
\draw[blue,very thick] (-0.8,0)--(-1.73,-0.63);
\node at (-0.1,-0.4) {\textcolor{blue}{$\Delta_A$}}; 
\node at (-1.6,0.1) {\textcolor{blue}{$\vdots$}}; 
\node at (-2.1,0) {\textcolor{blue}{$i_L$}}; 
\end{tikzpicture}}
}
 \end{matrix}
 \right)
 \left(
 \begin{matrix}
\text{
\scalebox{0.75}{
\begin{tikzpicture}
\draw[blue,thick,fill=blue] (0,0) ellipse (0.05cm and 0.05cm);
\draw[blue,very thick] (0,0)--(0.6,1.36);
\draw[blue,very thick] (0,0)--(-0.6,1.36);
\node at (0,1.1) {\textcolor{blue}{...}}; 
\node at (0,1.65) {\textcolor{blue}{$i_U$}}; 
\draw[blue,very thick] (1.4,0)--(-1.4,0);
\node at (-0.8,-0.4) {\textcolor{blue}{$\Delta_A$}};
\node at (0.85,-0.4) {\textcolor{blue}{$\Delta_B$}}; 
\end{tikzpicture}}
}
 \end{matrix}
 \right) 
   \left(
 \begin{matrix}
\text{
\scalebox{0.75}{
\begin{tikzpicture}
\draw[blue,thick,fill=blue] (0.8,0) ellipse (0.05cm and 0.05cm);
\draw[blue,very thick] (-0.6,0)--(0.8,0);
\draw[blue,very thick] (0.8,0)--(1.73,0.63);
\draw[blue,very thick] (0.8,0)--(1.73,-0.63);
\node at (0.1,-0.4) {\textcolor{blue}{$\Delta_B$}}; 
\node at (1.6,0.1) {\textcolor{blue}{$\vdots$}}; 
\node at (2.1,0) {\textcolor{blue}{$i_R$}}; 
\end{tikzpicture}}
} 
\end{matrix}
 \right)
 \cr 
 & \hspace{57mm} \times 
\left(
 \begin{matrix}
\text{
\scalebox{0.75}{
\begin{tikzpicture}
\draw[red,very thick] (-0.8-0.6,0)--(-1.73-0.6,0.63);
\draw[red,very thick] (-0.8-0.6,0)--(-1.73-0.6,-0.63);
\draw[red,very thick] (0.8+0.6,0)--(1.73+0.6,0.63);
\draw[red,very thick] (0.8+0.6,0)--(1.73+0.6,-0.63);
\node at (-1.6-0.6,0.1) {\textcolor{red}{$\vdots$}}; 
\node at (-2.1-0.6,0) {\textcolor{red}{$i_L$}}; 
\node at (1.6+0.6,0.1) {\textcolor{red}{$\vdots$}}; 
\node at (2.1+0.6,0) {\textcolor{red}{$i_R$}}; 
\draw[red,thick,fill=red] (0,0) ellipse (0.05cm and 0.05cm);
\draw[red,thick,fill=red] (1.4,0) ellipse (0.05cm and 0.05cm);
\draw[red,thick,fill=red] (-1.4,0) ellipse (0.05cm and 0.05cm);
\draw[red,very thick] (0,0)--(0.6,1.36);
\draw[red,very thick] (0,0)--(-0.6,1.36);
\node at (0,1.1) {\textcolor{red}{...}}; 
\node at (0,1.65) {\textcolor{red}{$i_U$}}; 
\draw[red,very thick] (1.4,0)--(-1.4,0);
\node at (-0.75,-0.4) {\textcolor{red}{$\Delta_A$}};
\node at (0.85,-0.4) {\textcolor{red}{$\Delta_B$}}; 
\node at (-0.75,0.3) {\textcolor{red}{$s_A$}};
\node at (0.85,0.3) {\textcolor{red}{$s_B$}}; 
\end{tikzpicture}}
}
 \end{matrix}
 \right)
 \cr
 & =\int\frac{dc_A}{2\pi i} \,f_{\Delta_A}(c_A)\,\int\frac{dc_B}{2\pi i} \,f_{\Delta_B}(c_B)
  \left(
 \begin{matrix}
\text{
\scalebox{0.75}{
\begin{tikzpicture}
\draw[blue,thick,fill=blue] (-0.8,0) ellipse (0.05cm and 0.05cm);
\draw[blue,very thick] (0.6,0)--(-0.8,0);
\draw[blue,very thick] (-0.8,0)--(-1.73,0.63);
\draw[blue,very thick] (-0.8,0)--(-1.73,-0.63);
\node at (-0.1,-0.4) {\textcolor{blue}{$h-c_A$}}; 
\node at (-1.6,0.1) {\textcolor{blue}{$\vdots$}}; 
\node at (-2.1,0) {\textcolor{blue}{$i_L$}}; 
\end{tikzpicture}}
}
 \end{matrix}
 \right)
 \left(
 \begin{matrix}
\text{
\scalebox{0.75}{
\begin{tikzpicture}
\draw[red,thick,fill=red] (-0.8,0) ellipse (0.05cm and 0.05cm);
\draw[red,thick,fill=red] (0.6,0) ellipse (0.05cm and 0.05cm);
\draw[red,very thick] (2.0,0)--(-0.8,0);
\draw[red,very thick] (-0.8,0)--(-1.73,0.63);
\draw[red,very thick] (-0.8,0)--(-1.73,-0.63);
\node at (-0.1,-0.4) {\textcolor{red}{$h-c_A$}};
\node at (-0.1,0.3) {\textcolor{red}{$s_A$}}; 
\node at (1.3,-0.4) {\textcolor{red}{$h+c_A$}}; 
\node at (-1.6,0.1) {\textcolor{red}{$\vdots$}}; 
\node at (-2.1,0) {\textcolor{red}{$i_L$}}; 
\end{tikzpicture}}
}
 \end{matrix}
 \right)
 \cr
 & \hspace{57mm}  \times
\left(
 \begin{matrix}
\text{
\scalebox{0.75}{
\begin{tikzpicture}
\draw[blue,thick,fill=blue] (0,0) ellipse (0.05cm and 0.05cm);
\draw[blue,very thick] (0,0)--(0.6,1.36);
\draw[blue,very thick] (0,0)--(-0.6,1.36);
\node at (0,1.1) {\textcolor{blue}{...}}; 
\node at (0,1.65) {\textcolor{blue}{$i_U$}}; 
\draw[blue,very thick] (1.4,0)--(-1.4,0);
\node at (-0.8,-0.4) {\textcolor{blue}{$h+c_A$}};
\node at (0.85,-0.4) {\textcolor{blue}{$h-c_B$}}; 
\end{tikzpicture}}
}
 \end{matrix}
 \right)
   \left(
 \begin{matrix}
\text{
\scalebox{0.75}{
\begin{tikzpicture}
\draw[red,thick,fill=red] (0,0) ellipse (0.05cm and 0.05cm);
\draw[red,thick,fill=red] (1.4,0) ellipse (0.05cm and 0.05cm);
\draw[red,thick,fill=red] (-1.4,0) ellipse (0.05cm and 0.05cm);
\draw[red,very thick] (0,0)--(0.6,1.36);
\draw[red,very thick] (0,0)--(-0.6,1.36);
\node at (0,1.1) {\textcolor{red}{...}}; 
\node at (0,1.65) {\textcolor{red}{$i_U$}}; 
\draw[red,very thick] (2.8,0)--(-2.8,0);
\node at (-0.75,-0.4) {\textcolor{red}{$h+c_A$}};
\node at (0.85,-0.4) {\textcolor{red}{$h-c_B$}}; 
\node at (-0.8,0.4) {\textcolor{red}{$s_A$}};
\node at (0.85,0.4) {\textcolor{red}{$s_B$}}; 
\node at (-2.2,-0.4) {\textcolor{red}{$h-c_A$}};
\node at (2.25,-0.4) {\textcolor{red}{$h+c_B$}}; 
\end{tikzpicture}}
} 
\end{matrix}
 \right)
\cr
& \hspace{57mm} \times
  \left(
 \begin{matrix}
\text{
\scalebox{0.75}{
\begin{tikzpicture}
\draw[blue,thick,fill=blue] (0.8,0) ellipse (0.05cm and 0.05cm);
\draw[blue,very thick] (-0.6,0)--(0.8,0);
\draw[blue,very thick] (0.8,0)--(1.73,0.63);
\draw[blue,very thick] (0.8,0)--(1.73,-0.63);
\node at (0.1,-0.4) {\textcolor{blue}{$h+c_B$}}; 
\node at (1.6,0.1) {\textcolor{blue}{$\vdots$}}; 
\node at (2.1,0) {\textcolor{blue}{$i_R$}}; 
\end{tikzpicture}}
} 
\end{matrix}
 \right)
   \left(
 \begin{matrix}
\text{
\scalebox{0.75}{
\begin{tikzpicture}
\draw[red,thick,fill=red] (0.8,0) ellipse (0.05cm and 0.05cm);
\draw[red,thick,fill=red] (-0.6,0) ellipse (0.05cm and 0.05cm);
\draw[red,very thick] (-2.0,0)--(0.8,0);
\draw[red,very thick] (0.8,0)--(1.73,0.63);
\draw[red,very thick] (0.8,0)--(1.73,-0.63);
\node at (0.1,-0.4) {\textcolor{red}{$h+c_B$}}; 
\node at (-1.3,-0.4) {\textcolor{red}{$h-c_B$}}; 
\node at (0.1,0.3) {\textcolor{red}{$s_B$}}; 
\node at (1.6,0.1) {\textcolor{red}{$\vdots$}}; 
\node at (2.1,0) {\textcolor{red}{$i_R$}}; 
\end{tikzpicture}}
}
 \end{matrix}
 \right).
 }
 \end{example} 

\clearpage

\begin{example}
The Mellin amplitude for the diagram with three internal lines arranged in series, \eno{tripleLineAmp}  decomposes into a pre-amplitude thus:
 
\eqn{tripleLinePreAmp}{
&
\hspace{5mm}
\begin{matrix}
\text{
\scalebox{0.75}{
\begin{tikzpicture}
\draw[thick] (0,0) ellipse (3.35cm and 1.45cm);
\node at (0,0.3) {$s_B$}; 
\node at (0,-0.4) {$\Delta_B$};
\draw[thick,fill=black] (-0.8,0) ellipse (0.05cm and 0.05cm);
\draw[thick,fill=black] (0.8,0) ellipse (0.05cm and 0.05cm);
\draw[thick,fill=black] (-2.1,0) ellipse (0.05cm and 0.05cm);
\draw[thick,fill=black] (2.1,0) ellipse (0.05cm and 0.05cm);
\draw[very thick] (-0.8,0)--(-1.4,1.32);
\draw[very thick] (-0.8,0)--(-0.2,1.44);
\node at (-0.8,1.1) {...}; 
\node at (-0.8,1.75) {$i_l$};
\draw[very thick] (0.8,0)--(1.4,1.32);
\draw[very thick] (0.8,0)--(0.2,1.44);
\node at (0.8,1.1) {...}; 
\node at (0.8,1.75) {$i_r$}; 
\draw[very thick] (2.1,0)--(-2.1,0);
\draw[very thick] (2.1,0)--(3.03,0.63);
\draw[very thick] (2.1,0)--(3.03,-0.63);
\draw[very thick] (-2.1,0)--(-3.03,0.63);
\draw[very thick] (-2.1,0)--(-3.03,-0.63);
\node at (-1.5,0.3) {$s_A$}; 
\node at (-1.5,-0.4) {$\Delta_A$};
\node at (1.5,0.3) {$s_C$}; 
\node at (1.5,-0.4) {$\Delta_C$}; 
\node at (2.9,0.1) {$\vdots$}; 
\node at (-2.9,0.1) {$\vdots$}; 
\node at (3.7,0) {$i_R$}; 
\node at (-3.65,0) {$i_L$}; 
\end{tikzpicture}}
}
 \end{matrix}
\cr
 =
\,
&
 \int\frac{dc_A}{2\pi i} \,f_{\Delta_A}(c_A)\,\int\frac{dc_B}{2\pi i} \,f_{\Delta_B}(c_B)\,
 \int\frac{dc_C}{2\pi i} \,f_{\Delta_B}(c_C)
 \cr
&
\times
  \left(
 \begin{matrix}
\text{
\scalebox{0.75}{
\begin{tikzpicture}
\draw[blue,thick,fill=blue] (-0.8,0) ellipse (0.05cm and 0.05cm);
\draw[blue,very thick] (0.6,0)--(-0.8,0);
\draw[blue,very thick] (-0.8,0)--(-1.73,0.63);
\draw[blue,very thick] (-0.8,0)--(-1.73,-0.63);
\node at (-0.1,-0.4) {\textcolor{blue}{$h-c_A$}}; 
\node at (-1.6,0.1) {\textcolor{blue}{$\vdots$}}; 
\node at (-2.1,0) {\textcolor{blue}{$i_L$}}; 
\end{tikzpicture}}
}
 \end{matrix}
 \right)
 \left(
 \begin{matrix}
\text{
\scalebox{0.75}{
\begin{tikzpicture}
\draw[red,thick,fill=red] (-0.8,0) ellipse (0.05cm and 0.05cm);
\draw[red,thick,fill=red] (0.6,0) ellipse (0.05cm and 0.05cm);
\draw[red,very thick] (2.0,0)--(-0.8,0);
\draw[red,very thick] (-0.8,0)--(-1.73,0.63);
\draw[red,very thick] (-0.8,0)--(-1.73,-0.63);
\node at (-0.1,-0.4) {\textcolor{red}{$h-c_A$}};
\node at (-0.1,0.3) {\textcolor{red}{$s_A$}}; 
\node at (1.3,-0.4) {\textcolor{red}{$h+c_A$}}; 
\node at (-1.6,0.1) {\textcolor{red}{$\vdots$}}; 
\node at (-2.1,0) {\textcolor{red}{$i_L$}}; 
\end{tikzpicture}}
}
 \end{matrix}
 \right)
 \cr
 &
 \times
\left(
 \begin{matrix}
\text{
\scalebox{0.75}{
\begin{tikzpicture}
\draw[blue,thick,fill=blue] (0,0) ellipse (0.05cm and 0.05cm);
\draw[blue,very thick] (0,0)--(0.6,1.36);
\draw[blue,very thick] (0,0)--(-0.6,1.36);
\node at (0,1.1) {\textcolor{blue}{...}}; 
\node at (0,1.65) {\textcolor{blue}{$i_l$}}; 
\draw[blue,very thick] (1.4,0)--(-1.4,0);
\node at (-0.8,-0.4) {\textcolor{blue}{$h+c_A$}};
\node at (0.85,-0.4) {\textcolor{blue}{$h-c_B$}}; 
\end{tikzpicture}}
}
 \end{matrix}
 \right)
   \left(
 \begin{matrix}
\text{
\scalebox{0.75}{
\begin{tikzpicture}
\draw[red,thick,fill=red] (0,0) ellipse (0.05cm and 0.05cm);
\draw[red,thick,fill=red] (1.4,0) ellipse (0.05cm and 0.05cm);
\draw[red,thick,fill=red] (-1.4,0) ellipse (0.05cm and 0.05cm);
\draw[red,very thick] (0,0)--(0.6,1.36);
\draw[red,very thick] (0,0)--(-0.6,1.36);
\node at (0,1.1) {\textcolor{red}{...}}; 
\node at (0,1.65) {\textcolor{red}{$i_l$}}; 
\draw[red,very thick] (2.8,0)--(-2.8,0);
\node at (-0.75,-0.4) {\textcolor{red}{$h+c_A$}};
\node at (0.85,-0.4) {\textcolor{red}{$h-c_B$}}; 
\node at (-0.8,0.4) {\textcolor{red}{$s_A$}};
\node at (0.85,0.4) {\textcolor{red}{$s_B$}}; 
\node at (-2.2,-0.4) {\textcolor{red}{$h-c_A$}};
\node at (2.25,-0.4) {\textcolor{red}{$h+c_B$}}; 
\end{tikzpicture}}
} 
\end{matrix}
 \right)
\cr
&
 \times
\left(
 \begin{matrix}
\text{
\scalebox{0.75}{
\begin{tikzpicture}
\draw[blue,thick,fill=blue] (0,0) ellipse (0.05cm and 0.05cm);
\draw[blue,very thick] (0,0)--(0.6,1.36);
\draw[blue,very thick] (0,0)--(-0.6,1.36);
\node at (0,1.1) {\textcolor{blue}{...}}; 
\node at (0,1.65) {\textcolor{blue}{$i_r$}}; 
\draw[blue,very thick] (1.4,0)--(-1.4,0);
\node at (-0.8,-0.4) {\textcolor{blue}{$h+c_B$}};
\node at (0.85,-0.4) {\textcolor{blue}{$h-c_C$}}; 
\end{tikzpicture}}
}
 \end{matrix}
 \right)
   \left(
 \begin{matrix}
\text{
\scalebox{0.75}{
\begin{tikzpicture}
\draw[red,thick,fill=red] (0,0) ellipse (0.05cm and 0.05cm);
\draw[red,thick,fill=red] (1.4,0) ellipse (0.05cm and 0.05cm);
\draw[red,thick,fill=red] (-1.4,0) ellipse (0.05cm and 0.05cm);
\draw[red,very thick] (0,0)--(0.6,1.36);
\draw[red,very thick] (0,0)--(-0.6,1.36);
\node at (0,1.1) {\textcolor{red}{...}}; 
\node at (0,1.65) {\textcolor{red}{$i_r$}}; 
\draw[red,very thick] (2.8,0)--(-2.8,0);
\node at (-0.75,-0.4) {\textcolor{red}{$h+c_B$}};
\node at (0.85,-0.4) {\textcolor{red}{$h-c_C$}}; 
\node at (-0.8,0.4) {\textcolor{red}{$s_B$}};
\node at (0.85,0.4) {\textcolor{red}{$s_C$}}; 
\node at (-2.2,-0.4) {\textcolor{red}{$h-c_A$}};
\node at (2.25,-0.4) {\textcolor{red}{$h+c_B$}}; 
\end{tikzpicture}}
} 
\end{matrix}
 \right)
 \cr
&
 \times
  \left(
 \begin{matrix}
\text{
\scalebox{0.75}{
\begin{tikzpicture}
\draw[blue,thick,fill=blue] (0.8,0) ellipse (0.05cm and 0.05cm);
\draw[blue,very thick] (-0.6,0)--(0.8,0);
\draw[blue,very thick] (0.8,0)--(1.73,0.63);
\draw[blue,very thick] (0.8,0)--(1.73,-0.63);
\node at (0.1,-0.4) {\textcolor{blue}{$h+c_C$}}; 
\node at (1.6,0.1) {\textcolor{blue}{$\vdots$}}; 
\node at (2.1,0) {\textcolor{blue}{$i_R$}}; 
\end{tikzpicture}}
} 
\end{matrix}
 \right)
   \left(
 \begin{matrix}
\text{
\scalebox{0.75}{
\begin{tikzpicture}
\draw[red,thick,fill=red] (0.8,0) ellipse (0.05cm and 0.05cm);
\draw[red,thick,fill=red] (-0.6,0) ellipse (0.05cm and 0.05cm);
\draw[red,very thick] (-2.0,0)--(0.8,0);
\draw[red,very thick] (0.8,0)--(1.73,0.63);
\draw[red,very thick] (0.8,0)--(1.73,-0.63);
\node at (0.1,-0.4) {\textcolor{red}{$h+c_C$}}; 
\node at (-1.3,-0.4) {\textcolor{red}{$h-c_C$}}; 
\node at (0.1,0.3) {\textcolor{red}{$s_C$}}; 
\node at (1.6,0.1) {\textcolor{red}{$\vdots$}}; 
\node at (2.1,0) {\textcolor{red}{$i_R$}}; 
\end{tikzpicture}}
}
 \end{matrix}
 \right).
 }
\end{example} 
\hspace{1mm}
\\
\\
\\
\begin{example}
As a final example, consider the Mellin amplitude for the diagram with three internal lines meeting at a center vertex, \eno{starAmp}. It exhibits the following decomposition into a contour integral over its pre-amplitude:

\clearpage

\eqn{starPreAmp}{
&
\hspace{5mm}
\begin{matrix}
\text{
\scalebox{0.75}{
\begin{tikzpicture}
\draw[thick] (0,-0.7) ellipse (2.8cm and 2.1cm);
\draw[thick,fill=black] (0,0) ellipse (0.05cm and 0.05cm);
\draw[thick,fill=black] (-1.4,0) ellipse (0.05cm and 0.05cm);
\draw[thick,fill=black] (1.4,0) ellipse (0.05cm and 0.05cm);
\draw[thick,fill=black] (0,-1.4) ellipse (0.05cm and 0.05cm);
\draw[very thick] (0,0)--(0,-1.4);
\draw[very thick] (0,0)--(0.6,1.36);
\draw[very thick] (0,0)--(-0.6,1.36);
\draw[very thick] (0,-1.4)--(0.6,-2.76);
\draw[very thick] (0,-1.4)--(-0.6,-2.76);
\node at (-0.5,-1.1) {$s_C$}; 
\node at (0.5,-1.1) {$\Delta_C$}; 
\node at (0,1.1) {...}; 
\node at (0,1.75) {$i_U$}; 
\node at (0,-2.5) {...}; 
\node at (0,-3.15) {$i_D$}; 
\draw[very thick] (1.4,0)--(-1.4,0);
\draw[very thick] (1.4,0)--(2.33,0.46);
\draw[very thick] (1.4,0)--(2.8,-0.63);
\draw[very thick] (-1.4,0)--(-2.33,0.46);
\draw[very thick] (-1.4,0)--(-2.8,-0.63);
\node at (-0.8,0.3) {$s_A$}; 
\node at (-0.8,-0.4) {$\Delta_A$};
\node at (0.8,0.3) {$s_B$}; 
\node at (0.8,-0.4) {$\Delta_B$}; 
\node at (2.2,0.1) {$\vdots$}; 
\node at (-2.2,0.1) {$\vdots$}; 
\node at (3.1,0) {$i_R$}; 
\node at (-3.05,0) {$i_L$}; 
\end{tikzpicture}}
}
\end{matrix}
 \cr
 =\,
 &
 \int\frac{dc_A}{2\pi i}\,f_{\Delta_A}(c_A)\, \int\frac{dc_B}{2\pi i}\,f_{\Delta_B}(c_B)\, \int\frac{dc_C}{2\pi i}\,f_{\Delta_C}(c_C)\,
 \cr
 &
 \times
  \left(
 \begin{matrix}
\text{
\scalebox{0.75}{
\begin{tikzpicture}
\draw[blue,thick,fill=blue] (-0.8,0) ellipse (0.05cm and 0.05cm);
\draw[blue,very thick] (0.6,0)--(-0.8,0);
\draw[blue,very thick] (-0.8,0)--(-1.73,0.63);
\draw[blue,very thick] (-0.8,0)--(-1.73,-0.63);
\node at (-0.1,-0.4) {\textcolor{blue}{$h-c_A$}}; 
\node at (-1.6,0.1) {\textcolor{blue}{$\vdots$}}; 
\node at (-2.1,0) {\textcolor{blue}{$i_L$}}; 
\end{tikzpicture}}
}
 \end{matrix}
 \right)
 \left(
 \begin{matrix}
\text{
\scalebox{0.75}{
\begin{tikzpicture}
\draw[red,thick,fill=red] (-0.8,0) ellipse (0.05cm and 0.05cm);
\draw[red,thick,fill=red] (0.6,0) ellipse (0.05cm and 0.05cm);
\draw[red,very thick] (2.0,0)--(-0.8,0);
\draw[red,very thick] (-0.8,0)--(-1.73,0.63);
\draw[red,very thick] (-0.8,0)--(-1.73,-0.63);
\node at (-0.1,-0.4) {\textcolor{red}{$h-c_A$}};
\node at (-0.1,0.3) {\textcolor{red}{$s_A$}}; 
\node at (1.3,-0.4) {\textcolor{red}{$h+c_A$}}; 
\node at (-1.6,0.1) {\textcolor{red}{$\vdots$}}; 
\node at (-2.1,0) {\textcolor{red}{$i_L$}}; 
\end{tikzpicture}}
}
 \end{matrix}
 \right)
 \cr
 &
 \times
  \left(
 \begin{matrix}
\text{
\scalebox{0.75}{
\begin{tikzpicture}
\draw[blue,thick,fill=blue] (0.8,0) ellipse (0.05cm and 0.05cm);
\draw[blue,very thick] (-0.6,0)--(0.8,0);
\draw[blue,very thick] (0.8,0)--(1.73,0.63);
\draw[blue,very thick] (0.8,0)--(1.73,-0.63);
\node at (0.1,-0.4) {\textcolor{blue}{$h+c_B$}}; 
\node at (1.6,0.1) {\textcolor{blue}{$\vdots$}}; 
\node at (2.1,0) {\textcolor{blue}{$i_R$}}; 
\end{tikzpicture}}
} 
\end{matrix}
 \right)
   \left(
 \begin{matrix}
\text{
\scalebox{0.75}{
\begin{tikzpicture}
\draw[red,thick,fill=red] (0.8,0) ellipse (0.05cm and 0.05cm);
\draw[red,thick,fill=red] (-0.6,0) ellipse (0.05cm and 0.05cm);
\draw[red,very thick] (-2.0,0)--(0.8,0);
\draw[red,very thick] (0.8,0)--(1.73,0.63);
\draw[red,very thick] (0.8,0)--(1.73,-0.63);
\node at (0.1,-0.4) {\textcolor{red}{$h+c_B$}}; 
\node at (-1.3,-0.4) {\textcolor{red}{$h-c_B$}}; 
\node at (0.1,0.3) {\textcolor{red}{$s_B$}}; 
\node at (1.6,0.1) {\textcolor{red}{$\vdots$}}; 
\node at (2.1,0) {\textcolor{red}{$i_R$}}; 
\end{tikzpicture}}
}
 \end{matrix}
 \right)
 \cr
& 
\times
\left(
\begin{matrix}
\text{
\scalebox{0.75}{
 \begin{tikzpicture}
\draw[blue,thick,fill=blue] (0,0) ellipse (0.05cm and 0.05cm);
\draw[blue,very thick] (0,0)--(0,-1.4);
\draw[blue,very thick] (0,0)--(0.6,1.36);
\draw[blue,very thick] (0,0)--(-0.6,1.36);
\node at (0.8,-1.1) {\textcolor{blue}{$h+c_C$}}; 
\node at (0,1.1) {\textcolor{blue}{...}}; 
\node at (0,1.75) {\textcolor{blue}{$i_U$}}; 
\draw[blue,very thick] (1.4,0)--(-1.4,0);
\node at (-0.8,-0.4) {\textcolor{blue}{$h+c_A$}};
\node at (0.8,-0.4) {\textcolor{blue}{$h-c_B$}}; 
\end{tikzpicture}}
}
\end{matrix}
\right)
\left(
\begin{matrix}
\text{
\scalebox{0.75}{
\begin{tikzpicture}
\draw[red,thick,fill=red] (0,0) ellipse (0.05cm and 0.05cm);
\draw[red,thick,fill=red] (-1.4,0) ellipse (0.05cm and 0.05cm);
\draw[red,thick,fill=red] (1.4,0) ellipse (0.05cm and 0.05cm);
\draw[red,thick,fill=red] (0,-1.5) ellipse (0.05cm and 0.05cm);
\draw[red,very thick] (0,0)--(0,-2.4);
\draw[red,very thick] (0,0)--(0.6,1.36);
\draw[red,very thick] (0,0)--(-0.6,1.36);
\node at (-0.5,-1.1) {\textcolor{red}{$s_C$}}; 
\node at (0.8,-1.1) {\textcolor{red}{$h+c_C$}}; 
\node at (0.8,-2.) {\textcolor{red}{$h-c_C$}}; 
\node at (0,1.1) {\textcolor{red}{...}}; 
\node at (0,1.75) {\textcolor{red}{$i_U$}}; 
\draw[red,very thick] (2.8,0)--(-2.8,0);
\node at (-0.8,0.3) {\textcolor{red}{$s_A$}}; 
\node at (-0.8,-0.4) {\textcolor{red}{$h+c_A$}};
\node at (-2.2,-0.4) {\textcolor{red}{$h-c_A$}};
\node at (0.8,0.3) {\textcolor{red}{$s_B$}}; 
\node at (0.8,-0.4) {\textcolor{red}{$h-c_B$}}; 
\node at (2.2,-0.4) {\textcolor{red}{$h+c_B$}}; 
\end{tikzpicture}}
}
\end{matrix}
\right)
 \cr
 &
 \times
 \left(
 \begin{matrix}
 \text{
 \scalebox{0.75}{
 \begin{tikzpicture}
\draw[blue,thick,fill=blue] (0,-1.4) ellipse (0.05cm and 0.05cm);
\draw[blue,very thick] (0,0)--(0,-1.4);
\draw[blue,very thick] (0,-1.4)--(0.6,-2.76);
\draw[blue,very thick] (0,-1.4)--(-0.6,-2.76);
\node at (0.8,-0.7) {\textcolor{blue}{$h-c_C$}}; 
\node at (0,-2.5) {\textcolor{blue}{...}}; 
\node at (0,-3.15) {\textcolor{blue}{$i_D$}}; 
\end{tikzpicture}}
 }
 \end{matrix}
 \right)
 \left(
 \begin{matrix}
 \text{
 \scalebox{0.75}{
 \begin{tikzpicture}
\draw[red,thick,fill=red] (0,0) ellipse (0.05cm and 0.05cm);
\draw[red,thick,fill=red] (0,-1.4) ellipse (0.05cm and 0.05cm);
\draw[red,very thick] (0,1.4)--(0,-1.4);
\draw[red,very thick] (0,-1.4)--(0.6,-2.76);
\draw[red,very thick] (0,-1.4)--(-0.6,-2.76);
\node at (-0.5,-0.7) {\textcolor{red}{$s_C$}}; 
\node at (0.8,-0.7) {\textcolor{red}{$h-c_C$}}; 
\node at (0.8,0.7) {\textcolor{red}{$h+c_C$}}; 
\node at (0,-2.5) {\textcolor{red}{...}}; 
\node at (0,-3.15) {\textcolor{red}{$i_D$}}; 
\end{tikzpicture}}
 }
 \end{matrix}
 \right) .
 }
\end{example}

\subsection{Integrating the pre-amplitude to obtain the amplitude}
\label{sec:intPreamp}

In section \ref{recursionSection} we showed how to write a Mellin amplitude as the product of vertex factors times a single undressed amplitude.
In sections~\ref{PreampRecursion}-\ref{PreampExamples} we showed how to obtain the pre-amplitude by multiplying together one vertex factor and one undressed amplitude with complex dimensions for each vertex in the bulk diagram. In this subsection we explain the diagrammatic interpretation of performing  contour integrals over the pre-amplitude to obtain the full Mellin amplitude: for each integral that is carried out, two undressed diagrams merge and become one, and after carrying out all the contour integrals, all undressed diagrams have merged into the single undressed amplitude.

As an example of how this, consider the Mellin amplitude $\mathcal{M}^{\text{3--int}}$ for the diagram with three internal legs arranged in series as given in equation \eno{tripleLinePreAmp}. Carrying out the $c_A$ integral gives the following: 
\eqn{}{
\mathcal{M}^{\text{3--int}}
\equiv &
\,
 \int \frac{dc_B}{2\pi i}\,f_{\Delta_B}(c_B)
 \int \frac{dc_C}{2\pi i}\,f_{\Delta_C}(c_C)
 \cr
 &
 \times
\left(
 \begin{matrix}
\text{
  \scalebox{0.75}{
\begin{tikzpicture}
\draw[blue,thick,fill=blue] (-0.8,0) ellipse (0.05cm and 0.05cm);
\draw[blue,very thick] (0.6,0)--(-0.8,0);
\draw[blue,very thick] (-0.8,0)--(-1.73,0.63);
\draw[blue,very thick] (-0.8,0)--(-1.73,-0.63);
\node at (-0.1,-0.4) {\textcolor{blue}{$\Delta_A$}}; 
\node at (-1.6,0.1) {\textcolor{blue}{$\vdots$}}; 
\node at (-2.1,0) {\textcolor{blue}{$i_L$}}; 
\end{tikzpicture}}
}
 \end{matrix}
 \right) 
 \left(
 \begin{matrix}
\text{
  \scalebox{0.75}{
\begin{tikzpicture}
\draw[red,very thick] (-0.8-0.6,0)--(-1.73-0.6,0.63);
\draw[red,very thick] (-0.8-0.6,0)--(-1.73-0.6,-0.63);
\node at (-1.6-0.6,0.1) {\textcolor{red}{$\vdots$}}; 
\node at (-2.1-0.6,0) {\textcolor{red}{$i_L$}}; 
\draw[red,thick,fill=red] (0,0) ellipse (0.05cm and 0.05cm);
\draw[red,thick,fill=red] (1.4,0) ellipse (0.05cm and 0.05cm);
\draw[red,thick,fill=red] (-1.4,0) ellipse (0.05cm and 0.05cm);
\draw[red,very thick] (0,0)--(0.6,1.36);
\draw[red,very thick] (0,0)--(-0.6,1.36);
\node at (0,1.1) {\textcolor{red}{...}}; 
\node at (0,1.65) {\textcolor{red}{$i_l$}}; 
\draw[red,very thick] (2.8,0)--(-1.4,0);
\node at (-0.75,-0.4) {\textcolor{red}{$\Delta_A$}};
\node at (0.85,-0.4) {\textcolor{red}{$h-c_B$}}; 
\node at (2.25,-0.4) {\textcolor{red}{$h+c_B$}}; 
\end{tikzpicture}}
}
 \end{matrix}
 \right)
 \left(
 \begin{matrix}
\text{
  \scalebox{0.75}{
\begin{tikzpicture}
\draw[blue,thick,fill=blue] (0,0) ellipse (0.05cm and 0.05cm);
\draw[blue,very thick] (0,0)--(0.6,1.36);
\draw[blue,very thick] (0,0)--(-0.6,1.36);
\node at (0,1.1) {\textcolor{blue}{...}}; 
\node at (0,1.65) {\textcolor{blue}{$i_l$}}; 
\draw[blue,very thick] (1.4,0)--(-1.4,0);
\node at (-0.8,-0.4) {\textcolor{blue}{$\Delta_A$}};
\node at (0.85,-0.4) {\textcolor{blue}{$h-c_B$}}; 
\end{tikzpicture}}
}
 \end{matrix}
 \right)
\cr
&
\times 
   \left(
 \begin{matrix}
\text{
  \scalebox{0.75}{
\begin{tikzpicture}
\draw[red,thick,fill=red] (0,0) ellipse (0.05cm and 0.05cm);
\draw[red,thick,fill=red] (1.4,0) ellipse (0.05cm and 0.05cm);
\draw[red,thick,fill=red] (-1.4,0) ellipse (0.05cm and 0.05cm);
\draw[red,very thick] (0,0)--(0.6,1.36);
\draw[red,very thick] (0,0)--(-0.6,1.36);
\node at (0,1.1) {\textcolor{red}{...}}; 
\node at (0,1.65) {\textcolor{red}{$i_r$}}; 
\draw[red,very thick] (2.8,0)--(-2.8,0);
\node at (-0.75,-0.4) {\textcolor{red}{$h+c_B$}};
\node at (0.85,-0.4) {\textcolor{red}{$h-c_C$}}; 
\node at (-2.2,-0.4) {\textcolor{red}{$h-c_B$}};
\node at (2.25,-0.4) {\textcolor{red}{$h+c_C$}}; 
\end{tikzpicture}}
}
 \end{matrix}
 \right)
   \left(
 \begin{matrix}
\text{
  \scalebox{0.75}{
\begin{tikzpicture}
\draw[blue,thick,fill=blue] (0,0) ellipse (0.05cm and 0.05cm);
\draw[blue,very thick] (0,0)--(0.6,1.36);
\draw[blue,very thick] (0,0)--(-0.6,1.36);
\node at (0,1.1) {\textcolor{blue}{...}}; 
\node at (0,1.65) {\textcolor{blue}{$i_r$}}; 
\draw[blue,very thick] (1.4,0)--(-1.4,0);
\node at (-0.8,-0.4) {\textcolor{blue}{$h+c_B$}};
\node at (0.85,-0.4) {\textcolor{blue}{$h-c_C$}}; 
\end{tikzpicture}}
}
 \end{matrix}
 \right)
   \left(
 \begin{matrix}
\text{
  \scalebox{0.75}{
\begin{tikzpicture}
\draw[red,thick,fill=red] (0.8,0) ellipse (0.05cm and 0.05cm);
\draw[red,thick,fill=red] (-0.6,0) ellipse (0.05cm and 0.05cm);
\draw[red,very thick] (-2.0,0)--(0.8,0);
\draw[red,very thick] (0.8,0)--(1.73,0.63);
\draw[red,very thick] (0.8,0)--(1.73,-0.63);
\node at (0.1,-0.4) {\textcolor{red}{$h+c_C$}}; 
\node at (-1.3,-0.4) {\textcolor{red}{$h-c_C$}}; 
\node at (1.6,0.1) {\textcolor{red}{$\vdots$}}; 
\node at (2.1,0) {\textcolor{red}{$i_R$}}; 
\end{tikzpicture}}
}
 \end{matrix}
 \right) 
 \cr 
& \times 
 \left(
 \begin{matrix}
\text{
  \scalebox{0.75}{
\begin{tikzpicture}
\draw[blue,thick,fill=blue] (0.8,0) ellipse (0.05cm and 0.05cm);
\draw[blue,very thick] (-0.6,0)--(0.8,0);
\draw[blue,very thick] (0.8,0)--(1.73,0.63);
\draw[blue,very thick] (0.8,0)--(1.73,-0.63);
\node at (0.1,-0.4) {\textcolor{blue}{$h+c_C$}}; 
\node at (1.6,0.1) {\textcolor{blue}{$\vdots$}}; 
\node at (2.1,0) {\textcolor{blue}{$i_R$}}; 
\end{tikzpicture}}
}
 \end{matrix}
 \right).
}
Carrying out the $c_B$ integral next, one finds that
\eqn{}{
\mathcal{M}^{\text{3--int}}
=&
\,
  \int \frac{dc_C}{2\pi i}\,f_{\Delta_C}(c_C)
\left(
 \begin{matrix}
\text{
  \scalebox{0.75}{
\begin{tikzpicture}
\draw[blue,thick,fill=blue] (-0.8,0) ellipse (0.05cm and 0.05cm);
\draw[blue,very thick] (0.6,0)--(-0.8,0);
\draw[blue,very thick] (-0.8,0)--(-1.73,0.63);
\draw[blue,very thick] (-0.8,0)--(-1.73,-0.63);
\node at (-0.1,-0.4) {\textcolor{blue}{$\Delta_A$}}; 
\node at (-1.6,0.1) {\textcolor{blue}{$\vdots$}}; 
\node at (-2.1,0) {\textcolor{blue}{$i_L$}}; 
\end{tikzpicture}}
}
 \end{matrix}
 \right) 
 \left(
 \begin{matrix}
\text{
  \scalebox{0.75}{
\begin{tikzpicture}
\draw[blue,thick,fill=blue] (0,0) ellipse (0.05cm and 0.05cm);
\draw[blue,very thick] (0,0)--(0.6,1.36);
\draw[blue,very thick] (0,0)--(-0.6,1.36);
\node at (0,1.1) {\textcolor{blue}{...}}; 
\node at (0,1.65) {\textcolor{blue}{$i_l$}}; 
\draw[blue,very thick] (1.4,0)--(-1.4,0);
\node at (-0.8,-0.4) {\textcolor{blue}{$\Delta_A$}};
\node at (0.85,-0.4) {\textcolor{blue}{$\Delta_B$}}; 
\end{tikzpicture}}
}
 \end{matrix}
 \right)
\cr
&
\times 
   \left(
 \begin{matrix}
\text{
  \scalebox{0.75}{
\begin{tikzpicture}
\draw[red,very thick] (-0.8-0.6,0)--(-1.73-0.6,0.63);
\draw[red,very thick] (-0.8-0.6,0)--(-1.73-0.6,-0.63);
\node at (-1.6-0.6,0.1) {\textcolor{red}{$\vdots$}}; 
\node at (-2.1-0.6,0) {\textcolor{red}{$i_L$}}; 
\draw[red,thick,fill=red] (0,0) ellipse (0.05cm and 0.05cm);
\draw[red,thick,fill=red] (2.0,0) ellipse (0.05cm and 0.05cm);
\draw[red,thick,fill=red] (-1.4,0) ellipse (0.05cm and 0.05cm);
\draw[red,thick,fill=red] (3.4,0) ellipse (0.05cm and 0.05cm);
\draw[red,very thick] (0,0)--(0.6,1.36);
\draw[red,very thick] (0,0)--(-0.6,1.36);
\draw[red,very thick] (2,0)--(2.6,1.36);
\draw[red,very thick] (2,0)--(1.4,1.36);
\draw[red,very thick] (4.8,0)--(-1.4,0);
\node at (0,1.1) {\textcolor{red}{...}}; 
\node at (0,1.65) {\textcolor{red}{$i_l$}}; 
\node at (2,1.1) {\textcolor{red}{...}}; 
\node at (2,1.65) {\textcolor{red}{$i_r$}}; 
\node at (-0.75,-0.4) {\textcolor{red}{$\Delta_A$}};
\node at (1.,-0.4) {\textcolor{red}{$\Delta_B$}}; 
\node at (-0.8,0.4) {\textcolor{red}{$s_A$}};
\node at (1.,0.4) {\textcolor{red}{$s_B$}}; 
\node at (2.75,-0.4) {\textcolor{red}{$h-c_C$}}; 
\node at (4.15,-0.4) {\textcolor{red}{$h+c_C$}}; 
\end{tikzpicture}}
}
 \end{matrix}
 \right)
   \left(
 \begin{matrix}
\text{
  \scalebox{0.75}{
\begin{tikzpicture}
\draw[blue,thick,fill=blue] (0,0) ellipse (0.05cm and 0.05cm);
\draw[blue,very thick] (0,0)--(0.6,1.36);
\draw[blue,very thick] (0,0)--(-0.6,1.36);
\node at (0,1.1) {\textcolor{blue}{...}}; 
\node at (0,1.65) {\textcolor{blue}{$i_r$}}; 
\draw[blue,very thick] (1.4,0)--(-1.4,0);
\node at (-0.8,-0.4) {\textcolor{blue}{$\Delta_B$}};
\node at (0.85,-0.4) {\textcolor{blue}{$h-c_C$}}; 
\end{tikzpicture}}
}
 \end{matrix}
 \right)
 \cr 
 & \times
   \left(
 \begin{matrix}
\text{
  \scalebox{0.75}{
\begin{tikzpicture}
\draw[red,thick,fill=red] (0.8,0) ellipse (0.05cm and 0.05cm);
\draw[red,thick,fill=red] (-0.6,0) ellipse (0.05cm and 0.05cm);
\draw[red,very thick] (-2.0,0)--(0.8,0);
\draw[red,very thick] (0.8,0)--(1.73,0.63);
\draw[red,very thick] (0.8,0)--(1.73,-0.63);
\node at (0.1,-0.4) {\textcolor{red}{$h+c_C$}}; 
\node at (-1.3,-0.4) {\textcolor{red}{$h-c_C$}}; 
\node at (1.6,0.1) {\textcolor{red}{$\vdots$}}; 
\node at (2.1,0) {\textcolor{red}{$i_R$}}; 
\end{tikzpicture}}
}
 \end{matrix}
 \right)
 \left(
 \begin{matrix}
\text{
  \scalebox{0.75}{
\begin{tikzpicture}
\draw[blue,thick,fill=blue] (0.8,0) ellipse (0.05cm and 0.05cm);
\draw[blue,very thick] (-0.6,0)--(0.8,0);
\draw[blue,very thick] (0.8,0)--(1.73,0.63);
\draw[blue,very thick] (0.8,0)--(1.73,-0.63);
\node at (0.1,-0.4) {\textcolor{blue}{$h+c_C$}}; 
\node at (1.6,0.1) {\textcolor{blue}{$\vdots$}}; 
\node at (2.1,0) {\textcolor{blue}{$i_R$}}; 
\end{tikzpicture}}
}
 \end{matrix}
 \right).
 }
 $\hspace{1mm}$
 \\
Finally, carrying out the $c_C$ integral yields 

\clearpage
\eqn{}{
\mathcal{M}^{\text{3--int}}
=&
\left(
\begin{matrix}
\text{
  \scalebox{0.75}{
\begin{tikzpicture}
\draw[blue,thick,fill=blue] (-0.8,0) ellipse (0.05cm and 0.05cm);
\draw[blue,very thick] (0.6,0)--(-0.8,0);
\draw[blue,very thick] (-0.8,0)--(-1.73,0.63);
\draw[blue,very thick] (-0.8,0)--(-1.73,-0.63);
\node at (-0.1,-0.4) {\textcolor{blue}{$\Delta_A$}}; 
\node at (-1.6,0.1) {\textcolor{blue}{$\vdots$}}; 
\node at (-2.1,0) {\textcolor{blue}{$i_L$}}; 
\end{tikzpicture}}
}
\end{matrix}
\right)
 \left(
 \begin{matrix}
\text{
  \scalebox{0.75}{
\begin{tikzpicture}
\draw[blue,thick,fill=blue] (0,0) ellipse (0.05cm and 0.05cm);
\draw[blue,very thick] (0,0)--(0.6,1.36);
\draw[blue,very thick] (0,0)--(-0.6,1.36);
\node at (0,1.1) {\textcolor{blue}{...}}; 
\node at (0,1.65) {\textcolor{blue}{$i_l$}}; 
\draw[blue,very thick] (1.4,0)--(-1.4,0);
\node at (-0.8,-0.4) {\textcolor{blue}{$\Delta_A$}};
\node at (0.85,-0.4) {\textcolor{blue}{$\Delta_B$}}; 
\end{tikzpicture}}
}
 \end{matrix}
 \right)
   \left(
 \begin{matrix}
\text{
  \scalebox{0.75}{
\begin{tikzpicture}
\draw[blue,thick,fill=blue] (0,0) ellipse (0.05cm and 0.05cm);
\draw[blue,very thick] (0,0)--(0.6,1.36);
\draw[blue,very thick] (0,0)--(-0.6,1.36);
\node at (0,1.1) {\textcolor{blue}{...}}; 
\node at (0,1.65) {\textcolor{blue}{$i_r$}}; 
\draw[blue,very thick] (1.4,0)--(-1.4,0);
\node at (-0.8,-0.4) {\textcolor{blue}{$\Delta_B$}};
\node at (0.85,-0.4) {\textcolor{blue}{$\Delta_C$}}; 
\end{tikzpicture}}
}
 \end{matrix}
 \right)
 \left(
 \begin{matrix}
\text{
  \scalebox{0.75}{
\begin{tikzpicture}
\draw[blue,thick,fill=blue] (0.8,0) ellipse (0.05cm and 0.05cm);
\draw[blue,very thick] (-0.6,0)--(0.8,0);
\draw[blue,very thick] (0.8,0)--(1.73,0.63);
\draw[blue,very thick] (0.8,0)--(1.73,-0.63);
\node at (0.1,-0.4) {\textcolor{blue}{$\Delta_C$}}; 
\node at (1.6,0.1) {\textcolor{blue}{$\vdots$}}; 
\node at (2.1,0) {\textcolor{blue}{$i_R$}}; 
\end{tikzpicture}}
}
 \end{matrix}
 \right)
\cr
&
\times 
   \left(
 \begin{matrix}
\text{
  \scalebox{0.75}{
\begin{tikzpicture}
\draw[red,very thick] (-0.8-0.6,0)--(-1.73-0.6,0.63);
\draw[red,very thick] (-0.8-0.6,0)--(-1.73-0.6,-0.63);
\node at (-1.6-0.6,0.1) {\textcolor{red}{$\vdots$}}; 
\node at (-2.1-0.6,0) {\textcolor{red}{$i_L$}}; 
\draw[red,thick,fill=red] (0,0) ellipse (0.05cm and 0.05cm);
\draw[red,thick,fill=red] (2.0,0) ellipse (0.05cm and 0.05cm);
\draw[red,thick,fill=red] (-1.4,0) ellipse (0.05cm and 0.05cm);
\draw[red,thick,fill=red] (3.4,0) ellipse (0.05cm and 0.05cm);
\draw[red,very thick] (0,0)--(0.6,1.36);
\draw[red,very thick] (0,0)--(-0.6,1.36);
\draw[red,very thick] (2,0)--(2.6,1.36);
\draw[red,very thick] (2,0)--(1.4,1.36);
\draw[red,very thick] (3.4,0)--(-1.4,0);
\node at (0,1.1) {\textcolor{red}{...}}; 
\node at (0,1.65) {\textcolor{red}{$i_l$}}; 
\node at (2,1.1) {\textcolor{red}{...}}; 
\node at (2,1.65) {\textcolor{red}{$i_r$}}; 
\node at (-0.75,-0.4) {\textcolor{red}{$\Delta_A$}};
\node at (1.,-0.4) {\textcolor{red}{$\Delta_B$}}; 
\node at (-0.8,0.4) {\textcolor{red}{$s_A$}};
\node at (1.,0.4) {\textcolor{red}{$s_B$}}; 
\node at (2.75,-0.4) {\textcolor{red}{$\Delta_C$}}; 
\draw[red,very thick] (3.4,0)--(4.33,0.63);
\draw[red,very thick] (3.4,0)--(4.33,-0.63);
\node at (4.2,0.1) {\textcolor{red}{$\vdots$}}; 
\node at (4.7,0) {\textcolor{red}{$i_R$}}; 
\end{tikzpicture}}
}
 \end{matrix}
 \right).
 }
We see that carrying out  contour integrals of the form above effectively glues together two of the undressed diagrams defined in section \ref{recursionSection}. This holds true generally: given a preamplitude,
\eqn{}{
\widetilde{\mathcal{M}}\equiv 
&\left(
\begin{matrix}
  \scalebox{0.75}{
\begin{tikzpicture}
\draw[blue,thick,fill=blue] (0,0) ellipse (0.05cm and 0.05cm);
\draw[blue,very thick] (0,0)--(0,1.4);
\draw[blue,very thick] (0,0)--(0.6,-1.36);
\draw[blue,very thick] (0,0)--(-0.6,-1.36);
\node at (-0.4,0.8) {\textcolor{blue}{$F$}}; 
\node at (0,-1.1) {\textcolor{blue}{...}}; 
\node at (0,-1.55) {\textcolor{blue}{$i_1$}}; 
\draw[blue,very thick] (1.4,0)--(-1.4,0);
\node at (-0.9,-0.4) {\textcolor{blue}{$A$}};
\node at (0.9,-0.4) {\textcolor{blue}{$h-c$}}; 
\end{tikzpicture}}
\end{matrix}
\right)
\left(
\begin{matrix}
\text{
  \scalebox{0.75}{
\begin{tikzpicture}
\draw[red,very thick] (-1.4,0)--(-2.5,0.75);
\node at (-2.5-1.1*0.8,0.75+0.75*0.8) {\textcolor{red}{.}};
\node at (-2.5-1.1*1.2,0.75+0.75*1.2) {\textcolor{red}{$i_4$}}; 
\node at (-2.5-1.1*0.8+0.7*0.11,0.75+0.75*0.8+1*0.11){\textcolor{red}{.}}; 
\node at (-2.5-1.1*0.8-0.7*0.11,0.75+0.75*0.8-1*0.11){\textcolor{red}{.}}; 
\draw[red,very thick] (-2.5,0.75)--(-2.5-0.79,0.75+1.26);
\draw[red,very thick] (-2.5,0.75)--(-2.5-1.46,0.75+0.27);
\draw[red,very thick] (-1.4,0)--(-2.5,-0.75);
\draw[red,very thick] (-2.5,-0.75)--(-2.5-0.79,-0.75-1.26);
\draw[red,very thick] (-2.5,-0.75)--(-2.5-1.46,-0.75-0.27);
\node at (-2.5-1.1*0.8,-0.75-0.75*0.8) {\textcolor{red}{.}}; 
\node at (-2.5-1.1*1.2,-0.75-0.75*1.2) {\textcolor{red}{$i_3$}}; 
\node at (-2.5-1.1*0.8+0.7*0.11,-0.75-0.75*0.8-1*0.11){\textcolor{red}{.}}; 
\node at (-2.5-1.1*0.8-0.7*0.11,-0.75-0.75*0.8+1*0.11){\textcolor{red}{.}}; 
\draw[red,thick,fill=red] (-2.5,0.75) ellipse (0.05cm and 0.05cm);
\draw[red,thick,fill=red] (-2.5,-0.75) ellipse (0.05cm and 0.05cm);
\node at (-1.8,0.75) {\textcolor{red}{$D$}};
\node at (-1.8,-0.75) {\textcolor{red}{$C$}}; 
\node at (-0.4,0.7) {\textcolor{red}{$E$}};
\node at (1.6,0.7) {\textcolor{red}{$F$}};
\draw[red,thick,fill=red] (0,0) ellipse (0.05cm and 0.05cm);
\draw[red,thick,fill=red] (2.0,0) ellipse (0.05cm and 0.05cm);
\draw[red,thick,fill=red] (0,1.4) ellipse (0.05cm and 0.05cm);
\draw[red,thick,fill=red] (2.0,1.4) ellipse (0.05cm and 0.05cm);
\draw[red,thick,fill=red] (-1.4,0) ellipse (0.05cm and 0.05cm);
\draw[red,thick,fill=red] (3.5,0) ellipse (0.05cm and 0.05cm);
\draw[red,very thick] (4.9,0)--(-1.4,0);
\draw[red,very thick] (0,0)--(0,1.4);
\draw[red,very thick] (2,0)--(2,1.4);
\draw[red,very thick] (0,1.4)--(0.6,2.76);
\draw[red,very thick] (0,1.4)--(-0.6,2.76);
\draw[red,very thick] (0,0)--(0.6,-1.36);
\draw[red,very thick] (0,0)--(-0.6,-1.36);
\node at (0,-1.1) {\textcolor{red}{...}}; 
\node at (0,-1.55) {\textcolor{red}{$i_2$}}; 
\draw[red,very thick] (2,0)--(2.6,-1.36);
\draw[red,very thick] (2,0)--(1.4,-1.36);
\node at (2,-1.1) {\textcolor{red}{...}}; 
\node at (2,-1.55) {\textcolor{red}{$i_1$}}; 
\draw[red,very thick] (2,1.4)--(2.6,2.76);
\draw[red,very thick] (2,1.4)--(1.4,2.76);
\node at (0,1.1+1.4) {\textcolor{red}{...}}; 
\node at (0,1.65+1.4) {\textcolor{red}{$i_5$}}; 
\node at (2,1.1+1.4) {\textcolor{red}{...}}; 
\node at (2,1.65+1.4) {\textcolor{red}{$i_6$}}; 
\node at (-0.85,-0.4) {\textcolor{red}{$B$}};
\node at (1.,-0.4) {\textcolor{red}{$A$}}; 
\node at (2.9,-0.4) {\textcolor{red}{$h-c$}}; 
\node at (2.8,0.4) {\textcolor{red}{$s$}}; 
\node at (4.3,-0.4) {\textcolor{red}{$h+c$}}; 
\end{tikzpicture}}
}
\end{matrix}
\right)
\cr
&
\times
\left(
\begin{matrix}
\text{
  \scalebox{0.75}{
\begin{tikzpicture}
\draw[red,thick,fill=red] (0.8,0) ellipse (0.05cm and 0.05cm);
\draw[red,thick,fill=red] (-0.6,0) ellipse (0.05cm and 0.05cm);
\draw[red,very thick] (-2.0,0)--(0.8,0);
\draw[red,very thick] (0.8,0)--(1.73,0.63);
\draw[red,very thick] (0.8,0)--(1.73,-0.63);
\node at (0.1,-0.4) {\textcolor{red}{$h+c$}};
\node at (0.1,0.4) {\textcolor{red}{$s$}};
\node at (-1.3,-0.4) {\textcolor{red}{$h-c$}}; 
\node at (1.6,0.1) {\textcolor{red}{$\vdots$}}; 
\node at (2.1,0) {\textcolor{red}{$i_7$}}; 
\end{tikzpicture}}
}\end{matrix}
\right)
\left(
\begin{matrix}
\vspace*{-0.2cm}
\text{
  \scalebox{0.75}{
\begin{tikzpicture}
\draw[blue,thick,fill=blue] (0.8,0) ellipse (0.05cm and 0.05cm);
\draw[blue,very thick] (-0.6,0)--(0.8,0);
\draw[blue,very thick] (0.8,0)--(1.73,0.63);
\draw[blue,very thick] (0.8,0)--(1.73,-0.63);
\node at (0.1,-0.4) {\textcolor{blue}{$h+c$}}; 
\node at (1.6,0.1) {\textcolor{blue}{$\vdots$}}; 
\node at (2.1,0) {\textcolor{blue}{$i_7$}}; 
\end{tikzpicture}}
}
\end{matrix}
\right),
}
the following identity holds,
\eqn{toBeProven}
{
&
\int \frac{dc}{2\pi i}\,f_\Delta(c)\,\widetilde{\mathcal{M}}
= \cr 
& 
\left(
\begin{matrix}
  \scalebox{0.75}{
\begin{tikzpicture}
\draw[blue,thick,fill=blue] (0,0) ellipse (0.05cm and 0.05cm);
\draw[blue,very thick] (0,0)--(0,1.4);
\draw[blue,very thick] (0,0)--(0.6,-1.36);
\draw[blue,very thick] (0,0)--(-0.6,-1.36);
\node at (-0.4,0.8) {\textcolor{blue}{$F$}}; 
\node at (0,-1.1) {\textcolor{blue}{...}}; 
\node at (0,-1.55) {\textcolor{blue}{$i_1$}}; 
\draw[blue,very thick] (1.4,0)--(-1.4,0);
\node at (-0.9,-0.4) {\textcolor{blue}{$A$}};
\node at (0.9,-0.4) {\textcolor{blue}{$\Delta$}}; 
\end{tikzpicture}}
\end{matrix}
\right)
\left(
\begin{matrix}
\text{
  \scalebox{0.75}{
\begin{tikzpicture}
\draw[red,very thick] (-1.4,0)--(-2.5,0.75);
\node at (-2.5-1.1*0.8,0.75+0.75*0.8) {\textcolor{red}{.}}; 
\node at (-2.5-1.1*1.2,0.75+0.75*1.2) {\textcolor{red}{$i_4$}}; 
\node at (-2.5-1.1*0.8+0.7*0.11,0.75+0.75*0.8+1*0.11){\textcolor{red}{.}}; 
\node at (-2.5-1.1*0.8-0.7*0.11,0.75+0.75*0.8-1*0.11){\textcolor{red}{.}}; 
\draw[red,very thick] (-2.5,0.75)--(-2.5-0.79,0.75+1.26);
\draw[red,very thick] (-2.5,0.75)--(-2.5-1.46,0.75+0.27);
\draw[red,very thick] (-1.4,0)--(-2.5,-0.75);
\draw[red,very thick] (-2.5,-0.75)--(-2.5-0.79,-0.75-1.26);
\draw[red,very thick] (-2.5,-0.75)--(-2.5-1.46,-0.75-0.27);
\node at (-2.5-1.1*0.8,-0.75-0.75*0.8) {\textcolor{red}{.}}; 
\node at (-2.5-1.1*1.2,-0.75-0.75*1.2) {\textcolor{red}{$i_3$}}; 
\node at (-2.5-1.1*0.8+0.7*0.11,-0.75-0.75*0.8-1*0.11){\textcolor{red}{.}}; 
\node at (-2.5-1.1*0.8-0.7*0.11,-0.75-0.75*0.8+1*0.11){\textcolor{red}{.}}; 
\draw[red,thick,fill=red] (-2.5,0.75) ellipse (0.05cm and 0.05cm);
\draw[red,thick,fill=red] (-2.5,-0.75) ellipse (0.05cm and 0.05cm);
\node at (-1.8,0.75) {\textcolor{red}{$D$}};
\node at (-1.8,-0.75) {\textcolor{red}{$C$}}; 
\node at (-0.4,0.7) {\textcolor{red}{$E$}};
\node at (1.6,0.7) {\textcolor{red}{$F$}};
\draw[red,thick,fill=red] (0,0) ellipse (0.05cm and 0.05cm);
\draw[red,thick,fill=red] (2.0,0) ellipse (0.05cm and 0.05cm);
\draw[red,thick,fill=red] (0,1.4) ellipse (0.05cm and 0.05cm);
\draw[red,thick,fill=red] (2.0,1.4) ellipse (0.05cm and 0.05cm);
\draw[red,thick,fill=red] (-1.4,0) ellipse (0.05cm and 0.05cm);
\draw[red,thick,fill=red] (3.4,0) ellipse (0.05cm and 0.05cm);
\draw[red,very thick] (3.4,0)--(-1.4,0);
\draw[red,very thick] (0,0)--(0,1.4);
\draw[red,very thick] (2,0)--(2,1.4);
\draw[red,very thick] (0,1.4)--(0.6,2.76);
\draw[red,very thick] (0,1.4)--(-0.6,2.76);
\draw[red,very thick] (0,0)--(0.6,-1.36);
\draw[red,very thick] (0,0)--(-0.6,-1.36);
\node at (0,-1.1) {\textcolor{red}{...}}; 
\node at (0,-1.55) {\textcolor{red}{$i_2$}}; 
\draw[red,very thick] (2,0)--(2.6,-1.36);
\draw[red,very thick] (2,0)--(1.4,-1.36);
\node at (2,-1.1) {\textcolor{red}{...}}; 
\node at (2,-1.55) {\textcolor{red}{$i_1$}}; 
\draw[red,very thick] (2,1.4)--(2.6,2.76);
\draw[red,very thick] (2,1.4)--(1.4,2.76);
\draw[red,very thick] (3.4,0)--(4.33,0.63);
\draw[red,very thick] (3.4,0)--(4.33,-0.63);
\node at (0,1.1+1.4) {\textcolor{red}{...}}; 
\node at (0,1.65+1.4) {\textcolor{red}{$i_5$}}; 
\node at (2,1.1+1.4) {\textcolor{red}{...}}; 
\node at (2,1.65+1.4) {\textcolor{red}{$i_6$}}; 
\node at (-0.85,-0.4) {\textcolor{red}{$B$}};
\node at (1.,-0.4) {\textcolor{red}{$A$}}; 
\node at (2.8,-0.4) {\textcolor{red}{$\Delta$}}; 
\node at (2.8,0.4) {\textcolor{red}{$s$}}; 
\node at (4.2,0.1) {\textcolor{red}{$\vdots$}}; 
\node at (4.7,0) {\textcolor{red}{$i_7$}}; 
\end{tikzpicture}}
}
\end{matrix}
\right)
\left(
\begin{matrix}
\vspace*{-0.2cm}
\text{
  \scalebox{0.75}{
\begin{tikzpicture}
\draw[blue,thick,fill=blue] (0.8,0) ellipse (0.05cm and 0.05cm);
\draw[blue,very thick] (-0.6,0)--(0.8,0);
\draw[blue,very thick] (0.8,0)--(1.73,0.63);
\draw[blue,very thick] (0.8,0)--(1.73,-0.63);
\node at (0.1,-0.4) {\textcolor{blue}{$\Delta$}}; 
\node at (1.6,0.1) {\textcolor{blue}{$\vdots$}}; 
\node at (2.1,0) {\textcolor{blue}{$i_7$}}; 
\end{tikzpicture}}
}
\end{matrix}
\right).
}
 At the real place, this fact follows from the Feynman rules for Mellin amplitudes and pre-amplitudes that already exist in the literature (see Refs.~\cite{Fitzpatrick:2011ia,Paulos:2011ie,Nandan:2011wc}), and which in this paper we have presented in a  manner which highlights the common features that hold true over both the reals and the $p$-adics. At the $p$-adic place, this fact requires a proof which is provided in appendix \ref{AmpProof}.

\section{On-Shell Recursion for Mellin Amplitudes}
\label{BCFW}
The recursive structure of $p$-adic Mellin amplitudes described in section \ref{recursionSectionPadic} is of a somewhat unusual kind: the amplitude of an ${\cal N}$-point bulk diagram is expressed in terms of amplitudes of ${\cal N}$-point diagrams with fewer internal lines. Such a recursion relation invokes interaction vertices of higher and higher degree, that is, interaction terms involving more and more fields. A natural question to ask is whether it is possible to set up a recursive prescription in a form more similar to the familiar on-shell recursions in flat-space, such as the BCFW recursion relation~\cite{Britto:2005fq}, where no new interaction vertices are introduced and a higher-point amplitude is expressed in terms of lower point amplitudes of the same theory. In this section we show that this is indeed possible, though we emphasize that the recursion we formulate in this section applies at the level of \emph{individual} bulk diagrams. 

We start in section \ref{Factorization} by describing the factorization property of undressed diagrams and explaining how this property allows us to reformulate the recursion relation \eno{decomp1} in a manner that will play as a precursor to the on-shell recursion of this section. Then, in section \ref{PadicBCFW}, we present what we refer to as the ``on-shell'' recursion relation, following up with a proof that establishes the equivalence between the recursive prescription~\ref{pres:padicMellin} of section \ref{recursionSection} and the on-shell prescription of this section. At the end of this section we comment on its connection with the Feynman rules for real Mellin amplitudes stated earlier as prescription~\ref{pres:Feyn}, and in appendix \ref{BCFWMOM}, with the help of two explicit examples, we demonstrate the similarities (and differences) between this prescription and flat-space BCFW recursion.

\subsection{Factorization property of undressed diagrams}
\label{Factorization}

Consider an arbitrary undressed diagram:
\eqn{exampleDiagram}{
\begin{matrix}
\includegraphics[height=19ex]{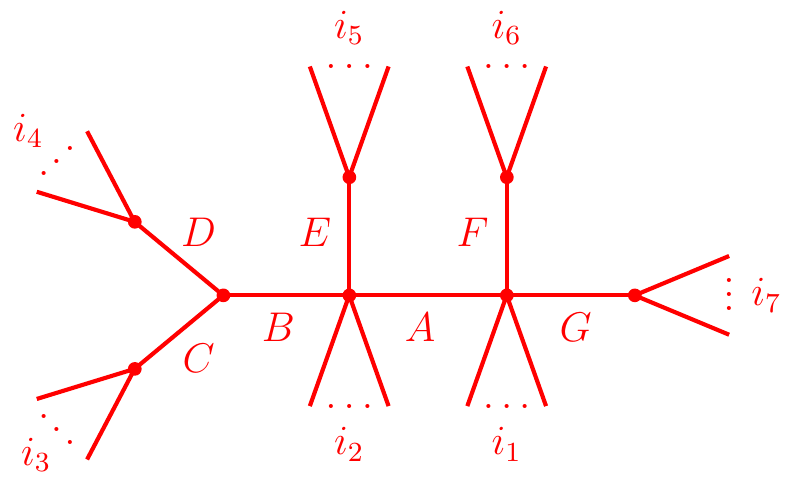}
 \end{matrix}\,.
}
It follows from the recursive prescription~\ref{pres:padicMellin} of section \ref{recursionSectionPadic} that the sum of all the terms of the undressed amplitude proportional to  the internal line factor $\zeta_p(s_G-\Delta_G)$ is itself given by an undressed diagram: 
\eqn{}{
-
\zeta_p(s_G-\Delta_G)
\left(
\begin{matrix}
\includegraphics[height=19ex]{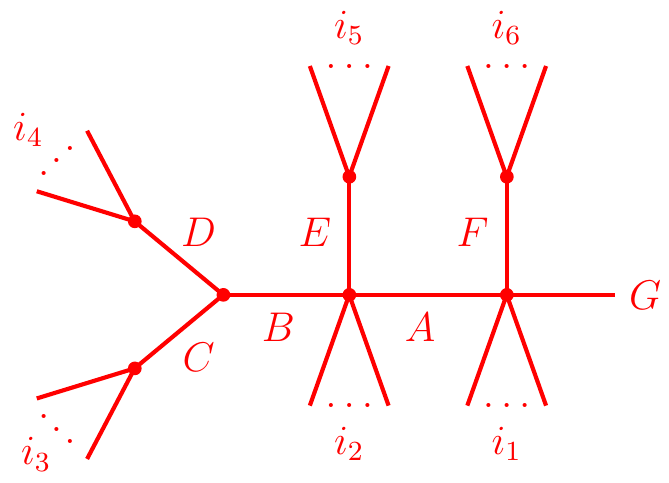}
 \end{matrix}
\right).
}
More generally, the coefficient of an internal line factor in an undressed diagram is given by the product of the left- and right-hand (undressed) diagrams obtained by cutting the diagram across the chosen internal line, and assigning the off-shell internal leg dimension to the newly generated external legs of the left- and right-hand undressed diagrams. For example, in the diagram \eno{exampleDiagram}, adding together all the terms containing the internal line factor $\zeta_p(s_A-\Delta_A)$ gives,
\eqn{}{
-
\zeta_p(s_A-\Delta_A)
\left(
\begin{matrix}
\includegraphics[height=19ex]{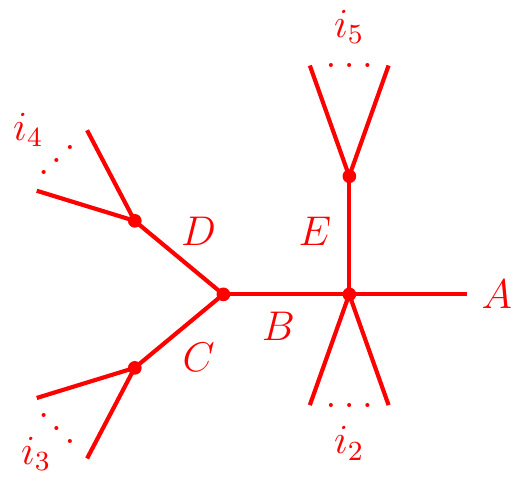}
 \end{matrix}
\right)
\left(
\begin{matrix}
\includegraphics[height=19ex]{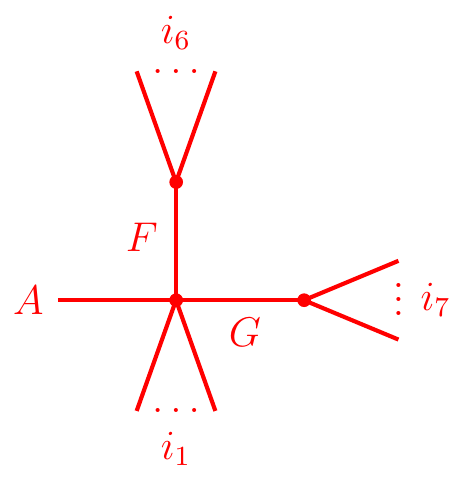}
 \end{matrix}
\right).
}
The procedure generalizes straightforwardly if one wants to pick out terms containing multiple propagator factors: 
\eqn{SelectTerms}{
& \big(\,\text{terms in undressed diagram $D$ containing propagator factors }\zeta_p(s_1-\Delta_1)...\zeta_p(s_f-\Delta_f)\,\big)
\cr
&=(-1)^f\left[\prod_{I=1}^f\zeta_p(s_I-\Delta_I)\right]\times
\begin{pmatrix*}[l]
\text{product of undressed diagrams resulting from} \\
\text{removing internal legs 1 to $f$ from $D$}
\end{pmatrix*}.
}
\hspace{1mm}
\\
For example, in diagram \eno{exampleDiagram}, the terms containing both factors $\zeta_p(s_A-\Delta_A)$ and $\zeta_p(s_B-\Delta_B)$ sum to
\clearpage
\eqn{}{
(-1)^2\zeta_p(s_A-\Delta_A)\zeta_p(s_B-\Delta_B)
\left(
\begin{matrix}
\includegraphics[height=19ex]{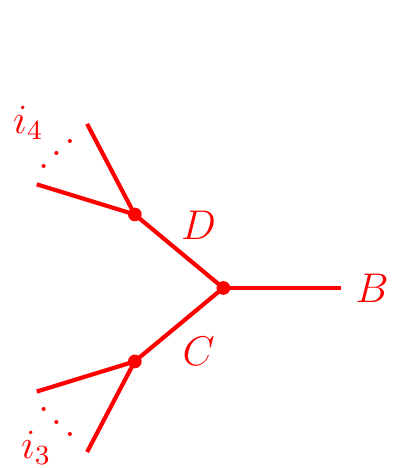}
 \end{matrix}
\right)
\left(
\begin{matrix}
\includegraphics[height=19ex]{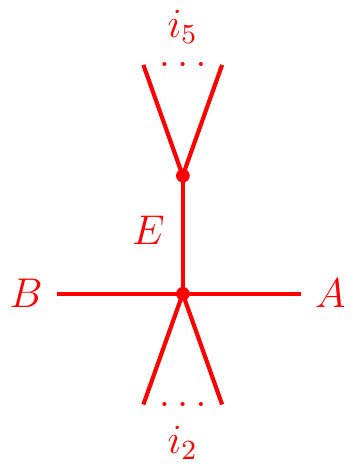}
 \end{matrix}
\right)
\left(
\begin{matrix}
\includegraphics[height=19ex]{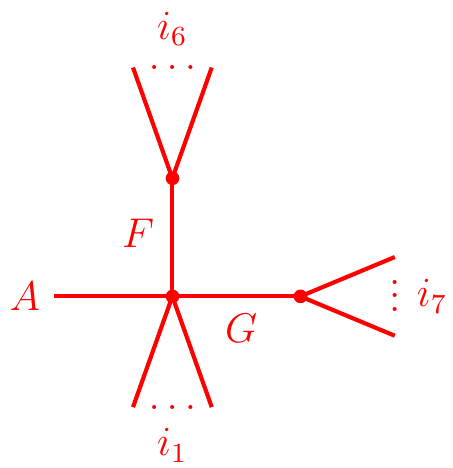}
 \end{matrix}
\right).
}
These observations reflect the meromorphic nature of arbitrary Mellin amplitudes, which have simple poles at values of Mandelstam-like invariants that put any internal leg ``on-shell''. For $p$-adic Mellin amplitudes it is clear from the above that such poles occur in the internal leg factor~\eno{internalPropagator} when the Mandelstam invariant of the associated leg equals the conformal dimension of the bulk field propagating along the leg. These poles correspond to the exchange of single-trace operators in the boundary theory. Thus $p$-adic Mellin amplitudes have the right factorization properties of (AdS) scattering amplitudes.

Another property of undressed diagrams that follows immediately from the recursive prescription~\ref{pres:padicMellin} is that the momentum independent terms (i.e.\ terms independent of any Mandelstam variable $s_I$ and thereby also independent of the Mellin variables $\gamma_{ij}$) are in one-to-one correspondence with the terms exhibiting momentum dependence. If for a given undressed diagram one takes all the momentum-dependent terms, sets all internal leg factors $\zeta_p(s-\Delta)$ to unity (which can be achieved by sending all Mandelstam variables $s$ to infinity), and multiplies with an overall factor of $(-1)\times\zeta_p(\Sigma)$, where $\Sigma$ is the sum of all external scaling dimensions minus $n=2h$, 
\eqn{}
{\Sigma \equiv \sum\Delta_{\text{external}}-2h\,,}
then one exactly recovers the momentum independent part of the undressed diagram:
\eqn{momIndep}
{
\big(\text{momentum independent terms}\big)
=-\zeta_p(\Sigma)\lim_{\forall s\rightarrow \infty}\big(\text{momentum dependent terms}\big)\,.
}
By applying the inclusion-exclusion principle to these two properties we can derive yet another recursive formula for undressed diagrams, as we now explain: As discussed above, given any diagram, we can, for each internal leg $I$, sum over the terms proportional to $\zeta_p(s_I-\Delta_I)$, the  sum of which gives us $-\zeta_p(s_I-\Delta_I)$ times a smaller diagram or a product of two smaller diagrams. Summing $-\zeta_p(s_I-\Delta_I)$ times these smaller diagrams over all $I$, we obtain a sum containing all momentum-dependent terms. But in doing so, terms containing two factors of $\zeta_p(s-\Delta)$ (i.e.\ terms proportional to $\zeta_p(s_I-\Delta_I)\zeta_p(s_J-\Delta_J)$ for $I\neq J$) have been double-counted. This over-counting can be compensated for by subtracting off all terms with the two factors of $\zeta_p(s-\Delta)$, which by the factorization property above, are also given in terms of smaller diagrams. But now terms containing three factors of $\zeta_p(s-\Delta)$ haven't been included at all, so we add those terms back in.  Carrying out this alternating sum until one reaches the one term that that contains the product of $\zeta_p(s_I-\Delta_I)$ over all $I$, we get an expression for all the momentum-dependent terms. (Note that the alternating sign of the inclusion-exclusion principle is cancelled by the $(-1)^f$ prefactor in \eno{SelectTerms} so that each term comes with a minus sign.) We then add in the momentum-independent terms to get an expression for the full undressed diagram under consideration, by adding $-\zeta_p(\Sigma)$ times the momentum-dependent terms, except with all Mandelstam variables set to infinity. 
Thus we arrive at the formula
\eqn{newFormula}
{
(\text{undressed diagram})=
-&\sum_{\ell=1}^{\#\text{ of internal legs}}\sum_{\text{all possible sets of $\ell$ internal legs } \{I_1,\ldots,I_\ell\}}
\cr
&
\times
\left[\prod_{i=1}^\ell\zeta_p(s_{I_i}-\Delta_{I_i})-\zeta_p(\Sigma)\lim_{\forall s\rightarrow \infty}\right]
\cr
&
\times
(\text{undressed diagrams left after removing legs $\{I_1,\ldots,I_\ell\}$})\,,
}
where by ``removing a leg'' we mean cutting or factorizing the diagram across the particular leg.
Concretely, writing out the first couple of terms on the right-hand side as well as the last term, this formula says that 
\eqn{}
{
(\text{undressed diagram})=
-&\sum_{I}
\left[\zeta_p(s_I-\Delta_I)-\zeta_p(\Sigma)\lim_{\forall s\rightarrow \infty}\right]
\cr
&\hspace{7mm} \times
(\text{undressed diagrams left after removing leg $I$})
\cr
-&\sum_{I<J}
\left[\zeta_p(s_I-\Delta_I)\zeta_p(s_J-\Delta_J)-\zeta_p(\Sigma)\lim_{\forall s\rightarrow \infty}\right]
\cr
&\hspace{7mm}\times
(\text{undressed diagrams left after removing legs $I,J$})
\cr
-&\sum_{I<J<K}
\left[\zeta_p(s_I-\Delta_I)\zeta_p(s_J-\Delta_J)\zeta_p(s_K-\Delta_K)-\zeta_p(\Sigma)\lim_{\forall s\rightarrow \infty}\right]
\cr
&\hspace{7mm}\times
(\text{undressed diagrams left after removing legs $I,J,K$})
\cr
\vdots\,&\hspace{26mm}\vdots\hspace{26mm}\vdots
\cr
-&
\left[\prod_{\text{internal legs $I$}}\zeta_p(s_I-\Delta_I)-\zeta_p(\Sigma)\right].
}

\clearpage
We now illustrate how to apply the formula  via an example:
\begin{example}
\eqn{}{
&
\hspace{20mm}
\left(
\begin{matrix}
\includegraphics[height=12ex]{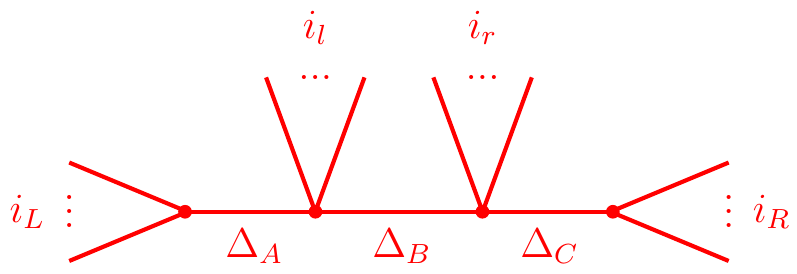}
 \end{matrix}
\right)
\cr
=&-
\left[\zeta_p(s_A-\Delta_A)-\zeta_p(\Sigma)\lim_{s_B,s_C\rightarrow \infty}\right]
\left(
\begin{matrix}
\includegraphics[height=12ex]{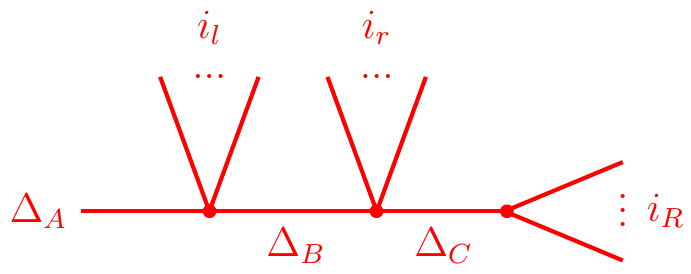}
 \end{matrix}
\right)
\cr
&
-
\left[\zeta_p(s_B-\Delta_B)-\zeta_p(\Sigma)\lim_{s_A,s_C\rightarrow \infty}\right]
\left(
\begin{matrix}
\includegraphics[height=12ex]{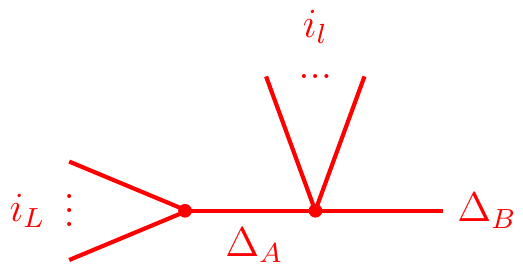}
 \end{matrix}
\right)
\left(
\begin{matrix}
\includegraphics[height=12ex]{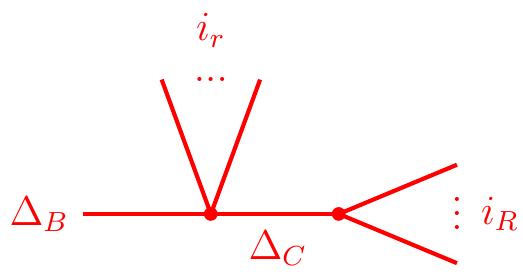}
 \end{matrix}
\right)
\cr
&
-
\left[\zeta_p(s_C-\Delta_C)-\zeta_p(\Sigma)\lim_{s_A,s_B\rightarrow \infty}\right]
\left(
\begin{matrix}
\includegraphics[height=12ex]{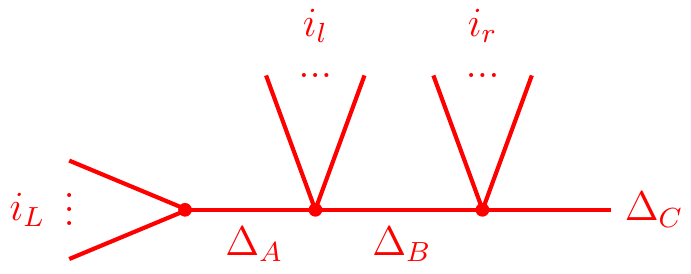}
 \end{matrix}
\right)
\cr
&
-\left[\zeta_p(s_A-\Delta_A)\zeta_p(s_B-\Delta_B)-\zeta_p(\Sigma)\lim_{s_C\rightarrow \infty}\right]
\left(
\begin{matrix}
\includegraphics[height=12ex]{figures/ThirdRecur4.pdf}
 \end{matrix}
\right)
\cr
&
-\left[\zeta_p(s_A-\Delta_A)\zeta_p(s_C-\Delta_C)-\zeta_p(\Sigma)\lim_{s_B\rightarrow \infty}\right]
\left(
\begin{matrix}
\includegraphics[height=12ex]{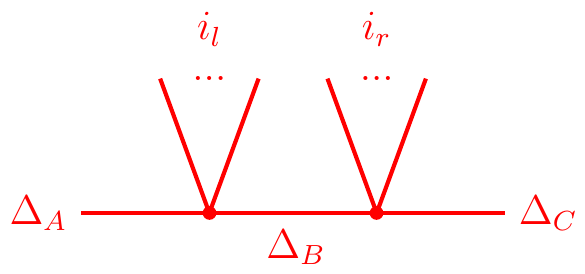}
 \end{matrix}
\right)
\cr
&
-\left[\zeta_p(s_B-\Delta_B)\zeta_p(s_C-\Delta_C)-\zeta_p(\Sigma)\lim_{s_A\rightarrow \infty}\right]
\left(
\begin{matrix}
\includegraphics[height=12ex]{figures/ThirdRecur3.pdf}
 \end{matrix}
\right)
\cr
&
-
\Big[\zeta_p(s_A-\Delta_A)\zeta_p(s_B-\Delta_B)\zeta_p(s_C-\Delta_C)-\zeta_p(\Sigma)\Big],
}
where for the last line we used the fact that the amplitude of an undressed diagram with no internal lines is precisely unity.
\end{example}

\subsection{$p$-adic on-shell recursion relation and proof}
\label{PadicBCFW}
Motivated by the factorization properties and the on-shell recursive structure observed for simple examples  in appendix \ref{BCFWMOM}, we propose the following recursive prescription for $p$-adic Mellin amplitudes:

\begin{prescription}[On-shell recursion for $p$-adic Mellin amplitudes]
\label{pres:BCFW}
The following recursive relation is obeyed by $p$-adic Mellin amplitudes, 
\eqn{mellinBCFW}
{
\mathcal{M}\big(\{s_I\}\big)
=-\!\!\!\! \sum_{\text{internal leg } I} \!\!\!\! \mathcal{M}_{\text{left}}\big(\{s_{\text{left}}\}\big) 
 \beta_p\big(s_I-\Delta_I,n-\sum_i \Delta_i\big)
\mathcal{M}_{\text{right}}\big(\{s_{\text{right}}\}\big)\bigg|_{\forall {J\neq I}:\: s_J \rightarrow s_J-s_I+\Delta_I}\!\!.
}
This formula is to be understood as follows. For an arbitrary tree-level bulk diagram:
\begin{itemize}
\item
For each internal leg $I$ with the associated Mandelstam invariant $s_I$, and which is propagating a scalar field of conformal dimension $\Delta_I$, include a \emph{``propagator''} factor of $\beta_p(s_I-\Delta_I,n-\sum_i \Delta_i)$ defined in \eno{BetaPDef}, times the Mellin amplitudes for the left and right sub-amplitudes separated by the internal leg $I$, where all Mandelstam invariants of the two sub-amplitudes have been shifted by $(-s_I+\Delta_I)$. In referring to the Mandelstam-dependence of the left (right) sub-amplitude as $\{s_{\text{left}}\}$ ($\{s_{\text{right}}\}$), we mean to say that the Mandelstam invariants of the sub-amplitude can be constructed entirely out of the Mellin variables $\gamma_{ij}$ and external scaling dimensions $\Delta_i$ indexed by numbers $i$ and $j$ that label external legs to the left (right) of the internal leg $I$ in question (and which, as described above, must be shifted further by $-s_I+\Delta_I$). 
\item We sum over all such terms (times an overall factor of minus one).
\end{itemize}
\end{prescription}

An alternate formulation of this prescription (in auxiliary momentum space) is provided in appendix~\ref{BCFWMOM}. Before demonstrating the equivalence between prescriptions~\ref{pres:BCFW} and~\ref{pres:padicMellin}, let us consider a few examples. The derivation of~\eno{mellinBCFW} for the case of four- and five-point diagrams is discussed in significant detail in appendix~\ref{BCFWMOM}, using the familiar technique of complexifying the amplitude and BCFW-shifting the (auxiliary) momenta.  We focus here on the final decomposition for two six-point examples to illustrate how to apply \eno{mellinBCFW}. 
\begin{example}
For the six-point series diagram, the on-shell recursion \eno{mellinBCFW} dictates
\eqn{6ptLinearPractical}{
&
\begin{matrix}
\includegraphics[height=14ex]{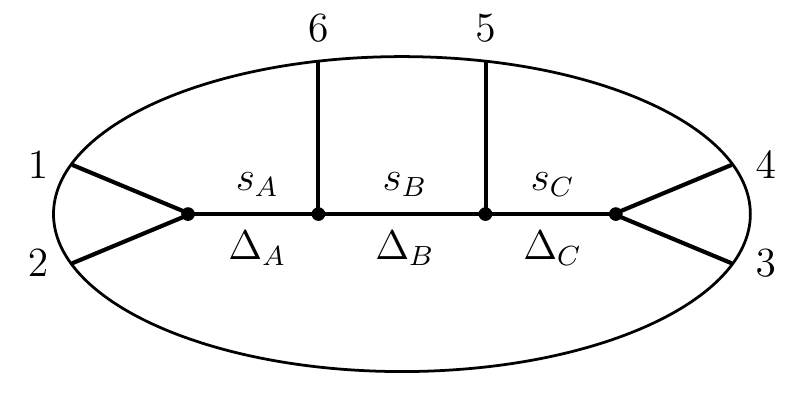}
 \end{matrix}
 =
 \cr
 &-
 \begin{matrix}
\includegraphics[height=11ex]{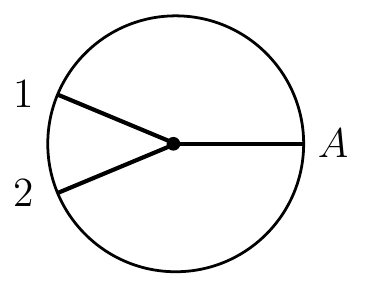}
 \end{matrix}
 \beta_p(s_A-\Delta_A,n-\Delta_\Sigma)
  \begin{matrix}\,
\includegraphics[height=14ex]{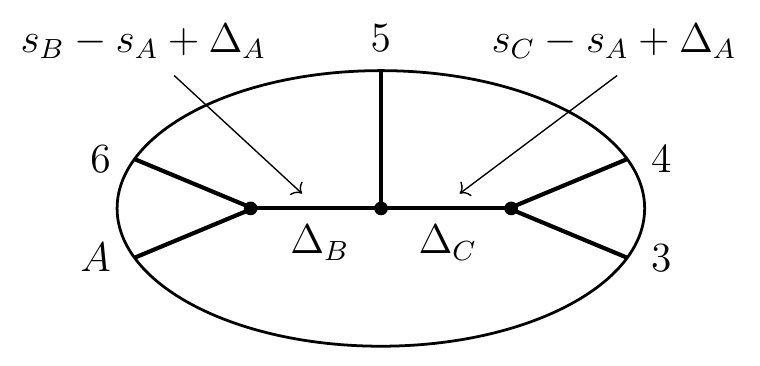}
 \end{matrix}
 \cr
 &-
 \begin{matrix}
\includegraphics[height=12ex]{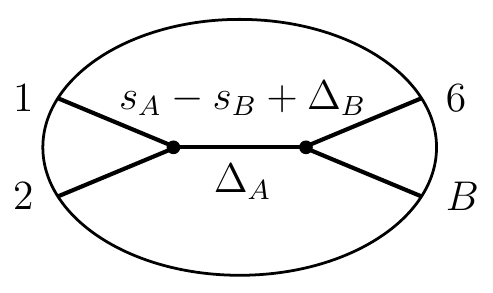}
 \end{matrix}\,
 \beta_p(s_B-\Delta_B,n-\Delta_\Sigma)\,
  \begin{matrix}
\includegraphics[height=12ex]{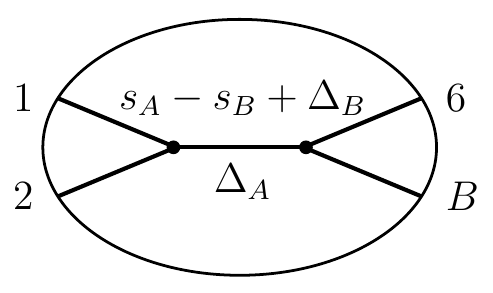}
 \end{matrix}
 \cr
 &-
 \begin{matrix}
\includegraphics[height=14ex]{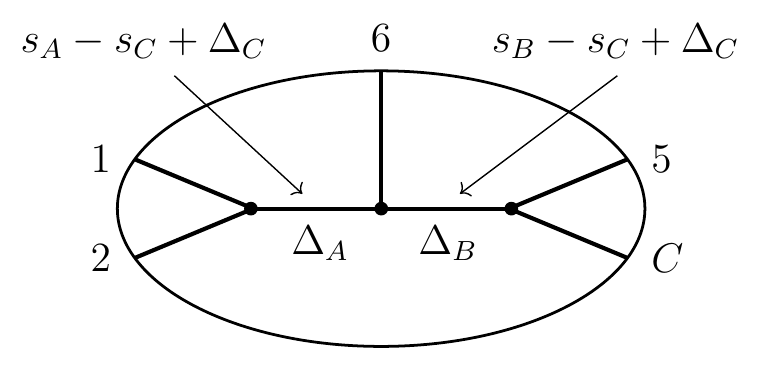}
 \end{matrix}\,
 \beta_p(s_C-\Delta_C,n-\Delta_\Sigma)\,
  \begin{matrix}
\includegraphics[height=11ex]{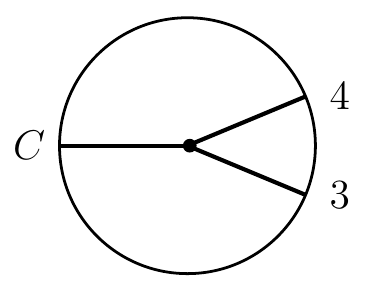}
 \end{matrix}\,,
 }
  where we have defined $\Delta_\Sigma \equiv \sum_{i=1}^6 \Delta_i$, 
and the Mandelstam variables are
\eqn{6ptLinearMand}{
s_A \equiv \Delta_1 + \Delta_2 - 2\gamma_{12}\,, \qquad s_C \equiv \Delta_3 + \Delta_4 - 2\gamma_{34}\,, \cr 
s_B \equiv s_A + \Delta_6  -2\gamma_{16}-2\gamma_{26} = s_C + \Delta_5 -2\gamma_{35}-2\gamma_{45}\,,
}
where the Mellin variables $\gamma_{ij}$ satisfy \eno{MellinVarConstraints} for ${\cal N}=6$.
Further, in \eno{6ptLinearPractical}, we have indicated the Mandelstam variables associated with the internal legs on top of the lines, and in the r.h.s.\ in the sub-amplitudes, we have explicitly displayed the shifted Mandelstam variables. 
\end{example}

\begin{example}
Likewise, for the six-point star diagram, \eno{mellinBCFW} gives

\clearpage
\eqn{6ptStarPractical}{
\begin{matrix}
\includegraphics[height=19ex]{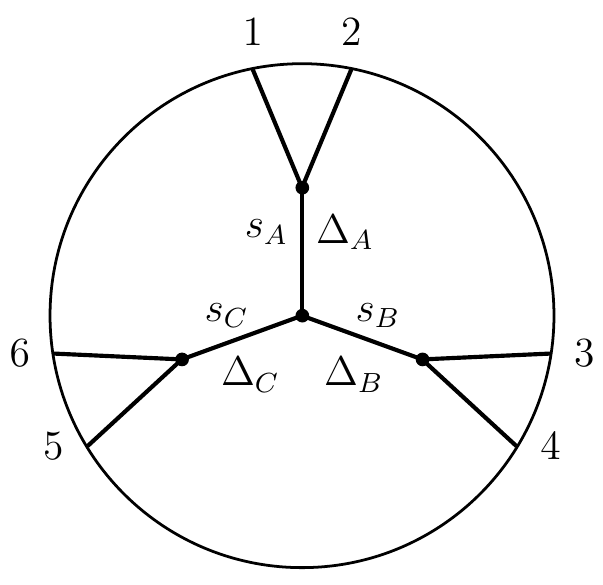}
 \end{matrix}
 =
 &-
 \begin{matrix}
\includegraphics[height=13ex]{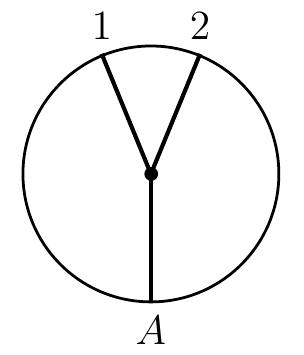}
 \end{matrix}
 \beta_p(s_A-\Delta_A,n-\Delta_\Sigma) \!\!\!
  \begin{matrix}
\includegraphics[height=14ex]{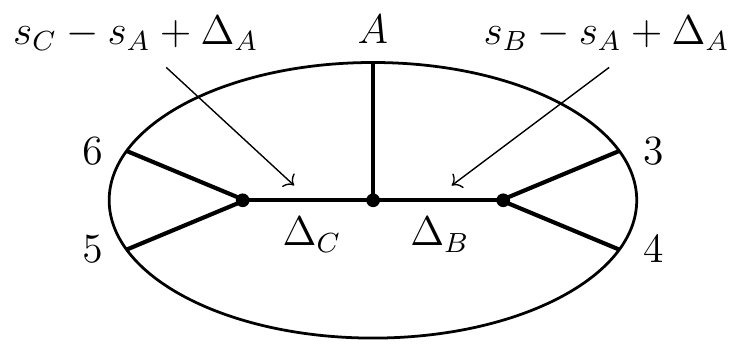}
 \end{matrix}
 \cr
 &-\!\!\!
  \begin{matrix}
\includegraphics[height=14ex]{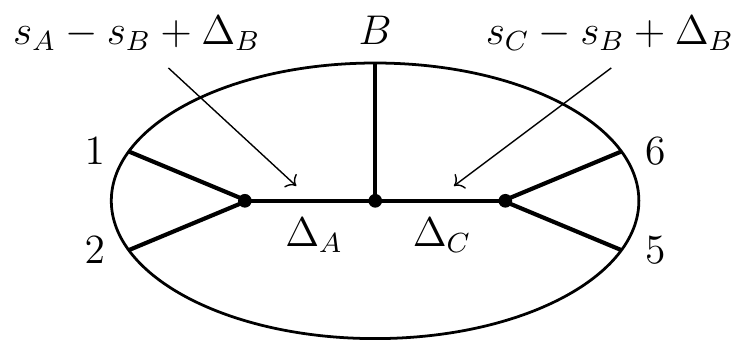}
 \end{matrix}
\beta_p(s_B-\Delta_B,n-\Delta_\Sigma)
 \begin{matrix}
\includegraphics[height=11ex]{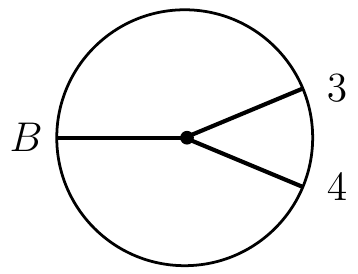}
 \end{matrix}
 \cr
 &-
 \begin{matrix}
\includegraphics[height=11ex]{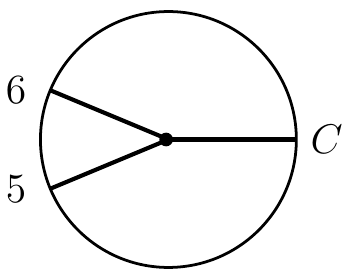}
 \end{matrix}\,
\beta_p(s_C-\Delta_C,n-\Delta_\Sigma)\!\!\!
  \begin{matrix}
\includegraphics[height=14ex]{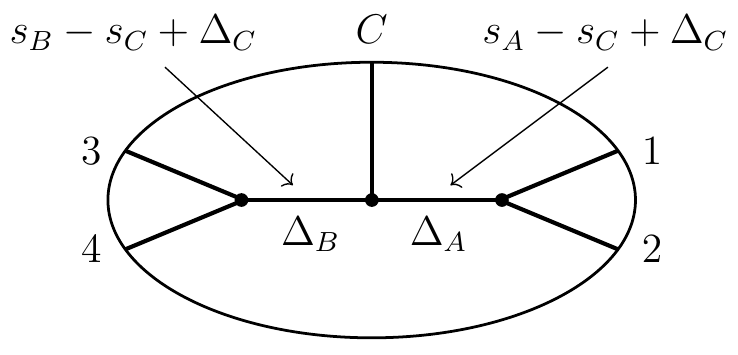}
 \end{matrix}\!\!\!,
 }
 where this time
 \eqn{6ptStarMand}{
 s_A \equiv \Delta_1+\Delta_2-2\gamma_{12}\,, \qquad s_B \equiv \Delta_{3}+\Delta_4-2\gamma_{34}\,, \qquad s_C \equiv \Delta_5+\Delta_6-2\gamma_{56}\,.
 }
\end{example} 
 Equations \eno{6ptLinearPractical} and \eno{6ptStarPractical} can be reduced further by repeated application of \eno{mellinBCFW} (see e.g.\ equations \eno{4ptPractical} and \eno{5ptPractical} in appendix \ref{BCFWMOM}) to expressions which involve only (in this case, three-point) contact interactions, whose Mellin amplitudes are given simply by the respective vertex factors \eno{vertexFactorDef}. It is straightforward to confirm that \eno{6ptLinearPractical} and \eno{6ptStarPractical} agree with the Mellin amplitudes for diagrams with three internal legs computed from first principles in Ref.~\cite{Jepsen:2018dqp} and rederived here in examples \ref{ex:sixptseries}-\ref{ex:sixptstar} using prescription~\ref{pres:padicMellin}.
 
 \vspace{1.5em}
 
We now prove that prescription \ref{pres:BCFW} is mathematically equivalent to the recursive prescription \ref{pres:padicMellin} described in section \ref{recursionSection}. (We will later prove prescription~\ref{pres:padicMellin} in appendix~\ref{AmpProof}.)
Starting with a Mellin amplitude $\mathcal{M}$ given in terms of the recursive prescription \ref{pres:padicMellin}, let $\mathcal{M}(z)$ be the complex deformation obtained by replacing each Mandelstam invariant $s_I$ with $s_I(z) \equiv s_I-z$. In that case the function, 
\eqn{Idef2}
{
\mathcal{I}(z) \equiv \log p\bigg[\zeta_p(z)-\zeta_p(\sum_i\Delta_i-n)\bigg] {\cal M}(z)\,,
}
is defined on a ``complex cylinder'' where the imaginary axis wraps around a circle of radius  $1/\log p$. In other words $z \in \mathbb{R} \times \left[-{\pi\over \log p}, {\pi \over \log p}\right)$. From here on, we will use the shorthand ${\cal M}={\cal M}(0)$. 
Consider now the sum
\eqn{contourSum}
{
S(r)\equiv
\int_{r-\frac{i\pi}{\log p}}^{r+\frac{i\pi}{\log p}} dz\,\mathcal{I}(z)+
\int_{-r+\frac{i\pi}{\log p}}^{-r-\frac{i\pi}{\log p}} dz\,\mathcal{I}(z),
}
where $r$ is some positive number. In the limit as $r$ tends to infinity, $S(r)$ tends to zero as we will now show. As the real part of $z$ tends to plus or minus infinity, the integrands in \eqref{contourSum} become constant, and the integrals evaluate to $2\pi /\log p$ (which is the circumference of the cylindrical manifold) times the constant integrand. This asymptotic behavior follows from the fact that $\mathcal{I}(z)$ depends on $z$ solely through the arguments of $p$-adic local zeta-functions, which themselves have the asymptotic behavior
\eqn{zetapAsymp}
{
\zeta_p(z) \rightarrow \begin{cases}
1 & \text{as } \Re [z]\rightarrow \infty\,,
\cr
0 & \text{as }\Re [z]\rightarrow -\infty\,.
\end{cases}
}
We see then that when $r \to \infty$, the first term in \eno{contourSum} gives
\eqn{firstContour}
{
\int_{\infty-\frac{i\pi}{\log p}}^{\infty+\frac{i\pi}{\log p}} dz\,\mathcal{I}(z)
&=2\pi\bigg[1-\zeta_p(\sum_i\Delta_i-n)\bigg]\mathcal{M}(z=\infty)
\cr
&=2\pi\bigg[1-\zeta_p(\sum_i\Delta_i-n)\bigg]\bigg(\text{the momentum-independent terms of }\mathcal{M}\bigg)\,.
}
For the second term in \eno{contourSum}, noting that the contour runs in the opposite direction, we get
\eqn{secondContour}
{
\int_{-\infty+\frac{i\pi}{\log p}}^{-\infty-\frac{i\pi}{\log p}} dz\,\mathcal{I}(z)
&=2\pi\, \zeta_p(\sum_i\Delta_i-n)\, \mathcal{M}(z=-\infty).
}
We split the expression \eqref{secondContour} into two parts:
\begin{enumerate}
\item A part proportional to the momentum-independent part of $\mathcal{M}$: this part exactly cancels the second term inside the square brackets in the second line of \eno{firstContour}.
\item A part proportional to the momentum-dependent part of $\mathcal{M}$: since $z$ is taken to minus infinity, all factors of $\zeta_p$ in ${\cal M}$ which carry Mandelstam variable dependence tend to unity.
Using equation \eno{momIndep}, this part is seen to exactly cancel with the first term inside square brackets in the second line of \eqref{firstContour}.
\end{enumerate}
We see then that $\lim_{r\rightarrow \infty}S(r)=0$. But from this it follows that the sum over all residues of ${\cal I}(z)$ defined in \eqref{Idef2} vanishes. For we may shift the contour of the first term in \eno{contourSum} from $\int_{r+\frac{i\pi}{\log p}}^{r-\frac{i\pi}{\log p}}$ to $\int_{-r+\frac{i\pi}{\log p}}^{-r-\frac{i\pi}{\log p}}$ so that this term cancels with the second term, provided we add in $2\pi i$ times the sum of all the residues in the strip $-r < \Re [z] < r$:
\begin{align}
 \begin{matrix}
 \vspace*{-1.cm}
\includegraphics[height=13ex]{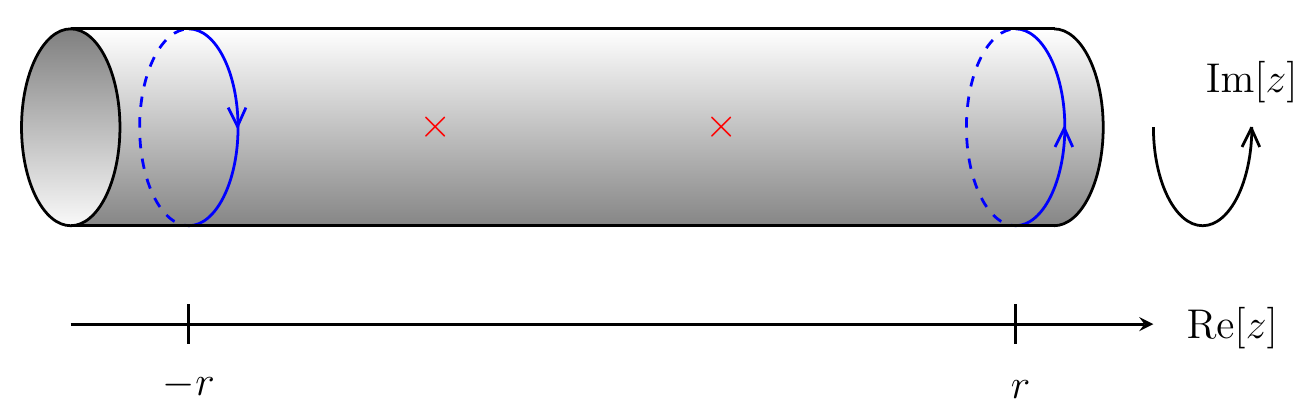}
 \end{matrix}
\,\,=
\,\,
 \begin{matrix}
 \vspace*{-1.cm}
\includegraphics[height=13ex]{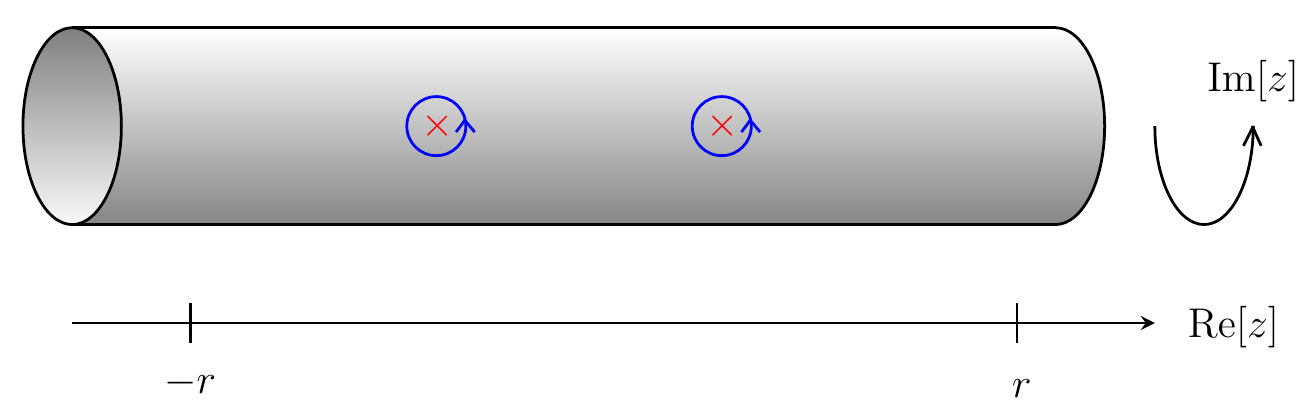}
 \end{matrix} \nonumber
\,.
\\ \nonumber
\end{align}
The residue of ${\cal I}(z)$ at $z=0$ equals ${\cal M} = \mathcal{M}(0)$, the original un-deformed amplitude. 
Since $\mathcal{M}(z)$ is obtained from ${\cal M}$  by replacing in $\mathcal{M}$ each factor of $\zeta_p(s_I-\Delta_I)$ with $\zeta_p(s_I-\Delta_I-z)$, we see that the remaining residues of $\mathcal{M}(z)$ occur at $z_\ast=s_I-\Delta_I$, exactly when the complex-shifted internal leg goes on shell, i.e.\  $s_I(z_\ast)=\Delta_I$. 
Furthermore, from~\eno{Idef2} we conclude that each of these residues is equal to a factor of $\bigg[\zeta_p(s_I-\Delta_I)-\zeta_p(\sum_i\Delta_i-n)\bigg] = \beta_p(s_I-\Delta_I,n-\sum_i \Delta_i)$ times the (on-shell) left and the right sub-amplitudes (evaluated at $z=z_\ast$), as follows from the factorization property described in section \ref{Factorization}.

Thus upon equating the residue of $\mathcal{I}(z)$ at $z=0$ with minus the sum of all remaining residues,  we recover the on-shell recursion formula \eqref{mellinBCFW}, proving prescription~\ref{pres:BCFW} assuming prescription~\ref{pres:padicMellin}. 

\vspace{1em}
Given the strong similarities between the structure of real and $p$-adic Mellin amplitudes as illuminated in this paper so far, it is natural to wonder if the on-shell recursion relations presented in this section also have a real counterpart. 
It turns out (a trivial restatement of) the Feynman rules for real Mellin amplitudes (as given in prescription~\ref{pres:Feyn}) provides the closest analog, with an important subtlety:
The on-shell recursion given in prescription~\ref{pres:BCFW} has a particularly simple form owing to the absence of descendant fields in the $p$-adic setup, which is no longer the case over the reals.
Thus~\eno{mellinBCFW} continues to hold for real Mellin amplitudes, provided we additionally assign to each internal leg $I$ an integer $m_I$, replace the propagator factor $\beta_p(\cdot,\cdot)$ above by the propagator factor $L(s_I,\Delta_I,m_I)$ defined in~\eno{lineFactorFeyn}, update the Mandelstam shift operation in~\eno{mellinBCFW} to be $s_J \to s_J -s_I+\Delta_I+2m_I$, and  finally include infinite sums over all integers $m_I \in [0,\infty)$ in~\eno{mellinBCFW}.
The fact that this prescription reduces to the Feynman rules trivially follows from the partial fraction identity,
\eqn{}
{
\frac{1}{D_1\ldots D_f}=\sum_{i=1}^f\frac{1}{D_i}\left[
\prod_{j \neq i}\frac{1}{D_j-D_i}
\right],
}
applied to the product of factors $1/(s_I-\Delta_I-2m_I)$ appearing in the propagator factors $L(s_I,\Delta_I,m_I)$.

\section{Outlook}
\label{DISCUSSION}

In this paper we completed the task, initiated in Ref.~\cite{Jepsen:2018dqp}, of computing  $p$-adic Mellin amplitudes for all tree-level, arbitrary-point bulk diagrams involving scalars. 
This is achieved by establishing various prescriptions for constructing Mellin amplitudes which overcome the need to perform bulk integrations altogether:
\begin{itemize}
    \item Prescription~\ref{pres:padicMellin} provides an effective computational strategy, recursive on the number of internal lines;
    \item Prescription~\ref{pres:BCFW} provides an even more effective ``on-shell'' strategy;
    \item Prescription~\ref{pres:Preamp} (along with~\eno{PreampRuleDiag}) provides a recipe for directly writing down the Mellin pre-amplitudes defined via Mellin-Barnes contour integrals in~\eno{preampDef}.
\end{itemize}
In fact, we showed further that the pre-amplitude prescription~\ref{pres:Preamp} of section~\ref{preAmpSec} applies simultaneously to both $p$-adic \emph{and} real Mellin pre-amplitudes, and provides a precise, unified recipe for obtaining the pre-amplitudes from full Mellin amplitudes.
This demonstrates the close ties between the real and $p$-adic formulations of AdS/CFT. 

While the real and $p$-adic Mellin amplitudes share many common properties, such as the recursion relations mentioned above, physical factorization properties, the analytic form of the pre-amplitudes, and the fact that they are meromorphic functions which carry simple poles corresponding to exchange of single-trace operators, the $p$-adic amplitudes are significantly simpler than the real amplitudes owing to the fact that the $p$-adic CFTs we are studying lack descendants~\cite{Melzer:1988he,Gubser:2017tsi}. 
The simplicity of $p$-adic Mellin amplitudes as compared to their real cousins is for example highlighted in the structure of the on-shell recursion relation~\ref{pres:BCFW} described in section~\ref{BCFW}, which is related to, though simpler than the
Feynman rules prescription~\ref{pres:Feyn} for real Mellin amplitudes. 
 Thus the $p$-adic formulation provides a promising, powerful computational tool for investigating the harder to work with real Mellin amplitudes.
 
Part of our motivation for studying $p$-adic Mellin space is the hope that the simplicity and tractability of the $p$-adic setting may lead to new inspiration and results for real Mellin amplitudes.
  We should like to think of the relation between real Mellin amplitudes and the corresponding real pre-amplitudes from section \ref{PreampRecursion} as an instantiation of this hope;  it may be worthwhile to also investigate whether such relations exist for spinning amplitudes and loop amplitudes. 
 
Another promising avenue is the study of $p$-adic Mellin amplitudes involving bulk fermions~\cite{Gubser:2018cha} and amplitudes at loop level. 
It is tempting to note that the real Mellin loop-amplitudes, especially in the Mellin-Barnes integral form for pre-amplitudes~\cite{Yuan:2017vgp,Yuan:2018qva}, are expressible in terms of the local zeta function $\zeta_\infty$,
which naturally suggests analogous candidates for $p$-adic loop amplitudes. It would be interesting to explore this connection more systematically.
We also wonder whether (possibly a generalization of) the recursion relations put forward in this paper extend to loop amplitudes. 

 In this paper we have restricted ourselves to computing Mellin amplitudes associated to individual bulk diagrams. 
 It remains to see whether it is possible to lift the on-shell recursion of section~\ref{BCFW}, which is applicable at the level of individual diagrams, to one which applies to total amplitudes, i.e.\  the sums over constitutive diagrams of a bulk theory, and to other bulk theories over reals. 
 Nevertheless, the techniques of this paper should still prove useful in extending the construction of the putative bulk dual of the  free $p$-adic $O(N)$ model, at least at tree-level~\cite{Gubser:2017tsi}.

On a closing note, we emphasize that at present the mounting number of remarkable similarities  between the conventional (real) and $p$-adic AdS/CFT formulations  are still mathematical curiosities which remain to be fully understood.
It is tempting to wonder whether one may think of the real and $p$-adic formulations  as ``local formulations'' of AdS/CFT at the place at infinity and the finite places, respectively. 
This immediately leads to the question~\cite{Freund:2017aqf}: Is there an adelic formulation of AdS/CFT, i.e.\ a ``global'' formulation which subsumes the real and $p$-adic formulations in a single field-independent (field in the mathematics sense) framework?
Such a unified formulation would constitute an important conceptual step forward in understanding as yet undiscovered mathematical aspects of AdS/CFT.

\section*{Acknowledgments}

C.\ B.\ J.\ and S.\ P.\ thank Steven S.\ Gubser, Matilde Marcolli, and Brian Trundy for useful discussions and encouragement. 
S.\ P.\ thanks Perimeter Institute for their kind hospitality while this work was in progress. 
The work of C.\ B.\ J. was supported in part by the Department of Energy under Grant No. DE-FG02-91ER40671, by the US NSF under Grant No. PHY-1620059, and by the Simons Foundation, Grant 511167 (SSG).
The work of S.\ P.\ was supported in part by Perimeter Institute for Theoretical Physics. Research at Perimeter Institute is supported by the Government of Canada through the Department of Innovation, Science and Economic Development and by the Province of Ontario through the Ministry of Research, Innovation and Science.

\appendix

\section{Useful Formulae and Conventions}
\label{NOTATION}
 
We now recall some useful functions and formulae from Ref.~\cite{Jepsen:2018dqp}, to which we refer for further details.
The characteristic function takes a $p$-adic argument, and is defined to be
\eqn{characteristicfn}{
\gamma_p(x)=\frac{2\log p}{2\pi i}\int_{\epsilon-\frac{i\pi}{2\log p}}^{\epsilon+\frac{i\pi}{2\log p}}d\gamma\,\frac{\zeta_p(2\gamma)}{|x|_p^{2\gamma}} = \begin{cases}
1 \quad \text{ for } |x|_p \leq 1, \\
0 \quad \text{ otherwise. }
\end{cases}
}
It plays the role of the Gaussian function for $p$-adic variables. 
If we let $\{S_1,...,S_{\mathcal{N}}\}$ denote a finite set of $p$-adic numbers and let $m$ be an index such that $|S_m|_p=\sup\{|S_1|_p,...,|S_\mathcal{N}|_p\}$, then it follows from the definition of the characteristic function that it satisfies the factorization property
\eqn{GammaFactorization}
{
\gamma_p(S_1)...\gamma_p(S_\mathcal{N}) = \gamma_p\big(S_m\big).
}
Furthermore, letting $\{x_1,...,x_{\mathcal{N}}\}$ denote another set of $p$-adic numbers of equal cardinality, it follows from the ultra-metricity of the $p$-adic norm that 
\eqn{prodgamma1}
{
\prod_{\substack{i=1\\i\neq m}}^\mathcal{N}\gamma_p\big(S_i (x_i-x_m)^2\big)=\prod_{1\leq i < j\leq \mathcal{N}}\gamma_p\bigg(\frac{S_iS_j}{S_m}(x_i-x_j)^2\bigg)\,.
}
An integral identity for characteristic functions that comes in handy in the context of $p$-adic Mellin amplitudes is
\eqn{characteristicIntegral}
{
\int_{\mathbb{Q}_{p^n}}dx\,\left[\prod_i\gamma_p\big(S_i(x-x_i)^2\big)\right]
=\frac{1}{|S_m|^{n/2}_p}\left[\prod_{i\neq m}\gamma_p\big(S_i(x_i-x_m)^2\big)\right].
}
Some useful integration formulae, which we collectively refer to as the ``$p$-adic Schwinger parameter trick'' are listed below:
\eqn{Schwinger}
{
\frac{1}{|x|_p^{\Delta}} = \frac{\zeta_p(n)}{\zeta_p(\Delta)} \int_{\mathbb{Q}_{p^n}}\frac{dS}{|S|^n_p}|S|_p^\Delta\gamma_p(xS)
\qquad {\rm for\ } x \in \mathbb{Q}_{p^n}\,,
}
\eqn{SchwingerSquare}
{
\frac{1}{|x|_p^{\Delta}}  = \frac{2}{|2|_p} \frac{\zeta_p(1)}{\zeta_p(2\Delta)} \int_{\mathbb{Q}^2_{p}}\frac{dS}{|S|_p}|S|_p^\Delta\gamma_p(xS)
\qquad \text{for }x\in \mathbb{Q}^2_{p}\,,
}
\eqn{SchwingerNotSquare}
{
\frac{1}{|x|_p^{\Delta}} = \frac{2}{|2|_p} \frac{\zeta_p(1)}{\zeta_p(2\Delta)} \int_{p\mathbb{Q}^2_{p}}\frac{dS}{|S|_p}|S|_p^\Delta\gamma_p(xS)
 \qquad \text{for }x\in p\mathbb{Q}^2_{p}\,,
}
 where
 \eqn{square}
{
\mathbb{Q}_p^2 \equiv \{x\in \mathbb{Q}_p \,\, : \,\, x=y^2 {\rm  \ for\ some\ } y\in\mathbb{Q}_p  \}\,.
}
\eqn{psquare}
{
p\mathbb{Q}_p^2 \equiv \{x\in \mathbb{Q}_p \,\, : \,\, x=p y^2 {\rm  \ for\ some\ } y\in\mathbb{Q}_p  \}\,.
}

The $p$-adic version of the Symanzik star integration formula takes the form,
\eqn{SymanzikStar}{
\int [d\gamma] \prod_{1\leq i<j\leq {\cal N}}\frac{\zeta_p(2\gamma_{ij})}{|x_{ij}|_p^{2\gamma_{ij}}} = 
   \sum_{a\in \{1,p\}}  \prod_{i=1}^{\cal N} \left( \frac{2\zeta_p(1)}{|2|_p} \int_{a\mathbb{Q}^2_p} \frac{ds_i}{|s_i|_p}|s_i|_p^{\Delta_i}\right) \prod_{1\leq i<j\leq {\cal N}}\gamma_p\bigg(s_is_jx_{ij}^2  \bigg).
}
An integral identity from which, with the help of the Symanzik formula, the contact amplitude may be extracted is,
\eqn{preMellin}{
\sum_{(z_0,z) \in {\cal T}_{p^n}} \prod_{i=1}^{\cal N} K_{\Delta_i}(z_0,z;x_i) =\sum_{a\in\{1,p\}}
\zeta_p\big(\sum_{i}\Delta_i-n\big)
\prod_{i=1}^{\cal N}\bigg(\frac{2\zeta_p(1)}{|2|_p}\int_{a\mathbb{Q}^2_p}\frac{ds_i}{|s_i|_p}|s_i|_p^{\Delta_i}\bigg)\prod_{i < j}\gamma_p\bigg(s_is_jx_{ij}^2  \bigg).
}
A $p$-adic integral that has appeared at some instances in this paper is
\eqn{supInt}{
\int_{\mathbb{Q}_p}\frac{dt}{|t|_p}|t|_p^a\left|1,t\right|^b_s=&
\sum_{n\in\mathbb{Z}}\int_{p^{-n}\mathbb{U}_p}\frac{dt}{|t|_p}|t|_p^a\left|1,t\right|^b_s
=\sum_{n=-\infty}^0\int_{p^{-n}\mathbb{U}_p}\frac{dt}{|t|_p}|t|_p^a
+\sum_{n=1}^\infty\int_{p^{-n}\mathbb{U}_p}\frac{dt}{|t|_p}|t|_p^{a+b}
\cr
=&\frac{1}{\zeta(1)}\bigg(\sum_{n=-\infty}^0p^{an} +\sum_{n=1}^\infty p^{(a+b)n} \bigg)=\frac{\zeta_p(a)-\zeta_p(a+b)}{\zeta_p(1)}\,.
}

\section{Inductive Proof of Prescription \ref{pres:Preamp} for Pre-Amplitudes}
\label{PreAmpProof}

\subsection{Outline of the proof}

In this section we prove that at tree-level any $p$-adic Mellin amplitude is given in terms of a multiple contour integral over a pre-amplitude  according to prescription \ref{pres:Preamp} of section \ref{preAmpSec}. Subsequently, in the next appendix we prove that the performing all the contour integrals over a pre-amplitude reproduces the Mellin amplitude given by the recursive prescription \ref{pres:padicMellin} presented in section \ref{recursionSectionPadic}.

The proof for the pre-amplitude prescription  is inductive. A possible inductive approach one might attempt to adopt is weak induction on the number of internal legs.
However, given the pre-amplitude for an arbitrary bulk diagram, it turns out to be technically difficult to find the pre-amplitude for a diagram obtained by inserting an extra internal leg at an arbitrary position in the diagram. 
It is significantly more tractable to attach extra internal legs to a special kind of bulk vertex, one that only has one internal leg attached to it. 
Thus our inductive strategy will be to assume that the pre-amplitude prescription~\ref{pres:Preamp} is satisfied by  an arbitrary bulk diagram $D_L$, and a bulk diagram $D_R$ built from $f$ internal lines all connected to the same vertex:
\eqn{LeftAndRight}{
D_L =
\begin{matrix}
\text{
\begin{tikzpicture}
\draw[thick,fill=black] (0,0) ellipse (0.05cm and 0.05cm);
\node at (0,0.4) {$V_1$}; 
\node at (1.3,0.4) {$R_0$};
\draw[thick,fill=black] (1.3,0) ellipse (0.05cm and 0.05cm);
\draw[very thick] (0,0)--(1.3,0);
\draw[thick,fill=black] (-0.93,1.82) ellipse (0.05cm and 0.05cm);
\draw[thick,fill=black] (-2.03,0.72) ellipse (0.05cm and 0.05cm);
\draw[very thick] (0,0)--(-0.93,0.72);
\draw[very thick] (-0.93,0.72)--(-0.93,1.82);
\draw[very thick] (-0.93,0.72)--(-2.03,0.72);
\draw[thick,fill=black] (-0.93,-1.82) ellipse (0.05cm and 0.05cm);
\draw[thick,fill=black] (-2.03,-0.72) ellipse (0.05cm and 0.05cm);
\draw[very thick] (0,0)--(-0.93,-0.72);
\draw[very thick] (-0.93,-0.72)--(-0.93,-1.82);
\draw[very thick] (-0.93,-0.72)--(-2.03,-0.72);
\draw[thick] (-0.5,0) ellipse (2.3cm and 2.3cm);
\draw (-0.609,0) circle (0.2mm)[fill=black];
\draw (-0.5684,-0.203) circle (0.2mm)[fill=black];
\draw (-0.5684,0.203) circle (0.2mm)[fill=black];
\draw[thick] (-0.93,-0.72) circle (1.5mm)[fill=light-gray];
\draw[thick] (-0.93,0.72) circle (1.5mm)[fill=light-gray];
\draw (-1.33,1.12) circle (0.2mm)[fill=black];
\draw (-1.13,1.25) circle (0.2mm)[fill=black];
\draw (-1.46,0.93) circle (0.2mm)[fill=black];
\draw (-1.33,-1.12) circle (0.2mm)[fill=black];
\draw (-1.13,-1.25) circle (0.2mm)[fill=black];
\draw (-1.46,-0.93) circle (0.2mm)[fill=black];
\end{tikzpicture}
}
\end{matrix}
\,,
\hspace{20mm}
D_R=
\begin{matrix}
\text{
\begin{tikzpicture}
\draw[thick,fill=black] (0,0) ellipse (0.05cm and 0.05cm);
\draw[thick,fill=black] (0.93,0.72) ellipse (0.05cm and 0.05cm);
\draw[thick,fill=black] (0.93,-0.72) ellipse (0.05cm and 0.05cm);
\draw[very thick] (0,0)--(0.93,0.72);
\draw[very thick] (0,0)--(0.93,-0.72);
\draw[thick] (0.6,0) ellipse (1.6cm and 1.6cm);
\draw (1.709-1,0) circle (0.2mm)[fill=black];
\draw (1.6684-1,-0.203) circle (0.2mm)[fill=black];
\draw (1.6684-1,0.203) circle (0.2mm)[fill=black];
\node at (-0.45,0) {$R_0$};
\node at (0.93,1.12) {$R_1$};
\node at (0.93,-1.17) {$R_f$};
\end{tikzpicture}
}
\end{matrix}
\,.
}
Here the grey circles each indicate an arbitrary number of vertices and internal lines that form part of the diagram $D_L$, and we have omitted drawing the external lines, of which an arbitrary number may be connected to any vertex. We will then consider the bulk diagram $D_M$ obtained by merging the two diagrams by fusing the two vertices labelled $R_0$ in \eno{LeftAndRight}, to get
\eqn{MergedDiagram}
{
D_M=
\begin{matrix}
\text{
\begin{tikzpicture}
\draw[thick] (0.2,0) ellipse (3.1cm and 2.3cm);
\draw[thick,fill=black] (0,0) ellipse (0.05cm and 0.05cm);
\node at (0,0.4) {$V_1$}; 
\node at (1.3,0.4) {$R_0$};
\draw[thick,fill=black] (1.3,0) ellipse (0.05cm and 0.05cm);
\draw[very thick] (0,0)--(1.3,0);
\draw[thick,fill=black] (-0.93,1.82) ellipse (0.05cm and 0.05cm);
\draw[thick,fill=black] (-2.03,0.72) ellipse (0.05cm and 0.05cm);
\draw[very thick] (0,0)--(-0.93,0.72);
\draw[very thick] (-0.93,0.72)--(-0.93,1.82);
\draw[very thick] (-0.93,0.72)--(-2.03,0.72);
\draw[thick,fill=black] (-0.93,-1.82) ellipse (0.05cm and 0.05cm);
\draw[thick,fill=black] (-2.03,-0.72) ellipse (0.05cm and 0.05cm);
\draw[very thick] (0,0)--(-0.93,-0.72);
\draw[very thick] (-0.93,-0.72)--(-0.93,-1.82);
\draw[very thick] (-0.93,-0.72)--(-2.03,-0.72);
\draw (-0.609,0) circle (0.2mm)[fill=black];
\draw (-0.5684,-0.203) circle (0.2mm)[fill=black];
\draw (-0.5684,0.203) circle (0.2mm)[fill=black];
\draw[thick] (-0.93,-0.72) circle (1.5mm)[fill=light-gray];
\draw[thick] (-0.93,0.72) circle (1.5mm)[fill=light-gray];
\draw (-1.33,1.12) circle (0.2mm)[fill=black];
\draw (-1.13,1.25) circle (0.2mm)[fill=black];
\draw (-1.46,0.93) circle (0.2mm)[fill=black];
\draw (-1.33,-1.12) circle (0.2mm)[fill=black];
\draw (-1.13,-1.25) circle (0.2mm)[fill=black];
\draw (-1.46,-0.93) circle (0.2mm)[fill=black];
\draw[thick,fill=black] (0.93+1.3,0.72) ellipse (0.05cm and 0.05cm);
\draw[thick,fill=black] (0.93+1.3,-0.72) ellipse (0.05cm and 0.05cm);
\draw[very thick] (0+1.3,0)--(0.93+1.3,0.72);
\draw[very thick] (0+1.3,0)--(0.93+1.3,-0.72);
\draw (1.709-1+1.3,0) circle (0.2mm)[fill=black];
\draw (1.6684-1+1.3,-0.203) circle (0.2mm)[fill=black];
\draw (1.6684-1+1.3,0.203) circle (0.2mm)[fill=black];
\node at (0.93+1.3,1.12) {$R_1$};
\node at (0.93+1.3,-1.17) {$R_f$};
\end{tikzpicture}
}
\end{matrix}
\,,
}
where the merged diagram has a set of external legs which is the union of the set of external legs for $D_L$ and $D_R$.
Using the inductive assumptions we will demonstrate that this bulk-diagram has a Mellin amplitude given by the pre-amplitude prescription~\ref{pres:Preamp}, thereby completing the inductive proof.

\subsection{Details of the proof}
 
\paragraph{The ``left-hand diagram'' $D_L$.}

As a first step in the proof, we will recast the pre-amplitude for $D_L$ in a more useful form. To this end, it is useful to first consider a smaller diagram $D_l$, namely the one obtained by removing the vertex $R_0$ and the internal line connected to it (i.e.\ collapsing the internal line joining $R_0$ to $V_1$ in $D_L$). We keep the number and dimensions of external legs attached to vertex $V_1$ in $D_l$ arbitrary for now but fix them later. The smaller diagram takes the following form, where this time we show a subset of the external legs,
\eqn{recurdia1}
{
D_l =
 \begin{matrix}
\includegraphics[height=42ex]{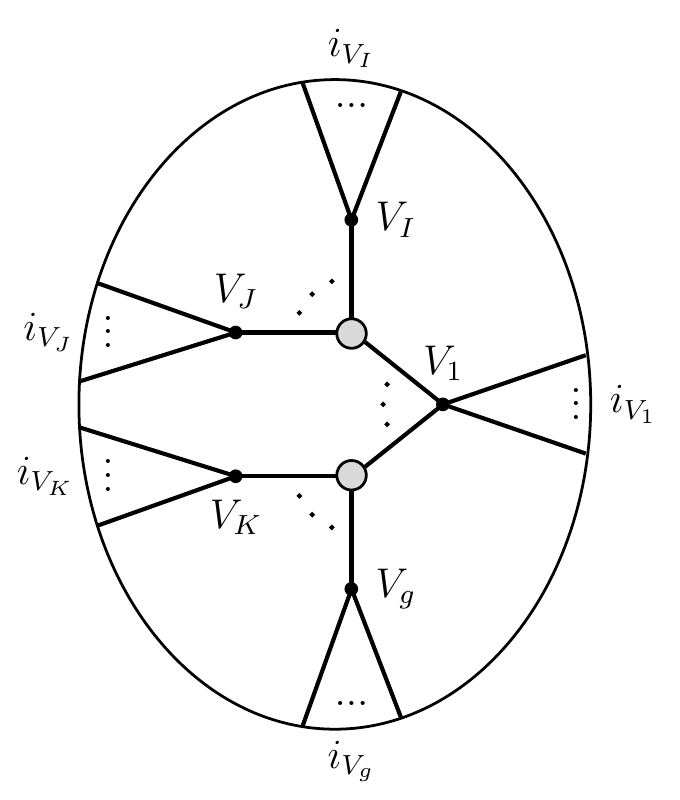}
 \end{matrix}.
 }
The vertices of this diagram 
are labelled $V_1$, $V_2$, \ldots, $V_g$, and the external legs are labelled by the lower-case alphabet $i$ or $j$, where we sometimes put a subscript on them to indicate which vertex they are incident on. Specifically the label $i_{V_I}$ runs over the external legs incident on vertex $V_I$, and  $i_V$ runs over all labels $i_{V_1}$, $i_{V_2}$, \ldots, $i_{V_g}$. We posit that the position space pre-amplitude $\widetilde{\mathcal{A}}_l$ for $D_l$, denoted with a tilde and defined via 
\eqn{}{ 
{\cal A}(x) \equiv \left(\prod_I \int {dc \over 2\pi i} f_{\Delta_I}(c_I) \right) \widetilde{\cal A}(x)\,,
}
 can be written in the form\footnote{We are allowing for the possibility of vertices $V_I$ with no external legs incident to them. Such vertices have no associated variables $s_{i_{V_I}}$ as $i_{V_I}$ runs over the empty set. But the total number of $t_L$ variables remains $2g-2$ where $g$ is the total number of vertices, with or without incident external legs.}
\eqn{inductionstart}
{
\widetilde{\mathcal{A}}_l=&\mathcal{L} \left[\prod_{L=1}^{2g-2}\int_{\mathbb{Q}_p}\frac{dt_L}{|t_L|_p}
 \right]\mathcal{G}(\{t_L\}) 
\sum_{a\in\{1,p\}} \left[ \prod_{i_V} \frac{2\zeta_p(1)}{|2|_p}\int_{a\mathbb{Q}_p^2}\frac{ds_{i_V}}{|s_{i_V}|_p}|s_{i_V}|_p^{\sum\Delta_{i_V}} \right]
\cr 
&\quad \times 
  \left[ \prod_{1\leq I \leq J \leq g}\prod_{i_{V_I},i_{V_J}}\gamma_p\bigg(f_{IJ}(\{t_L\})s_{i_{V_I}}s_{i_{V_J}}x_{i_{V_I}i_{V_J}}^2\bigg) \right] \,
}
where $\mathcal{G}(\{t_L\})$ and $f_{IJ}(\{t_L\})$ are, respectively, real and $p$-adic valued functions whose exact form we will not need, except for the fact that the functions $f_{IJ}$ are valued in $\mathbb{Q}_p^2$. $\mathcal{L}$ and  $\mathcal{G}(\{t_L\})$ also depend on external scaling dimensions $\Delta_{i_V}$, the degree of the field extension $n$, and complex variables $c_I$, but we suppress such dependencies.
The fact that the $\widetilde{\mathcal{A}}_l$ has the form \eno{inductionstart} can be seen inductively. For the initial step, we note that equations (4.34) and (4.48) in Ref.~\cite{Jepsen:2018dqp} show that the diagrams with one and two internal legs have a position space pre-amplitude of the form \eno{inductionstart}. We will now assume \eno{inductionstart} and show that the diagram $D_L$ (with one additional internal leg) has a pre-amplitude of the same form as well. 

Applying the split representation \eno{Gsplit} to the internal leg connecting vertices $V_1$ and $R_0$ in $D_L$, 
\eqn{}{ 
D_L =
\int {dc \over 2\pi i} f_{\Delta}(c) \int_{\mathbb{P}^1(\mathbb{Q}_{p^n})} dx 
\begin{matrix}
\text{
\begin{tikzpicture}
\draw[very thick] (0,0)--(0,-2.6);
\draw[very thick] (0,0)--(0.8,-2.5);
\node at (0.4,0.) {$V_1$}; 
\node at (1.3,0) {$R_0$};
\node at (0.35,-2.3)  {$...$};
\node at (0.5,-2.9)  {$i_{V_1}$};
\draw[thick] (0.,0) ellipse (2.8cm and 2.6cm);
\draw[very thick] (0,0)--(1.75/2,2.48);
\draw[very thick] (1.78,0)--(1.75/2,2.48);
\draw[thick,fill=black] (1.75,0) ellipse (0.05cm and 0.05cm);
\draw[very thick] (1.78,0)--(2.71,0.63);
\draw[very thick] (1.78,0)--(2.71,-0.63);
\node at (2.5,0.1)  { \vdots};
\node at (1.75/2+0.05,2.78) {$x$};
\node at (3.2,0)  {$i_{R_0}$};
\node at (-0.11,1.3) {{\small $h-c$}}; 
\node at (1.88,1.3) {{\small $h+c$}};
\draw[thick,fill=black] (0,0) ellipse (0.05cm and 0.05cm);
\draw[thick,fill=black] (-0.93,1.82) ellipse (0.05cm and 0.05cm);
\node at (-1.,2.8) {$i_{V_I}$};
\node at (-0.95,-2.8) {$i_{V_g}$};
\draw[thick,fill=black] (-2.03,0.72) ellipse (0.05cm and 0.05cm);
\node at (-3.1,0.72) {$i_{V_J}$};
\node at (-3.1,-0.72) {$i_{V_K}$};
\node at (-2.46,0.8)  {{\small \vdots}};
\node at (-0.93,2.22)  {{\small ...}};
\draw[very thick] (-0.93,1.82)--(-0.93+0.63*0.8,1.82+0.93*0.8);
\draw[very thick] (-0.93,1.82)--(-0.93-0.63*0.53,1.82+0.93*0.53);
\draw[very thick] (-2.03,0.72)--(-2.03-0.93*0.56,0.72+0.63*0.56);
\draw[very thick] (-2.03,0.72)--(-2.03-0.93*0.82,0.72-0.63*0.82);
\draw[thick,fill=black] (-0.93,-1.82) ellipse (0.05cm and 0.05cm);
\draw[thick,fill=black] (-2.03,-0.72) ellipse (0.05cm and 0.05cm);
\node at (-2.46,-0.62)  {{\small \vdots}};
\node at (-0.93,-2.22)  {{\small ...}};
\draw[very thick] (-0.93,-1.82)--(-0.93+0.63*0.8,-1.82-0.93*0.8);
\draw[very thick] (-0.93,-1.82)--(-0.93-0.63*0.53,-1.82-0.93*0.53);
\draw[very thick] (-2.03,-0.72)--(-2.03-0.93*0.56,-0.72-0.63*0.56);
\draw[very thick] (-2.03,-0.72)--(-2.03-0.93*0.82,-0.72+0.63*0.82);
\draw[very thick] (0,0)--(-0.93,0.72);
\draw[very thick] (-0.93,0.72)--(-0.93,1.82);
\draw[very thick] (-0.93,0.72)--(-2.03,0.72);
\draw[very thick] (0,0)--(-0.93,-0.72);
\draw[very thick] (-0.93,-0.72)--(-0.93,-1.82);
\draw[very thick] (-0.93,-0.72)--(-2.03,-0.72);
\draw (-0.609,0) circle (0.2mm)[fill=black];
\draw (-0.5684,-0.203) circle (0.2mm)[fill=black];
\draw (-0.5684,0.203) circle (0.2mm)[fill=black];
\draw[thick] (-0.93,-0.72) circle (1.5mm)[fill=light-gray];
\draw[thick] (-0.93,0.72) circle (1.5mm)[fill=light-gray];
\draw (-1.33,1.12) circle (0.2mm)[fill=black];
\draw (-1.13,1.25) circle (0.2mm)[fill=black];
\draw (-1.46,0.93) circle (0.2mm)[fill=black];
\draw (-1.33,-1.12) circle (0.2mm)[fill=black];
\draw (-1.13,-1.25) circle (0.2mm)[fill=black];
\draw (-1.46,-0.93) circle (0.2mm)[fill=black];
\end{tikzpicture}
}
\end{matrix}\,,
}
we find that the position space pre-amplitude $\widetilde{\mathcal{A}}_L$ is given by a boundary integral over the product of 1)~the pre-amplitude $\widetilde{\cal A}_l$ as written in \eno{inductionstart} with appropriately chosen external dimensions incident at $V_1$, and 2)~a contact amplitude associated to the vertex $R_0$. 
We fix the dimension of one of the external legs incident at $V_1$ in $D_l$ to a complex-shifted value consistent with the split representation, and assign the rest of the external legs the dimensions of the external operators originally incident on $V_1$ in $D_L$. Label the complex-shifted dimension $\Delta_1$.
 We define 
\eqn{}
{
\widetilde{\mathcal{L}}=\mathcal{L}\big|_{\Delta_1 \rightarrow h-c}, \hspace{10mm}
\widetilde{\mathcal{G}}(\{t_L\})=\mathcal{G}(\{t_L\})\big|_{\Delta_1 \rightarrow h-c}\,, 
}
and rename the integration variable $s_1$ associated to this external leg to $t$.
Furthermore, we will now take the set $\{i_{V_1}\}$ to not include the index $i=1$ associated to this leg. For the contact amplitude associated to the vertex $R_0$, we apply the identity \eno{preMellin}. Again we must assign a complex value to the dimension of an external leg consistent with the split representation, and we  call the integration variable associated with this external leg $u$, while the other external legs (with real scaling dimensions) are labelled by an index $i_{R_0}$. The split representation then leads to
\eqn{}
{
\widetilde{\mathcal{A}}_L= 
\int_{\mathbb{Q}_{p^n}} &dx\,
\widetilde{\mathcal{L}}
\bigg[\prod_{L=1}^{2g-2}\int_{\mathbb{Q}_p}\frac{dt_L}{|t_L|_p}\bigg]\widetilde{\mathcal{G}}(\{t_L\}) 
\cr
&\times 
\sum_{a\in\{1,p\}}
\bigg[\prod_{i_V}\frac{2\zeta_p(1)}{|2|_p}\int_{a\mathbb{Q}_p^2}\frac{ds_{i_V}}{|s_{i_V}|_p}|s_{i_V}|_p^{\sum\Delta_{i_V}}\bigg]
\frac{2\zeta_p(1)}{|2|_p}\int_{a\mathbb{Q}_p^2}\frac{dt}{|t|_p}|t|_p^{h-c}
\cr
&\times 
\bigg[\prod_{1\leq I \leq J \leq g}\prod_{i_{V_I},i_{V_J}}\gamma_p\bigg(f_{IJ}(t_L)s_{i_{V_I}}s_{i_{V_J}}x_{i_{V_I}i_{V_J}}^2\bigg)\bigg]
\cr
&\times
\bigg[\prod_{1\leq I \leq g}\prod_{i_{V_I}}\gamma_p\big(f_{1I}(t_L)ts_{i_{V_I}}(x-x_{i_{V_I}})^2\big)\bigg]
\cr
&\times
\zeta_p\big(\sum\Delta_{i_{R_0}}-h+c\big)
\cr
&\times
\sum_{b\in\{1,p\}} 
\bigg[\prod_{i_{R_0}}\frac{2\zeta_p(1)}{|2|_p}\int_{b\mathbb{Q}_p^2}\frac{ds_{i_{R_0}}}{|s_{i_{R_0}}|_p}|s_{i_{R_0}}|_p^{\sum\Delta_{i_{R_0}}}\bigg]
\frac{2\zeta_p(1)}{|2|_p}\int_{b\mathbb{Q}_p^2}\frac{du}{|u|_p}|u|_p^{h+c}
\cr
&\times
\bigg[\prod_{i_{R_0}<j_{R_0}}\gamma_p(s_{i_{R_0}}s_{j_{R_0}}x_{i_{R_0}j_{R_0}}^2)\bigg]
\bigg[\prod_{i_{R_0}}\gamma_p\big(us_{i_{R_0}}(x-x_{i_{R_0}})^2\big)\bigg].
}
Performing the changes of variables $s_{i_{R_0}} \rightarrow \displaystyle \frac{s_{i_{R_0}}}{u}$ and $s_{i_{V_I}} \rightarrow \displaystyle \frac{s_{i_{V_I}}}{f_{1I}(t_L)\,t}$ and
letting $m$ denote an index such that $|s_m|_p=\text{sup}(|s_{i_{V_1}}|_p,|s_{i_{V_2}}|_p,\ldots|s_{i_{V_g}}|_p,|s_{i_{R_0}}|_p)$, we can carry out the $x$ integral with the integral formula \eno{characteristicIntegral}. Applying this formula, and re-expressing the domains of integration of $t$ and $u$ to all of $\mathbb{Q}_p$ at the cost of two factors of $\frac{2}{|2|_p}$, we find that 
\eqn{}
{
\widetilde{\mathcal{A}}_L&=
\widetilde{\mathcal{L}}\bigg[\prod_{L=1}^{2g-2}\int_{\mathbb{Q}_p}\frac{dt_L}{|t_L|_p}\bigg]\widetilde{\mathcal{G}}(\{t_L\})
\cr
&\quad\times
\bigg[\prod_{i_V}\frac{2\zeta_p(1)}{|2|_p}\int_{\mathbb{Q}_p^2}\frac{ds_{i_V}}{|s_{i_V}|_p}|s_{i_V}|_p^{\sum\Delta_{i_V}}\bigg]
\zeta_p(1)\int_{\mathbb{Q}_p}\frac{dt}{|t|_p}|t|_p^{-\sum\Delta_{i_V}+h-c}
\cr
&\quad\times
\bigg[\prod_{1\leq I \leq J \leq g}\prod_{i_{V_I},i_{V_J}}\gamma_p\bigg(\frac{s_{i_{V_I}}s_{i_{V_J}}f_{IJ}(t_L)}{t^2f_{1I}(t_L)f_{1J}(t_L)}x_{i_{V_I}i_{V_J}}^2\bigg)\bigg]
\cr
&\quad\times
\bigg(\prod_{1\leq I \leq g}|f_{1I}(t_L)|_p^{-\sum\Delta_{i_{V_I}}}\bigg[\prod_{i_{V_I}}\gamma_p\big(s_{i_{V_I}}(x_m-x_{i_{V_I}})^2\big)\bigg]\bigg)
\cr
&\quad\times
\zeta_p\big(\sum\Delta_{i_{R_0}}-h+c\big)
\cr
&\quad\times
\bigg[\prod_{i_{R_0}}\frac{2\zeta_p(1)}{|2|_p}\int_{\mathbb{Q}_p^2}\frac{ds_{i_{R_0}}}{|s_{i_{R_0}}|_p}|s_{i_{R_0}}|_p^{\sum\Delta_{i_{R_0}}}\bigg]
\zeta(1)\int_{\mathbb{Q}_p}\frac{du}{|u|_p}|u|_p^{-\sum\Delta_{i_{R_0}}+h+c}
\cr
&\quad\times
\bigg[\prod_{i_{R_0}<j_{R_0}}
\gamma_p\left(\frac{s_{i_{R_0}}s_{j_{R_0}}}{u^2}x_{i_{R_0}j_{R_0}}^2\right)\bigg]
\bigg[\prod_{i_{R_0}}\gamma_p\big(s_{i_{R_0}}(x_m-x_{R_0})^2\big)\bigg]
|s_m|_p^{-h}.
}
Next,  we change variables $t\rightarrow t \sqrt{s_m}$ and $u\rightarrow u \sqrt{s_m}$, and then change variables $s_i \rightarrow s_i \sqrt{s_m}$ for all $i\neq m$, and lastly change variable $s_m \rightarrow s_m^2$: 
\eqn{utrolig}
{
\widetilde{\mathcal{A}}_L&=
\widetilde{\mathcal{L}}
\bigg[\prod_{L=1}^{2g-2}\int_{\mathbb{Q}_p}\frac{dt_L}{|t_L|_p}\bigg]
\widetilde{\mathcal{G}}(\{t_L\})
\cr
&\quad\times
\sum_{a\in\{1,p\}}
\bigg[\prod_{i_V}\frac{2\zeta_p(1)}{|2|_p}\int_{a\mathbb{Q}_p^2}\frac{ds_{i_V}}{|s_{i_V}|_p}|s_{i_V}|_p^{\sum\Delta_{i_V}}\bigg]
\zeta_p(1)\int_{\mathbb{Q}_p}\frac{dt}{|t|_p}|t|_p^{-\sum\Delta_{i_V}+h-c}
\cr
&\quad\times
\bigg(\prod_{1\leq I \leq J \leq g}
\bigg[\prod_{i_{V_I},i_{V_J}}\gamma_p\bigg(\frac{s_{i_{V_I}}s_{i_{V_J}}f_{IJ}(t_L)}{t^2f_{1I}(t_L)f_{1J}(t_L)}x_{i_{V_I}i_{V_J}}^2\bigg)\bigg]\bigg)
\cr
&\quad\times
\bigg(\prod_{1\leq I \leq g}|f_{1I}(t_L)|_p^{-\sum\Delta_{i_{V_I}}}
\bigg[\prod_{i_{V_I}}\gamma_p\big(s_{i_{V_I}}s_m(x_m-x_{i_{V_I}})^2\big)\bigg]\bigg)
\cr
&\quad\times
\zeta_p\big(\sum\Delta_{i_{R_0}}-h+c\big)
\cr
&\quad\times
\bigg[\prod_{i_{R_0}}\frac{2\zeta_p(1)}{|2|_p}\int_{a\mathbb{Q}_p^2}\frac{ds_{i_{R_0}}}{|s_{i_{R_0}}|_p}|s_{i_{R_0}}|_p^{\sum\Delta_{i_{R_0}}}\bigg]
\zeta_p(1)\int_{\mathbb{Q}_p}\frac{du}{|u|_p}|u|_p^{-\sum\Delta_{i_{R_0}}+h+c}
\cr
&\quad\times
\bigg[\prod_{i_{R_0}<j_{R_0}}
\gamma_p\left(\frac{s_{i_{R_0}}s_{j_{R_0}}}{u^2}x_{i_{R_0}j_{R_0}}^2\right)\bigg]
\bigg[\prod_{i_{R_0}}\gamma_p\big(s_{i_{R_0}}s_m(x_m-x_{R_0})^2\big)\bigg].
}
The above expression for $\widetilde{\mathcal{A}}_L$ exhibits an explicit dependence on $s_m$, which is inexpedient for further manipulations. But let $i$ denote an index that runs over all the indices $i_{V_I}$ and $i_{R_0}$. Using \eno{prodgamma1}, one can show that 
\eqn{gammaReduce}
{
\bigg[\prod_{1\leq I \leq g}\prod_{i_{V_I}}\gamma_p\big(s_{i_{V_I}}s_m(x_m-x_{i_{V_I}})^2\big)\bigg]
\bigg[\prod_{i_{R_0}}\gamma_p\big(s_{i_{R_0}}s_m(x_m-x_{R_0})^2\big)\bigg]
=
\prod_{i,j}\gamma_p\big(s_is_jx_{ij}^2\big)\,.
}
After using this formula, we note that equation \eno{utrolig} is of the same form as equation \eno{inductionstart},  and so, as promised, we have demonstrated inductively that the position space pre-amplitude $\widetilde{A}_L$ could indeed be cast in the form \eno{inductionstart}. This completes the inductive proof that any arbitrary tree-level diagram, in particular the diagram in~\eno{recurdia1} corresponding to the position space pre-amplitude $\widetilde{A}_l$, takes the form described in~\eno{inductionstart}.

Applying the  $p$-adic Symanzik star integration formula \eno{SymanzikStar} to \eno{utrolig} after invoking \eno{gammaReduce}, we obtain
\eqn{leftdiagram}
{
\widetilde{\mathcal{A}}_L&=
\widetilde{\mathcal{L}}\,
\zeta_p\big(\sum\Delta_{i_{R_0}}-h+c\big)
\bigg[\prod_{L=1}^{2g-2}\int_{\mathbb{Q}_p}\frac{dt_L}{|t_L|_p}\bigg]
\widetilde{\mathcal{G}}(\{t_L\})
\bigg[\prod_{1\leq I \leq g}|f_{1I}(t_L)|_p^{-\sum\Delta_{i_{V_I}}}\bigg]
\cr
&\quad\times
\int[d\gamma]
\bigg[
\prod_{i< j}\frac{\zeta_p(2\gamma_{ij})}{|x_{ij}|_p^{2\gamma_{ij}}}\bigg]\zeta_p(1)\int_{\mathbb{Q}_p}\frac{du}{|u|_p}|u|_p^{-\sum\Delta_{i_{R_0}}+h+c}
\bigg[\prod_{i_{R_0}<j_{R_0}}\left|1,\frac{1}{u}\right|_s^{-2\gamma_{i_{R_0}j_{R_0}}}\bigg]
\cr
&\quad\times
\zeta_p(1)\int_{\mathbb{Q}_p}\frac{dt}{|t|_p}|t|_p^{-\sum\Delta_{i_V}+h-c}
\bigg[\prod_{1\leq I \leq J \leq g}\prod_{i_{V_I},i_{V_J}}\left|1,\frac{f_{IJ}(t_L)}{t^2f_{1I}(t_L)f_{1J}(t_L)}\right|_s^{-\gamma_{i_{V_I}i_{V_J}}}\bigg].
}
The factor of $\zeta_p\big(\sum\Delta_{i_{R_0}}-h+c\big)$ gives us the vertex factor \eno{vertexFactorDef} associated with the vertex $R_0$ in the diagram $D_L$. The $u$ integral evaluates to  (see \eno{supInt} in appendix~\ref{NOTATION}) 
\eqn{}{
\frac{-1}{\zeta_p(1)}\bigg[
\zeta_p\big(\sum\Delta_{R_0}-2\sum\gamma_{i_{R_0},j_{R_0}}-h-c\big)
-\zeta_p\big(\sum\Delta_{i_{R_0}}-h-c\big)
\bigg],
}
which is the undressed diagram associated with the vertex $R_0$ (recall identity~\eno{PreampRuleDiag}). We are making the inductive assumption that expression \eqref{leftdiagram} is equal to the position space pre-amplitude for diagram $D_L$ according to prescription~\ref{pres:Preamp} of section \ref{preAmpSec}, and this assumption implies that the remaining part of the right-hand side of \eno{leftdiagram} is equal to the product of all the vertex factors and undressed diagrams associated with the remaining vertices of $D_L$, that is,
\eqn{partAns}
{
& \widetilde{\mathcal{L}}\,
\bigg[\prod_{L=1}^{2g-2}\int_{\mathbb{Q}_p}\frac{dt_L}{|t_L|_p}\bigg]
\widetilde{\mathcal{G}}(\{t_L\})
\bigg[\prod_{1\leq I \leq g}|f_{1I}(t_L)|_p^{-\sum\Delta_{i_{V_I}}}\bigg]
\cr
&\quad \times
\zeta_p(1)\int_{\mathbb{Q}_p}\frac{dt}{|t|_p}|t|_p^{-\sum\Delta_{i_V}+h-c}
\bigg[\prod_{1\leq I \leq J \leq g}\prod_{i_{V_I},i_{V_J}}\left|1,\frac{f_{IJ}(t_L)}{t^2f_{1I}(t_L)f_{1J}(t_L)}\right|_s^{-\gamma_{i_{V_I}i_{V_J}}}\bigg]
\cr
 &= \bigg[\prod_{I=1}^g \text{(vertex factor for vertex $V_I$)(undressed diagram for vertex $V_I$)}\bigg].
}

\paragraph{The ``right-hand diagram'' $D_R$.}

Now we turn our attention to the position space pre-amplitude $\widetilde{\mathcal{A}}_R$ for diagram $D_R$ in \eno{LeftAndRight}. According to prescription~\ref{pres:Preamp} and \eno{PreampRuleDiag} of section~\ref{preAmpSec}, this Mellin pre-amplitude is given by the following product of vertex factors and undressed diagrams: 
\eqn{diagrams}
{
\widetilde{\mathcal{M}}_R&=
\left( 
 \begin{matrix}
\text{
\begin{tikzpicture}
\draw[blue,thick,fill=blue] (0,0) ellipse (0.05cm and 0.05cm);
\draw[blue,very thick] (0,0)--(-0.93,0.63);
\draw[blue,very thick] (0,0)--(-0.93,-0.63);
\draw[blue,very thick] (0,0)--(0.93,0.63);
\draw[blue,very thick] (0,0)--(0.93,-0.63);
\node at (-0.8,0.1) {\textcolor{blue}{$\vdots$}}; 
\node at (0.8,0.1) {\textcolor{blue}{$\vdots$}}; 
\node at (-1.3,0) {\textcolor{blue}{$i_{R_0}$}}; 
\node at (1.65,0.7) {\textcolor{blue}{$h+c_1$}}; 
\node at (1.65,-0.7) {\textcolor{blue}{$h+c_f$}}; 
\end{tikzpicture}
}
 \end{matrix}
\right)
\left( 
 \begin{matrix}
\text{
\begin{tikzpicture}
\draw[red,thick,fill=red] (0,0) ellipse (0.05cm and 0.05cm);
\draw[red,thick,fill=red] (1.2*0.93,1.2*0.63) ellipse (0.05cm and 0.05cm);
\draw[red,thick,fill=red] (1.2*0.94,-1.2*0.63) ellipse (0.05cm and 0.05cm);
\draw[red,very thick] (0,0)--(-0.93,0.63);
\draw[red,very thick] (0,0)--(-0.93,-0.63);
\draw[red,very thick] (0,0)--(2.2*0.93,2.2*0.63);
\draw[red,very thick] (0,0)--(2.2*0.93,-2.2*0.63);
\node at (-0.8,0.1) {\textcolor{red}{$\vdots$}}; 
\node at (0.8,0.1) {\textcolor{red}{$\vdots$}}; 
\node at (-1.3,0) {\textcolor{red}{$i_{R_0}$}}; 
\node at (0.2,0.8) {\textcolor{red}{$h+c_1$}}; 
\node at (0.2+0.93,0.8+0.63) {\textcolor{red}{$h-c_1$}}; 
\node at (0.2,-0.75) {\textcolor{red}{$h+c_f$}}; 
\node at (0.2+0.93,-0.75-0.63) {\textcolor{red}{$h-c_f$}};
\end{tikzpicture}
}
 \end{matrix}
\right)
\cr
&\quad\times
\prod_{I=1}^f
\Bigg[
\left( 
 \begin{matrix}
\text{
\begin{tikzpicture}
\draw[blue,thick,fill=blue] (0.8,0) ellipse (0.05cm and 0.05cm);
\draw[blue,very thick] (-0.6,0)--(0.8,0);
\draw[blue,very thick] (0.8,0)--(1.73,0.63);
\draw[blue,very thick] (0.8,0)--(1.73,-0.63);
\node at (0.1,-0.4) {\textcolor{blue}{$h-c_I$}}; 
\node at (1.6,0.1) {\textcolor{blue}{$\vdots$}}; 
\node at (2.1,0) {\textcolor{blue}{$i_{R_I}$}}; 
\end{tikzpicture}
}
 \end{matrix}
\right)
\left( 
 \begin{matrix}
\text{
\begin{tikzpicture}
\draw[red,thick,fill=red] (0.8,0) ellipse (0.05cm and 0.05cm);
\draw[red,thick,fill=red] (-0.6,0) ellipse (0.05cm and 0.05cm);
\draw[red,very thick] (-2.0,0)--(0.8,0);
\draw[red,very thick] (0.8,0)--(1.73,0.63);
\draw[red,very thick] (0.8,0)--(1.73,-0.63);
\node at (0.1,-0.4) {\textcolor{red}{$h-c_I$}}; 
\node at (-1.3,-0.4) {\textcolor{red}{$h+c_I$}}; 
\node at (1.6,0.1) {\textcolor{red}{$\vdots$}}; 
\node at (2.1,0) {\textcolor{red}{$i_{R_I}$}}; 
\end{tikzpicture}
}
 \end{matrix}
\right)
\Bigg]
.
}
From this diagrammatic decomposition, we can obtain an expression for the Mellin amplitude by applying to it equations \eno{vertexFactorDef} and \eno{toShowPadic}-\eno{recipint}. We find it useful, however, to recast the integral expression \eno{toShowPadic}-\eno{recipint} for undressed diagrams by changing variables to $u_i \equiv \frac{1}{\sqrt{x_i}}$:
\eqn{anotherUndressed}{ 
 &\hspace{25mm}
 \begin{matrix}
 \text{
 \begin{tikzpicture}
\draw[red,thick,fill=red] (0,0) ellipse (0.05cm and 0.05cm);
\draw[red,thick,fill=red] (1.2,0.5) ellipse (0.05cm and 0.05cm);
\draw[red,thick,fill=red] (1.2,-0.5) ellipse (0.05cm and 0.05cm);
\draw[red,very thick] (2*1.2,2*0.5) to (0,0);
\draw[red,very thick] (2*1.2,-2*0.5) to (0,0);
\draw[red,very thick] (-1.1*1.2,-1.1*0.5) to (0,0);
\draw[red,very thick] (-1.1*1.2,1.1*0.5) to (0,0);
\node at (-1.5*1.2,-1.4*0.5) {\textcolor{red}{$\Delta_{l+1}$}};
\node at (-1.5*1.2,1.4*0.5) {\textcolor{red}{$\Delta_{L}$}};
\node at (1,0.1) {\textcolor{red}{$\vdots$}}; 
\node at (-1.1,0.1) {\textcolor{red}{$\vdots$}}; 
\node at (0.5,1.2) {\textcolor{red}{$s_1,$}}; 
\node at (0.5,0.75) {\textcolor{red}{$h+c_1$}}; 
\node at (0.5,-0.7) {\textcolor{red}{$h+c_l,$}}; 
\node at (0.5,-1.25) {\textcolor{red}{$s_l$}}; 
\node at (1.8,1.27) {\textcolor{red}{$h-c_1$}}; 
\node at (1.8,-1.23) {\textcolor{red}{$h-c_l$}}; 
\end{tikzpicture}
 }
 \end{matrix}
 \cr
&
=
 \bigg[\prod_{i=1}^l \zeta_p(1)\int_{\mathbb{Q}_p^2}\frac{du_i}{|u_i|_p}\,|u_i|_p^{-2c_i} |1,u_i|^{s_i+c_i-h}_s\bigg] \left|1,\frac{1}{u_1},\ldots,\frac{1}{u_l}\right|_s^{2h-\sum_{i=1}^l(c_i+h)-\sum_{i=l+1}^L\Delta_i}\,.
}
Writing out the vertex factors in terms of their $p$-adic zeta functions and using the above integral form for the undressed diagrams, \eqref{diagrams} can be rewritten as 
\eqn{lyser}
{
\widetilde{\mathcal{A}}_R&=
\zeta_p\big(\sum\Delta_{I_{R_0}}+\sum_{I=1}^f(h+c_I)-2h\big)
\int [d\gamma]
\bigg[\prod_{i<j}\frac{\zeta_p(2\gamma_{ij})}{|x_{ij}|_p^{2\gamma_{ij}}}\bigg]
\cr
&\quad\times
\bigg[\prod_{I=1}^f\zeta_p(\sum\Delta_{i_{R_I}}-h-c_I)\,\zeta_p(1)\int_{\mathbb{Q}_p}\frac{dv_I}{|v_I|}|v_I|_p^{\sum\Delta_{i_{R_I}}-h+c_I} |1,v_I|_s^{-2\sum\gamma_{i_{R_I}j_{R_I}}}\bigg]
\cr
&\quad\times
\bigg[\prod_{I=1}^f
\zeta_p(1)\int_{\mathbb{Q}_p}\frac{du_I}{|u_I|_p}|u_I|_p^{-2c_I}|1,u_I|_s^{\sum\Delta_{i_{R_I}}-h+c_I-2\sum\gamma_{i_{R_I}j_{R_I}}}
\bigg]
\cr
&\quad\times
\left|
1,\frac{1}{u_1},\frac{1}{u_2},\ldots,\frac{1}{u_f}
\right|_s^{2h-\sum\Delta_{i_{R_0}}-\sum_{I=1}^f(c_I+h)}.
}
where $i,j$ run over all oxternal leg indices $i_{R_0},i_{R_1},\ldots,i_{R_f}$.
Lumping together the vertex factors into one symbol
\eqn{}{
\mathcal{R} \equiv
\zeta_p\big(\sum\Delta_{I_{R_0}}+\sum_{I=1}^f(h+c_I)-2h\big)
\bigg[\prod_{I=1}^f\zeta_p\big(\sum\Delta_{i_{R_I}}-h-c_I\big)\bigg],
}
and using the Symanzik star integration formula \eno{SymanzikStar}, one finds that the position space pre-amplitude for diagram $D_R$ is given by 
\eqn{rightexpression}
{
\widetilde{\mathcal{A}}_R
&=
\mathcal{R}\,
\sum_{b\in\{1,p\}}
\bigg[\prod_{i_R}\frac{2\zeta_p(1)}{|2|_p}\int_{b\mathbb{Q}_p^2}\frac{ds_{i_R}}{|s_{i_R}|_p}|s_{i_R}|_p^{\sum\Delta_{i_R}}\bigg]
\cr
&\quad\times
\bigg[ \prod_{I=1}^f \zeta_p(1)\int_{\mathbb{Q}_p}\frac{du_I}{|u_I|_p}|u_I|_p^{-2c_I}|1,u_I|_s^{\sum\Delta_{i_{R_I}}-h+c_I}\zeta_p(1)\int_{\mathbb{Q}_p}\frac{dv_I}{|v_I|_p}|v_I|_p^{\sum\Delta_{i_{R_I}}-h+c_I}\bigg]
\cr
&\quad\times
\left|1,\frac{1}{u_1},\frac{1}{u_2},\ldots,\frac{1}{u_f}\right|_s^{2h-\sum\Delta_{i_{R_0}}-\sum_{I=1}^f(c_I+h)}
\bigg[\prod_{i_R,j_R}\gamma_p(s_{i_R}s_{j_R}x_{i_Rj_R}^2)\bigg]
\cr
&\quad\times
\bigg[
\prod_{I=1}^f\prod_{i_{R_I}<j_{R_I}} \gamma_p\big(s_{i_{R_I}}s_{j_{R_I}}(1,u_I)^2_s(1,v_I)^2_sx_{i_{R_I}j_{R_I}}^2\big)
\bigg],
}
 where the index $i_R$ runs over $i_{R_0}, i_{R_1},\ldots,i_{R_f}$ and
 \eqn{}
{ (1,u)_s \equiv \begin{cases}
1, \qquad \text{when }|u|_p \leq 1\,,
\\
u, \qquad \text{when }|u|_p > 1\,.
\end{cases}
}

\paragraph{The ``merged diagram'' $D_M$.}

By strong induction we are making the assumption that diagrams $D_l$, $D_L$, and $D_R$ satisfy prescription~\ref{pres:Preamp}. The inductive step we shall undertake in order to prove the pre-amplitude prescription is that under this assumption, the position space pre-amplitude of diagram $D_M$, which we call $\widetilde{\mathcal{A}}_M$, evaluates to the appropriate product of vertex factors and undressed diagrams as prescribed by prescription~\ref{pres:Preamp}. 

Applying the split representation to the internal leg connecting vertices $V_1$ and $R_0$ in $D_M$, we can express $\widetilde{\mathcal{A}}_M$ as a boundary integral over the pre-amplitudes for the left- and right-hand diagrams, for which we may use equations \eqref{inductionstart} and \eqref{rightexpression}. However, in using each of these two equations, we once again need to assign a complex scaling dimension to one of the external legs. For the left-hand part of the diagram we assign a complex value $h-c$ to the scaling dimension of an external leg connected to vertex $V_1$ and once again rename the integration variable $s_{i_{V_1}}$ as $t$. For the right-hand part of the diagram, we assign the value $h+c$ to the scaling dimension, call it $\Delta_0$, of an external leg connected to vertex $R_0$. Furthermore, we rename the integration variable $s_{i_{R_0}}$ as $u$ and define $\widetilde{\mathcal{R}}\equiv \mathcal{R}\big|_{\Delta_0=h+c}$. We arrive, then, at the following expression for the position space pre-amplitude of $D_M$: 
\eqn{}
{
\widetilde{\mathcal{A}}_M=
\int_{\mathbb{Q}_{p^n}} & dx \,
\widetilde{\mathcal{L}}\bigg[\prod_{L=1}^{2g-2}\int_{\mathbb{Q}_p}\frac{dt_L}{|t_L|_p}\bigg]
\widetilde{\mathcal{G}}(\{t_L\})
\cr
&\times
\sum_{a\in\{1,p\}}
\bigg[\prod_{i_V}\frac{2\zeta_p(1)}{|2|_p}\int_{a\mathbb{Q}_p^2}\frac{ds_{i_V}}{|s_{i_V}|_p}|s_{i_V}|_p^{\sum\Delta_{i_V}}\bigg]
\frac{2\zeta_p(1)}{|2|_p}\int_{a\mathbb{Q}_p^2}\frac{dt}{|t|_p}|t|_p^{h-c}
\cr
&\times
\bigg[\prod_{1\leq I \leq J \leq g}\prod_{i_{V_I},i_{V_J}}\gamma_p\big(f_{IJ}(t_L)s_{i_{V_I}}s_{i_{V_J}}x_{i_{V_I}i_{V_J}}^2\big)\bigg]
\cr
&\times
\bigg[\prod_{1\leq I \leq g}\prod_{i_{V_I}}\gamma_p\big(f_{1I}(t_L)ts_{i_{V_I}}(x-x_{i_{V_I}})^2\big)\bigg]
\cr
&\times
\widetilde{\mathcal{R}}\,
\sum_{b\in\{1,p\}}
\bigg[\prod_{i_R}\frac{2\zeta_p(1)}{|2|_p}\int_{b\mathbb{Q}_p^2}\frac{ds_{i_R}}{|s_{i_R}|_p}|s_{i_R}|_p^{\sum\Delta_{i_R}}\bigg]
\cr
&\times
\frac{2\zeta_p(1)}{|2|_p}\int_{b\mathbb{Q}_p^2}\frac{du}{|u|_p}|u|_p^{h+c}
\bigg[\prod_{i_{R}}\gamma_p\big(us_{i_R}(x-x_{i_{R}})^2\big)\bigg]
\cr
&\times
 \bigg[\prod_{I=1}^f\zeta_p(1)\int_{\mathbb{Q}_p}\frac{du_I}{|u_I|_p}|u_I|_p^{-2c_I}|1,u_I|_s^{\sum\Delta_{i_{R_I}}-h+c_I}
 \zeta_p(1)\int_{\mathbb{Q}_p}\frac{dv_I}{|v_I|_p}|v_I|_p^{\sum\Delta_{i_{R_I}}-h+c_I}\bigg]
\cr
&\times 
 \left|1,\frac{1}{u_1},\frac{1}{u_2},\ldots,\frac{1}{u_f}\right|_s^{2h-\sum\Delta_{i_{R_0}}-\sum_{I=1}^f(c_I+h)-h-c}
\bigg[\prod_{i_R,j_R}\gamma_p(s_{i_R}s_{j_R}x_{i_Rj_R}^2)\bigg]
\cr
&\times
\bigg[
\prod_{I=1}^f 
\prod_{i_{R_I}<j_{R_I}} \gamma_p\big(s_{i_{R_I}}s_{j_{R_I}}(1,u_I)^2_s(1,v_I)^2_sx_{i_{R_I}j_{R_I}}^2\big)
\bigg].
}
$\hspace{1mm}$
\\
Changing variables, $s_{i_{V_I}}\rightarrow \displaystyle \frac{s_{i_{V_I}}}{tf_{1I}(t_L)}$ and $s_{i_R} \rightarrow \displaystyle\frac{s_{i_R}}{u}$,
re-expressing the domains of the $t$ and $u$ integrals, letting $m$ denote an index such that $|s_m|_p=\text{sup}_i|s_i|_p$, and carrying out the $x$ integral, we get
\clearpage
\eqn{}
{
\widetilde{\mathcal{A}}_M=
&
\widetilde{\mathcal{L}}
\bigg[\prod_{L=1}^{2g-2}\int_{\mathbb{Q}_p}\frac{dt_L}{|t_L|_p}\bigg]
\widetilde{\mathcal{G}}(t_L)
\cr
&
\times
\bigg[\prod_{i_V}\frac{2\zeta_p(1)}{|2|_p}\int_{\mathbb{Q}_p^2}\frac{ds_{i_V}}{|s_{i_V}|_p}|s_{i_V}|_p^{\sum\Delta_{i_V}}\bigg]
\zeta_p(1)\int_{\mathbb{Q}_p}\frac{dt}{|t|_p}|t|_p^{-\sum\Delta_{i_{V}}+h-c}
\cr
&
\times
\bigg[\prod_{1\leq I \leq J \leq g}\prod_{i_{V_I},i_{V_J}}\gamma_p\bigg(\frac{f_{IJ}(t_L)s_{i_{V_I}}s_{i_{V_J}}}{t^2f_{1I}(t_L)f_{1J}(t_L)}x_{i_{V_I}i_{V_J}}^2\bigg)\bigg]
\cr
&
\times
\bigg(
\prod_{1\leq I \leq g}|f_{1I}(t_L)|_p^{-\sum\Delta_{i_{V_I}}}\bigg[\prod_{i_{V_I}}\gamma_p\bigg(s_{i_{V_I}}(x_m-x_{i_{V_I}})^2\bigg)\bigg]
\bigg)
\cr
&
\times
\widetilde{\mathcal{R}}\bigg[\prod_{i_R}\frac{2\zeta_p(1)}{|2|_p}\int_{\mathbb{Q}_p^2}\frac{ds_{i_R}}{|s_{i_R}|_p}|s_{i_R}|_p^{\sum\Delta_{i_R}}\bigg]
\cr
&
\times
\zeta_p(1)\int_{\mathbb{Q}_p}\frac{du}{|u|_p}|u|_p^{-\sum\Delta_{i_R}+h+c}
\bigg[\prod_{i_{R}}\gamma_p\big(s_{i_R}(x_m-x_{i_{R}})^2\big)\bigg]|s_m|_p^{-h}
\cr
&
\times
 \left[\prod_{I=1}^f \zeta_p(1)\int_{\mathbb{Q}_p}\frac{du_I}{|u_I|_p}|u_I|_p^{-2c_I}|1,u_I|_s^{\sum\Delta_{i_{R_I}}-h+c_I}
 \zeta_p(1)
 \int_{\mathbb{Q}_p}\frac{dv_I}{|v_I|_p}|v_I|_p^{\sum\Delta_{i_{R_I}}-h+c_I}\right]
\cr
&
\times
 \left|1,\frac{1}{u_1},\frac{1}{u_2},\ldots,\frac{1}{u_f}\right|_s^{2h-\sum\Delta_{i_{R_0}}-\sum_{I=1}^f(c_I+h)-h-c}
\bigg[\prod_{i_R,j_R}\gamma_p\bigg(\frac{s_{i_R}s_{j_R}}{u^2}x_{i_Rj_R}^2\bigg)\bigg]
\cr
&
\times
\bigg[
\prod_{I=1}^f 
\prod_{i_{R_I}<j_{R_I}} \gamma_p\bigg(\frac{s_{i_{R_I}}s_{j_{R_I}}}{u^2}(1,u_I)^2_s(1,v_I)^2_sx_{i_{R_I}j_{R_I}}^2\bigg)
\bigg]\,.
}
$\hspace{1mm}$
\\
\\
Changing variables $t \rightarrow \sqrt{s_m}t$, $u \rightarrow \sqrt{s_m}u$, and $s_i \rightarrow \sqrt{s_m}s_i$ for $i\neq m$, and then changing variable $s_i \rightarrow s_i^2$ gives: 

\clearpage

\eqn{}
{
\widetilde{\mathcal{A}}_M=
&
\widetilde{\mathcal{L}}\bigg[\prod_{L=1}^{2g-2}\int_{\mathbb{Q}_p}\frac{dt_L}{|t_L|_p}\bigg]
\widetilde{\mathcal{G}}(t_L)
\cr
&
\times
\sum_{a\in \{1,p\}}
\bigg[\prod_{i_V}\frac{2\zeta_p(1)}{|2|_p}\int_{a\mathbb{Q}_p^2}\frac{ds_{i_V}}{|s_{i_V}|_p}|s_{i_V}|_p^{\sum\Delta_{i_V}}\bigg]
\zeta_p(1)\int_{\mathbb{Q}_p}\frac{dt}{|t|_p}|t|_p^{-\sum\Delta_{i_{V}}+h-c}
\cr
&
\times
\bigg[\prod_{1\leq I \leq J \leq g}\prod_{i_{V_I},i_{V_J}}\gamma_p\bigg(\frac{f_{IJ}(t_L)s_{i_{V_I}}s_{i_{V_J}}}{t^2f_{1I}(t_L)f_{1J}(t_L)}x_{i_{V_I}i_{V_J}}^2\bigg)\bigg]
\cr
&
\times
\bigg(
\prod_{1\leq I \leq g}
|f_{1I}(t_L)|_p^{-\sum\Delta_{i_{V_I}}}\bigg[\prod_{i_{V_I}}\gamma_p\big(s_ms_{i_{V_I}}(x_m-x_{i_{V_I}})^2\big)\bigg]\bigg)
\cr
&
\times
\widetilde{\mathcal{R}}\bigg[\prod_{i_R}\frac{2\zeta_p(1)}{|2|_p}\int_{a\mathbb{Q}_p^2}\frac{ds_{i_R}}{|s_{i_R}|_p}|s_{i_R}|_p^{\sum\Delta_{i_R}}\bigg]
\cr
&
\times
\zeta_p(1)\int_{\mathbb{Q}_p}\frac{du}{|u|_p}|u|_p^{-\sum\Delta_{i_R}+h+c}
\bigg[\prod_{i_{R}}\gamma_p\big(s_ms_{i_R}(x_m-x_{i_{R}})^2\big)\bigg]
\cr
&
\times
\bigg[
 \prod_{I=1}^f\zeta_p(1)\int_{\mathbb{Q}_p}\frac{du_I}{|u_I|_p}|u_I|_p^{-2c_I}|1,u_I|_s^{\sum\Delta_{i_{R_I}}-h+c_I}
 \zeta_p(1)\int_{\mathbb{Q}_p}\frac{dv_I}{|v_I|_p}|v_I|_p^{\sum\Delta_{i_{R_I}}-h+c_I}\bigg]
 \cr
 &
 \times
 \left|1,\frac{1}{u_1},\frac{1}{u_2},\ldots,\frac{1}{u_f}\right|_s^{2h-\sum\Delta_{i_{R_0}}-\sum_{I=1}^f(c_I+h)-h-c}
\bigg[\prod_{i_R,j_R}\gamma_p\bigg(\frac{s_{i_R}s_{j_R}}{u^2}x_{i_Rj_R}^2\bigg)\bigg]
\cr
&
\times
\bigg[
\prod_{I=1}^f 
\prod_{i_{R_I}<j_{R_I}} \gamma_p\bigg(\frac{s_{i_{R_I}}s_{j_{R_I}}}{u^2}(1,u_I)^2_s(1,v_I)^2_sx_{i_{R_I}j_{R_I}}^2\bigg)
\bigg]
\,.
}
$\hspace{1mm}$
\\
\\
Introducing redundant characteristic functions to remove the explicit $s_m$ dependence like in \eno{gammaReduce}, and then invoking the Symanzik star integration formula \eno{SymanzikStar}, we obtain
\clearpage
\eqn{dancer}
{
\widetilde{\mathcal{A}}_M=
&
\int[d\gamma]
\bigg[\prod_{i< j}\frac{\zeta(2\gamma_{ij})}{|x_{ij}|_p^{2\gamma_{ij}}}\bigg]
\cr
&
\times
\widetilde{\mathcal{L}}
\bigg[
\prod_{L=1}^{2g-2}\int_{\mathbb{Q}_p}\frac{dt_L}{|t_L|_p}\bigg]
\widetilde{\mathcal{G}}(t_L)
\bigg[
\prod_{1\leq I \leq g}|f_{1I}(t_L)|_p^{-\sum\Delta_{i_{V_I}}}\bigg]
\cr
&
\times
\zeta_p(1)\int_{\mathbb{Q}_p}\frac{dt}{|t|_p}|t|_p^{-\sum\Delta_{i_V}+h-c}
\bigg[
\prod_{1\leq I \leq J \leq g}
\prod_{i_{V_I},i_{V_J}}\left|1,\frac{f_{IJ}(t_L)}{t^2f_{1I}(t_L)f_{1J}(t_L)}\right|_s^{-\gamma_{i_{V_I}i_{V_J}}}\bigg]
\cr
&
\times
\widetilde{\mathcal{R}} \,
\zeta_p(1)\int_{\mathbb{Q}_p}\frac{du}{|u|_p}|u|_p^{-\sum\Delta_{i_{R}}+h+c}
\left|1,\frac{1}{u}\right|_s^{-2\sum\gamma_{i_{R_0}j_{R_0}}}
\bigg[
\prod_{0\leq I < J \leq f}\left|1,\frac{1}{u}\right|_s^{-2\sum\gamma_{i_{R_I}i_{R_J}}}
\bigg]
\cr
&
\times
\left[
\prod_{I=1}^f\zeta_p(1)\int_{\mathbb{Q}_p}\frac{du_I}{|u_I|_p}|u_I|_p^{-2c_I}|1,u_I|_p^{\sum\Delta_{i_{R_I}}-h+c_I}
\zeta_p(1)
\int_{\mathbb{Q}_p}\frac{dv_I}{|v_I|_p}|v_I|_p^{\sum\Delta_{i_{R_I}}-h+c_I}\right]
\cr
&
\times
\left|1,\frac{1}{u_1},\frac{1}{u_2},\ldots,\frac{1}{u_f}\right|_s^{2h-\sum\Delta_{i_{R_0}}-\sum_{I=1}^f(c_I+h)-h-c}
\bigg[
\prod_{I=1}^f\left|1,\frac{(1,u_I)_s(1,v_I)_s}{u}\right|_s^{-2 \sum\gamma_{i_{R_I}j_{R_I}}}\bigg],
}
where by the sum $\sum\gamma_{i_{R_I}j_{R_I}}$ we mean $\sum_{i_{R_i} < j_{R_j}}\gamma_{i_{R_I}j_{R_I}}$, that is, no double-counting. Comparing the second and third lines of \eno{dancer} with \eqref{partAns}, we see that the Mellin amplitude of $D_M$ by the inductive assumption includes the vertex factors and undressed diagrams of all the vertices $V_I$ on the left-hand side of diagram $D_M$. What remains for us to show is that the bottom three lines of \eqref{dancer}, denote them $\widetilde{M}_R$, equal the vertex factors and undressed diagrams of the vertices $R_I$ of the right-hand diagram. Changing the variable $u$ to its reciprocal, the bottom three lines can be rewritten as
\eqn{}
{
\widetilde{M}_R=\,
&
\widetilde{\mathcal{R}}\,
\zeta_p(1)\int_{\mathbb{Q}_p}\frac{du}{|u|_p}|u|_p^{\sum\Delta_{i_{R}}-h-c}
\left|1,u\right|_s^{-2\sum\gamma_{i_{R_0}j_{R_0}}}
\bigg[\prod_{0\leq I < J \leq f}\left|1,u\right|_s^{-2\sum\gamma_{i_{R_I}i_{R_J}}}
\bigg]
\cr
&
\times
\bigg[\prod_{I=1}^f\zeta_p(1)\int_{\mathbb{Q}_p}\frac{du_I}{|u_I|_p}|u_I|_p^{-2c_I}|1,u_I|_p^{\sum\Delta_{i_{R_I}}-h+c_I}\zeta_p(1)\int_{\mathbb{Q}_p}\frac{dv_I}{|v_I|_p}|v_I|_p^{\sum\Delta_{i_{R_I}}-h+c_I}\bigg]
\cr
&
\times
\left|1,\frac{1}{u_1},\frac{1}{u_2},\ldots,\frac{1}{u_f}\right|_s^{2h-\sum\Delta_{i_{R_0}}-\sum_{I=1}^f(c_I+h)-h-c}
\bigg[\prod_{I=1}^f\left|1,u,uu_I,u(1,u_I)_sv_I\right|_s^{-2\sum \gamma_{i_{R_I}j_{R_I}}}\bigg]
.
}
Noting that $\sum\Delta_{i_R}=\sum\Delta_{i_{R_0}}+\sum_{I=1}^f\sum\Delta_{i_{R_I}}$ and changing variables $v_I\rightarrow \displaystyle \frac{(1,u,uu_I)_s}{u(1,u_I)_s}v_I$, we obtain
\eqn{}
{
\widetilde{M}_R=\,&
\widetilde{\mathcal{R}}\,
\zeta_p(1)\int_{\mathbb{Q}_p}\frac{du}{|u|_p}|u|_p^{\sum\Delta_{i_{R_0}}+\sum_{I=1}^f(h-c_I)-h-c}
\left|1,u\right|_s^{-2\sum\gamma_{i_{R_0}j_{R_0}}}
\bigg[
\prod_{0\leq I < J \leq f}\left|1,u\right|_s^{-2\sum\gamma_{i_{R_I}i_{R_J}}}
\bigg]
\cr
&
\times
\bigg[\prod_{I=1}^f\zeta_p(1)\int_{\mathbb{Q}_p}\frac{du_I}{|u_I|_p}|u_I|_p^{-2c_I}
\left|1,u,uu_I\right|_s^{\sum\Delta_{i_{R_I}}-2\sum\gamma_{i_{R_I}j_{R_I}}-h+c_I}
\bigg]
\cr
&
\times
\bigg[
\zeta_p(1)\int_{\mathbb{Q}_p}\frac{dv_I}{|v_I|_p}|v_I|_p^{\sum\Delta_{i_{R_I}}-h+c_I}\left|1,v_I\right|_s^{-2\sum\gamma_{i_{R_I}j_{R_I}}}\bigg]
\cr
&
\times
\left|1,\frac{1}{u_1},\frac{1}{u_2},\ldots,\frac{1}{u_f}\right|_s^{2h-\sum\Delta_{i_{R_0}}-\sum_{I=1}^f(c_I+h)-h-c}
\,.
}
Carrying out the $v_I$ integrals using \eno{supInt} and changing variables $u_I \rightarrow \displaystyle \frac{(1,u)_s}{u} u_I$, the pre-amplitude simplifies to
\eqn{almostThere}
{
\widetilde{M}_R=\,&
\widetilde{\mathcal{R}}\,
\bigg[\prod_I(-1)\bigg(\zeta_p(\sum\Delta_{i_{R_I}}-2\sum\gamma_{i_{R_I}j_{R_I}}-h+c_I)-\zeta_p(\sum\Delta_{i_{R_I}}-h+c_I)\bigg)\bigg]
\cr
&
\times
\zeta_p(1)\int_{\mathbb{Q}_p}\frac{du}{|u|_p}|u|_p^{-2c}
|1,u|_s^{\sum\Delta_{i_R}-2\sum\gamma_{i_{R_0}j_{R_0}}-h+c}
\bigg[\prod_{0\leq I < J \leq f}\left|1,u\right|_s^{-2\sum\gamma_{i_{R_I}i_{R_J}}}\bigg]
\cr
&
\times
\bigg[\prod_{I=1}^f\zeta_p(1)\int_{\mathbb{Q}_p}\frac{du_I}{|u_I|_p}|u_I|_p^{-2c_I}
\left|1,u_I\right|_s^{\sum\Delta_{i_{R_I}}-2\sum\gamma_{i_{R_I}j_{R_I}}-h+c_I}
|1,u|_s^{-2\sum\gamma_{i_{R_I}j_{R_I}}}\bigg]
\cr
&
\times
\left|1,\frac{1}{u},\frac{1}{u_1},\frac{1}{u_2},\ldots,\frac{1}{u_f}\right|_s^{2h-\sum\Delta_{i_{R_0}}-\sum_{I=1}^f(c_I+h)-h-c}
\,.
}
Let $s_I$ denote the Mandelstam invariant associated with the internal leg connecting vertices $R_0$ and $R_I$, and let $s_0$ denote the Mandelstam invariant associated with the internal leg connecting vertices $V_1$ and $R_1$. Then we have that 
\eqn{}
{
s_I &= \sum \Delta_{i_{R_I}} - 2\sum \gamma_{i_{R_I}j_{R_I}}\,, \cr
s_0 &= \sum\Delta_{i_R}-2\sum_{0\leq I < J \leq f}\sum\gamma_{i_{R_I}i_{R_J}}-2\sum\gamma_{i_{R_0}j_{R_0}}-2\sum_{I=1}^f\sum\gamma_{i_{R_I}j_{R_I}}\,,
}
using which equation \eqref{almostThere} can be rewritten as
\eqn{There}
{
\widetilde{M}_R=\,&
\widetilde{\mathcal{R}}\,
\bigg[\prod_I(-1)\bigg(\zeta_p(s_I-h+c_I)-\zeta_p(\sum\Delta_{i_{R_I}}-h+c_I)\bigg)\bigg]
\cr
&
\times
\zeta_p(1)
\int_{\mathbb{Q}_p}\frac{du}{|u|_p}|u|_p^{-2c}\left|1,u\right|_s^{s_0-h+c}
\bigg[\prod_{I=1}^f\zeta_p(1)\int_{\mathbb{Q}_p}\frac{du_I}{|u_I|_p}|u_I|_p^{-2c_I}
\left|1,u_I\right|_s^{s_I-h+c_I}
\bigg]
\cr
&
\times
\left|1,\frac{1}{u},\frac{1}{u_1},\frac{1}{u_2},\ldots,\frac{1}{u_f}\right|_s^{2h-\sum\Delta_{i_{R_0}}-\sum_{I=1}^f(c_I+h)-h-c}.
}
In the first line of this equation, $\widetilde{\mathcal{R}}$ is the product of vertex factors associated with the vertices $R_0$, $R_1$, \ldots, $R_f$ of diagram $D_M$ and the product over $I$ gives the product of undressed diagrams associated with vertices $R_1$, $R_2$, \ldots, $R_f$. Comparing with \eno{anotherUndressed}, we recognize in lines two and three of \eqref{There} the integral form of the undressed diagram associated with vertex $R_0$. We conclude that $\widetilde{\mathcal{M}}_R$ is equal to the product of vertex factors and undressed diagrams associated with vertices $R_0$, $R_1$, \ldots, $R_f$, and so the full pre-amplitude for $D_M$ is equal to the product of  vertex factors and undressed diagrams associated with all the vertices of this diagram. This completes the proof of prescription~\ref{pres:Preamp}.

\section{Inductive Proof of Prescription \ref{pres:padicMellin} for Mellin Amplitudes}
\label{AmpProof}

In order to obtain a $p$-adic Mellin amplitude from a pre-amplitude one must carry out a contour integral around a cylindrical manifold for each internal leg of the diagram. In section~\ref{sec:intPreamp} we proved that such contour integrals over pre-amplitudes, given according to prescription~\ref{pres:Preamp} of section \ref{preAmpSec}, correctly reproduce Mellin amplitudes as defined in \eno{pMel}. What remains to be shown to demonstrate that the recursion relations of section~\ref{recursionSectionPadic} are true is that carrying out the contour integrals over the pre-amplitudes exactly reproduces the Mellin amplitudes given by prescription~\ref{pres:padicMellin}. In this appendix we prove this for arbitrary tree-level bulk diagrams.

Consider first the exchange amplitude. The fact that carrying out the contour integral over its  pre-amplitude yields the appropriate Mellin amplitude (see~\eqref{singlePreAmp}) relies mathematically on the following identity: 
\eqn{firstContourId}
{
& \quad 
\frac{\log p}{2\pi i}\int_{-\frac{i\pi}{\log p}}^{\frac{i\pi}{\log p}}dc
\,
\frac{\zeta_p(A+c)\zeta_p(A-c)\zeta_p(B-c)\zeta_p(C+c)}{\zeta_p(2c)\zeta_p(-2c)}
\cr 
& \quad  \times
\bigg(\zeta_p(D-c)+\zeta_p(B+c)-1\bigg)\bigg(\zeta_p(D+c)+\zeta_p(C-c)-1\bigg)
\cr
&= 
2\,\zeta_p(A+B)\zeta_p(A+C)\bigg(\zeta_p(B+C)+\zeta_p(A+D)-1\bigg),
}
where $A,B,C,D>0$. This identity is a rewriting of equation~(4.39) in Ref.~\cite{Jepsen:2018dqp}.

For the bulk diagram with two internal lines, the fact that carrying out the contour integral over the pre-amplitude yields the desired Mellin amplitude (see~\eqref{doublePreAmp}) follows from \eqref{firstContourId} and the following identity:
\eqn{secondContourId}
{
&
\frac{\log p}{2\pi i}\int_{-\frac{i\pi}{\log p}}^{\frac{i\pi}{\log p}}dc
\,
\frac{\zeta_p(A+c)\zeta_p(A-c)\zeta_p(B-c)\zeta_p(C+c)}{\zeta_p(2c)\zeta_p(-2c)}
\bigg(\zeta_p(D-c)+\zeta_p(B+c)-1\bigg) 
\cr 
&
\times \bigg(\zeta_p(E+c)\zeta_p(D+c)+\zeta_p(C-c)\zeta_p(E-c)+\zeta_p(E+c)\zeta_p(E-c) 
 -\zeta_p(E+c)-\zeta_p(E-c)\bigg)
\cr
&=
2\zeta_p(A+B)\zeta_p(A+C)  \bigg(\zeta_p(B+C)\zeta_p(B+E)+\zeta_p(A+E)\zeta_p(A+D)
\cr
&
\qquad\qquad\qquad\qquad\qquad +\zeta_p(A+E)\zeta_p(B+E)-\zeta_p(A+E)-\zeta_p(B+E)\bigg).
}
In fact in general, the fact that the pre-amplitudes given by prescription~\ref{pres:Preamp} of section~\ref{preAmpSec} integrate to the appropriate Mellin amplitudes given by prescription~\ref{pres:padicMellin} of section~\ref{recursionSectionPadic} hinges mathematically on an infinite tower of increasingly convoluted contour integral identities for the local zeta function $\zeta_p$.\footnote{The factors of two on the r.h.s sides of equations \eqref{firstContourId} and \eqref{secondContourId} arise non-trivially: If one computes the contour integrals on the left-hand sides by summing over the residues, it is only the full sum of residues that conspire to yield two times a sum of zeta functions. This holds true for all identities in the previously mentioned infinite tower of identities.}  In the diagrammatic notation we have employed in this paper, it turns out these integral identities have a simple interpretation,  expressed in the form of identity~\eno{toBeProven}. 
In fact such a diagrammatic interpretation holds true for both real as well as $p$-adic Mellin amplitudes. In this appendix we will prove identity~\eno{toBeProven} over the $p$-adics.

\subsection{The ``leg adding'' operation}

An important part of the proof involves showing there exists a mathematical operation that adds an extra internal line to an undressed diagram, except  the operation should only output terms that do not depend on the Mandelstam variable of the added internal line. The previous statement will become  clear after considering a few examples. In the simplest example the starting point is the undressed diagram with no internal lines, which by definition equals unity, and after the ``leg adding'' operation one obtains an undressed diagram in a particular limit:
\eqn{}
{
&
\sum _{ z^\ast\in\{\Delta,\Delta_{i_R}\}}
\res_{z^\ast}
\Bigg[
\zeta_p\big(\Delta_{i_L}+z-n\big) \times (1) \times
\zeta_p\big(\Delta_{i_R}-z\big)
\zeta_p\big(\Delta-z\big)
\Bigg]
\cr
&
=-
\zeta_p\big(\Delta+\Delta_{i_L}-n\big)
\lim_{s\rightarrow -\infty}
\left(
\begin{matrix}
\vspace*{-0.2cm}
\includegraphics[height=7ex]{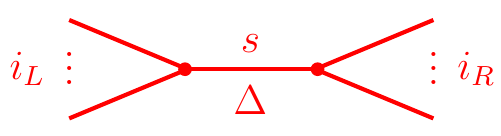}
\end{matrix}
\right).
}
Here as elsewhere, we have shortened notation by omitting a summation symbol that is implied, eg. $\Delta_{i_R} \equiv \sum_{i_R}\Delta_{i_R}$.
The limit $s \rightarrow -\infty$ kills off the momentum dependent factor $\zeta_p(s-\Delta)$ of the undressed diagram. The l.h.s.\ above is symmetric w.r.t.\ $\Delta$ and $\Delta_{i_R}$, and so the same is true for the r.h.s., although we have chosen to write it in a way that does not make this symmetry manifest.

The above example is almost too simple to be informative, so let us consider the action of the ``leg adding operation'' on a diagram that has one internal line to begin with: 
\eqn{}
{
&
\sum _{z^\ast\in\{\Delta_R,\Delta_{i_R}\}}
\res_{z^\ast}
\Bigg[
\zeta_p\big(\Delta_{i_U}+\Delta_L+z-n\big)
\left(
\begin{matrix}
\vspace*{-0.2cm}
\includegraphics[height=12ex]{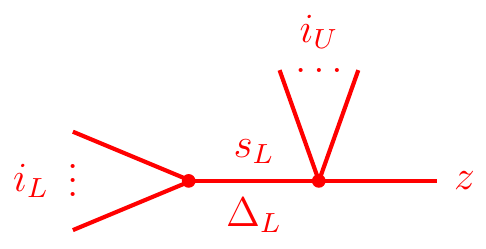}
\end{matrix}
\right)
\zeta_p\big(\Delta_{i_R}-z\big)
\zeta_p\big(\Delta_R-z\big)
\Bigg]
\cr
&
=-
\zeta_p\big(\Delta_L+\Delta_R+\Delta_{i_U}-n\big)
\lim_{s_R\rightarrow -\infty}
\left(
\begin{matrix}
\vspace*{-0.2cm}
\includegraphics[height=12ex]{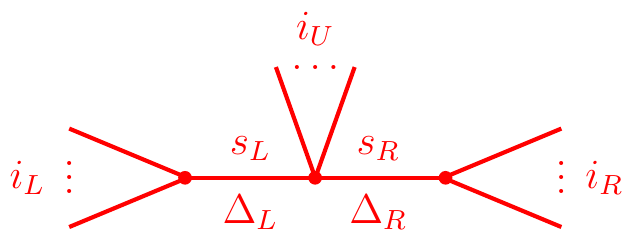}
\end{matrix}
\right).
}
Consider one final example, where we now attach an extra internal line onto a vertex that already has two internal lines incident on it:
\eqn{}
{
&
\sum _{z^\ast\in\{\Delta_C,\Delta_{i_D}\}}
\res_{z^\ast}
\Bigg[
\zeta_p\big(\Delta_{i_U}+\Delta_A+\Delta_B+z-n\big)
\left(
\begin{matrix}
\vspace*{-0.2cm}
\includegraphics[height=19ex]{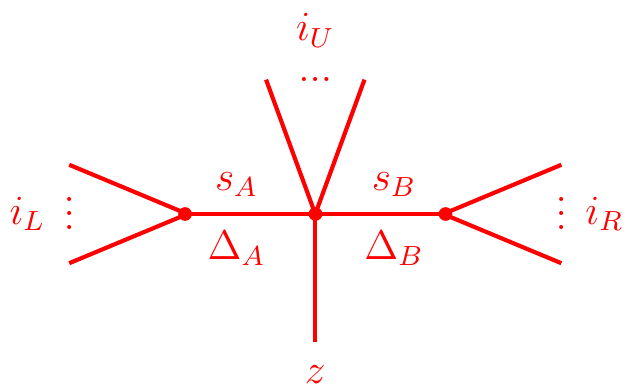}
\end{matrix}
\right)
\cr
&
\hspace{31mm}
\times \zeta_p\big(\Delta_{i_D}-z\big)
\zeta_p\big(\Delta_C-z\big)
\Bigg]
\cr
&
=-
\zeta_p\big(\Delta_A+\Delta_B+\Delta_C+\Delta_{i_U}-n\big)
\lim_{s_C\rightarrow -\infty}
\left(
\begin{matrix}
\vspace*{-0.2cm}
\includegraphics[height=25ex]{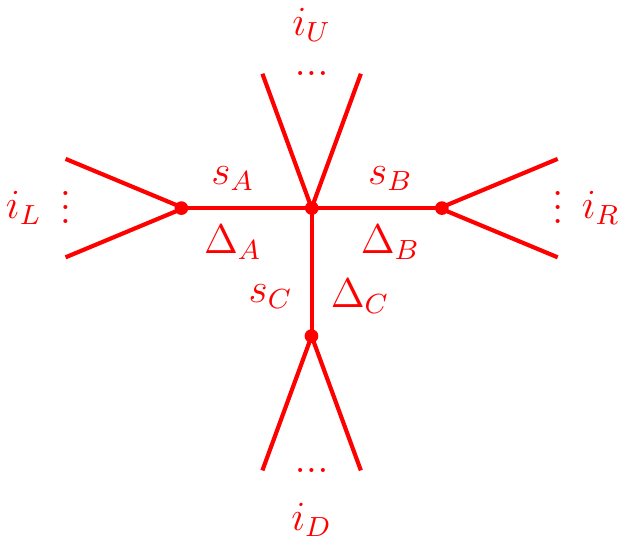}
\end{matrix}
\right).
}
These examples  demonstrate how the operation of ``adding an extra internal line'' works. Assuming that such an operation exists for any undressed diagram, it is easy to  prove that the type of contour integral displayed in the preceding subsection can be employed to glue together any two undressed diagrams. This proof is carried out in the next subsection. Then, in the subsection after the next, we justify the above-mentioned assumption by proving inductively that the ``internal leg adding operation'' can be applied to any undressed diagram, and in so doing we complete the proof of the recursive prescription~\ref{pres:padicMellin}.

\subsection{Integrating pre-amplitudes using ``leg addition''}

Given an arbitrary undressed diagram parametrized by the complex number $c$,
\eqn{}
{
\mathcal{D}\equiv
\left(
\begin{matrix}
\text{
\scalebox{0.75}{
\begin{tikzpicture}
\draw[red,very thick] (-1.4,0)--(-2.5,0.75);
\node at (-2.5-1.1*0.8,0.75+0.75*0.8) {\textcolor{red}{.}}; 
\node at (-2.5-1.1*1.2,0.75+0.75*1.2) {\textcolor{red}{$i_4$}}; 
\node at (-2.5-1.1*0.8+0.7*0.11,0.75+0.75*0.8+1*0.11){\textcolor{red}{.}}; 
\node at (-2.5-1.1*0.8-0.7*0.11,0.75+0.75*0.8-1*0.11){\textcolor{red}{.}}; 
\draw[red,very thick] (-2.5,0.75)--(-2.5-0.79,0.75+1.26);
\draw[red,very thick] (-2.5,0.75)--(-2.5-1.46,0.75+0.27);
\draw[red,very thick] (-1.4,0)--(-2.5,-0.75);
\draw[red,very thick] (-2.5,-0.75)--(-2.5-0.79,-0.75-1.26);
\draw[red,very thick] (-2.5,-0.75)--(-2.5-1.46,-0.75-0.27);
\node at (-2.5-1.1*0.8,-0.75-0.75*0.8) {\textcolor{red}{.}}; 
\node at (-2.5-1.1*1.2,-0.75-0.75*1.2) {\textcolor{red}{$i_3$}}; 
\node at (-2.5-1.1*0.8+0.7*0.11,-0.75-0.75*0.8-1*0.11){\textcolor{red}{.}}; 
\node at (-2.5-1.1*0.8-0.7*0.11,-0.75-0.75*0.8+1*0.11){\textcolor{red}{.}}; 
\draw[red,thick,fill=red] (-2.5,0.75) ellipse (0.05cm and 0.05cm);
\draw[red,thick,fill=red] (-2.5,-0.75) ellipse (0.05cm and 0.05cm);
\node at (-1.8,0.75) {\textcolor{red}{$D$}};
\node at (-1.8,-0.75) {\textcolor{red}{$C$}}; 
\node at (-0.4,0.7) {\textcolor{red}{$E$}};
\node at (1.6,0.7) {\textcolor{red}{$F$}};
\draw[red,thick,fill=red] (0,0) ellipse (0.05cm and 0.05cm);
\draw[red,thick,fill=red] (2.0,0) ellipse (0.05cm and 0.05cm);
\draw[red,thick,fill=red] (0,1.4) ellipse (0.05cm and 0.05cm);
\draw[red,thick,fill=red] (2.0,1.4) ellipse (0.05cm and 0.05cm);
\draw[red,thick,fill=red] (-1.4,0) ellipse (0.05cm and 0.05cm);
\draw[red,thick,fill=red] (3.5,0) ellipse (0.05cm and 0.05cm);
\draw[red,very thick] (4.9,0)--(-1.4,0);
\draw[red,very thick] (0,0)--(0,1.4);
\draw[red,very thick] (2,0)--(2,1.4);
\draw[red,very thick] (0,1.4)--(0.6,2.76);
\draw[red,very thick] (0,1.4)--(-0.6,2.76);
\draw[red,very thick] (0,0)--(0.6,-1.36);
\draw[red,very thick] (0,0)--(-0.6,-1.36);
\node at (0,-1.1) {\textcolor{red}{...}}; 
\node at (0,-1.55) {\textcolor{red}{$i_2$}}; 
\draw[red,very thick] (2,0)--(2.6,-1.36);
\draw[red,very thick] (2,0)--(1.4,-1.36);
\node at (2,-1.1) {\textcolor{red}{...}}; 
\node at (2,-1.55) {\textcolor{red}{$i_1$}}; 
\draw[red,very thick] (2,1.4)--(2.6,2.76);
\draw[red,very thick] (2,1.4)--(1.4,2.76);
\node at (0,1.1+1.4) {\textcolor{red}{...}}; 
\node at (0,1.65+1.4) {\textcolor{red}{$i_5$}}; 
\node at (2,1.1+1.4) {\textcolor{red}{...}}; 
\node at (2,1.65+1.4) {\textcolor{red}{$i_6$}}; 
\node at (-0.85,-0.4) {\textcolor{red}{$B$}};
\node at (1.,-0.4) {\textcolor{red}{$A$}}; 
\node at (2.9,-0.4) {\textcolor{red}{$h-c$}}; 
\node at (2.8,0.4) {\textcolor{red}{$s$}}; 
\node at (4.3,-0.4) {\textcolor{red}{$h+c$}}; 
\end{tikzpicture}}
}
\end{matrix}
\right),
}
we need to show that if it appears as part of a pre-amplitude as follows:
\eqn{}{
\widetilde{\mathcal{M}}=
\left(
\begin{matrix}
\text{
\scalebox{0.75}{
\begin{tikzpicture}
\draw[blue,thick,fill=blue] (0,0) ellipse (0.05cm and 0.05cm);
\draw[blue,very thick] (0,0)--(0,1.4);
\draw[blue,very thick] (0,0)--(0.6,-1.36);
\draw[blue,very thick] (0,0)--(-0.6,-1.36);
\node at (-0.4,0.8) {\textcolor{blue}{$F$}}; 
\node at (0,-1.1) {\textcolor{blue}{...}}; 
\node at (0,-1.55) {\textcolor{blue}{$i_1$}}; 
\draw[blue,very thick] (1.4,0)--(-1.4,0);
\node at (-0.9,-0.4) {\textcolor{blue}{$A$}};
\node at (0.9,-0.4) {\textcolor{blue}{$h-c$}}; 
\end{tikzpicture}}
}
\end{matrix}
\right)
\left(
\begin{matrix}
\text{
\scalebox{0.75}{
\begin{tikzpicture}
\draw[red,thick,fill=red] (0.8,0) ellipse (0.05cm and 0.05cm);
\draw[red,thick,fill=red] (-0.6,0) ellipse (0.05cm and 0.05cm);
\draw[red,very thick] (-2.0,0)--(0.8,0);
\draw[red,very thick] (0.8,0)--(1.73,0.63);
\draw[red,very thick] (0.8,0)--(1.73,-0.63);
\node at (0.1,-0.4) {\textcolor{red}{$h+c$}}; 
\node at (-1.3,-0.4) {\textcolor{red}{$h-c$}}; 
\node at (0.1,0.3) {\textcolor{red}{$s$}}; 
\node at (1.6,0.1) {\textcolor{red}{$\vdots$}}; 
\node at (2.1,0) {\textcolor{red}{$i_7$}}; 
\end{tikzpicture}}
}\end{matrix}
\right)
\left(
\begin{matrix}
\text{
\scalebox{0.75}{
\begin{tikzpicture}
\draw[blue,thick,fill=blue] (0.8,0) ellipse (0.05cm and 0.05cm);
\draw[blue,very thick] (-0.6,0)--(0.8,0);
\draw[blue,very thick] (0.8,0)--(1.73,0.63);
\draw[blue,very thick] (0.8,0)--(1.73,-0.63);
\node at (0.1,-0.4) {\textcolor{blue}{$h+c$}}; 
\node at (1.6,0.1) {\textcolor{blue}{$\vdots$}}; 
\node at (2.1,0) {\textcolor{blue}{$i_7$}}; 
\end{tikzpicture}}
}\end{matrix}
\right)
\mathcal{D},
}
then the contour integral over $c$ as described below effectively glues together the two undressed diagrams. That is, 
\eqn{ToShowGlue}
{
&\quad{\cal M} = 
\int_{-\frac{i\pi}{\log p}}^{\frac{i\pi}{\log p}}\, \frac{dc}{2\pi i}\,f_\Delta(c)\,\widetilde{\mathcal{M}}
  \cr 
  =
\left(
\begin{matrix}
\text{
\scalebox{0.75}{
\begin{tikzpicture}
\draw[blue,thick,fill=blue] (0,0) ellipse (0.05cm and 0.05cm);
\draw[blue,very thick] (0,0)--(0,1.4);
\draw[blue,very thick] (0,0)--(0.6,-1.36);
\draw[blue,very thick] (0,0)--(-0.6,-1.36);
\node at (-0.4,0.8) {\textcolor{blue}{$F$}}; 
\node at (0,-1.1) {\textcolor{blue}{...}}; 
\node at (0,-1.55) {\textcolor{blue}{$i_1$}}; 
\draw[blue,very thick] (1.4,0)--(-1.4,0);
\node at (-0.9,-0.4) {\textcolor{blue}{$A$}};
\node at (0.9,-0.4) {\textcolor{blue}{$\Delta$}}; 
\end{tikzpicture}}}
\end{matrix}
\right)
&
\left(
\begin{matrix}
\text{
\scalebox{0.75}{
\begin{tikzpicture}
\draw[red,very thick] (-1.4,0)--(-2.5,0.75);
\node at (-2.5-1.1*0.8,0.75+0.75*0.8) {\textcolor{red}{.}}; 
\node at (-2.5-1.1*1.2,0.75+0.75*1.2) {\textcolor{red}{$i_4$}}; 
\node at (-2.5-1.1*0.8+0.7*0.11,0.75+0.75*0.8+1*0.11){\textcolor{red}{.}}; 
\node at (-2.5-1.1*0.8-0.7*0.11,0.75+0.75*0.8-1*0.11){\textcolor{red}{.}}; 
\draw[red,very thick] (-2.5,0.75)--(-2.5-0.79,0.75+1.26);
\draw[red,very thick] (-2.5,0.75)--(-2.5-1.46,0.75+0.27);
\draw[red,very thick] (-1.4,0)--(-2.5,-0.75);
\draw[red,very thick] (-2.5,-0.75)--(-2.5-0.79,-0.75-1.26);
\draw[red,very thick] (-2.5,-0.75)--(-2.5-1.46,-0.75-0.27);
\node at (-2.5-1.1*0.8,-0.75-0.75*0.8) {\textcolor{red}{.}}; 
\node at (-2.5-1.1*1.2,-0.75-0.75*1.2) {\textcolor{red}{$i_3$}}; 
\node at (-2.5-1.1*0.8+0.7*0.11,-0.75-0.75*0.8-1*0.11){\textcolor{red}{.}}; \node at (-2.5-1.1*0.8-0.7*0.11,-0.75-0.75*0.8+1*0.11){\textcolor{red}{.}}; \draw[red,thick,fill=red] (-2.5,0.75) ellipse (0.05cm and 0.05cm);
\draw[red,thick,fill=red] (-2.5,-0.75) ellipse (0.05cm and 0.05cm);
\node at (-1.8,0.75) {\textcolor{red}{$D$}};
\node at (-1.8,-0.75) {\textcolor{red}{$C$}}; 
\node at (-0.4,0.7) {\textcolor{red}{$E$}};
\node at (1.6,0.7) {\textcolor{red}{$F$}};
\draw[red,thick,fill=red] (0,0) ellipse (0.05cm and 0.05cm);
\draw[red,thick,fill=red] (2.0,0) ellipse (0.05cm and 0.05cm);
\draw[red,thick,fill=red] (0,1.4) ellipse (0.05cm and 0.05cm);
\draw[red,thick,fill=red] (2.0,1.4) ellipse (0.05cm and 0.05cm);
\draw[red,thick,fill=red] (-1.4,0) ellipse (0.05cm and 0.05cm);
\draw[red,thick,fill=red] (3.4,0) ellipse (0.05cm and 0.05cm);
\draw[red,very thick] (3.4,0)--(-1.4,0);
\draw[red,very thick] (0,0)--(0,1.4);
\draw[red,very thick] (2,0)--(2,1.4);
\draw[red,very thick] (0,1.4)--(0.6,2.76);
\draw[red,very thick] (0,1.4)--(-0.6,2.76);
\draw[red,very thick] (0,0)--(0.6,-1.36);
\draw[red,very thick] (0,0)--(-0.6,-1.36);
\node at (0,-1.1) {\textcolor{red}{...}}; 
\node at (0,-1.55) {\textcolor{red}{$i_2$}}; 
\draw[red,very thick] (2,0)--(2.6,-1.36);
\draw[red,very thick] (2,0)--(1.4,-1.36);
\node at (2,-1.1) {\textcolor{red}{...}}; 
\node at (2,-1.55) {\textcolor{red}{$i_1$}}; 
\draw[red,very thick] (2,1.4)--(2.6,2.76);
\draw[red,very thick] (2,1.4)--(1.4,2.76);
\draw[red,very thick] (3.4,0)--(4.33,0.63);
\draw[red,very thick] (3.4,0)--(4.33,-0.63);
\node at (0,1.1+1.4) {\textcolor{red}{...}}; 
\node at (0,1.65+1.4) {\textcolor{red}{$i_5$}}; 
\node at (2,1.1+1.4) {\textcolor{red}{...}}; 
\node at (2,1.65+1.4) {\textcolor{red}{$i_6$}}; 
\node at (-0.85,-0.4) {\textcolor{red}{$B$}};
\node at (1.,-0.4) {\textcolor{red}{$A$}}; 
\node at (2.8,-0.4) {\textcolor{red}{$\Delta$}}; 
\node at (2.8,0.4) {\textcolor{red}{$s$}}; 
\node at (4.2,0.1) {\textcolor{red}{$\vdots$}}; 
\node at (4.7,0) {\textcolor{red}{$i_7$}}; 
\end{tikzpicture}}
}
\end{matrix}
\right)
\left(
\begin{matrix}
\text{
\scalebox{0.75}{
\begin{tikzpicture}
\draw[blue,thick,fill=blue] (0.8,0) ellipse (0.05cm and 0.05cm);
\draw[blue,very thick] (-0.6,0)--(0.8,0);
\draw[blue,very thick] (0.8,0)--(1.73,0.63);
\draw[blue,very thick] (0.8,0)--(1.73,-0.63);
\node at (0.1,-0.4) {\textcolor{blue}{$\Delta$}}; 
\node at (1.6,0.1) {\textcolor{blue}{$\vdots$}}; 
\node at (2.1,0) {\textcolor{blue}{$i_7$}}; 
\end{tikzpicture}}
}
\end{matrix}
\right),
}
where $f_\Delta(c)$ is given by \eno{fDef}. 
We will show this inductively,  by first assuming that the gluing of an extra internal line onto a diagram works for any diagram with fewer internal lines than the arbitrary diagram ${\cal D}$ we started with.

Now, split the undressed diagram into two parts: $\mathcal{D}=\mathcal{D}_1+\mathcal{D}_2$, where $\mathcal{D}_1$ consists of all the terms in $\mathcal{D}$ that are proportional to an internal line factor, say $\zeta_p(s_E-\Delta_E)$, while $\mathcal{D}_2$ consists of all the remaining terms of $\mathcal{D}$. That is,
\eqn{}
{
\mathcal{D}_1=-
\zeta_p(s_E-\Delta_E)
\left(
\begin{matrix}
\text{
\scalebox{0.75}{
\begin{tikzpicture}
\draw[red,very thick] (-1.4,0)--(-2.5,0.75);
\node at (-2.5-1.1*0.8,0.75+0.75*0.8) {\textcolor{red}{.}}; 
\node at (-2.5-1.1*1.2,0.75+0.75*1.2) {\textcolor{red}{$i_4$}}; 
\node at (-2.5-1.1*0.8+0.7*0.11,0.75+0.75*0.8+1*0.11){\textcolor{red}{.}}; 
\node at (-2.5-1.1*0.8-0.7*0.11,0.75+0.75*0.8-1*0.11){\textcolor{red}{.}}; 
\draw[red,very thick] (-2.5,0.75)--(-2.5-0.79,0.75+1.26);
\draw[red,very thick] (-2.5,0.75)--(-2.5-1.46,0.75+0.27);
\draw[red,very thick] (-1.4,0)--(-2.5,-0.75);
\draw[red,very thick] (-2.5,-0.75)--(-2.5-0.79,-0.75-1.26);
\draw[red,very thick] (-2.5,-0.75)--(-2.5-1.46,-0.75-0.27);
\node at (-2.5-1.1*0.8,-0.75-0.75*0.8) {\textcolor{red}{.}}; 
\node at (-2.5-1.1*1.2,-0.75-0.75*1.2) {\textcolor{red}{$i_3$}}; 
\node at (-2.5-1.1*0.8+0.7*0.11,-0.75-0.75*0.8-1*0.11){\textcolor{red}{.}}; 
\node at (-2.5-1.1*0.8-0.7*0.11,-0.75-0.75*0.8+1*0.11){\textcolor{red}{.}}; 
\draw[red,thick,fill=red] (-2.5,0.75) ellipse (0.05cm and 0.05cm);
\draw[red,thick,fill=red] (-2.5,-0.75) ellipse (0.05cm and 0.05cm);
\node at (-1.8,0.75) {\textcolor{red}{$D$}};
\node at (-1.8,-0.75) {\textcolor{red}{$C$}}; 
\node at (0,1.8) {\textcolor{red}{$E$}};
\node at (1.6,0.7) {\textcolor{red}{$F$}};
\draw[red,thick,fill=red] (0,0) ellipse (0.05cm and 0.05cm);
\draw[red,thick,fill=red] (2.0,0) ellipse (0.05cm and 0.05cm);
\draw[red,thick,fill=red] (2.0,1.4) ellipse (0.05cm and 0.05cm);
\draw[red,thick,fill=red] (-1.4,0) ellipse (0.05cm and 0.05cm);
\draw[red,thick,fill=red] (3.5,0) ellipse (0.05cm and 0.05cm);
\draw[red,very thick] (4.9,0)--(-1.4,0);
\draw[red,very thick] (0,0)--(0,1.4);
\draw[red,very thick] (2,0)--(2,1.4);
\draw[red,very thick] (0,0)--(0.6,-1.36);
\draw[red,very thick] (0,0)--(-0.6,-1.36);
\node at (0,-1.1) {\textcolor{red}{...}}; 
\node at (0,-1.55) {\textcolor{red}{$i_2$}}; 
\draw[red,very thick] (2,0)--(2.6,-1.36);
\draw[red,very thick] (2,0)--(1.4,-1.36);
\node at (2,-1.1) {\textcolor{red}{...}}; 
\node at (2,-1.55) {\textcolor{red}{$i_1$}}; 
\draw[red,very thick] (2,1.4)--(2.6,2.76);
\draw[red,very thick] (2,1.4)--(1.4,2.76);
\node at (2,1.1+1.4) {\textcolor{red}{...}}; 
\node at (2,1.65+1.4) {\textcolor{red}{$i_6$}}; 
\node at (-0.85,-0.4) {\textcolor{red}{$B$}};
\node at (1.,-0.4) {\textcolor{red}{$A$}}; 
\node at (2.9,-0.4) {\textcolor{red}{$h-c$}}; 
\node at (2.8,0.4) {\textcolor{red}{$s$}}; 
\node at (4.3,-0.4) {\textcolor{red}{$h+c$}}; 
\end{tikzpicture}}
}
\end{matrix}
\right),
}
where we used the factorization property of undressed diagrams discussed in section \ref{Factorization}, and 
\eqn{}
{
\mathcal{D}_2
&= \lim_{s_E\rightarrow -\infty} {\cal D}
= \lim_{s_E\rightarrow -\infty}
\left(
\begin{matrix}
\scalebox{0.75}{
\text{\begin{tikzpicture}
\draw[red,very thick] (-1.4,0)--(-2.5,0.75);
\node at (-2.5-1.1*0.8,0.75+0.75*0.8) {\textcolor{red}{.}}; 
\node at (-2.5-1.1*1.2,0.75+0.75*1.2) {\textcolor{red}{$i_4$}}; 
\node at (-2.5-1.1*0.8+0.7*0.11,0.75+0.75*0.8+1*0.11){\textcolor{red}{.}}; 
\node at (-2.5-1.1*0.8-0.7*0.11,0.75+0.75*0.8-1*0.11){\textcolor{red}{.}}; 
\draw[red,very thick] (-2.5,0.75)--(-2.5-0.79,0.75+1.26);
\draw[red,very thick] (-2.5,0.75)--(-2.5-1.46,0.75+0.27);
\draw[red,very thick] (-1.4,0)--(-2.5,-0.75);
\draw[red,very thick] (-2.5,-0.75)--(-2.5-0.79,-0.75-1.26);
\draw[red,very thick] (-2.5,-0.75)--(-2.5-1.46,-0.75-0.27);
\node at (-2.5-1.1*0.8,-0.75-0.75*0.8) {\textcolor{red}{.}}; 
\node at (-2.5-1.1*1.2,-0.75-0.75*1.2) {\textcolor{red}{$i_3$}}; 
\node at (-2.5-1.1*0.8+0.7*0.11,-0.75-0.75*0.8-1*0.11){\textcolor{red}{.}}; 
\node at (-2.5-1.1*0.8-0.7*0.11,-0.75-0.75*0.8+1*0.11){\textcolor{red}{.}}; 
\draw[red,thick,fill=red] (-2.5,0.75) ellipse (0.05cm and 0.05cm);
\draw[red,thick,fill=red] (-2.5,-0.75) ellipse (0.05cm and 0.05cm);
\node at (-1.8,0.75) {\textcolor{red}{$D$}};
\node at (-1.8,-0.75) {\textcolor{red}{$C$}}; 
\node at (-0.4,0.7) {\textcolor{red}{$E$}};
\node at (1.6,0.7) {\textcolor{red}{$F$}};
\draw[red,thick,fill=red] (0,0) ellipse (0.05cm and 0.05cm);
\draw[red,thick,fill=red] (2.0,0) ellipse (0.05cm and 0.05cm);
\draw[red,thick,fill=red] (0,1.4) ellipse (0.05cm and 0.05cm);
\draw[red,thick,fill=red] (2.0,1.4) ellipse (0.05cm and 0.05cm);
\draw[red,thick,fill=red] (-1.4,0) ellipse (0.05cm and 0.05cm);
\draw[red,thick,fill=red] (3.5,0) ellipse (0.05cm and 0.05cm);
\draw[red,very thick] (4.9,0)--(-1.4,0);
\draw[red,very thick] (0,0)--(0,1.4);
\draw[red,very thick] (2,0)--(2,1.4);
\draw[red,very thick] (0,1.4)--(0.6,2.76);
\draw[red,very thick] (0,1.4)--(-0.6,2.76);
\draw[red,very thick] (0,0)--(0.6,-1.36);
\draw[red,very thick] (0,0)--(-0.6,-1.36);
\node at (0,-1.1) {\textcolor{red}{...}}; 
\node at (0,-1.55) {\textcolor{red}{$i_2$}}; 
\draw[red,very thick] (2,0)--(2.6,-1.36);
\draw[red,very thick] (2,0)--(1.4,-1.36);
\node at (2,-1.1) {\textcolor{red}{...}}; 
\node at (2,-1.55) {\textcolor{red}{$i_1$}}; 
\draw[red,very thick] (2,1.4)--(2.6,2.76);
\draw[red,very thick] (2,1.4)--(1.4,2.76);
\node at (0,1.1+1.4) {\textcolor{red}{...}}; 
\node at (0,1.65+1.4) {\textcolor{red}{$i_5$}}; 
\node at (2,1.1+1.4) {\textcolor{red}{...}}; 
\node at (2,1.65+1.4) {\textcolor{red}{$i_6$}}; 
\node at (-0.85,-0.4) {\textcolor{red}{$B$}};
\node at (1.,-0.4) {\textcolor{red}{$A$}}; 
\node at (2.9,-0.4) {\textcolor{red}{$h-c$}}; 
\node at (2.8,0.4) {\textcolor{red}{$s$}}; 
\node at (4.3,-0.4) {\textcolor{red}{$h+c$}}; 
\end{tikzpicture}}
}
\end{matrix}
\right)
\cr
&=
-
\sum _{z^\ast\in\{\Delta_E,\Delta_{i_2}\}}
\res_{z^\ast}
\Bigg[
\zeta_p\big(\Delta_A+\Delta_B+\Delta_{i_2}+z-n\big)
\left(
\begin{matrix}
\text{
\scalebox{0.75}{
\begin{tikzpicture}
\draw[red,very thick] (-1.4,0)--(-2.5,0.75);
\node at (-2.5-1.1*0.8,0.75+0.75*0.8) {\textcolor{red}{.}}; 
\node at (-2.5-1.1*1.2,0.75+0.75*1.2) {\textcolor{red}{$i_4$}}; 
\node at (-2.5-1.1*0.8+0.7*0.11,0.75+0.75*0.8+1*0.11){\textcolor{red}{.}}; 
\node at (-2.5-1.1*0.8-0.7*0.11,0.75+0.75*0.8-1*0.11){\textcolor{red}{.}}; 
\draw[red,very thick] (-2.5,0.75)--(-2.5-0.79,0.75+1.26);
\draw[red,very thick] (-2.5,0.75)--(-2.5-1.46,0.75+0.27);
\draw[red,very thick] (-1.4,0)--(-2.5,-0.75);
\draw[red,very thick] (-2.5,-0.75)--(-2.5-0.79,-0.75-1.26);
\draw[red,very thick] (-2.5,-0.75)--(-2.5-1.46,-0.75-0.27);
\node at (-2.5-1.1*0.8,-0.75-0.75*0.8) {\textcolor{red}{.}}; 
\node at (-2.5-1.1*1.2,-0.75-0.75*1.2) {\textcolor{red}{$i_3$}}; 
\node at (-2.5-1.1*0.8+0.7*0.11,-0.75-0.75*0.8-1*0.11){\textcolor{red}{.}}; 
\node at (-2.5-1.1*0.8-0.7*0.11,-0.75-0.75*0.8+1*0.11){\textcolor{red}{.}}; 
\draw[red,thick,fill=red] (-2.5,0.75) ellipse (0.05cm and 0.05cm);
\draw[red,thick,fill=red] (-2.5,-0.75) ellipse (0.05cm and 0.05cm);
\node at (-1.8,0.75) {\textcolor{red}{$D$}};
\node at (-1.8,-0.75) {\textcolor{red}{$C$}}; 
\node at (0,1.8) {\textcolor{red}{$z$}};
\node at (1.6,0.7) {\textcolor{red}{$F$}};
\draw[red,thick,fill=red] (0,0) ellipse (0.05cm and 0.05cm);
\draw[red,thick,fill=red] (2.0,0) ellipse (0.05cm and 0.05cm);
\draw[red,thick,fill=red] (2.0,1.4) ellipse (0.05cm and 0.05cm);
\draw[red,thick,fill=red] (-1.4,0) ellipse (0.05cm and 0.05cm);
\draw[red,thick,fill=red] (3.5,0) ellipse (0.05cm and 0.05cm);
\draw[red,very thick] (4.9,0)--(-1.4,0);
\draw[red,very thick] (0,0)--(0,1.4);
\draw[red,very thick] (2,0)--(2,1.4);
\draw[red,very thick] (0,0)--(0.6,-1.36);
\draw[red,very thick] (0,0)--(-0.6,-1.36);
\node at (0,-1.1) {\textcolor{red}{...}}; 
\node at (0,-1.55) {\textcolor{red}{$i_2$}}; 
\draw[red,very thick] (2,0)--(2.6,-1.36);
\draw[red,very thick] (2,0)--(1.4,-1.36);
\node at (2,-1.1) {\textcolor{red}{...}}; 
\node at (2,-1.55) {\textcolor{red}{$i_1$}}; 
\draw[red,very thick] (2,1.4)--(2.6,2.76);
\draw[red,very thick] (2,1.4)--(1.4,2.76);
\node at (2,1.1+1.4) {\textcolor{red}{...}}; 
\node at (2,1.65+1.4) {\textcolor{red}{$i_6$}}; 
\node at (-0.85,-0.4) {\textcolor{red}{$B$}};
\node at (1.,-0.4) {\textcolor{red}{$A$}}; 
\node at (2.9,-0.4) {\textcolor{red}{$h-c$}}; 
\node at (2.8,0.4) {\textcolor{red}{$s$}}; 
\node at (4.3,-0.4) {\textcolor{red}{$h+c$}}; 
\end{tikzpicture}}
}
\end{matrix}
\right)
\cr
&
\hspace{31mm}
\times \zeta_p\big(\Delta_{i_5}-z\big)
\zeta_p\big(\Delta_E-z\big)
\Bigg]\frac{1}{\zeta_p\big(\Delta_A+\Delta_B+\Delta_E+\Delta_{i_2}-n\big)}
\,,
}
where we used the ``leg adding operation'' from the previous subsection to write the final equality above. (The validity of such a leg adding operation is proven in full generality in the next subsection.)

Essentially, the point of this exercise was to re-express $\mathcal{D}_1$ and $\mathcal{D}_2$ (and thus in turn ${\cal D}$)  in terms of a diagram with one fewer internal line than the original undressed diagram ${\cal D}$.
Having done that,
we can now carry out the contour integral in \eno{ToShowGlue} by breaking $\widetilde{{\cal M}}$ into two terms, 
 each of which, using the assumption of our inductive setup,
admits an extra internal line to be glued on.

Explicitly, splitting up the integrated amplitude $\mathcal{M}=\mathcal{M}_1+\mathcal{M}_2$ into the parts coming from $\mathcal{D}_1$ and $\mathcal{D}_2$, respectively, the inductive assumption allows us to straightforwardly evaluate the two parts. The first part is given by 

\clearpage
\eqn{}
{\mathcal{M}_1\equiv
&\,
\int_{-\frac{i\pi}{\log p}}^{\frac{i\pi}{\log p}}\, \frac{dc}{2\pi i}\,f_\Delta(c)\,\left(
\begin{matrix}
\text{
\scalebox{0.75}{
\begin{tikzpicture}
\draw[blue,thick,fill=blue] (0,0) ellipse (0.05cm and 0.05cm);
\draw[blue,very thick] (0,0)--(0,1.4);
\draw[blue,very thick] (0,0)--(0.6,-1.36);
\draw[blue,very thick] (0,0)--(-0.6,-1.36);
\node at (-0.4,0.8) {\textcolor{blue}{$F$}}; 
\node at (0,-1.1) {\textcolor{blue}{...}}; 
\node at (0,-1.55) {\textcolor{blue}{$i_1$}}; 
\draw[blue,very thick] (1.4,0)--(-1.4,0);
\node at (-0.9,-0.4) {\textcolor{blue}{$A$}};
\node at (0.9,-0.4) {\textcolor{blue}{$h-c$}}; 
\end{tikzpicture}}
}
\end{matrix}
\right)
\left(
\begin{matrix}
\text{
\scalebox{0.75}{
\begin{tikzpicture}
\draw[red,thick,fill=red] (0.8,0) ellipse (0.05cm and 0.05cm);
\draw[red,thick,fill=red] (-0.6,0) ellipse (0.05cm and 0.05cm);
\draw[red,very thick] (-2.0,0)--(0.8,0);
\draw[red,very thick] (0.8,0)--(1.73,0.63);
\draw[red,very thick] (0.8,0)--(1.73,-0.63);
\node at (0.1,-0.4) {\textcolor{red}{$h+c$}}; 
\node at (-1.3,-0.4) {\textcolor{red}{$h-c$}}; 
\node at (0.1,0.3) {\textcolor{red}{$s$}}; 
\node at (1.6,0.1) {\textcolor{red}{$\vdots$}}; 
\node at (2.1,0) {\textcolor{red}{$i_R$}}; 
\end{tikzpicture}}
}
\end{matrix}
\right)
\left(
\begin{matrix}
\text{
\scalebox{0.75}{
\begin{tikzpicture}
\draw[blue,thick,fill=blue] (0.8,0) ellipse (0.05cm and 0.05cm);
\draw[blue,very thick] (-0.6,0)--(0.8,0);
\draw[blue,very thick] (0.8,0)--(1.73,0.63);
\draw[blue,very thick] (0.8,0)--(1.73,-0.63);
\node at (0.1,-0.4) {\textcolor{blue}{$h+c$}}; 
\node at (1.6,0.1) {\textcolor{blue}{$\vdots$}}; 
\node at (2.1,0) {\textcolor{blue}{$i_7$}}; 
\end{tikzpicture}}
}
\end{matrix}
\right)
\mathcal{D}_1
\cr
=&
-
\left(
\begin{matrix}
\text{
\scalebox{0.75}{
\begin{tikzpicture}
\draw[blue,thick,fill=blue] (0,0) ellipse (0.05cm and 0.05cm);
\draw[blue,very thick] (0,0)--(0,1.4);
\draw[blue,very thick] (0,0)--(0.6,-1.36);
\draw[blue,very thick] (0,0)--(-0.6,-1.36);
\node at (-0.4,0.8) {\textcolor{blue}{$F$}}; 
\node at (0,-1.1) {\textcolor{blue}{...}}; 
\node at (0,-1.55) {\textcolor{blue}{$i_1$}}; 
\draw[blue,very thick] (1.4,0)--(-1.4,0);
\node at (-0.9,-0.4) {\textcolor{blue}{$A$}};
\node at (0.9,-0.4) {\textcolor{blue}{$\Delta$}}; 
\end{tikzpicture}}
}
\end{matrix}
\right)
\zeta_p(s_E-\Delta_E)
\left(
\begin{matrix}
\text{
\scalebox{0.75}{
\begin{tikzpicture}
\draw[red,very thick] (-1.4,0)--(-2.5,0.75);
\node at (-2.5-1.1*0.8,0.75+0.75*0.8) {\textcolor{red}{.}}; 
\node at (-2.5-1.1*1.2,0.75+0.75*1.2) {\textcolor{red}{$i_4$}}; 
\node at (-2.5-1.1*0.8+0.7*0.11,0.75+0.75*0.8+1*0.11){\textcolor{red}{.}}; 
\node at (-2.5-1.1*0.8-0.7*0.11,0.75+0.75*0.8-1*0.11){\textcolor{red}{.}}; 
\draw[red,very thick] (-2.5,0.75)--(-2.5-0.79,0.75+1.26);
\draw[red,very thick] (-2.5,0.75)--(-2.5-1.46,0.75+0.27);
\draw[red,very thick] (-1.4,0)--(-2.5,-0.75);
\draw[red,very thick] (-2.5,-0.75)--(-2.5-0.79,-0.75-1.26);
\draw[red,very thick] (-2.5,-0.75)--(-2.5-1.46,-0.75-0.27);
\node at (-2.5-1.1*0.8,-0.75-0.75*0.8) {\textcolor{red}{.}}; 
\node at (-2.5-1.1*1.2,-0.75-0.75*1.2) {\textcolor{red}{$i_3$}}; 
\node at (-2.5-1.1*0.8+0.7*0.11,-0.75-0.75*0.8-1*0.11){\textcolor{red}{.}}; 
\node at (-2.5-1.1*0.8-0.7*0.11,-0.75-0.75*0.8+1*0.11){\textcolor{red}{.}}; 
\draw[red,thick,fill=red] (-2.5,0.75) ellipse (0.05cm and 0.05cm);
\draw[red,thick,fill=red] (-2.5,-0.75) ellipse (0.05cm and 0.05cm);
\node at (-1.8,0.75) {\textcolor{red}{$D$}};
\node at (-1.8,-0.75) {\textcolor{red}{$C$}}; 
\node at (0,1.7) {\textcolor{red}{$E$}};
\node at (1.6,0.7) {\textcolor{red}{$F$}};
\draw[red,thick,fill=red] (0,0) ellipse (0.05cm and 0.05cm);
\draw[red,thick,fill=red] (2.0,0) ellipse (0.05cm and 0.05cm);
\draw[red,thick,fill=red] (2.0,1.4) ellipse (0.05cm and 0.05cm);
\draw[red,thick,fill=red] (-1.4,0) ellipse (0.05cm and 0.05cm);
\draw[red,thick,fill=red] (3.4,0) ellipse (0.05cm and 0.05cm);
\draw[red,very thick] (3.4,0)--(-1.4,0);
\draw[red,very thick] (0,0)--(0,1.4);
\draw[red,very thick] (2,0)--(2,1.4);
\draw[red,very thick] (0,0)--(0.6,-1.36);
\draw[red,very thick] (0,0)--(-0.6,-1.36);
\node at (0,-1.1) {\textcolor{red}{...}}; 
\node at (0,-1.55) {\textcolor{red}{$i_2$}}; 
\draw[red,very thick] (2,0)--(2.6,-1.36);
\draw[red,very thick] (2,0)--(1.4,-1.36);
\node at (2,-1.1) {\textcolor{red}{...}}; 
\node at (2,-1.55) {\textcolor{red}{$i_1$}}; 
\draw[red,very thick] (2,1.4)--(2.6,2.76);
\draw[red,very thick] (2,1.4)--(1.4,2.76);
\draw[red,very thick] (3.4,0)--(4.33,0.63);
\draw[red,very thick] (3.4,0)--(4.33,-0.63);
\node at (2,1.1+1.4) {\textcolor{red}{...}}; 
\node at (2,1.65+1.4) {\textcolor{red}{$i_6$}}; 
\node at (-0.85,-0.4) {\textcolor{red}{$B$}};
\node at (1.,-0.4) {\textcolor{red}{$A$}}; 
\node at (2.8,-0.4) {\textcolor{red}{$\Delta$}}; 
\node at (2.8,0.4) {\textcolor{red}{$s$}}; 
\node at (4.2,0.1) {\textcolor{red}{$\vdots$}}; 
\node at (4.7,0) {\textcolor{red}{$i_7$}}; 
\end{tikzpicture}}
}
\end{matrix}
\right) \cr 
& \times 
\left(
\begin{matrix}
\text{
\scalebox{0.75}{
\begin{tikzpicture}
\draw[blue,thick,fill=blue] (0.8,0) ellipse (0.05cm and 0.05cm);
\draw[blue,very thick] (-0.6,0)--(0.8,0);
\draw[blue,very thick] (0.8,0)--(1.73,0.63);
\draw[blue,very thick] (0.8,0)--(1.73,-0.63);
\node at (0.1,-0.4) {\textcolor{blue}{$\Delta$}}; 
\node at (1.6,0.1) {\textcolor{blue}{$\vdots$}}; 
\node at (2.1,0) {\textcolor{blue}{$i_7$}}; 
\end{tikzpicture}}
}
\end{matrix}
\right).
}
And the second part is given by 
\eqn{}
{
\mathcal{M}_2\equiv &
\,
\int_{-\frac{i\pi}{\log p}}^{\frac{i\pi}{\log p}}\, \frac{dc}{2\pi i}\,f_\Delta(c)
\,\left(
\begin{matrix}
\text{
\scalebox{0.75}{
\begin{tikzpicture}
\draw[blue,thick,fill=blue] (0,0) ellipse (0.05cm and 0.05cm);
\draw[blue,very thick] (0,0)--(0,1.4);
\draw[blue,very thick] (0,0)--(0.6,-1.36);
\draw[blue,very thick] (0,0)--(-0.6,-1.36);
\node at (-0.4,0.8) {\textcolor{blue}{$F$}}; 
\node at (0,-1.1) {\textcolor{blue}{...}}; 
\node at (0,-1.55) {\textcolor{blue}{$i_1$}}; 
\draw[blue,very thick] (1.4,0)--(-1.4,0);
\node at (-0.9,-0.4) {\textcolor{blue}{$A$}};
\node at (0.9,-0.4) {\textcolor{blue}{$h-c$}}; 
\end{tikzpicture}}
}
\end{matrix}
\right)
\left(
\begin{matrix}
\text{
\scalebox{0.75}{
\begin{tikzpicture}
\draw[red,thick,fill=red] (0.8,0) ellipse (0.05cm and 0.05cm);
\draw[red,thick,fill=red] (-0.6,0) ellipse (0.05cm and 0.05cm);
\draw[red,very thick] (-2.0,0)--(0.8,0);
\draw[red,very thick] (0.8,0)--(1.73,0.63);
\draw[red,very thick] (0.8,0)--(1.73,-0.63);
\node at (0.1,-0.4) {\textcolor{red}{$h+c$}}; 
\node at (-1.3,-0.4) {\textcolor{red}{$h-c$}}; 
\node at (0.1,0.3) {\textcolor{red}{$s$}}; 
\node at (1.6,0.1) {\textcolor{red}{$\vdots$}}; 
\node at (2.1,0) {\textcolor{red}{$i_7$}}; 
\end{tikzpicture}}
}
\end{matrix}
\right)
\left(
\begin{matrix}
\text{
\scalebox{0.75}{
\begin{tikzpicture}
\draw[blue,thick,fill=blue] (0.8,0) ellipse (0.05cm and 0.05cm);
\draw[blue,very thick] (-0.6,0)--(0.8,0);
\draw[blue,very thick] (0.8,0)--(1.73,0.63);
\draw[blue,very thick] (0.8,0)--(1.73,-0.63);
\node at (0.1,-0.4) {\textcolor{blue}{$h+c$}}; 
\node at (1.6,0.1) {\textcolor{blue}{$\vdots$}}; 
\node at (2.1,0) {\textcolor{blue}{$i_7$}}; 
\end{tikzpicture}}
}
\end{matrix}
\right)
\mathcal{D}_2\,,
}
which, using the leg adding operation, we may re-express as
\clearpage
\eqn{}
{
\mathcal{M}_2
=&-\sum _{z^\ast\in\{\Delta_E,\Delta_{i_5}\}}
\res_{z^\ast}
\Bigg[
\zeta_p\big(\Delta_A+\Delta_B+\Delta_{i_2}+z-n\big)\,
\int_{-\frac{i\pi}{\log p}}^{\frac{i\pi}{\log p}}\, \frac{dc}{2\pi}\,f_\Delta(c) 
\left(
\begin{matrix}
\text{
\scalebox{0.75}{
\begin{tikzpicture}
\draw[blue,thick,fill=blue] (0,0) ellipse (0.05cm and 0.05cm);
\draw[blue,very thick] (0,0)--(0,1.4);
\draw[blue,very thick] (0,0)--(0.6,-1.36);
\draw[blue,very thick] (0,0)--(-0.6,-1.36);
\node at (-0.4,0.8) {\textcolor{blue}{$F$}}; 
\node at (0,-1.1) {\textcolor{blue}{...}}; 
\node at (0,-1.55) {\textcolor{blue}{$i_1$}}; 
\draw[blue,very thick] (1.4,0)--(-1.4,0);
\node at (-0.9,-0.4) {\textcolor{blue}{$A$}};
\node at (0.9,-0.4) {\textcolor{blue}{$h-c$}}; 
\end{tikzpicture}}
}
\end{matrix}
\right)
\cr
&
\hspace{34mm}
\times 
\left(
\begin{matrix}
\text{
\scalebox{0.75}{
\begin{tikzpicture}
\draw[red,thick,fill=red] (0.8,0) ellipse (0.05cm and 0.05cm);
\draw[red,thick,fill=red] (-0.6,0) ellipse (0.05cm and 0.05cm);
\draw[red,very thick] (-2.0,0)--(0.8,0);
\draw[red,very thick] (0.8,0)--(1.73,0.63);
\draw[red,very thick] (0.8,0)--(1.73,-0.63);
\node at (0.1,-0.4) {\textcolor{red}{$h+c$}}; 
\node at (-1.3,-0.4) {\textcolor{red}{$h-c$}}; 
\node at (0.1,0.3) {\textcolor{red}{$s$}}; 
\node at (1.6,0.1) {\textcolor{red}{$\vdots$}}; 
\node at (2.1,0) {\textcolor{red}{$i_7$}}; 
\end{tikzpicture}}
}
\end{matrix}
\right)
\left(
\begin{matrix}
\text{
\scalebox{0.75}{
\begin{tikzpicture}
\draw[blue,thick,fill=blue] (0.8,0) ellipse (0.05cm and 0.05cm);
\draw[blue,very thick] (-0.6,0)--(0.8,0);
\draw[blue,very thick] (0.8,0)--(1.73,0.63);
\draw[blue,very thick] (0.8,0)--(1.73,-0.63);
\node at (0.1,-0.4) {\textcolor{blue}{$h+c$}}; 
\node at (1.6,0.1) {\textcolor{blue}{$\vdots$}}; 
\node at (2.1,0) {\textcolor{blue}{$i_7$}}; 
\end{tikzpicture}}
}
\end{matrix}
\right)
\cr
&
\hspace{34mm}
\times 
\left(
\begin{matrix}
\text{
\scalebox{0.75}{
\begin{tikzpicture}
\draw[red,very thick] (-1.4,0)--(-2.5,0.75);
\node at (-2.5-1.1*0.8,0.75+0.75*0.8) {\textcolor{red}{.}};
\node at (-2.5-1.1*1.2,0.75+0.75*1.2) {\textcolor{red}{$i_4$}}; 
\node at (-2.5-1.1*0.8+0.7*0.11,0.75+0.75*0.8+1*0.11){\textcolor{red}{.}}; 
\node at (-2.5-1.1*0.8-0.7*0.11,0.75+0.75*0.8-1*0.11){\textcolor{red}{.}}; 
\draw[red,very thick] (-2.5,0.75)--(-2.5-0.79,0.75+1.26);
\draw[red,very thick] (-2.5,0.75)--(-2.5-1.46,0.75+0.27);
\draw[red,very thick] (-1.4,0)--(-2.5,-0.75);
\draw[red,very thick] (-2.5,-0.75)--(-2.5-0.79,-0.75-1.26);
\draw[red,very thick] (-2.5,-0.75)--(-2.5-1.46,-0.75-0.27);
\node at (-2.5-1.1*0.8,-0.75-0.75*0.8) {\textcolor{red}{.}}; 
\node at (-2.5-1.1*1.2,-0.75-0.75*1.2) {\textcolor{red}{$i_3$}}; 
\node at (-2.5-1.1*0.8+0.7*0.11,-0.75-0.75*0.8-1*0.11){\textcolor{red}{.}}; 
\node at (-2.5-1.1*0.8-0.7*0.11,-0.75-0.75*0.8+1*0.11){\textcolor{red}{.}}; 
\draw[red,thick,fill=red] (-2.5,0.75) ellipse (0.05cm and 0.05cm);
\draw[red,thick,fill=red] (-2.5,-0.75) ellipse (0.05cm and 0.05cm);
\node at (-1.8,0.75) {\textcolor{red}{$D$}};
\node at (-1.8,-0.75) {\textcolor{red}{$C$}}; 
\node at (0,1.8) {\textcolor{red}{$z$}};
\node at (1.6,0.7) {\textcolor{red}{$F$}};
\draw[red,thick,fill=red] (0,0) ellipse (0.05cm and 0.05cm);
\draw[red,thick,fill=red] (2.0,0) ellipse (0.05cm and 0.05cm);
\draw[red,thick,fill=red] (2.0,1.4) ellipse (0.05cm and 0.05cm);
\draw[red,thick,fill=red] (-1.4,0) ellipse (0.05cm and 0.05cm);
\draw[red,thick,fill=red] (3.5,0) ellipse (0.05cm and 0.05cm);
\draw[red,very thick] (4.9,0)--(-1.4,0);
\draw[red,very thick] (0,0)--(0,1.4);
\draw[red,very thick] (2,0)--(2,1.4);
\draw[red,very thick] (0,0)--(0.6,-1.36);
\draw[red,very thick] (0,0)--(-0.6,-1.36);
\node at (0,-1.1) {\textcolor{red}{...}}; 
\node at (0,-1.55) {\textcolor{red}{$i_2$}}; 
\draw[red,very thick] (2,0)--(2.6,-1.36);
\draw[red,very thick] (2,0)--(1.4,-1.36);
\node at (2,-1.1) {\textcolor{red}{...}}; 
\node at (2,-1.55) {\textcolor{red}{$i_1$}}; 
\draw[red,very thick] (2,1.4)--(2.6,2.76);
\draw[red,very thick] (2,1.4)--(1.4,2.76);
\node at (2,1.1+1.4) {\textcolor{red}{...}}; 
\node at (2,1.65+1.4) {\textcolor{red}{$i_6$}}; 
\node at (-0.85,-0.4) {\textcolor{red}{$B$}};
\node at (1.,-0.4) {\textcolor{red}{$A$}}; 
\node at (2.9,-0.4) {\textcolor{red}{$h-c$}}; 
\node at (2.8,0.4) {\textcolor{red}{$s$}}; 
\node at (4.3,-0.4) {\textcolor{red}{$h+c$}}; 
\end{tikzpicture}}
}
\end{matrix}
\right)
\cr
&
\hspace{34mm}
\times 
\zeta_p\big(\Delta_{i_5}-z\big)
\zeta_p\big(\Delta_E-z\big)
\Bigg]\frac{1}{\zeta_p\big(\Delta_A+\Delta_B+\Delta_E+\Delta_{i_2}-n\big)}
\,.
}
Here we have made use of the fact that the leg adding operation commutes with the contour integral so that we may choose to first carry out the contour integral, which is easily done using the inductive assumption:
\eqn{}
{
\mathcal{M}_2
=
&
-\sum _{z^\ast\in\{\Delta_E,\Delta_{i_5}\}}
\res_{z^\ast}
\Bigg[
\zeta_p\big(\Delta_A+\Delta_B+\Delta_{i_2}+z-n\big)\,
\cr
&
\quad \times \left(
\begin{matrix}
\vspace*{-0.2cm}
\includegraphics[height=13ex]{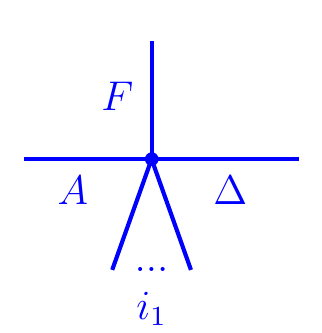}
\end{matrix}
\right)
\left(
\begin{matrix}
\vspace*{-0.2cm}
\includegraphics[height=18ex]{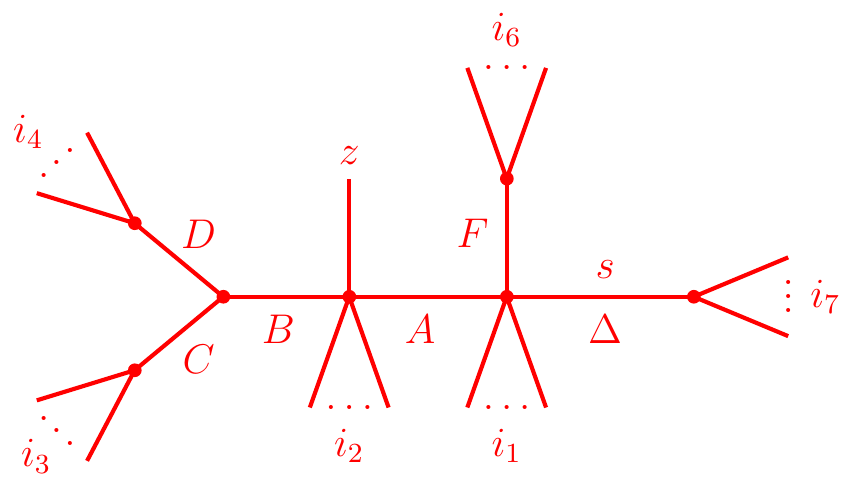}
\end{matrix}
\right)
\left(
\begin{matrix}
\vspace*{-0.2cm}
\includegraphics[height=6ex]{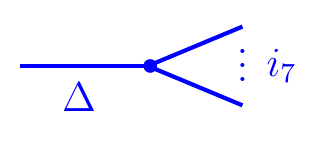}
\end{matrix}
\right)
\cr
&
\quad \times 
\zeta_p\big(\Delta_{i_5}-z\big)
\zeta_p\big(\Delta_E-z\big)
\Bigg]\frac{1}{\zeta_p\big(\Delta_A+\Delta_B+\Delta_E+\Delta_{i_2}-n\big)}\,.
}
At this point we can use the leg adding operation as described in the previous subsection to re-express $\mathcal{M}_2$ in terms of a larger diagram: 
\eqn{}
{
\mathcal{M}_2
=
&
\left(
\begin{matrix}
\vspace*{-0.2cm}
\includegraphics[height=13ex]{figures/glue5.pdf}
\end{matrix}
\right)
\lim_{s_E\rightarrow -\infty}
\left(
\begin{matrix}
\vspace*{-0.2cm}
\includegraphics[height=18ex]{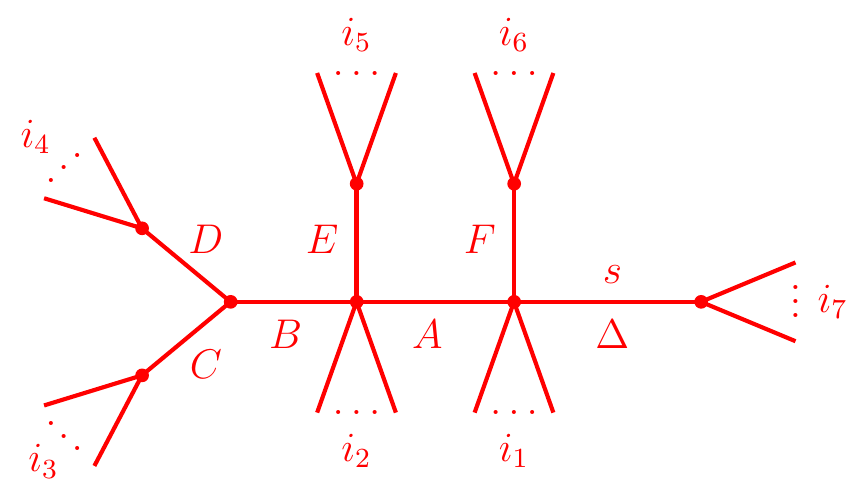}
\end{matrix}
\right)
\left(
\begin{matrix}
\vspace*{-0.2cm}
\includegraphics[height=6ex]{figures/glue7.pdf}
\end{matrix}
\right).
}
Comparing the final expressions for $\mathcal{M}_1$ and $\mathcal{M}_2$, we see that up to a common pre-factor, $\mathcal{M}_2$ is given by the $s_E$-independent terms of a diagram, and  $\mathcal{M}_1$ is given by the $s_E$-dependent terms of the same diagram. We thus conclude that their sum is equal to 
\eqn{}
{
\mathcal{M}=
\left(
\begin{matrix}
\vspace*{-0.2cm}
\includegraphics[height=13ex]{figures/glue5.pdf}
\end{matrix}
\right)
\left(
\begin{matrix}
\vspace*{-0.2cm}
\includegraphics[height=18ex]{figures/glue6.pdf}
\end{matrix}
\right)
\left(
\begin{matrix}
\vspace*{-0.2cm}
\includegraphics[height=6ex]{figures/glue7.pdf}
\end{matrix}
\right),
}
which is precisely what we wanted to show in \eno{ToShowGlue}.

\subsection{Proof of the general ``leg adding'' operation}

The last task that remains to be accomplished to have a complete proof of the recursive prescription for $p$-adic Mellin amplitudes is to show that the leg adding operation works generally. 
That is, given an arbitrary undressed diagram
\eqn{}
{
\mathfrak{D}\equiv
\left(
\begin{matrix}
\vspace*{-0.2cm}
\includegraphics[height=18ex]{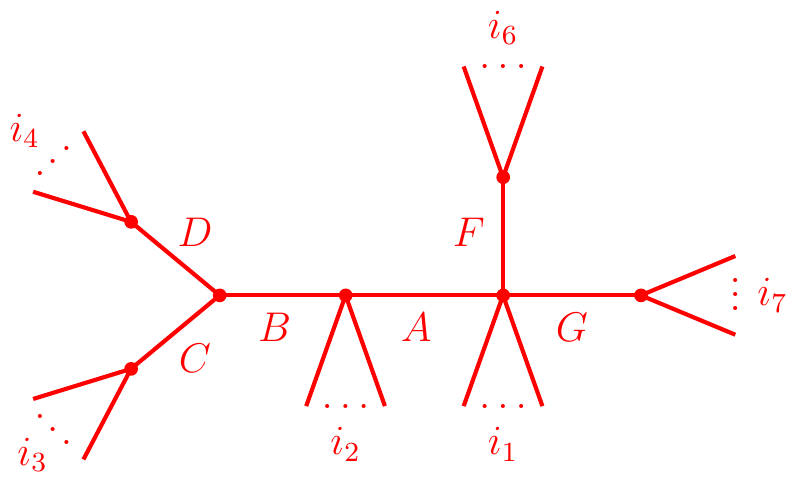}
\end{matrix}
\right),
}
we need to show that we can obtain a diagram with an extra internal leg (more precisely, the part of the diagram that does not depend on a particular Mandelstam variable), by taking an appropriate sum of residues of $\mathfrak{D}$.
Concretely, we need to show that the following sum of residues,
\eqn{Bdef}
{
B &\equiv
\sum _{z^\ast\in\{\Delta_E,\Delta_{i_5}\}}
\res_{z^\ast}
\Bigg[
\zeta_p\big(\Delta_A+\Delta_B+\Delta_{i_2}+z-n\big)
\left(
\begin{matrix}
\vspace*{-0.2cm}
\includegraphics[height=18ex]{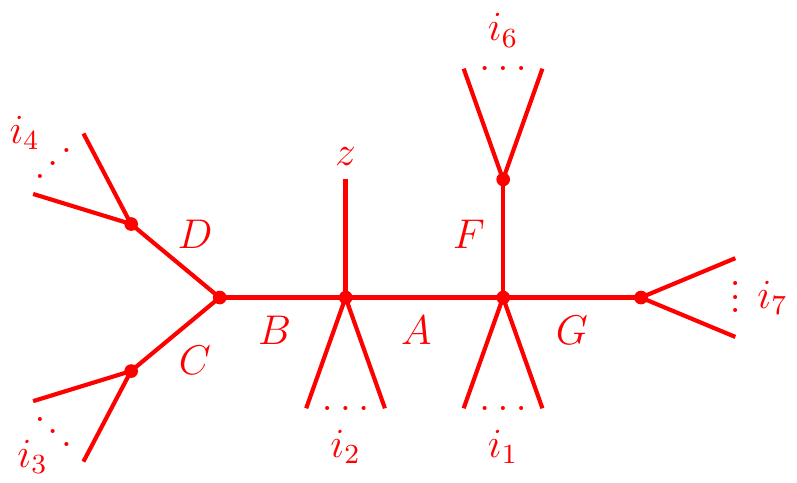}
\end{matrix}
\right) \cr 
& \qquad  \times 
\zeta_p\big(\Delta_{i_5}-z\big)
\zeta_p\big(\Delta_E-z\big)
\Bigg]
\,
}
is equal to the following limit, 
\eqn{toShow}
{
C\equiv -\zeta_p\big(\Delta_A+\Delta_B+\Delta_E+\Delta_{i_2}-n\big)
\lim_{s_E\rightarrow -\infty}
\left(
\begin{matrix}
\vspace*{-0.2cm}
\includegraphics[height=18ex]{figures/Factorization1.pdf}
\end{matrix}
\right).
}
We will show this by strong induction on the number of internal lines. The idea is to apply the formula \eno{newFormula} to the undressed diagram in equation \eno{Bdef} in order to re-cast it in terms of diagrams with fewer internal lines and then use induction to evaluate the contribution to $B$ from these smaller diagrams. Specifically, applying the formula \eno{newFormula} to the diagram in \eno{Bdef}, we may decompose $B$ as 
\eqn{Bdecompose}
{
B=&
-B_A-B_B-B_C-B_D-B_F-B_G
\cr
&-B_{AB}-B_{AC}-B_{AD}-\cdots-B_{FG}
\cr
&-B_{ABC}-B_{ABD}-B_{ABF}-\cdots-B_{DFG}
\cr 
&\hspace{2mm}\vdots
\cr
& -B_{ABCDFG}\,,
}
where $B_{I_{i_1}\ldots I_{i_f}}$ represents the contribution to $B$ arising from the term in the sum in \eno{newFormula} which has $l=f$ and $I_1=I_{i_1},\ldots,I_f=I_{i_f}$. For example,
\eqn{Bsplit}
{
B_A =&
\sum _{z^\ast\in\{\Delta_E,\Delta_{i_5}\}}
\res_{z^\ast}
\Bigg[
\zeta_p\big(\Delta_A+\Delta_B+\Delta_{i_2}+z-n\big)
\zeta_p\big(\Delta_{i_5}-z\big)
\zeta_p\big(\Delta_E-z\big)
\cr
& \times
\left[\zeta_p(s_A-\Delta_A)-\zeta_p(\Delta_{i_1 i_2 i_3 i_4 i_6 i_7,}+z-n)\lim_{\forall s \rightarrow\infty}\right]
\left(
\begin{matrix}
\vspace*{-0.2cm}
\includegraphics[height=16ex]{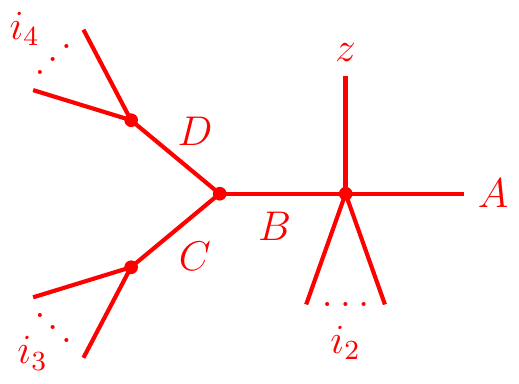}
\end{matrix}
\right) \cr 
& \times 
\left(
\begin{matrix}
\vspace*{-0.2cm}
\includegraphics[height=17ex]{figures/Factorization4.pdf}
\end{matrix}
\right)\Bigg],
}
where we are using the convention,
\eqn{DeltaijkComma}
{
\Delta_{i_1\ldots i_k,i_{k+1}\ldots i_{\ell}}\equiv\sum_{j=1}^k\Delta_{i_j}-\sum_{j=k+1}^\ell\Delta_{i_j}\,.
}
so that
\eqn{}
{
\Delta_{i_1 i_2 i_3 i_4 i_6 i_7,} = \Delta_{i_1}+\Delta_{i_2}+\Delta_{i_3}+\Delta_{i_4}+\Delta_{i_6}+\Delta_{i_7}
\,.
}
and summation is implied over each of the terms on the r.h.s.
It may be remarked right away that we can decompose $C$ in a similar manner: 
\eqn{Cdecompose}
{
C=&
-C_A-C_B-C_C-C_D-C_E-C_F-C_G
\cr
&-C_{AB}-C_{AC}-C_{AD}-\cdots-C_{FG}
\cr
&-C_{ABC}-C_{ABD}-C_{ABE}-\cdots-C_{EFG}
\cr 
&\hspace{2mm}\vdots
\cr
& -C_{ABCDEFG}\,,
}
where, introducing the definition
\eqn{}
{
\Sigma\equiv \Delta_{i_1i_2i_3i_4i_5i_6i_7,} -n 
\,,
}
the term $C_A$, for example, is given by
\eqn{}{
C_A=
-&
\zeta_p(\Delta_{ABEi_2,}-n)
\lim_{s_E \rightarrow -\infty}
\big[\zeta_p(s_A-\Delta_A)
-\zeta_p(\Sigma)\lim_{\forall s\rightarrow \infty}\big]
\cr
&
\left(
\begin{matrix}
\includegraphics[height=19ex]{figures/Factorization3.pdf}
 \end{matrix}
\right)
\left(
\begin{matrix}
\includegraphics[height=19ex]{figures/Factorization4.pdf}
 \end{matrix}
\right).
}
The two limits $s_E \rightarrow -\infty$ and $\forall s\rightarrow \infty$ do not commute, since any term containing a factor of $\zeta_p(s_E-\Delta_E)$ is killed off if the $s_E \rightarrow -\infty$ is taken first but not if it is taken last. We can therefore commute the two limits only if we add back in these terms. That is, we may write $C_A=C_{A,1}+C_{A,2}$, where
\eqn{}{
C_{A,1}=
-&
\zeta_p(\Delta_{ABEi_2,}-n)
\big[\zeta_p(s_A-\Delta_A)
-\zeta_p(\Sigma)\lim_{\forall s\rightarrow \infty}\big]
\lim_{s_E \rightarrow -\infty}
\cr
&
\left(
\begin{matrix}
\includegraphics[height=19ex]{figures/Factorization3.pdf}
 \end{matrix}
\right)
\left(
\begin{matrix}
\includegraphics[height=19ex]{figures/Factorization4.pdf}
 \end{matrix}
\right),
}
\eqn{}{
C_{A,2}=
-&
\zeta_p(\Delta_{ABEi_2,}-n)\zeta_p(\Sigma)
\lim_{\forall s\rightarrow \infty}
\left(
\begin{matrix}
\vspace*{-0.2cm}
\includegraphics[height=16ex]{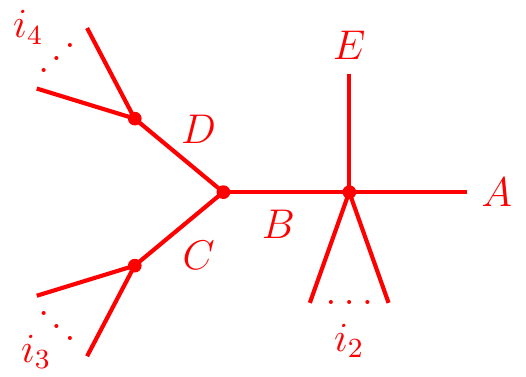}
\end{matrix}
\right)
\left(
\begin{matrix}
\includegraphics[height=19ex]{figures/Factorization4.pdf}
 \end{matrix}
\right).
}

In equation \eno{Cdecompose} we get extra terms as compared to \eno{Bdecompose} because the diagram in \eno{toShow} has an extra internal line, but we should remember that the $s_E\rightarrow -\infty$ limit in \eno{toShow} kills off the part of any term $C_{I_1\ldots I_f}$ that contains a factor of $\zeta_p(s_E-\Delta_E)$.

Let us now look more closely at the contribution $B_A$. It is straightforward to check that 
\eqn{}
{
\zeta_p(\Delta_{i_5}-z)\zeta_p( \Delta_{i_1i_2i_3i_4i_6i_7,}+z-n)=\zeta_p(\Sigma)
\bigg(
\zeta_p(\Delta_{i_5}-z)+\zeta_p(\Delta_{i_1i_2i_3i_4i_6i_7,}+z-n)
-1
\bigg)\,.
}
Using this, we write $B_A=B_{A,1}+B_{A,2}$, where the first part is given by
\eqn{}
{
B_{A,1} &=
\sum _{z^\ast\in\{\Delta_E,\Delta_{i_5}\}}
\res_{z^\ast}
\Bigg[
\zeta_p\big(\Delta_A+\Delta_B+\Delta_{i_2}+z-n\big)
\zeta_p\big(\Delta_{i_5}-z\big)
\zeta_p\big(\Delta_E-z\big)
\cr
& 
\times
\left[\zeta_p(s_A-\Delta_A)-\zeta_p(\Sigma)\lim_{\forall s \rightarrow\infty}\right]
\left(
\begin{matrix}
\vspace*{-0.2cm}
\includegraphics[height=16ex]{figures/legProof3.pdf}
\end{matrix}
\right)
\left(
\begin{matrix}
\vspace*{-0.2cm}
\includegraphics[height=17ex]{figures/Factorization4.pdf}
\end{matrix}
\right)\Bigg],
}
and the second part of $B_A$ is given by
\eqn{BA2}
{
B_{A,2} &= -
\sum _{z^\ast\in\{\Delta_E,\Delta_{i_5}\}}
\res_{z^\ast}
\Bigg[
\zeta_p\big(\Delta_A+\Delta_B+\Delta_{i_2}+z-n\big)
\zeta_p\big(\Delta_E-z\big)
\cr
& 
\times
\zeta_p(\Sigma)\bigg(\zeta_p(\Delta_{i_1i_2i_3i_4i_6i_7,}+z-n)-1\bigg)\lim_{\forall s \rightarrow\infty}
\left(
\begin{matrix}
\vspace*{-0.2cm}
\includegraphics[height=16ex]{figures/legProof3.pdf}
\end{matrix}
\right)
\left(
\begin{matrix}
\vspace*{-0.2cm}
\includegraphics[height=17ex]{figures/Factorization4.pdf}
\end{matrix}
\right)\Bigg].
}
Using the fact that the $s\rightarrow \infty$ limit commutes with the sum over residues and invoking the inductive assumption, the first part is seen to equal 
\eqn{CA1}
{
B_{A,1} =&-
\zeta_p(\Delta_A+\Delta_B+\Delta_E+\Delta_{i_2}-n)
\cr
&
\times
\left[\zeta_p(s_A-\Delta_A)-\zeta_p(\Sigma)\lim_{\forall s \rightarrow\infty}\right]
\lim_{s_E\rightarrow -\infty}
\left(
\begin{matrix}
\vspace*{-0.2cm}
\includegraphics[height=17ex]{figures/Factorization3.pdf}
\end{matrix}
\right)
\left(
\begin{matrix}
\vspace*{-0.2cm}
\includegraphics[height=17ex]{figures/Factorization4.pdf}
\end{matrix}
\right).
}
We notice that $B_{A,1}$ is exactly equal to $C_{A,1}$. And this observation does not depend on any special property of $B_A$  not shared with any of the other terms on the r.h.s.\ of \eno{Bdecompose}: we may decompose any term $B_{I_1\ldots I_f}$ on the r.h.s.\ of \eno{Bdecompose} into two parts, called $B_{I_1\ldots I_f,1}$ and $B_{I_1\ldots I_f,2}$, such that $B_{I_1\ldots I_f,1}=C_{I_1\ldots I_f,1}$. 
So we have now matched some of \eno{Bdef} with some of \eno{toShow}. 

What remains to be shown is that the remaining part of $B$ is equal to the remaining part of $C$. The remaining part of $C$ consists of 1) terms $C_{I_1\ldots I_f,2}$ where $I_i \neq$ for all $i\in \{1,2,\ldots,f\}$ and, 2) terms $C_{I_1\ldots I_f}$ for which $I_i=E$ for some $i\in \{1,2,\ldots,f\}$. 
However, note that the sum of all these terms admits a simple diagrammatic interpretation, as we now explain.
It was pointed out in section \ref{Factorization} that for any undressed diagram, the momentum-independent terms are in one-to-one correspondence with the terms proportional to internal line factors $\zeta_p(s-\Delta)$. Thus in \eno{toShow} when we take the limit $s_E \rightarrow -\infty$, although we kill off the terms  proportional to  $\zeta_p(s_E-\Delta_E)$, we do not kill off the momentum-independent terms which are in one-to-one correspondence with these killed-off terms. 
It is exactly these surviving, unpartnered momentum-independent terms that we have yet to account for. But as these remaining terms stand in one-to-one correspondence with the killed-off terms containing the propagator factor $\zeta_p(s_E-\Delta)$, the factorization property of undressed diagrams from section \ref{Factorization}
allows us to express these remaining unaccounted terms of $C$ as 
\eqn{unAccounted}
{
\zeta_p\big(\Delta_A+\Delta_B+\Delta_E+\Delta_{i_2}-n\big)\zeta_p(\Sigma)
\lim_{\forall s \rightarrow \infty}
\left(
\begin{matrix}
\vspace*{-0.2cm}
\includegraphics[height=18ex]{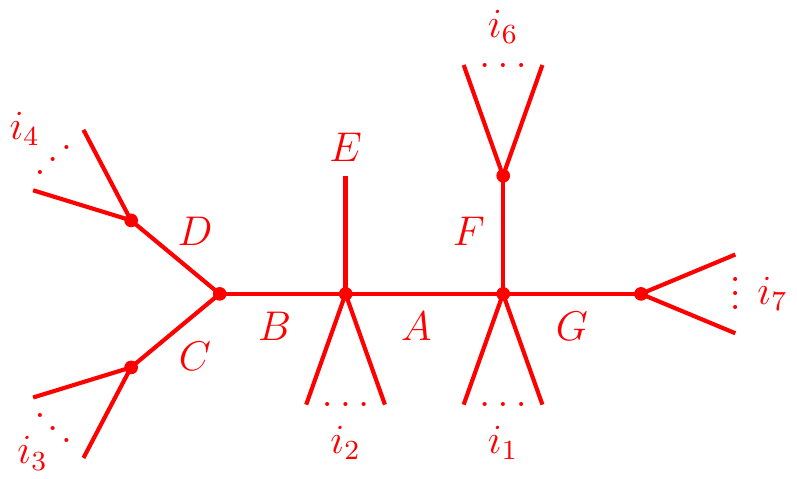}
\end{matrix}
\right).
}

Let's return to the second part of $B_A$. 
We see in the expression for  $B_{A,2}$ in \eno{BA2} that  there is no longer a pole at $z=\Delta_{i_5}$, so that $B_{A,2}$ is simply equal to the residue at $\Delta_E$. Thus, we have
\eqn{}
{
B_{A,2} =&
-\zeta_p\big(\Delta_A+\Delta_B+\Delta_E+\Delta_{i_2}-n\big)\: \zeta_p(\Sigma)
\cr
& 
\times
\bigg(1-\zeta_p(\Delta_{i_1i_2i_3i_4i_6i_7,}+\Delta_E-n)\bigg)\lim_{\forall s \rightarrow\infty}
\left(
\begin{matrix}
\vspace*{-0.2cm}
\includegraphics[height=14ex]{figures/legProof4.pdf}
\end{matrix}
\right)
\left(
\begin{matrix}
\vspace*{0.4cm}
\includegraphics[height=17ex]{figures/Factorization4.pdf}
\end{matrix}
\right).
}
We can rewrite this expression in a more suggestive form as
\eqn{}
{
B_{A,2} &=
-\zeta_p\big(\Delta_A+\Delta_B+\Delta_E+\Delta_{i_2}-n\big)\:\zeta_p(\Sigma)
\cr
& 
\times
\lim_{\forall s \rightarrow\infty}\!\! \bigg(\zeta_p(s_A-\Delta_A)-\zeta_p(\Delta_{i_1i_2i_3i_4i_6i_7,}+\Delta_E-n)\bigg)\!\!
\left(
\begin{matrix}
\vspace*{-0.2cm}
\includegraphics[height=16ex]{figures/legProof4.pdf}
\end{matrix}
\right)\cr 
& \times
\left(
\begin{matrix}
\vspace*{-0.2cm}
\includegraphics[height=17ex]{figures/Factorization4.pdf}
\end{matrix}
\right),
}
where now the symbol $\forall s\to \infty$ also includes the limit $s_A \to \infty$.
We recognize the above as exactly the contribution from the $l=1, I=A$ term if we apply the formula \eno{newFormula} to the undressed diagram in \eno{unAccounted}. Again there is nothing special about the $B_A$ term. In general when we split a term $B_{I_{i_1}\ldots I_{i_f}}$ on the right-hand side of \eno{Bsplit} into $B_{I_{i_1}\ldots I_{i_f},1}$ and $B_{I_{i_1}\ldots I_{i_f},2}$, the second term $B_{I_{i_1}\ldots I_{i_f},2}$ exactly reproduces the term with $l=f$, $I_1=I_{i_1},\ldots,I_f=I_{i_f}$ on applying the formula \eno{newFormula} to \eno{unAccounted}. We conclude that the sum over these contributions coming from all the terms on the right-hand side of \eno{Bdecompose}, that is, the sum
\eqn{}
{
&
-B_{A,2}-B_{B,2}-B_{C,2}-B_{D,2}-B_{F,2}-B_{G,2}
\cr
&-B_{AB,2}-B_{AC,2}-B_{AD,2}-\cdots-B_{FG,2}
\cr
&-B_{ABC,2}-B_{ABD,2}-B_{ABF,2}-\cdots-B_{DFG,2}
\cr 
&\hspace{2mm}\vdots
\cr
& -B_{ABCDFG,2}\,
}
is exactly equal to \eno{unAccounted}, which shows that $B$ exactly reproduces all of $C$. 

This concludes the proof of the leg adding operation, and in turn proves the recursion prescription~\ref{pres:padicMellin} from section~\ref{recursionSection}.

\section{BCFW-Shifts of Auxiliary Momenta}
\label{BCFWMOM}

In this appendix, we provide an alternative but equivalent form of the recursive prescription \eno{mellinBCFW} from section \ref{BCFW}, in auxiliary  momentum space.
The $p$-adic Mellin amplitude of a tree-level bulk diagram with external legs labeled $\{1, \ldots {\cal N}\}$ and auxiliary momenta (defined in section~\ref{setup}) labeled $\{k_1,\ldots,k_{\cal N}\}$, can be written as
\eqn{BCFWschematic}{
{\cal M} = -\sum_{{\rm partitions\ } I} {\cal M}(I, -\hat{k}_I)(z_I) \times {\rm (propagator)} \times {\cal M}(I^c, \hat{k}_I)(z_I)\,,
}
where the partition of external legs, $I$ is a subset $I \subset \{1,\ldots {\cal N}\}$ such that it can be obtained by splitting the diagram into two by cutting across any chosen internal line, and $I^c$ is the complement of $I$. The hatted momenta are defined to be $\hat{k}_I \equiv -\sum_{a \in I} k_a(z)$, where $k_a(z)$ are the complex-shifted external momenta which are on-shell at all values of $z \in \mathbb{C}$, while $\hat{k}_I$ goes on-shell only at $z=z_I$.

The amplitude ${\cal M}(I,-\hat{k}_I)(z)$ is one of the $z$-dependent sub-amplitudes obtained by splitting the original diagram into two by cutting across the chosen internal line, and the factor of ``${\rm (propagator)}$'' is the Mellin space ``propagator'' $\beta_p(-{k}_I^2-\Delta_I,n-\sum_{i=1}^{\cal N}\Delta_{i})$ defined in \eno{BetaPDef}, where the unhatted momentum $k_I = -\sum_{a \in I} k_a$ is constructed out of {\it un-shifted} external momentum variables, and $\Delta_I$ is the conformal dimension of the bulk field propagating along the chosen internal line.\footnote{Here we have made the implicit assumption that {\it all} internal lines get complex-shifted -- to ensure this is true in general, a ``multi-line BCFW-shift'' of the external momenta may be required. The final Mellin amplitude ${\cal M}$ will be independent of the choice of the particular BCFW-shift employed, as long as all internal lines develop a $z$-dependence and consequently can be put on-shell at particular values of $z$.}

From here on, we will sometimes refer to the on-shell recursion above as BCFW-type recursion because of the similarity of~\eno{BCFWschematic} with the usual BCFW decomposition, but we stress that we are working at the level of individual diagrams. We note that the recursion in~\eno{BCFWschematic} is identical to the recursion relation~\eno{mellinBCFW} except~\eno{mellinBCFW} is written without any direct reference to auxiliary momenta.
We now explain this recursive prescription in detail with the help of two illustrative examples.

\subsection{Four-point exchange diagram}
\label{4PTBCFW}

We start by demonstrating the BCFW-type recursion \eno{BCFWschematic} for the simplest case, the four-point function. 
The full tree-level four-point bulk (Witten) diagram for a theory with cubic and quartic interaction vertices is given by
\begin{align}
\label{FourBlob}
\begin{matrix}
\includegraphics[height=18ex]{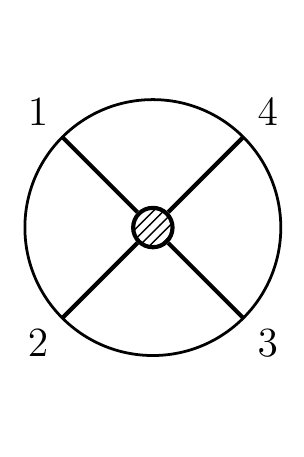}
 \end{matrix}
 \,\,\,
 \sim
 \,\,\,
 \begin{matrix}
\includegraphics[height=18ex]{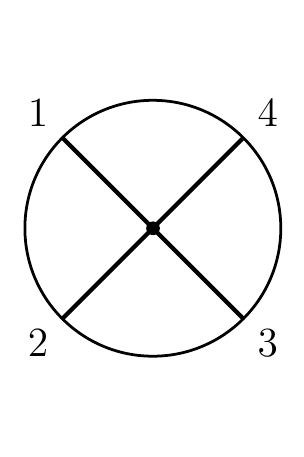}
 \end{matrix}
+
\begin{matrix}
\includegraphics[height=18ex]{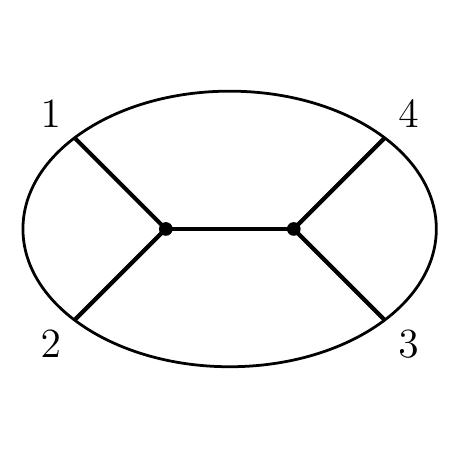}
 \end{matrix}
 +
 \begin{matrix}
\includegraphics[height=15ex]{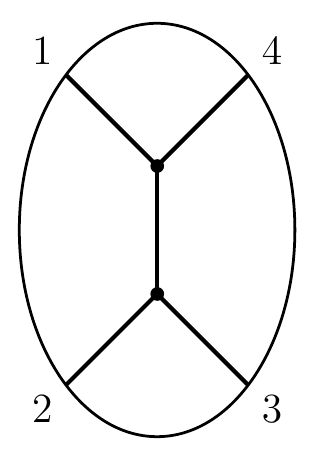}
 \end{matrix}
 +
 \begin{matrix}
\includegraphics[height=14ex]{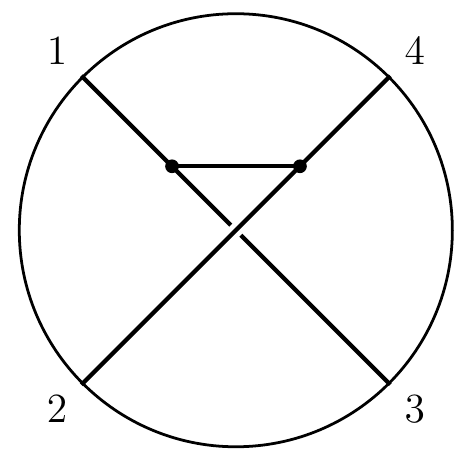}
 \end{matrix}\,,
 \end{align}
where we must sum over all fields which can propagate in the intermediate channels in the $s$-, $t$- and $u$-channel exchange diagrams. 
We will be suppressing coupling constants and any symmetry factors associated with the diagrams. 
These details depend on the specifics of the bulk theory, particularly the bulk Lagrangian. 
We now show that {\it each individual} diagram in \eno{FourBlob} in Mellin space admits a BCFW-type recursion relation, with the basic building blocks of the recursion being the contact diagrams. 
Consequently, we'll find the four-point contact diagram in \eno{FourBlob} is already in its `maximally reduced' form.

Let's focus on the $s$-channel exchange where a scalar field dual to a (single-trace) operator of conformal dimension $\Delta_A$ propagates along the internal leg.\footnote{A similar analysis can be done for the $t$- and $u$-channel exchange diagrams.}
It turns out the form of the Mellin amplitude for the ($s$-channel) four-point exchange diagram as written in example~\ref{ex:exch} is, rather trivially, already in the desired BCFW-type on-shell recursion form. 
However in this subsection, we will take the time to describe in detail the various ingredients which go into recognizing the BCFW-type recursion structure, as they  illuminate how this works with more complicated higher-point diagrams.

Define
\eqn{4ptschannel}{
{\cal M}^{1-{\rm int}}_{\Delta_{12,};\Delta_{34,}}(s_A,\Delta_A) \equiv
\begin{matrix}
\includegraphics[height=11ex]{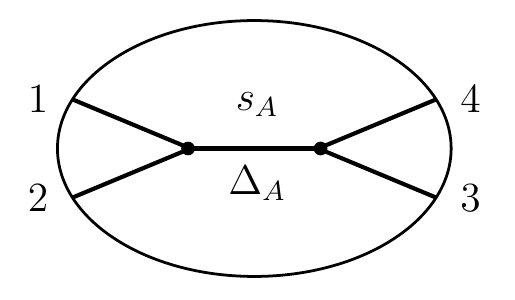}
 \end{matrix} 
 =
- V_{\Delta_{12A,}}\, V_{\Delta_{34A,}}\, \beta_p\left(s_A-\Delta_A, n- \sum_{i=1}^4 \Delta_i\right), 
}
where the vertex factor $V_{\Delta_{1 2 A,}}$, defined  using the convention \eno{DeltaijkComma}, via \eno{vertexFactorDef}, 
\eqn{vertexFactorDef2}{ 
V(\Delta_{1\ldots f,}) \equiv \zeta_p(\Delta_{1\ldots f,}-n)\,,
}
is the Mellin amplitude associated with the $3$-point contact diagram between operators of dimensions $\Delta_1$, $\Delta_2$ and $\Delta_A$. 
Further, we have defined 
\eqn{sADef}{
s_A \equiv \Delta_1 + \Delta_2 - 2\gamma_{12} = \Delta_3 + \Delta_4 - 2\gamma_{34}\,,
}
and $\beta_p$ is defined in \eno{BetaPDef}.
To go to the auxiliary momentum space, described in section \ref{setup}, we assign to each external leg of the four-point $s$-channel exchange diagram an auxiliary momentum, as shown below:  
\eqn{4ptMomentum}{
 {\cal M}^{1-{\rm int}}_{\Delta_{12,};\Delta_{34,}}(s_A,\Delta_A) =  
\begin{matrix}
\includegraphics[height=12ex]{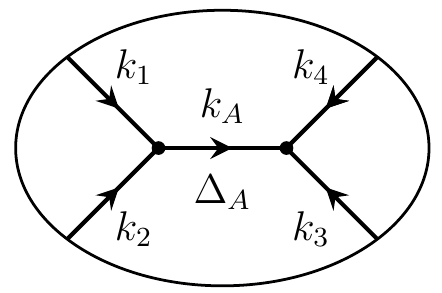}
 \end{matrix} \equiv {\cal M}^{1-{\rm int}}_{-k_1^2-k_2^2;-k_3^2-k_4^2}\left(-k_A^2,\Delta_A\right) ,
}
where the external momenta are on-shell,
\eqn{4ptOnshell}{
-k_i^2 = \Delta_i \qquad i=1,\ldots, 4\,,
}
while the internal momentum is not, and momentum is conserved at each vertex.
We are suppressing the Lorentz indices on the momentum variables. 
To obtain a BCFW-type relation for the exchange diagram, we begin by complexifying some of the external momenta in such a way that the complex-shifted external legs remain on-shell and still satisfy momentum conservation at each vertex for all values of the complex parameter, and as a result of the shift the momentum along the internal leg gets complex-shifted as well. For instance, we may choose the following ``two-line shift'' that preserves momentum conservation,
\eqn{4ptBCFWshift}{
k_1 \to k_1(z) \equiv k_1 + z q\,, \qquad k_4 \to k_4(z) \equiv k_4 - z q\,,
}
where $z \in \mathbb{C}$, and $q$ is a constrained momentum variable which will in general be complex-valued. The constraints on $q$ arise from the following on-shell requirements,
\eqn{4ptComplexOnshell}{
-k_1(z)^2 = \Delta_1\,, \qquad -k_4(z)^2 = \Delta_4\,, \qquad \forall z \in \mathbb{C}\,.
}
As desired, the momentum running through the internal leg gets complex-shifted as well,
\eqn{4ptInternalShift}{
k_A \to k_A(z) \equiv k_1(z)+k_2\,.
}
The conditions \eno{4ptComplexOnshell}
are met if $q$ satisfies
\eqn{4ptqConditions}{
q^2 = 0 \qquad q \cdot k_1 = 0 \qquad q \cdot k_4 =0\,.
}
Momentum conservation along with the constraints above then imply
\eqn{4ptqImplies}{
q \cdot k_2 = - q \cdot k_3\,.
}
Owing to the fact that the auxiliary momenta $k_i$ are not null, it is in general not possible to specify in a Lorentz covariant manner the momentum variable $q$ satisfying \eno{4ptqConditions}.\footnote{The momentum variables $k_i$ and $q$ are respectively, real and complex valued $(n+1)$-dimensional Lorentz vectors with a Lorentzian  inner product.
If $(n+1)=3$, then only $q=0$ satisfies \eno{4ptqConditions} for arbitrary $k_1, k_4$. Thus we need at least $(n+1) \geq 4 ={\cal N}$ to admit a non-vanishing complex-deformation $z q$. This requirement is consistent with the inequality mentioned in section \ref{setup}.} 
However, the explicit form for $q$ will not be important in the following; $q$ will enter in the calculations via its dot-product with  various momenta and it will be sufficient for us to inquire how these dot-products are related to each other without establishing the explicit form for any of them.

The complex-shifts \eno{4ptBCFWshift} transform the Mandelstam invariant $s_A$ as well,
\eqn{4ptsShift}{
s_A \to s_A(z) \equiv -(k_1(z)+k_2)^2 = s_A -2z q \cdot k_2 = -2 q \cdot k_2 (z-z_A) + \Delta_A\,,}
where we have defined
\eqn{zADef}{
z_A \equiv {s_A -\Delta_A \over 2q \cdot k_2}\,.
}
From this it is clear that the complex-shifted momentum running through the internal leg, $k_A(z)$ goes on-shell at $z=z_A$.
Consequently the complexified Mellin amplitude is given by\footnote{Note that the complex Mellin amplitude ${\cal M}(z)$ in this appendix is different from the one defined in section \ref{BCFW} above \eno{Idef2}.} 
\eqn{4ptMellinC}{
{\cal M}(z) \equiv - V_{\Delta_{12A,}}\, V_{\Delta_{34A,}}\, \beta_p\left(s_A(z) -\Delta_A, n- \sum_{i=1}^4 \Delta_i\right),
}
so that the original Mellin amplitude is recovered upon setting $z=0$,
\eqn{4ptRecoverOrigin}{
{\cal M}(0) = {\cal M}^{1-{\rm int}}_{\Delta_{12,};\Delta_{34,}}(s_A,\Delta_A)\,.
}

Consider now the integral 
\eqn{4ptIDef}{
I \equiv \oint_{\mathsf{C}} {dz \over 2\pi i} {{\cal M}(z) \over z}\,,
}
where the contour $\mathsf{C}$ is chosen to be a circle of infinite radius centered at origin. The integrand has simple poles at
\eqn{4ptMzPoles}{
z = 0\,, \qquad z= z_A + {1 \over 2 q\cdot k_2} {2\pi i m \over \log p} \qquad m \in \mathbb{Z}\,,
}
where the infinite sequence of poles arises from the factor of $\zeta_p(s_A(z)-\Delta_A)$ in the $\beta_p$ function.
We will now apply the residue theorem
to obtain an expression for the original amplitude in terms of the residues from the remaining poles.
It is clear that the residue at $z=0$ is precisely the original amplitude ${\cal M}(0)$. The residue at the remaining poles takes the form 
\eqn{4ptResiduezA}{
\underset{\substack{z =z_A \\  \hspace{12mm}+ {1 \over 2 q\cdot k_2} {2\pi i m \over \log p}}}{\rm Res} \left({{\cal M}(z) \over z}\right) =  { V_{\Delta_{12A,}} V_{\Delta_{34A,}} \over (s_A - \Delta_A)\log p +  2\pi i m } \qquad m \in \mathbb{Z}\,.
}
Thus we conclude that
\eqn{4ptIagain2}{
I = {\cal M}(0) + V_{\Delta_{12A,}} V_{\Delta_{34A,}} \sum_{m=-\infty}^{+\infty} { 1 \over (s_A - \Delta_A)\log p +  2\pi i m }\,.
}
At this point, one hopes that ${\cal M}(z)$ falls off sufficiently fast at large $z$, so that the contour integral $I=0$, i.e.\ the boundary term vanishes. 
The infinite sum in \eno{4ptIagain2} evaluates to give
\eqn{4ptSumRes}{
 \sum_{m=-\infty}^{+\infty} { 1 \over (s_A - \Delta_A)\log p +  2\pi i m } = \zeta_p(s_A-\Delta_A) - {1 \over 2}\,,
}
so the claim that $I=0$ becomes
\eqn{4ptClaim2}{
{\cal M}(0) &\stackrel{?}{=} -  V_{\Delta_{12A,}}  \left( \zeta_p(s_A-\Delta_A) - {1 \over 2} \right) V_{\Delta_{34A,}} \cr 
  &\stackrel{?}{=} -  {\cal M}(k_1(z),k_2,-k_1(z)-k_2)\Big|_{z=z_A}  \left( \zeta_p(s_A-\Delta_A) - {1 \over 2} \right) {\cal M}(k_1(z)+k_2,k_3,k_4(z))\Big|_{z=z_A} ,
}
where in the second line above we have defined the sub-amplitude
\eqn{4ptMDef}{
{\cal M}(k_a,k_b,k_c) \equiv V_{-k_a^2,-k_b^2,-k_c^2}
}
with $k_a+k_b+k_c=0$.
Consider the sub-amplitude 
\eqn{Moffshell}{
 {\cal M}(k_1(z),k_2,-k_1(z)-k_2) = V_{\Delta_1,\Delta_2, s_A(z)}\,.
 }
It represents an ``off-shell'' three-point contact interaction with incoming momenta $k_1(z),k_2$ and $-k_1(z)-k_2$. At generic values of $z$, two of the three external legs are on-shell, while the third one is off-shell, but the third leg goes on-shell at $z=z_A$, 
 and we write
\eqn{Monshell}{
{\cal M}(k_1(z_A),k_2,-k_1(z_A)-k_2) = \begin{matrix}
\includegraphics[height=17ex]{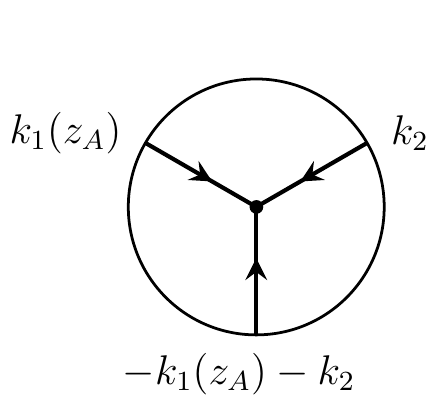}
 \end{matrix} \,.
}
Thus if it is true that $I=0$, we can interpret the second line of  \eno{4ptClaim2} as the claim that we have rewritten the original Mellin amplitude for the $s$-channel four-point exchange diagram as a product of two on-shell sub-amplitudes obtained by cutting across the internal leg of the original diagram along which the field dual to an operator of dimension $\Delta_A$ propagates, times a putative ``propagator'' running across the internal leg given by the expression $\zeta_p(s_A-\Delta_A) - 1/2$.

However, it can be explicitly checked  that the boundary contribution in \eno{4ptIagain2} does {\it not} vanish, as the complex Mellin amplitude ${\cal M}(z)$ does not vanish at infinity. Instead, we find
\eqn{4ptBdyTerm}{
I &= {1\over 2} V_{\Delta_{12A,}} V_{\Delta_{34A,}} V_{\Delta_{1234,}} (1+p^{n-\Delta_{1234,}}) \cr 
 &= {\cal M}(k_1(z),k_2,-k_1(z)-k_2)\Big|_{z=z_A}  \left( {V_{\Delta_{1234,}} (1+p^{n-\Delta_{1234,}}) \over 2} \right)  {\cal M}(k_1(z)+k_2,k_3,k_4(z))\Big|_{z=z_A}\,,
}
where the second line is merely a suggestive rewriting of the first line.
Taking the boundary term into account, the corrected version of \eno{4ptClaim2} is then given by
\eqn{4ptCorrect0}{
& {\cal M}(0) \cr &= -  {\cal M}(k_1(z),k_2,-k_1(z)-k_2)\Big|_{z=z_A}  \left( \zeta_p(s_A-\Delta_A) - {1 \over 2} \right) {\cal M}(k_1(z)+k_2,k_3,k_4(z))\Big|_{z=z_A} \cr
   &\quad +  {\cal M}(k_1(z),k_2,-k_1(z)-k_2)\Big|_{z=z_A}  \left( {V_{\Delta_{1234,}} (1+p^{n-\Delta_{1234,}}) \over 2} \right)  {\cal M}(k_1(z)+k_2,k_3,k_4(z))\Big|_{z=z_A},
}
which simplifies to the compact expression
\eqn{4ptCorrect}{
& {\cal M}(0) \cr 
   &= - {\cal M}(k_1(z),k_2,-k_1(z)-k_2)\Big|_{z=z_A}  \beta_p(s_A-\Delta_A, n- \Delta_{1234,})  {\cal M}(k_1(z)+k_2,k_3,k_4(z))\Big|_{z=z_A}.
}
The amplitude in \eno{4ptCorrect} is, indeed, mathematically identical to the starting point \eno{4ptschannel}, and in fact is a special case of \eno{BCFWschematic}.
 More importantly, we see the boundary contribution does not destroy the nice decomposition of the amplitude first observed in \eno{4ptClaim2}; instead it combines nicely with the residue from the poles of the complexified Mellin amplitude in such a way that the original amplitude is still given by a product of two on-shell sub-amplitudes obtained by cutting across the internal leg, times a propagator running across the same internal leg, which we refer to as the ``Mellin space propagator'' joining the left and right sub-amplitudes. We comment on the appearance of non-vanishing boundary terms at the end of this appendix.
  
Diagrammatically, we may write \eno{4ptCorrect} as 
\eqn{4ptDiagrammatic}{
\begin{matrix}
\includegraphics[height=17ex]{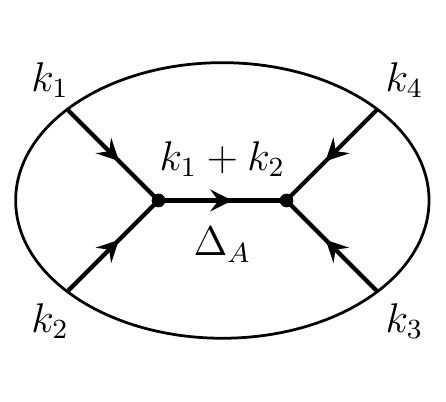}
 \end{matrix}  = -\begin{matrix}
\includegraphics[height=17ex]{figures/4ptMom3.pdf}
 \end{matrix} 
 \beta_p(s_A-\Delta_A,n-\Delta_{1234,})
 \begin{matrix}
\includegraphics[height=17ex]{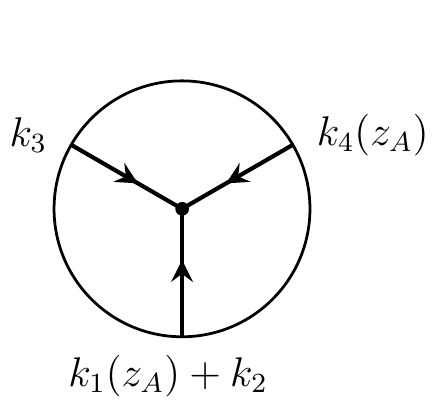}
 \end{matrix}\,.
}
We may also equivalently depict the identity \eno{4ptCorrect} as
\eqn{4ptPractical}{
\begin{matrix}
\includegraphics[height=12ex]{figures/4ptDia.pdf}
 \end{matrix}
 =&
- \begin{matrix}
\includegraphics[height=12ex]{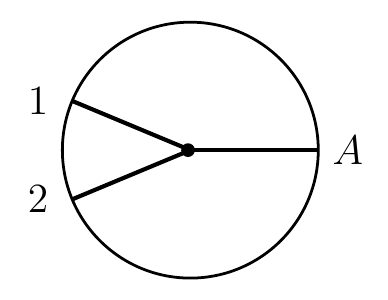}
 \end{matrix}
 \beta_p(s_A-\Delta_A, n-\Delta_{1234,})
\begin{matrix}
\includegraphics[height=12ex]{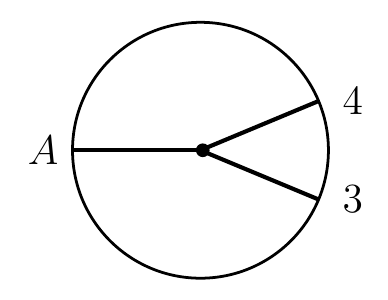}
 \end{matrix}\,,
}
which avoids reference to the auxiliary momentum space and any BCFW-shifts altogether, and takes the form presented in \eno{mellinBCFW}.

Admittedly, the example of the four-point exchange diagram is quite special in that the sub-amplitudes turn out to be contact diagrams and thus have no Mandelstam dependence. 
Nevertheless, we demonstrate in the next subsection that the BCFW-type decomposition \eno{BCFWschematic} holds, rather non-trivially, for the Mellin amplitude of the five-point diagram with two internal exchanges.

\subsection{Five-point diagram with two internal lines}
\label{5PTBCFW}

\begin{figure}
\centering
\begin{subfigure}[b]{0.35\textwidth}
        \includegraphics[width=\textwidth]{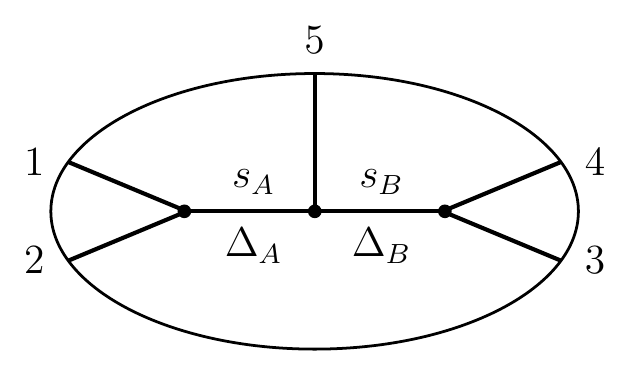}
        \caption{}
         \label{fig:5ptCubica}
\end{subfigure}
\hspace{20mm}
\begin{subfigure}[b]{0.27\textwidth}
        \includegraphics[width=\textwidth]{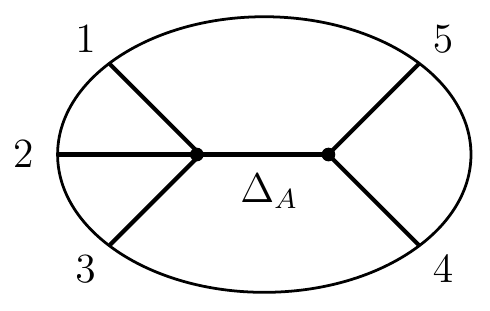}
        \caption{}
         \label{fig:5ptCubicb}
\end{subfigure}

\caption{(a) A five-point diagram built out of cubic interaction vertices. The Mandelstam variables $s_A$ and $s_B$ are defined in \eno{sAsBDef}. (b) A five-point diagram built out of cubic and quartic interaction vertices.}
\label{fig:5ptCubic}
\end{figure}
In a bulk theory with cubic interaction vertices, five-point diagrams of the kind shown in figure \ref{fig:5ptCubica}, with two internal legs, are allowed. Further, if the theory contains quartic interaction vertices as well, exchange diagrams of the kind shown in figure \ref{fig:5ptCubicb} are also possible. However, exchange diagrams like the one in figure \ref{fig:5ptCubicb} fall in the category of diagrams discussed in the previous subsection, and results obtained there extend trivially to arbitrary exchange diagrams (that is, bulk diagrams with exactly one internal line).

So in this subsection, we focus on the tree-level five-point diagram built solely from cubic vertices. A BCFW-type recursion can be set up in each channel {\it individually}, so just like in the previous subsection, we will restrict attention to a particular channel, which will be the one shown in figure \ref{fig:5ptCubica}; an identical analysis will hold in all other channels.

As with the four-point exchange diagram, we start with the Mellin amplitude for the diagram in figure \ref{fig:5ptCubica}, which is a special case of the diagram evaluated in example~\ref{ex:double},
\eqn{5ptchannel}{
{\cal M}^{2-{\rm int}}_{\Delta_{12,}; \Delta_5; \Delta_{34,}}(s_A,\Delta_A; s_B, \Delta_B) &\equiv
\begin{matrix}
\includegraphics[height=15ex]{figures/5ptDia.pdf}
 \end{matrix} \cr 
&= V_{\Delta_{12A,}}\, V_{\Delta_{A5B,}} \, V_{\Delta_{34B,}} 
\bigg[
\zeta_p(s_A-\Delta_A) \zeta_p(s_B-\Delta_B) \cr 
&\quad -V_{\Delta_{345A,}}\, \beta_p(s_A-\Delta_A, n-\Delta_{\Sigma}) 
 -V_{\Delta_{125B,}} \, \beta_p(s_B-\Delta_B, n-\Delta_{\Sigma}) \cr &\quad  -V_{\Delta_\Sigma} 
\bigg]\,,
}
where the contact amplitudes $V_{\Delta_{i_1 \ldots i_f\,,}}$ were defined in \eno{vertexFactorDef2}, $\beta_p$ was defined in \eno{BetaPDef}, and we are using the shorthand
\eqn{DeltaSigmaDef5pt}{
\Delta_\Sigma \equiv \sum_{i=1}^5 \Delta_i = \Delta_{12345,}\,.
}
Furthermore, with the help of the Mellin variable constraints \eno{MellinVarConstraints} for ${\cal N}=5$, we have defined
\eqn{sAsBDef}{
s_A &\equiv \Delta_1 + \Delta_2 - 2\gamma_{12} = \Delta_3 + \Delta_4 + \Delta_5 - 2\gamma_{34} -2\gamma_{35}-2\gamma_{45} \cr 
s_B & \equiv \Delta_3 + \Delta_4 -2 \gamma_{34} = \Delta_1 + \Delta_2 + \Delta_5 - 2\gamma_{12} - 2\gamma_{15} -2\gamma_{25}\,.
}
Unlike the case of the four-point exchange diagram, it is not immediately clear that the amplitude in \eno{5ptchannel} admits a BCFW-type decomposition; in the rest of this appendix we show precisely how this works out. 

We start by passing to the auxiliary momentum space, with the momentum assignments as shown: 
\eqn{5ptMomentum}{
{\cal M}^{2-{\rm int}}_{\Delta_{12,}; \Delta_5; \Delta_{34,}}(s_A,\Delta_A; s_B, \Delta_B) = 
\begin{matrix}
\includegraphics[height=18ex]{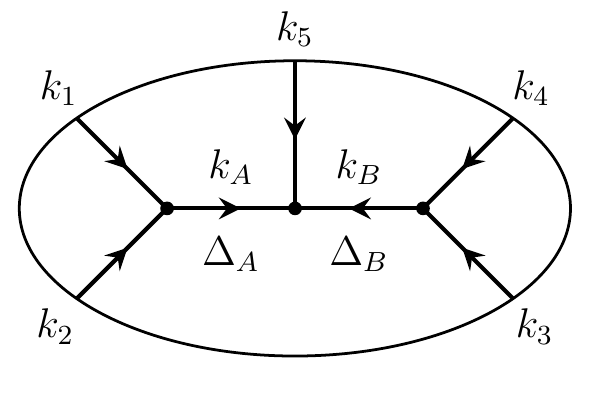}
 \end{matrix}   \,,
}
where, like before, the external momenta are on-shell,
\eqn{5ptOnshell}{
-k_i^2 = \Delta_i \qquad i=1,\ldots, 5\,,
}
and momentum is conserved at each vertex.
Once again we employ a complex-shift of momenta; specifically we apply the two-line BCFW-shift \eno{4ptBCFWshift} subject to the on-shell constraints \eno{4ptComplexOnshell}. 
The on-shell conditions lead us to constraints on $q$ as written down in \eno{4ptqConditions}. Momentum conservation along with constraints \eno{4ptqConditions} further implies
\eqn{5ptqImplied}{
q \cdot (k_2 + k_3 + k_5)=0\,.
}
Looking ahead, it will be convenient to set
\eqn{5ptqNewCond}{
q \cdot k_5 = 0\,.
}
We are free to make this choice and still have (at least one) non-vanishing solution for the complex momentum variable $q$, as long as $(n+1) \geq 5$, since \eno{4ptqConditions} and \eno{5ptqNewCond} amount to $4$ conditions on the $(n+1)$-component vector $q$. This proviso is consistent with the requirement $(n+1) \geq {\cal N}=5$ which ensures there are precisely ${\cal N}({\cal N}-3)/2=5$ independent Mandelstam variables.

 Applying momentum conservation for the complex-shifted momenta vertex by vertex, the shifted internal momenta are given by
\eqn{5ptInternalShift}{
k_A \to k_A(z) \equiv k_1(z)+k_2\,, \qquad k_B \to k_B(z) \equiv k_3 + k_4(z)\,,
}
which allows the possibility of setting the internal legs on-shell at specific (non-zero) values of the complex parameter $z$. In particular, 
\eqn{kAkBOnshell}{
s_A(z) &\equiv -k_A(z)^2 = s_A -2z q \cdot k_2 = -2q\cdot k_2 (z-z_A) + \Delta_A  \cr 
s_B(z) &\equiv -k_B(z)^2 = s_B +2z q \cdot k_3 = 2q\cdot k_3 (z-z_B) + \Delta_B \,,
}
where we have defined
\eqn{zAzBDef}{
z_A \equiv {s_A - \Delta_A \over 2q \cdot k_2}\,, \qquad z_B \equiv -{s_B - \Delta_B \over 2q \cdot k_3}\,.
}
We conclude that $k_A(z)$ goes on-shell at $z=z_A$, while $k_B(z)$ goes on-shell at $z=z_B$. Equation \eno{kAkBOnshell} also suggests  what the complex-shifted Mellin amplitude should look like,
\eqn{5ptMellinC}{
{\cal M}(z) \equiv M^{2-{\rm int}}_{\Delta_{12,}; \Delta_5; \Delta_{34,}}(s_A(z),\Delta_A; s_B(z), \Delta_B) \,,
}
that is, we simply promote the Mandelstam variables $s_A$ and $s_B$ in \eno{5ptchannel} to their complex-shifted versions. 

Like in the previous subsection, consider now the contour integral
\eqn{5ptIDef}{
I \equiv \oint_{\mathsf{C}} {dz\over 2\pi i} {{\cal M}(z) \over z}\,,
}
where the contour $\mathsf{C}$ is a circle of infinite radius centered at origin. It is clear from the explicit form in \eno{5ptchannel} that the integrand in \eno{5ptIDef} has simple poles at
\eqn{5ptMzPoles}{
z= 0\,, \qquad z= z_A + {1 \over 2 q\cdot k_2} {2\pi i m \over \log p}  \qquad z= z_B + {1 \over 2 q\cdot k_3} {2\pi i m \over \log p} \qquad m \in \mathbb{Z}\,.
}
We will first apply the residue theorem to evaluate \eno{5ptIDef}. The residue at $z=0$ reproduces the original amplitude ${\cal M}(0)$ which we are interested in, while the residues at the remaining poles evaluate to
\eqn{5ptResidues}{
\underset{\substack{z =z_A \\ \hspace{12mm} + {1 \over 2 q\cdot k_2} {2\pi i m \over \log p}}}{\rm Res} \left({{\cal M}(z) \over z}\right) &=  -{V_{\Delta_{12A,}} V_{\Delta_{A5B,}} V_{\Delta_{34B,}} \, \beta_p(s_B-\Delta_B -s_A + \Delta_A,n-\Delta_{345A,})   \over (s_A - \Delta_A)\log p +  2\pi i m }  \cr 
\underset{\substack{z =z_B \\ \hspace{12mm} - {1 \over 2 q\cdot k_2} {2\pi i m \over \log p}}}{\rm Res} \left({{\cal M}(z) \over z}\right) &=  -{ V_{\Delta_{12A,}}V_{\Delta_{A5B,}}   V_{\Delta_{34B,}} \, \beta_p(s_A-\Delta_A -s_B + \Delta_B,n-\Delta_{125B,})   \over (s_B - \Delta_B)\log p +  2\pi i m }  \,,
}
for all $m \in \mathbb{Z}$, where we made important use of the identity $q \cdot k_2 = -q \cdot k_3$ which follows from \eno{5ptqImplied}-\eno{5ptqNewCond}. Using \eno{4ptSumRes} to sum up the residues, we obtain
\eqn{5ptClaim0}{
& {\cal M}(0) \cr 
&=  -  {\cal M}(k_1(z),k_2,-k_1(z)-k_2)\Big|_{z=z_A}  \left( \zeta_p(s_A-\Delta_A) - {1 \over 2} \right) {\cal M}_{\Delta_B}(k_1(z)+k_2,k_5,k_3,k_4(z))\Big|_{z=z_A} \cr 
 &\quad-  {\cal M}_{\Delta_A}(k_1(z),k_2,k_5,k_3+k_4(z))\Big|_{z=z_B}  \left( \zeta_p(s_B-\Delta_B) - {1 \over 2} \right) {\cal M}(-k_3-k_4(z),k_3,k_4(z))\Big|_{z=z_B} \cr 
 &\quad + I\,,
}
where we have used~\eno{4ptMDef} and have defined the (off-shell) sub-amplitude
\eqn{5ptMDeltaDef}{
{\cal M}_{\Delta}(k_a,k_b,k_c,k_d) &\equiv -V_{-k_a^2,-k_b^2,\Delta} \,V_{-k_c^2,-k_d^2,\Delta} \cr 
&\quad \times \beta_p\left(-(k_a+k_b)^2-\Delta,n+k_a^2+k_b^2+k_c^2+k_d^2\right) 
}
with $k_a + k_b +k_c +k_d= 0$.
The  sub-amplitude
\eqn{MoffshellAgain}{
 {\cal M}(k_1(z),k_2,-k_1(z)-k_2) = V_{\Delta_1,\Delta_2, s_A(z)}
 }
in \eno{5ptClaim0} is on-shell at $z=z_A$, i.e.\ all external momenta go on-shell at $z=z_A$. Moreover, at this value of $z$, it precisely takes the form of an (on-shell) three-point contact Mellin amplitude.
Similarly, the sub-amplitude,
\eqn{MDeltaOffshell}{
{\cal M}_{\Delta_B}(k_1(z)+k_2,k_5,k_3,k_4(z)) &= -V_{s_A(z),\Delta_5,\Delta_B}\, V_{\Delta_3,\Delta_4,\Delta_B}  \cr 
&\quad \times \beta_p\left(s_B(z)-\Delta_B,n-\Delta_3-\Delta_4-\Delta_5-s_A(z)\right)
}
is on-shell at $z=z_A$, whence $s_A(z_A)=\Delta_A$, $s_B(z_A) =-(k_3+k_4(z_A))^2= s_B - s_A + \Delta_A$, and we recognize from the previous subsection that \eno{MDeltaOffshell} takes precisely the form of an (on-shell) four-point exchange diagram (see, e.g.\ \eno{4ptMomentum}): 
\eqn{MDeltaOnshell}{
 {\cal M}_{\Delta_B}(k_1(z_A)+k_2,k_5,k_3,k_4(z_A)) &=\begin{matrix}
\includegraphics[height=18ex]{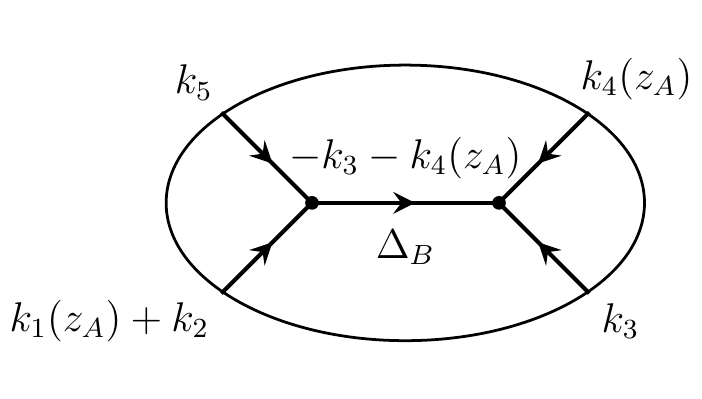}  
 \end{matrix} \cr 
 &= {\cal M}^{1-{\rm int}}_{-k_5^2-(k_1(z_A)+k_2)^2; -k_3^2-k_4(z_A)^2}\left(-(k_3+k_4(z_A))^2,\Delta_B\right) \cr 
 &= {\cal M}^{1-{\rm int}}_{\Delta_5+\Delta_A; \Delta_3+\Delta_4}(s_B-s_A+\Delta_A,\Delta_B)\,.
}
A similar interpretation can be given to the sub-amplitudes in the second term on the r.h.s.\ of \eno{5ptClaim0}, which are to be evaluated at $z=z_B$.

We still need to take into account the boundary contribution in \eno{5ptClaim0}, denoted by $I$. Just like in the previous subsection, such a contribution is non-vanishing and can  be directly computed by evaluating the contour integral at infinity in \eno{5ptIDef}. We omit the details of the computation, but point out that just like in the previous subsection, the boundary contribution combines with the sum of residues at all (non-zero) poles to give
\eqn{5ptFinal}{
& {\cal M}(0) \cr 
&=  -  {\cal M}(k_1(z),k_2,-k_1(z)-k_2)\Big|_{z=z_A}  \beta_p(s_A-\Delta_A,n-\Delta_\Sigma) \, {\cal M}_{\Delta_B}(k_1(z)+k_2,k_5,k_3,k_4(z))\Big|_{z=z_A} \cr 
 &\quad-  {\cal M}_{\Delta_A}(k_1(z),k_2,k_5,k_3+k_4(z))\Big|_{z=z_B} \beta_p(s_B-\Delta_B,n-\Delta_\Sigma) \, {\cal M}(-k_3-k_4(z),k_3,k_4(z))\Big|_{z=z_B} \,.
}
We may write this diagrammatically as

\eqn{5ptDiagrammatic}{
& \begin{matrix}
\includegraphics[height=18ex]{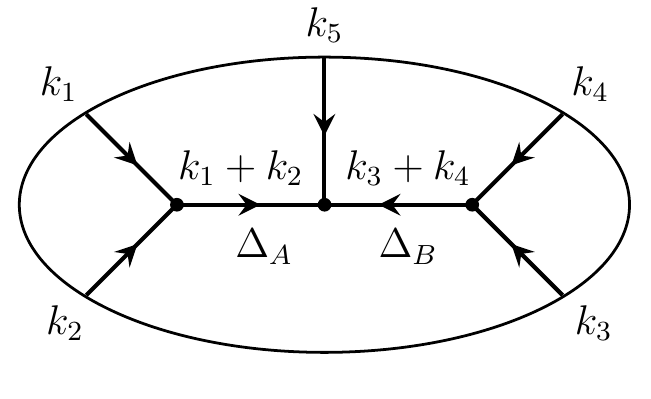}
 \end{matrix} \cr 
 &= -\begin{matrix}
\includegraphics[height=18ex]{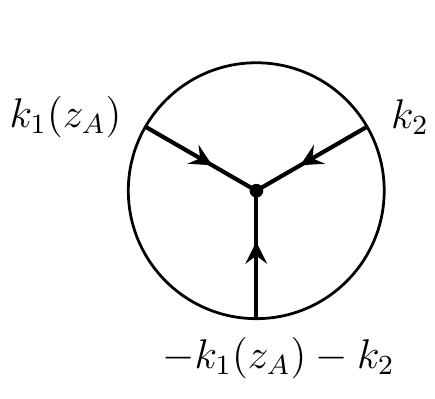}
 \end{matrix} 
 \beta_p(s_A-\Delta_A,n-\Delta_{\Sigma})
 \begin{matrix}
\includegraphics[height=18ex]{figures/5ptMom3.pdf}
 \end{matrix} \cr 
 & \quad - \begin{matrix}
\includegraphics[height=18ex]{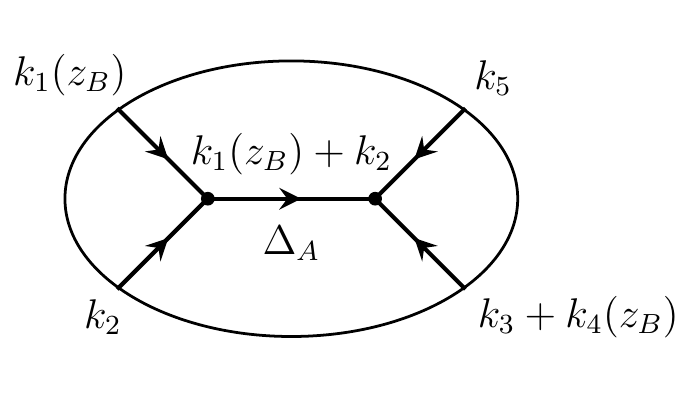}
 \end{matrix} 
 \beta_p(s_B-\Delta_B,n-\Delta_{\Sigma})
 \begin{matrix}
\includegraphics[height=18ex]{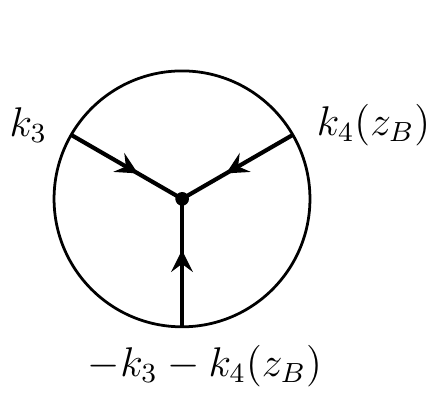}
 \end{matrix}\,.
} 
Equivalently, the result \eno{5ptFinal} may also  be recast in the form of \eno{mellinBCFW} as
\eqn{5ptPractical}{
\begin{matrix}
\includegraphics[height=14ex]{figures/5ptDia.pdf}
 \end{matrix}
 =&
- \begin{matrix}
\includegraphics[height=11ex]{figures/12A.pdf}
 \end{matrix}
 \beta_p(s_A-\Delta_A,n-\Delta_\Sigma)
\begin{matrix}
\includegraphics[height=11ex]{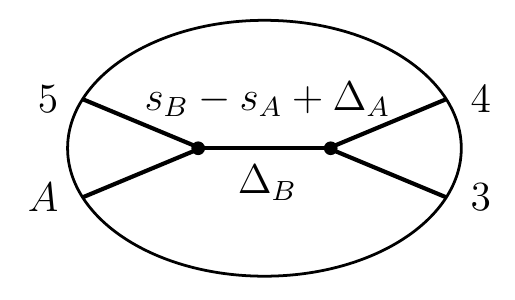}
 \end{matrix}
\cr
&
- \begin{matrix}
\includegraphics[height=11ex]{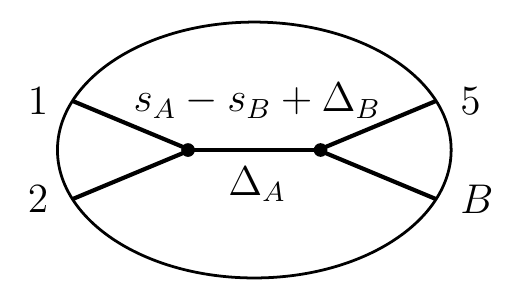}
 \end{matrix}
 \beta_p(s_B-\Delta_B,n-\Delta_\Sigma)
\begin{matrix}
\includegraphics[height=11ex]{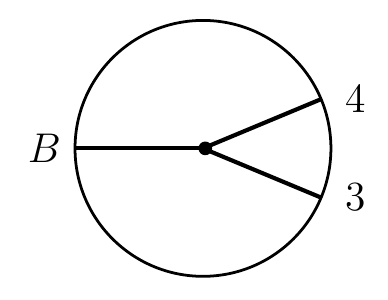}
 \end{matrix}\,,
}
which avoids any reference to momentum variables.  Further, we may use the decomposition \eno{4ptDiagrammatic}-\eno{4ptPractical} from the previous subsection to further reduce \eno{5ptDiagrammatic}-\eno{5ptPractical} in such a way that the r.h.s.\ is given entirely in terms of the three-point (contact) Mellin amplitudes.

\vspace{1em}

The proof presented in section \ref{BCFW} (for the on-shell recursive prescription~\ref{pres:BCFW}) sheds light on the appearance of non-vanishing boundary terms in the examples considered in this appendix, where we apply Cauchy's theorem in the complexified auxiliary momentum space to obtain an on-shell recursion relation: Had we chosen instead the complex-cylindrical manifold for the complex variable $z$ like we do in section \ref{BCFW}, the boundary terms would not have made an appearance in the intermediate steps.

\bibliographystyle{ssg}
\bibliography{mellin}

\end{document}